\documentclass[11pt]{article}
\pdfoutput=1
\usepackage[utf8x]{inputenc}	%for german Umlaute
\usepackage[english]{babel}
\usepackage{amssymb}
\usepackage{mathtools}		%for definition ":="
\usepackage{dsfont}		%for math modes
\usepackage{stmaryrd}		%GIT quotient
\usepackage{cite}		%for better citing
\usepackage[svgnames]{xcolor}
\usepackage{enumerate}
\usepackage{setspace}
\usepackage{booktabs}
\usepackage{tikz}
\usetikzlibrary{shapes,snakes}
\usepackage{todonotes}
\usepackage[font={small,it},labelfont=bf]{caption} % make captions ``small'' and 
\usepackage{subcaption}
\usepackage[pdftex,breaklinks,colorlinks=true,urlcolor=RoyalBlue,linkcolor=blue,
citecolor=blue]{hyperref}		%for links on refs etc.
\catcode`@=11
\renewcommand{\section}{\@startsection{section}{1}{0pt}{\medskipamount}
{\medskipamount}{\Large\bf}}
\numberwithin{equation}{section}
\catcode`@=12

\newcommand{\im}{\mathrm{i}}
\def\for{\qquad\textrm{for}\quad}
\def\with{\qquad\textrm{with}\quad}
\def\und{\quad\textrm{and}\quad}
\def\st{\quad\textrm{such that}\quad}
\def\rank{\mathrm{rk}}
\def\det{\mathrm{det}}
\def\deg{\mathrm{deg}}
\def\dim{\mathrm{dim}}
\def\diag{\mathrm{diag}}
\newcommand{\tr}{\mathrm{Tr}}
\newcommand{\diff}{\mathrm{d}}
\newcommand{\HS}{\mathrm{HS}}
\newcommand{\PL}{\mathrm{PL}}
\newcommand{\C}{\mathbb C}
\newcommand{\R}{\mathbb R}
\newcommand{\NN}{\mathbb N}  
\newcommand{\Z}{\mathbb Z}

\newcommand{\MCoulomb}{\mathcal{M}_C}
\newcommand{\MHiggs}{\mathcal{M}_H}
\newcommand{\Ncal}{\mathcal{N}}
\newcommand{\Zcal}{\mathcal{Z}}
\newcommand{\Ufrak}{\mathfrak{U}}

\newcommand{\spanZ}{\mathrm{Span}_{\Z}}
\newcommand{\cone}{\mathrm{Cone}}
\newcommand{\HilbS}{\mathrm{H}}

\newcommand{\colvec}[3]{\begin{pmatrix}#1 \\ #2 \\ #3  \end{pmatrix}}
\newcommand{\Casi}[1]{\mathcal{C}_{#1}}
  
\newcommand{\gfrak}{\mathfrak{g}}  
\newcommand{\hfrak}{\mathfrak{h}}
\newcommand{\tfrak}{\mathfrak{t}}
\newcommand{\Hcal}{\mathcal{H}}  
\newcommand{\Wcal}{\mathcal{W}}
\newcommand{\su}{{{\rm SU}(2)}}

\newcommand{\sut}{{{\rm SU}(3)}}
\newcommand{\uo}{{{\rm U}(1)}}

\newcommand{\sorm}{{{\rm SO}}}
\newcommand{\orm}{{{\rm O}}}

\newcommand{\ut}{{{\rm U}(3)}}
\newcommand{\utwo}{{{\rm U}(2)}}

\newcommand{\Gtwo}{{\rm G}_2}
\newcommand{\USp}{{\rm USp}(4)}
\newcommand{\Spin}{{\rm Spin}}
\newcommand{\G}{\mathrm{G}}
\newcommand{\Hh}{\mathrm{H}}
\newcommand{\T}{\mathrm{T}}
\newcommand{\GNOG}{\widehat{\G}}
\newcommand{\GNOfrak}{\widehat{\mathfrak{g}}}
\newcommand{\Roots}{\boldsymbol{\Phi}}
\newcommand{\posRoots}{\boldsymbol{\Phi}_+}
\newcommand{\negRoots}{\boldsymbol{\Phi}_-}
\newcommand{\simRoots}{\boldsymbol{\Phi}_s}

\allowdisplaybreaks

\begin{document}
\begin{titlepage}
\setcounter{page}{0}
\begin{flushright}
Imperial/TP/16/AH/03\\
ITP--UH--09/16 
\end{flushright}

\vskip 2cm

\begin{center}

{\Large\bf Coulomb branches for rank 2 gauge groups \\ in $3d$ $\mathcal{N}=4$ 
gauge theories
}

\vspace{15mm}

{\large Amihay Hanany${}^{1}$} , \ {\large Marcus Sperling${}^{2}$} 
\\[5mm]

\noindent ${}^1${\em Theoretical Physics Group, Imperial College London\\
Prince Consort Road, London, SW7 2AZ, UK}\\
{Email: {\tt a.hanany@imperial.ac.uk}}
\\[5mm]
\noindent ${}^{2}${\em Institut f\"ur Theoretische Physik, Leibniz Universit\"at 
Hannover}\\
{\em Appelstra\ss e 2, 30167 Hannover, Germany}\\
Email: {\tt marcus.sperling@itp.uni-hannover.de}
\\[5mm]

\vspace{15mm}

\begin{abstract}
The Coulomb branch of $3$-dimensional $\Ncal=4$ gauge theories is the space of 
bare and dressed BPS monopole operators. We 
utilise the conformal dimension to define a fan which, upon 
intersection with the weight lattice of a GNO-dual group, gives rise to a 
collection of semi-groups. It turns out that the unique Hilbert bases of these 
semi-groups are a sufficient, finite set of monopole operators which 
generate the entire chiral ring. 
Moreover, the knowledge of the properties of the minimal generators is enough 
to compute the Hilbert series explicitly. The techniques of this paper allow an 
efficient evaluation of the Hilbert series for general rank gauge groups. As an 
application, we provide 
various examples for all rank two gauge groups to demonstrate the novel 
interpretation.
\end{abstract}

\end{center}

\end{titlepage}

{\baselineskip=12pt
\tableofcontents
}

%%%%%%%%%%%%%%%%%%%%%%%%%%%%%%%%%%%%%%%%%%%%%%%%%%%%%%%%%%%%%%%%%%%%%%%%%%%%%%%%
  \section{Introduction}
The moduli spaces of supersymmetric gauge theories with $8$ supercharges have 
generically two branches: the Higgs and the Coulomb branch. In this paper we 
focus on $3$-dimensional $\Ncal=4$ gauge theories, for which both branches are 
hyper-Kähler spaces. Despite this fact, the branches are fundamentally 
different.

The Higgs branch $\MHiggs$ is understood as hyper-Kähler quotient
\begin{equation}
 \MHiggs = \R^{4N}/\!/\!/\,\G\, \; ,
\end{equation}
in which the vanishing locus of the $\Ncal=4$ F-terms is quotient by the 
complexified gauge group. The F-term equations play the role of complex 
hyper-Kähler moment maps, while the transition to the complexified gauge group 
eliminates the necessity to impose the D-term constraints. Moreover, this 
classical description is sufficient as the Higgs branch is protected from 
quantum corrections. The explicit quotient construction can be supplemented by 
the study of the Hilbert series, which allows to gain further understanding of 
$\MHiggs$ as a complex space.

Classically, the Coulomb branch $\MCoulomb$ is the hyper-Kähler space 
\begin{equation}
 \MCoulomb \approx (\R^3 \times S^1)^{\rank(\G)}\slash \Wcal_\G \; ,
\end{equation}
where $\Wcal_\G $ is the Weyl group of $\G$ and $\rank(\G)$ denotes the rank of 
$\G$. However, the geometry and topology 
of $\MCoulomb$ are affected by quantum corrections. Recently, the 
understanding of the Coulomb branch has been subject of active research 
from various viewpoints: the authors   
of~\cite{Bullimore:2015lsa} aim to provide a description for the 
quantum-corrected Coulomb 
branch of any $3d$ $\Ncal=4$ gauge theory, with particular emphasis on the 
full Poisson algebra of the chiral ring $\C[\MCoulomb]$. In contrast, a 
rigorous mathematical definition of the Coulomb branch itself lies at the heart 
of the attempts presented 
in~\cite{Nakajima:2015txa,Nakajima:2015gxa,Braverman:2016wma}.
In this paper, we take the perspective centred around the \emph{monopole 
formula} proposed in~\cite{Cremonesi:2013lqa}; that is, the computation of the 
Hilbert series for the Coulomb branch allows to gain information on $\MCoulomb$ 
as a complex space.

Let us briefly recall the set-up. Select an $\Ncal =2$ subalgebra in the 
$\Ncal=4$ algebra, which implies a decomposition of the $\Ncal=4$ vector 
multiplet into an $\Ncal=2$ vector multiplet (containing a gauge field $A$ and 
a real adjoint scalar $\sigma$) and an $\Ncal=2$ chiral multiplet (containing a 
complex adjoint scalar $\Phi$) which transforms in the adjoint representation 
of the gauge group $\G$. In addition, the selection of an $\Ncal=2$ subalgebra 
is equivalent to the choice of a complex structure on $\MCoulomb$ and 
$\MHiggs$, which is the reason why one studies the branches only as complex
and not as hyper-Kähler spaces.

The description of the Coulomb branch relies on 't~Hooft monopole 
operators~\cite{tHooft:1977hy}, which are local disorder 
operators~\cite{Borokhov:2002ib} defined by specifying a Dirac monopole 
singularity
\begin{equation}
 A_{\pm} \sim \frac{m}{2} \left(\pm1 -\cos \theta \right) \diff \varphi
\end{equation}
for the gauge field, where $m\in \gfrak=\mathrm{Lie}(\G)$ and 
$(\theta,\varphi)$ are coordinates on the $2$-sphere around the insertion point.
An important consequence is that the \emph{generalised Dirac quantisation 
condition}~\cite{Englert:1976ng}
\begin{equation}
 \exp\left(2 \pi \im m \right) = \mathds{1}_\G
 \label{eqn:general_Dirac}
\end{equation}
has to hold. As proven in~\cite{Goddard:1976qe}, the set of solutions 
to~\eqref{eqn:general_Dirac} equals the weight lattice $\Lambda_w(\GNOG)$ of 
the GNO (or Langlands) dual group $\GNOG$, which is uniquely associated to the 
gauge group $\G$.

For Coulomb branches of supersymmetric gauge theories, the monopole operators 
need to be supersymmetric as well, see for instance~\cite{Borokhov:2002cg}. In 
a pure $\Ncal =2$ theory, the supersymmetry condition amounts to the singular 
boundary condition 
\begin{equation}
 \sigma \sim \frac{m}{2r} \for r \to \infty \; ,
\end{equation}
for the real adjoint scalar in the $\Ncal=2 $ vector multiplet.
Moreover, an $\Ncal=4$ theory also allows for a non-vanishing vacuum 
expectation value of the complex adjoint scalar $\Phi$ of the adjoint-valued 
chiral multiplet. Compatibility with supersymmetry requires $\Phi$ to take 
values in the stabiliser $\Hh_m$ of the ``magnetic weight'' $m$ in $\G$. This 
phenomenon gives rise to dressed monopole operators. 

Dressed monopole operators and $\G$-invariant functions of $\Phi$ are believed 
to generate the entire chiral ring $\C[\MCoulomb]$. The corresponding Hilbert 
series allows for two points of view: seen via the \emph{monopole formula}, 
each operator is precisely counted once in the Hilbert series --- no 
over-counting appears. Evaluating the Hilbert series as rational function, 
however, provides an over-complete set of generators that, in general, satisfies 
relations. In order 
to count polynomials in the chiral ring, a notion of degree or dimension is 
required. Fortunately, in a CFT one employs the conformal dimension $\Delta$, 
which for BPS states agrees with the $\su_R$ highest weight. 
Following~\cite{Borokhov:2002cg,Gaiotto:2008ak,Benna:2009xd,Bashkirov:2010kz}, 
the conformal dimension of a BPS bare monopole operator of GNO-charge m is 
given by
\begin{equation}
 \Delta(m)= \frac{1}{2} \sum_{i=1}^{n} \sum_{\rho \in \mathcal{R}_i} \left| 
\rho(m)\right| - \sum_{\alpha \in \Phi_+} \left| 
\alpha(m)\right|\; ,
\label{eqn:Def_ConfDim}
\end{equation}
where $\mathcal{R}_i$ denotes the set of all weights $\rho$ of the 
$\G$-representation in 
which the $i$-th flavour of $\Ncal=4$ hypermultiplets transform. Moreover,  
$\Phi_+$ denotes the set of positive roots $\alpha$ of the Lie algebra $\gfrak$ 
and provides the contribution of the $\Ncal=4$ vector multiplet. Bearing in 
mind the proposed classification of $3d$ $\Ncal=4$ theories 
by~\cite{Gaiotto:2008ak}, we restrict ourselves to ``good'' theories (i.e.\
$\Delta > \tfrac{1}{2}$ for all BPS monopoles).

If the centre $\Zcal(\GNOG)$ is non-trivial, then the monopole operators can be 
charged under this topological symmetry group and one can refine the counting 
on the chiral ring.

Putting all the pieces together, the by now well-established \emph{monopole 
formula} of~\cite{Cremonesi:2013lqa} reads
\begin{equation}
 \HS_{\G}(t,z) = \sum_{m \in \Lambda_w(\GNOG) \slash \mathcal{W}_{\GNOG}} 
z^{J(m)} t^{\Delta(m)} P_{\G}(t,m) \; .
\label{eqn:HS_refined}
\end{equation}
Here, the fugacity $t$ counts the $\su_R$-spin, while the (multi-)fugacity $z$ 
counts the quantum numbers $J(m)$ of the topological symmetry $\Zcal(\GNOG)$.

This paper serves three purposes: firstly, we provide a geometric derivation of 
a sufficient set of monopole operators, called the \emph{Hilbert basis}, that 
generates the entire chiral ring. Secondly, employing the Hilbert basis 
allows an explicit summation of~\eqref{eqn:HS_refined}, which we demonstrate 
for 
$\rank(\G)=2$ explicitly. Thirdly, we provide various examples for all rank two 
gauge groups and display how the knowledge of the Hilbert basis completely 
determines the Hilbert series. 

The remainder of this paper is organised as follows: 
Sec.~\ref{sec:general_idea} is devoted to the exposition of our main 
points: after recapitulating basics on (root and weight) lattices and 
rational polyhedral cones in Subsec.~\ref{subsec:preliminaries}, we explain in 
Subsec.~\ref{subsec:Hilbert basis} how the conformal dimension decomposes the 
Weyl chamber of $\GNOG$ into a fan. Intersecting the fan with the weight lattice 
$\Lambda_w(\GNOG)$ introduces affine semi-groups, which are finitely generated 
by a unique set of irreducible elements --- called the Hilbert basis. Moving on 
to Subsec.~\ref{subsec:Dressings_as_HS}, we collect mathematical results that 
interpret the dressing factors $P_{\G}(t,m)$ as Poincar\'{e} series for the set 
of $\Hh_m$-invariant polynomials on the Lie algebra $\hfrak_m$. Finally, we 
explicitly sum the unrefined Hilbert series in 
Subsec.~\ref{subsec:summation_HS_unrefined} and the refined Hilbert series 
in~\ref{subsec:summation_HS_refined} utilising the knowledge about the Hilbert 
basis.
After establishing the generic results, we provide a comprehensive collection 
of examples for all rank two gauge groups in 
Sec.~\ref{sec:U1xU1}-\ref{sec:SU3}. Lastly, Sec.~\ref{sec:conclusions} 
concludes.

Before proceeding to the details, we present our main 
result~\eqref{eqn:HS_generic_refined} already at 
this stage: the refined Hilbert series for any rank two gauge group $\G$.
\begin{align}
 \HS_{\G}(t,z)&= \frac{P_{\G}(t,0) }{\prod_{p=0}^{L} 
\left(1-z^{J(x_p)} 
t^{\Delta(x_{p})}\right)}
\Bigg\{ 
\prod_{q=0}^{L} \left( 1-z^{J(x_q)}t^{\Delta(x_{q})} \right) 
\label{eqn:HS_generic_intro}\\
&\phantom{= \frac{P_{\G}(t,0) }{\prod_{p=0}^{L} \left(1-z^{J(x_p)} 
t^{\Delta(x_{p})}\right)}}
+ \sum_{q=0}^{L} \frac{P_{\G}(t,x_q)}{P_{\G}(t,0)} 
z^{J(x_q)} t^{\Delta(x_q)} 
\prod_{r=0\atop r\neq q}^{L} 
\left( 1-z^{J(x_r)} t^{\Delta(x_{r})} \right) 
\notag \\
&\phantom{= \frac{P_{\G}(t,0) }{\prod_{p=0}^{L} \left(1-z^{J(x_p)} 
t^{\Delta(x_{p})}\right)}}
+\sum_{q=1}^{L} \frac{P_{\G}(t,C_q^{(2)})}{P_{\G}(t,0)} 
\bigg[z^{J(x_{q-1})+J(x_q)} t^{\Delta(x_{q-1}) +\Delta(x_q)} 
\notag\\*
&\phantom{= \frac{P_{\G}(t,0) }{\prod_{p=0}^{L} \left(1-z^{J(x_p)} 
t^{\Delta(x_{p})}\right)}}
\qquad  \qquad
+  \sum_{s \in \mathrm{Int}( \mathcal{P}(C_q^{(2)}))} 
z^{J(s)} t^{\Delta(s)} \bigg] \prod_{r=0\atop{r\neq q-1,q}}^L 
\left(1- z^{J(x_r)} t^{\Delta(x_{r})}\right) 
\Bigg\} \notag \; ,
\end{align}
where the ingredients can be summarised as follows:
\begin{itemize}
 \item A fan $F_\Delta=\{C_p^{(2)}  \, , \, p =1,\ldots,L\}$, and each 
$2$-dimensional 
cone satisfies $\partial C_p^{(2)} = C_{p-1}^{(1)} \cup C_p^{(1)}$ and 
$C_{p-1}^{(1)} \cap C_p^{(1)} = \{0\} $.
\item The Hilbert basis for $C_p^{(2)}$ comprises the ray generators $x_{p-1}$, 
$x_p$ as well as other minimal generators $\{u_{\kappa}^p\}$.
\item The $x_{p-1}$, $x_p$ generate a fundamental parallelotope 
$\mathcal{P}(C_p^{(2)})$, where the discriminant counts the number of lattice 
points in the interior $\mathrm{Int}( \mathcal{P}(C_p^{(2)}))$ via $d(C_p^{(2)}) 
-1 = \# \text{pts.}\ \left( \mathrm{Int}( \mathcal{P}(C_p^{(2)})\right)$.
\end{itemize}
The form of~\eqref{eqn:HS_generic_intro} is chosen to emphasis that the terms 
within the curly bracket represent the numerator of the Hilbert series as 
rational function, i.e.\ the curly bracket is a proper polynomial in $t$ 
without 
poles. On the other hand, the first fraction represents the denominator of the 
rational function, which is again a proper polynomial by construction.
%%%%%%%%%%%%%%%%%%%%%%%%%%%%%%%%%%%%%%%%%%%%%%%%%%%%%%%%%%%%%%%%%%%%%%%%%%%%%%%%
   \section{Hilbert basis for monopole operators}
\label{sec:general_idea}
\subsection{Preliminaries}
\label{subsec:preliminaries}
Let us recall some basic properties of Lie algebras, 
c.f.~\cite{Humphreys:1972}, 
and combine them with the description of \emph{strongly convex rational 
polyhedral cones} and \emph{affine semi-groups}, c.f.~\cite{Cox:2011}. 
Moreover, we recapitulate the definition and properties of the GNO-dual group, 
which can be found in~\cite{Goddard:1976qe,Kapustin:2005py}.
\paragraph{Root and weight lattices of $\gfrak$}
Let $\G$ be a Lie group with semi-simple Lie algebra $\gfrak$ and 
$\rank(\G)=r$. Moreover, $\widetilde{\G}$ is the universal covering group of 
$\G$, i.e.\ the unique simply connected Lie group with Lie algebra $\gfrak$. 
Choose a maximal torus $\T \subset \G$ and the corresponding Cartan subalgebra 
$\tfrak \subset \gfrak$. Denote by 
$\Roots$ the set of all roots $\alpha \in \tfrak^*$.
By the choice of a 
hyperplane, one divides the root space into positive $\posRoots$ and negative 
roots $\negRoots$. In the half-space of positive roots one introduces the simple 
positive roots as irreducible basis elements and denotes their set by 
$\simRoots$.
The roots span a lattice $\Lambda_r(\gfrak)\subset \tfrak^*$, the \emph{root 
lattice}, with basis $\simRoots$.

Besides roots, one can always choose a basis in the complexified Lie algebra 
that gives rise to the notion of coroots $\alpha^\vee \in \tfrak$ which satisfy 
$\alpha \left( \beta^\vee \right) \in \Z$ for any $\alpha,\beta\in 
\mathbf{\Phi}$. Define $\alpha^\vee$ to be a simple 
coroot if and only if $\alpha$ is a simple root.
Then the coroots span a lattice $\Lambda^\vee_r(\gfrak)$ in $\tfrak$ ---  
called the \emph{coroot lattice} of $\gfrak$.

The dual lattice $\Lambda_w(\gfrak)$ of the coroot lattice is the set of points 
$\mu \in \tfrak^*$ for which $\mu (\alpha^\vee) \in \Z$ for all $\alpha \in 
\boldsymbol{\Phi}$. This lattice is called \emph{weight lattice} of $\gfrak$.
Choosing a basis $\boldsymbol{B}$ of simple coroots
\begin{equation}
 \boldsymbol{B}\coloneqq \left\{ \alpha^{\vee} \ , \ \alpha\in \simRoots 
\right\} \subset \tfrak \; ,
\end{equation}
one readily defines a basis for the dual space via
\begin{equation}
 \boldsymbol{B^*} \coloneqq \left\{ \lambda_{\alpha} \ , \ \alpha \in \simRoots 
\right\} \subset \tfrak^*\for \lambda_{\alpha} \left(\beta^{\vee}\right) = 
\delta_{\alpha,\beta} \; , \; \forall \alpha,\beta \in \simRoots \; .
\end{equation}
The basis elements $\lambda_{\alpha} $ are precisely the fundamental weights of 
$\gfrak$ (or $\widetilde{\G}$) and they are a basis for the weight lattice.

Analogous, the dual lattice $\Lambda_{mw}(\gfrak) \subset \tfrak$ of the root 
lattice is the set of points $m \in \tfrak$ such that $\alpha(m) \in \Z$ for all 
$\alpha \in \boldsymbol{\Phi}$. In particular, the coroot lattice is a 
sublattice of $\Lambda_{mw}(\gfrak)$.

As a remark, the lattices defined so far solely depend on the Lie algebra 
$\gfrak$, or equivalently on $\widetilde{G}$, but not on $\G$. Because any group 
defined via $\widetilde{\G}\slash \Gamma$ for $\Gamma\subset \Zcal(\G)$ has the 
same Lie algebra.
\paragraph{Weight and coweight lattice of $\G$}
The weight lattice of the group $\G$ is the lattice of the infinitesimal 
characters, i.e.\ a character $\chi: \T \to \uo$ is a homomorphism, which is 
then uniquely determined by the derivative at the identity. Let $X \in \tfrak$ 
then $\chi(\exp{(X)}) = \exp{(i \mu(X))}$, wherein $\mu \in \tfrak^*$ is an 
\emph{infinitesimal character} or weight of $\G$. The weights form then a 
lattice $\Lambda_w(\G)\subset \tfrak^*$, because the exponential map translates 
the multiplicative structure of the character group into an additive structure. 
Most importantly, the following inclusion of lattices holds:
\begin{equation}
 \Lambda_r(\gfrak) \subset \Lambda_w(\G) \subset \Lambda_w(\gfrak) \;. 
\label{eqn:inclusion_root_weight_weight}
\end{equation}
Note that the weight lattice $\Lambda_w$ of $\gfrak$ equals the weight lattice 
of the universal cover $\widetilde{\G}$.

As before, the dual lattice for $\Lambda_w(\G)$ in $\tfrak$ is readily defined
\begin{equation}
 \Lambda_w^*(\G) \coloneqq \mathrm{Hom}\left( \Lambda_w(\G),\Z\right) = \ker 
\left\{ \begin{matrix} \tfrak & \to & \T \\ X & \mapsto & \exp(2\pi \im X) 
\end{matrix} \right\} \; .
\end{equation}
As we see, the \emph{coweight lattice} $\Lambda_w^*(\G)$ is precisely the set of 
solutions to the generalised Dirac quantisation 
condition~\eqref{eqn:general_Dirac} for $\G$. In addition, an inclusion of 
lattices holds
\begin{equation}
 \Lambda_{r}^\vee(\gfrak) \subset \Lambda_w^*(\G)  \subset \Lambda_{mw}(\gfrak)
\; ,
\end{equation}
which follows from dualising~\eqref{eqn:inclusion_root_weight_weight}.
\paragraph{GNO-dual group and algebra}
Following~\cite{Goddard:1976qe,Kapustin:2005py}, a Lie algebra $\GNOfrak$ is 
the \emph{magnetic dual} of $\gfrak$ if its roots coincide with the coroots of 
$\gfrak$. Hence, the Weyl groups of $\gfrak$ and $\GNOfrak$ agree. The 
\emph{magnetic dual group} $\GNOG$ is, by definition, the unique Lie 
group with Lie algebra $\GNOfrak$ and weight lattice $\Lambda_w(\GNOG)$ equal to 
$\Lambda_w^*(\G)$. In physics, $\GNOG$ is called the GNO-dual group; while in 
mathematics, it is known under Langlands dual group.
\paragraph{Polyhedral cones}
A \emph{rational convex polyhedral cone} in $\tfrak$ is a set $\sigma_B$ of the 
form
\begin{equation}
 \sigma_B \equiv \cone(\boldsymbol{B}) = \left\{ \sum_{\alpha^\vee \in 
\boldsymbol{B}}  f_{\alpha^\vee}\ \alpha^\vee \ | \  f_{\alpha^\vee} \geq 
0 \right\} \subseteq \tfrak
\end{equation}
where $\boldsymbol{B} \subseteq \Lambda^\vee_r$, the basis of simple coroots, 
is finite. 
Moreover, we note that $\sigma_B$ is a \emph{strongly convex} cone, i.e.\ 
$\{0\}$ is a face of the cone, and of \emph{maximal dimension}, i.e.\ $\dim( 
\sigma_B) =r$.
Following~\cite{Cox:2011}, such cones $\sigma_B$ are generated 
by the \emph{ray generators} of their edges, where the ray generators in this 
case are precisely the simple coroots of $\gfrak$.

For a polyhedral cone $\sigma_B\subseteq \tfrak$ one naturally defines the 
\emph{dual cone}
\begin{equation}
 \sigma_B^\vee = \left\{m \in \tfrak^* \ | \ m(u)\geq0 \ \text{for all} \ u \in 
 \sigma_B \right\} \subseteq \tfrak^* \; .
\end{equation}
One can prove that $\sigma_B^\vee$ equals the rational convex 
polyhedral cone generated by  $\boldsymbol{B^*}$, i.e.
\begin{equation}
 \sigma_B^\vee = \sigma_{B^*} = \cone(\boldsymbol{B^*}) = \left\{ 
\sum_{\lambda \in 
\boldsymbol{B^*}}  g_{\lambda}\ \lambda \ | \  g_{\lambda} \geq 
0 \right\} \subseteq \tfrak^* \; ,
\end{equation}
which is well-known under the name \emph{(closed) principal Weyl chamber}. By 
the very same arguments as above, the cone $ \sigma_{B^*}$ is generated by its 
ray generators, which are the fundamental weights of $\gfrak$.

For any $m\in \tfrak$ and $d\geq0$, let us define an \emph{affine hyperplane} 
$H_{m,d}$ and \emph{closed linear 
half-spaces} $H_{m,d}^{\pm}$ in $\tfrak^*$ via
\begin{subequations}
\begin{align}
 H_{m,d} &\coloneqq \left\{ \mu \in \tfrak^* \ | \ \mu(m)=d \right\} \subseteq 
\tfrak^* \; ,\\
H_{m,d}^{\pm} &\coloneqq \left\{ \mu \in \tfrak^* \ | \ \mu(m)\geq \pm d 
\right\} \subseteq 
\tfrak^* \;.
\end{align}
\end{subequations}
If $d=0$ then $H_{m,0}$ is hyperplane through the origin, sometimes denoted as 
\emph{central} affine hyperplane. A theorem~\cite{Ziegler:1995} then states: a 
cone $\sigma\subset \R^n$ is finitely generated if and only if 
it is the finite intersection of closed linear half spaces. 

This result allows to make contact with the usual definition of the Weyl 
chamber. Since we know that $\sigma_{B^*}$ is finitely generated by the 
fundamental weights $\{ \lambda_{\alpha} \}$ and the dual basis is 
$\{ \alpha^{\vee} \}$, one arrives at $\sigma_{B^*} = \cap_{\alpha \in 
\simRoots} H_{\alpha^{\vee},0}^+ $; thus, the dominant Weyl chamber is 
obtained by cutting the root space along the hyperplanes orthogonal to some root 
and selecting the cone which has only positive entries.
\paragraph{Remark}
Consider the group $\su$, then the fundamental weight is simply $\tfrac{1}{2}$ 
such that $\Lambda_w^{\su}= \spanZ(\tfrac{1}{2})= \Z \cup \{\Z +\tfrac{1}{2} 
\}$. Moreover, the corresponding cone (Weyl chamber) will be denoted by 
$\sigma_{\boldsymbol{B^*}}^{\su}=\cone(\tfrac{1}{2})$.
%
%%%%%%%%%%%%%%%%%%%%%%%%%%%%%%%%%%%%%%%%%%%%%%%%%%%%%%%%%%%%%%%%%%%%%%%%%%
%%%%%%%%%%%%%%%%%%%%%%%%%%%%%%%%%%%%%%%%%%%%%%%%%%%%%%%%%%%%%%%%%%%%%%%%%%
%
\subsection{Effect of conformal dimension}
\label{subsec:Hilbert basis}
Next, while considering the conformal dimension $\Delta(m)$ as map between two 
Weyl chambers we will stumble across the notion of \emph{affine semi-groups}, 
which are known to constitute the combinatorial background for toric 
varieties~\cite{Cox:2011}.
\paragraph{Conformal dimensions --- revisited}
Recalling the conformal dimension $\Delta$ to be interpreted as the highest 
weight under $\su_R$, it can be understood as the following map
\begin{equation}
 \Delta : \begin{matrix}
           \sigma_{B^*}^{\GNOG} \cap \Lambda_w(\GNOG) &  \to & 
\sigma_{\boldsymbol{B^*}}^{\su} \cap \Lambda_w(\su) \\
	      m & \mapsto & \Delta(m)
          \end{matrix} \; .
\end{equation}
Where $\sigma_{B^*}^{\GNOG}$ is the cone spanned by the fundamental 
weights of $\GNOfrak$, i.e.\ the dual basis of the simple roots 
$\boldsymbol{\Phi}_s$ of $\gfrak$. Likewise, $\sigma_{\boldsymbol{B^*}}^{\su}$ 
is the Weyl chamber for $\su_R$.
Upon continuation, $\Delta$ becomes a map between the dominant Weyl 
chamber of $\GNOG$ and $\su_R$
\begin{equation}
 \Delta : \begin{matrix}
           \sigma_{B^*}^{\GNOG}  &  \to & \sigma_{\boldsymbol{B^*}}^{\su} \\
	      m & \mapsto & \Delta(m) \end{matrix} \; .
\end{equation}
By definition, the conformal dimension~\eqref{eqn:Def_ConfDim} has two types of 
contributions: firstly, a 
positive contribution $|\rho(m)|$ for a weight $\rho \in \Lambda_w(\G)\subset 
\tfrak^*$ and a magnetic weight $m \in \Lambda_w(\GNOG)\subset 
\widehat{\tfrak}^*$. By definition $\Lambda_w(\GNOG)= \Lambda_w^*(\G)$; thus, 
$m$ is a coweight of $\G$ and $\rho(m)$ is the duality paring. Secondly, a 
negative contribution $-|\alpha(m)|$ for a positive root $\alpha \in 
\boldsymbol{\Phi}_+$ of $\gfrak$. By the same arguments, $\alpha(m)$ is the 
duality pairing of weights and coweights. The paring is also well-defined on 
the entire the cone.
\paragraph{Fan generated by conformal dimension}
The individual absolute values in $\Delta$ allow for another interpretation; we 
use them to associate a collection of affine central hyperplanes and closed 
linear half-spaces
\begin{align}
 H_{\mu,0}^{\pm} = \left\{ m \in  \tfrak \  \big| \ \pm \mu(m) 
\geq 0 \right\} \subset \tfrak \und H_{\mu,0} = \left\{ m \in  
\tfrak  \  \big| \  \mu(m) =  0 \right\} \subset \tfrak\; .
\end{align}
Here, $\mu$ ranges over all weights $\rho$ and all positive roots $\alpha$ 
appearing in the theory.
If two weights $\mu_1$, $\mu_2$ are (integer) multiples of each other, 
then $ H_{\mu_1,0} = H_{\mu_2,0}$ and we can reduce the number of relevant 
weights. From now on, denote by $\Gamma$ the set of weights $\rho$ and positive 
roots $\alpha$ which are not multiples of one another. Then the conformal 
dimension contains $Q \coloneqq |\Gamma| \in \NN$ distinct hyperplanes such 
that there exist $2^Q$ different finitely generates cones
\begin{align}
 \sigma_{\epsilon_1,\epsilon_2,\ldots,\epsilon_Q} \coloneqq 
H_{\mu_1,0}^{\epsilon_1} \cap H_{\mu_2,0}^{\epsilon_2} \cap \cdots \cap 
H_{\mu_Q,0}^{\epsilon_Q} \subset \tfrak \with \epsilon_i = \pm \for 
i=1,\ldots, Q \; .
\end{align}
By construction, each cone $\sigma_{\epsilon_1,\epsilon_2,\ldots,\epsilon_Q}$ 
is a strongly convex rational polyhedral cone of dimension $r$, for 
non-trivial cones, and $0$, for trivial intersections. Consequently, 
each cone is generated by its ray generators and these can be chosen to be 
lattice points of $\Lambda_w(\GNOG)$.
Moreover, 
the restriction of $\Delta$ to any
$\sigma_{\epsilon_1,\epsilon_2,\ldots,\epsilon_Q}$ yields a linear function, 
because we effectively resolved the absolute values by defining these cones.

It is, however, sufficient to restrict the considerations to the Weyl chamber 
of $\GNOG$; 
hence, we simply intersect the cones with the hyperplanes defining 
$\sigma_{B^*}^{\GNOG}$, i.e.
\begin{equation}
C_p \equiv C_{\epsilon_1,\epsilon_2,\ldots,\epsilon_Q} \coloneqq 
\sigma_{\epsilon_1,\epsilon_2,\ldots, \epsilon_Q} \cap \sigma_{B^*}^{\GNOG}  
\with p =(\epsilon_1,\epsilon_2,\ldots,\epsilon_Q) \; .
\end{equation}
Naturally, we would like to know for which $\mu \in \Lambda_w(\G)$ the 
hyperplane $H_{\mu,0}$ intersects the Weyl chamber $\sigma_{B^*}^{\GNOG}$ 
non-trivially, i.e.\ not only in the origin. Let us emphasis the differences of 
the Weyl chamber (and their dual cones) of $\G$ and $\GNOG$:
\begin{subequations}
\begin{alignat}{3}
 \sigma_{B^*}^{\G} &= \cone\left( \lambda_\alpha \ | \ \lambda_\alpha 
(\beta^\vee) =\delta_{\alpha,\beta} \ , \forall \alpha ,\beta \in \simRoots 
\right) \subset \tfrak^* & \; &\xleftrightarrow{\ * \ } \; &
\sigma_{B}^{\G} &= \cone \left( \alpha^\vee \ | \ \forall \alpha \in \simRoots 
\right)  \subset \tfrak \; ,
\\
\sigma_{B^*}^{\GNOG} &= \cone \left( m_\alpha \ | \ \beta(m_\alpha) = 
\delta_{\alpha,\beta}  \ , \forall \alpha ,\beta \in \simRoots  \right) \subset 
\tfrak  & \; &\xleftrightarrow{\ * \ } \; &
\sigma_{B}^{\GNOG} &= \cone \left( \alpha  \ | \ \forall \alpha \in \simRoots  
\right) \subset \tfrak^* \; .
\end{alignat}
\end{subequations}
It is possible to prove the following statements:
\begin{enumerate}
 \item If $\mu \in \mathrm{Int}\left( \sigma_{B}^{\GNOG} \cup  
(-\sigma_{B}^{\GNOG}) \right) $, i.e.\ $\mu = \sum_{\alpha \in \simRoots} 
g_\alpha \alpha $ where either all $g_\alpha >0$ \emph{or} all $g_\alpha < 0$ , 
then $H_{\mu,0} \cap \sigma_{B^*}^{\GNOG} = \{0\}$.
  \item If $\mu \in \partial \left(  \sigma_{B}^{\GNOG}\cup  
(-\sigma_{B}^{\GNOG})  \right)$ and 
$\mu\neq0$, i.e.\ $\mu = 
\sum_{\alpha \in \simRoots} g_\alpha \alpha $ where at least one $g_\alpha =0$, 
then $H_{\mu,0}$ intersects $\sigma_{B^*}^{\GNOG}$ at one of its boundary faces.
\item If $\mu \notin \sigma_{B}^{\GNOG}\cup  
(-\sigma_{B}^{\GNOG}) $, i.e.\ $\mu = 
\sum_{\alpha \in \simRoots} g_\alpha \alpha $ with at least one $g_\alpha >0$ 
\emph{and} at least one $g_\beta <0$, then  $\left(H_{\mu,0} \cap 
\sigma_{B^*}^{\GNOG}\right) \setminus\{0\} \neq \emptyset $.
\end{enumerate}
\textbf{Consequently, a weight $\boldsymbol{\mu \in \Lambda_w(\G)}$ appearing 
in $\boldsymbol{\Delta}$ leads to a hyperplane intersecting the Weyl chamber of 
$\GNOG$ non-trivially if and only if neither $\boldsymbol{\mu}$ nor 
$\boldsymbol{-\mu}$ lies in the rational cone spanned by the simple roots 
$\simRoots$ of $\boldsymbol{\G}$.} \par \noindent
Therefore, the contributions $- |\alpha(m) |$, for $\alpha \in \posRoots$, 
of the vector multiplet never yield a relevant hyperplane.
From now on, 
assume that trivial cones $C_p$ are omitted in the index set $I$ for $p$. The 
appropriate geometric object to consider is then the \emph{fan} $F_\Delta 
\subset \tfrak$ defined by the family $F_\Delta= \left\{ C_{p}\ , \ p\in 
I\right\}$ in $\tfrak$. A fan $F$ is a family of non-empty polyhedral 
cones such that (i) every non-empty face of a cone in $F$ is a cone in 
$F$ and (ii) the intersection of any two cones in $F$ is a face of 
both. In addition, the fan $F_\Delta$ defined above is a \emph{pointed 
fan}, because $\{0\}$ is a cone in $F_\Delta$ (called the trivial cone).
\paragraph{Semi-groups}
Although we already know the cone generators for the fan $F_\Delta$, we have to 
distinguish them from the generators of $F_\Delta \cap \Lambda_w(\GNOG)$, i.e.\ 
we need to restrict to the weight lattice of $\GNOG$.
The first observation is that
\begin{equation}
 S_p\coloneqq C_p \cap \Lambda_w(\GNOG)  \for p\in I
\end{equation}
are semi-groups, i.e.\ sets with an associative binary operation. This is 
because 
the addition of elements is commutative, but there is no inverse 
defined as ``subtraction'' would lead out of the cone. Moreover, the $S_p$ 
satisfy further properties, which we now simply collect, see for 
instance~\cite{Ziegler:1995}. Firstly, the $S_p$ are affine semi-groups, which 
are semi-groups that can be embedded in $\Z^n$ for some $n$. Secondly, every 
$S_p$ possesses an identity element, here $m=0$, and such semi-groups are 
called \emph{monoids}. Thirdly, the $S_p$ are \emph{positive} because the only 
invertible element is $m=0$.

Now, according to \emph{Gordan's Lemma}~\cite{Ziegler:1995,Cox:2011}, we know 
that every $S_p$ is finitely generated, because all $C_p$'s are finitely 
generated, rational polyhedral cones. Even more is true, since the division into 
the $C_p$ is realised via affine hyperplanes $H_{\mu_i,0}$ passing through 
the origin, the $C_p$ are strongly convex rational cones of maximal dimension. 
Then~\cite[Prop.~1.2.22.]{Cox:2011} holds and we know that there exist a 
unique minimal generating set for $S_p$, which is called \emph{Hilbert basis}.

The Hilbert basis $\Hcal(S_p)$ is defined via
\begin{equation}
 \Hcal(S_p) \coloneqq \left\{ m \in S_p \ | \ m  \ \text{is 
irreducible}  \right\} \; ,
\end{equation}
where an element is called \emph{irreducible} if and only if $m = x 
+ y$ for $x,y\in S_p$ implies $x=0$ or $y=0$.
The importance of the Hilbert basis is that it is a unique, finite, 
minimal set of irreducible elements that generate 
$S_p$. Moreover, $\Hcal(S_p)$ always contains the ray generators of the 
edges of $C_p$. The elements of $\Hcal(S_p)$ are sometimes called \emph{minimal 
generators}.

As a remark, there exist various algorithms for computing the Hilbert basis, 
which are, for example, discussed in~\cite{Miller:2005,Sturmfels:1995}. For the 
computations presented in this paper, we used the \texttt{Sage} module 
\emph{Toric 
varieties} programmed by A.~Novoseltsev and V.~Braun as well as the 
\texttt{Macaulay2} package \emph{Polyhedra} written by René Birkner.

After the exposition of the idea to employ the conformal dimension to define a 
fan in the Weyl chamber of $\GNOG$, for which the intersection with the weight 
lattice leads to affine semi-groups, we now state the main consequence:  \par 
\noindent
\textbf{The collection $\boldsymbol{\{\Hcal(S_p) \; , 
p \in I \}}$ of all Hilbert bases is the set of necessary (bare) monopole 
operators for a theory with conformal dimension $\boldsymbol{\Delta}$.} \par
At this stage we did not include the Casimir invariance described by the 
dressing factors $P_{G}(t,m)$. For a generic situation, the bare and dressed 
monopole operators for a GNO-charge $m \in \Hcal(S_p)$ for some $p$ are all 
necessary generators for the chiral ring $\C[\MCoulomb]$. However, there will 
be scenarios for which there exists a further reduction of the number of 
generators. For those cases, we will comment and explain the cancellations.
% 
%%%%%%%%%%%%%%%%%%%%%%%%%%%%%%%%%%%%%%%%%%%%%%%%%%%%%%%%%%%%%%%%%%%%
%%%%%%%%%%%%%%%%%%%%%%%%%%%%%%%%%%%%%%%%%%%%%%%%%%%%%%%%%%%%%%%%%%%%
%
\subsection{Dressing of monopole operators}
\label{subsec:Dressings_as_HS}
One crucial ingredient of the monopole formula of~\cite{Cremonesi:2013lqa} 
are the dressing factors $P_{\G}(t,m)$ and this section provides an algebraic 
understanding. We refer 
to~\cite{Humphreys:1972,Varadarajan:1984,Humphreys:1990} for the exposition of 
the 
mathematical details used here.

It is known that in $\Ncal=4$ the $\Ncal =2$ BPS-monopole operator $V_{m}$ is 
compatible with a constant background of the $\Ncal=2$ adjoint complex scalar 
$\Phi$, provided $\Phi$ takes values on the Lie algebra $\hfrak_m $ of the 
residual gauge group $\Hh_m \subset \G$, i.e.\ the stabiliser of $m$ in $\G$. 
Consequently, each bare monopole operator $V_{m}$ is compatible with \emph{any} 
$\Hh_m$-invariant polynomial on $\hfrak_m$. We will now argue that the dressing 
factors $P_{\G}(t,m)$ are to be understood as Hilbert (or Poincar\'{e}) series 
for this so-called \emph{Casimir-invariance}.
\paragraph{Chevalley-Restriction Theorem}
Let $\G$ be a Lie group of rank $l$ with a semi-simple Lie algebra 
$\gfrak$ over $\C$ and $\G$ acts via the adjoint representation on $\gfrak$. 
Denote by $\mathfrak{P}(\gfrak)$ the algebra of all polynomial functions on 
$\gfrak$. The action of $\G$ extends to $\mathfrak{P}(\gfrak)$ and 
$\mathfrak{I}(\gfrak)^\G$ denotes the set of $\G$-invariant polynomials in 
$\mathfrak{P}(\gfrak)$. In addition, denote by $\mathfrak{P}(\hfrak)$ the 
algebra of all polynomial functions on $\hfrak$. The Weyl group $\Wcal_\G$, 
which acts naturally on $\hfrak$, acts also on $\mathfrak{P}(\hfrak)$ and 
$\mathfrak{I}(\hfrak)^{\Wcal_\G}$ denotes the Weyl-invariant polynomials on 
$\hfrak$. The \emph{Chevalley-Restriction Theorem} now states
\begin{equation}
 \mathfrak{I}(\gfrak)^\G \cong \mathfrak{I}(\hfrak)^{\Wcal_\G} \; ,
\end{equation}
where the isomorphism is given by the restriction map $p \mapsto p |_{\hfrak}$ 
for $p \in \mathfrak{I}(\gfrak)^\G$.

Therefore, the study of $\Hh_m$-invariant polynomials on $\hfrak_m$ is reduced 
to $\Wcal_{\Hh_m}$-invariant polynomials on a Cartan subalgebra $\tfrak_m 
\subset \hfrak_m$.
\paragraph{Finite reflection groups}
It is due to a theorem by Chevalley~\cite{Chevalley:1955}, in the context of 
\emph{finite reflection groups}, that there exist $l$ algebraically independent 
homogeneous elements $p_1,\ldots, p_l$ of positive degrees $d_i$, for 
$i=1,\ldots,l$, such that 
\begin{equation}
 \mathfrak{I}(\hfrak)^{\Wcal_\G}  = \C\left[p_1,\ldots, p_l\right] \; .
\end{equation}
In addition, the degrees $d_i$ satisfy
\begin{align}
 \left| \Wcal_\G \right| = \prod_{i=1}^l d_i \und \sum_{i=1}^d (d_i -1) = 
\text{number of reflections in $\Wcal_\G$} \;.
\end{align}
The degrees $d_i$ are unique~\cite{Humphreys:1990} and tabulated for all Weyl 
groups, see for instance~\cite[Sec. 3.7]{Humphreys:1990}. However, the 
generators $p_i$ are themselves not uniquely determined.
\paragraph{Poincar\'{e} or Molien series}
On the one hand, the Poincar\'{e} series for the 
$\mathfrak{I}(\hfrak)^{\Wcal_\G}$ is simply given by
\begin{equation}
 P_{\mathfrak{I}(\hfrak)^{\Wcal_\G}}(t)= \prod_{i=1}^l \frac{1}{1-t^{d_i}} \; .
\end{equation}
On the other hand, since $\hfrak$ is a $l$-dimensional complex vector space and 
$\Wcal_\G$ a finite group, the generating function for the invariant 
polynomials is known as Molien series~\cite{Molien:1897}
\begin{equation}
 P_{\mathfrak{I}(\hfrak)^{\Wcal_\G}}(t)= \frac{1}{\left| \Wcal_\G \right|} 
\sum_{g \in  \Wcal_\G } \frac{1}{\det\left(\mathds{1}-t \ g\right)} \; .
\end{equation}
Therefore, the dressing factors $P_{\G}(t,m)$ in the Hilbert 
series~\eqref{eqn:HS_refined} for the Coulomb branch are the Poincar\'{e} 
series for graded algebra of $\Hh_m$-invariant polynomials on $\hfrak_m$.
\paragraph{Harish-Chandra isomorphism}
In~\cite{Cremonesi:2013lqa}, the construction of the $P_\G(t,m)$ is based on 
Casimir invariants of $\G$ and $\Hh_m$; hence, we need to make contact with that 
idea. Casimir invariants live in the centre $\Zcal(\Ufrak(\gfrak))$ of the 
universal enveloping algebra $\Ufrak(\gfrak)$ of $\gfrak$. Fortunately, the 
Harish-Candra isomorphism~\cite{Harish:Chandra} provides us with
\begin{equation}
 \Zcal(\Ufrak(\gfrak)) \cong \mathfrak{I}(\hfrak)^{\Wcal_\G} \; .
\end{equation}
Consequently, $\Zcal(\Ufrak(\gfrak))$ is a polynomial algebra with $l$ 
algebraically independent homogeneous elements that have the same positive 
degrees $d_i$ as the generators of $\mathfrak{I}(\hfrak)^{\Wcal_\G}$. It is 
known that for semi-simple groups $\G$ these generators can be chosen to be the 
$\rank(\G)$ Casimir invariants; i.e.\ the space of Casimir-invariants 
is freely generated by $l$ generators (together with the unity).
\paragraph{Conclusions}
So far, $\G$ (and $\Hh_m$) had been restricted to be semi-simple. However, in 
most cases $\Hh_m$ is a direct product group of semi-simple Lie groups and 
$\uo$-factors. We proceed in two steps: firstly, $\uo$ acts trivially on its 
Lie-algebra $\cong \R$, thus all polynomials are invariant and we obtain
\begin{equation}
 \mathfrak{I}(\R)^{\uo} = \R[x] \und P_{\uo}(t) =\frac{1}{1-t} \; .
\end{equation}
Secondly, each factor $\G_i$ of a direct product $\G_1 \times \cdots 
\times \G_M$ acts via the adjoint representation on on its own Lie algebra 
$\gfrak_i$ and trivially on all other $\gfrak_j$ for $j\neq i$. Hence, the space 
of $\G_1 \times \cdots \times \G_M$-invariant polynomials on $\gfrak_1 \oplus 
\cdots \oplus \gfrak_M$ factorises into the product of the 
$\mathfrak{I}(\gfrak_i)^{\G_i}$ such that
\begin{equation}
 \mathfrak{I}(\oplus_i \gfrak_i)^{\prod_i \G_i} = \prod_i 
\mathfrak{I}(\gfrak_i)^{\G_i} \und P_{\mathfrak{I}(\oplus_i \gfrak_i)^{\prod_i 
\G_i} } (t) = \prod_i  P_{\mathfrak{I}(\gfrak_i)^{\G_i} } (t) \; .
\end{equation}
 For abelian groups $\G$, the Hilbert series for the Coulomb branch factorises 
in the Poincar\'{e} series $\G$-invariant polynomials on $\gfrak$ times the 
contribution of the (bare) monopole operators. In contrast, the Hilbert series 
does not factorise for non-abelian groups $\G$ as the stabiliser $\Hh_m\subset 
\G$ depends on $m$.
% 
%%%%%%%%%%%%%%%%%%%%%%%%%%%%%%%%%%%%%%%%%%%%%%%%%%%%%%%%%%%%%%%%%%%%
%%%%%%%%%%%%%%%%%%%%%%%%%%%%%%%%%%%%%%%%%%%%%%%%%%%%%%%%%%%%%%%%%%%%
%
\subsection{Consequences for unrefined Hilbert series}
\label{subsec:summation_HS_unrefined}
The aforementioned dissection of the Weyl chamber $\sigma_{B^*}^{\GNOG}$ into a 
fan, induced by the conformal dimension $\Delta$, and the subsequent collection 
of semi-groups in $\Lambda_w(\GNOG) \slash \Wcal_{\GNOG} $ provides an 
immediate consequence for the unrefined Hilbert series.
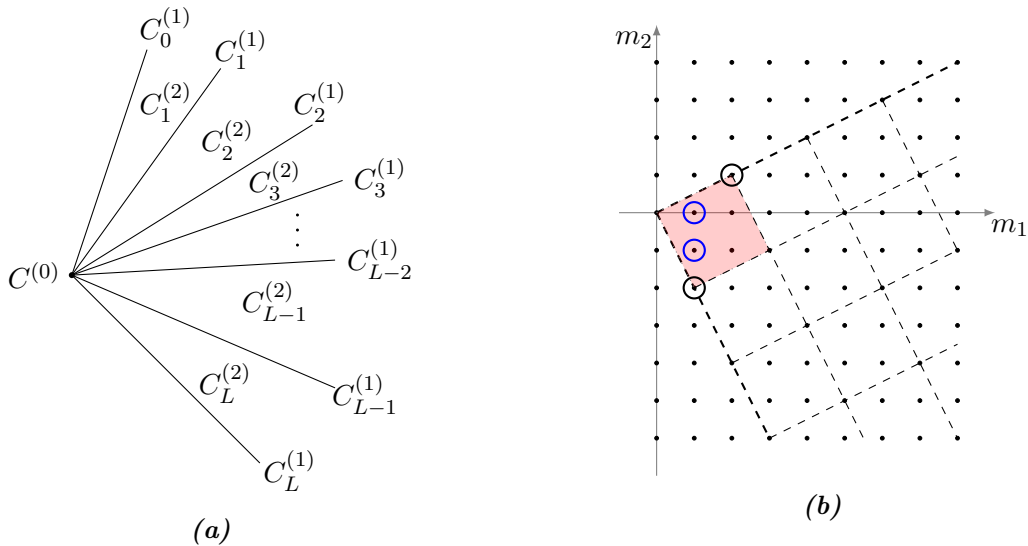
\begin{figure}[h]
\begin{subfigure}{0.485\textwidth}
\centering
 \begin{tikzpicture}
  \draw (0,0) node[circle,inner sep=0.8pt,fill,black] {};
  \draw[black] (0,0) -- (1,3);
  \draw[black] (0,0) -- (1.1*1.8,1.1*2.5);
  \draw[black] (0,0) -- (3.2,2);
  \draw[black] (0,0) -- (0.9*4,0.9*1.4);
  \draw (3,0.8) node[circle,inner sep=0.4pt,fill,black] {};
  \draw (3,0.6) node[circle,inner sep=0.4pt,fill,black] {};
  \draw (3,0.4) node[circle,inner sep=0.4pt,fill,black] {};
  \draw[black] (0,0) -- (3.5,0.2);
  \draw[black] (0,0) -- (3.5,-1.5);
  \draw[black] (0,0) -- (2.5,-2.5);
  \draw (1.25,2.3) node {$C_1^{(2)}$};
  \draw (0.9*2.3,0.9*2) node {$C_2^{(2)}$};
  \draw (0.9*3,0.9*1.4) node {$C_3^{(2)}$};
  \draw (0.9*3,-0.4*0.9) node {$C_{L-1}^{(2)}$};
  \draw (0.85*2.4,-1.8*0.8) node {$C_L^{(2)}$};
  \draw (1.25,3.3) node {$C_0^{(1)}$};
  \draw (2.25,3.0) node {$C_1^{(1)}$};
  \draw (3.3,2.3) node {$C_2^{(1)}$};
  \draw (4.1,1.3) node {$C_3^{(1)}$};
  \draw (4.1,0.2) node {$C_{L-2}^{(1)}$};
  \draw (3.9,-1.6) node {$C_{L-1}^{(1)}$};
  \draw (2.9,-2.6) node {$C_L^{(1)}$};
   \draw (-0.5,0) node {$C^{(0)}$};
 \end{tikzpicture}
 \caption{}
 \label{Fig:Rep_Fan}
  \end{subfigure}
  \begin{subfigure}{0.485\textwidth}
   \centering
   \begin{tikzpicture}
  \coordinate (Origin)   at (0,0);
  \coordinate (XAxisMin) at (-0.5,0);
  \coordinate (XAxisMax) at (4.5,0);
  \coordinate (YAxisMin) at (0,-3.5);
  \coordinate (YAxisMax) at (0,2.5);
  \draw [thin, gray,-latex] (XAxisMin) -- (XAxisMax);%
  \draw [thin, gray,-latex] (YAxisMin) -- (YAxisMax);%
  \draw (4.7,-0.2) node {$m_1$};
  \draw (-0.3,2.3) node {$m_2$};
     \foreach \x in {0,1,...,8}{%
      \foreach \y in {-6,-5,...,4}{%
        \node[draw,circle,inner sep=0.5pt,fill,black] at (0.5*\x,0.5*\y) {};
            }
            }
   \draw[black,dashed,thick] (0,0) -- (4,2) ;
   \draw[black,dashed,thick] (0,0) -- (1.5,-3) ;   
   \filldraw[dotted,fill=red, fill opacity=0.2, draw=gray] (Origin) -- (1,1/2) 
-- (3/2,-1/2) -- (1/2,-1) -- cycle ;
  \draw[black,dashed] (1/2,-1) -- (8/2,2/2-1/4) ;   
  \draw[black,dashed] (1,1/2) -- (5.5/2,-6/2) ;   
  \draw[black,dashed] (1,-2) -- (4,-1/2) ;   
  \draw[black,dashed] (2,1) -- (4,-3) ;   
  \draw[black,dashed] (3/2,-3) -- (4,-7/4) ;   
  \draw[black,dashed] (3,3/2) -- (4,-1/2) ;   
  \draw[black,thick] (1,1/2) circle (4pt);
  \draw[black,thick] (1/2,-1) circle (4pt);
  \draw[blue,thick] (1/2,0) circle (4pt);
  \draw[blue,thick] (1/2,-1/2) circle (4pt);
  \end{tikzpicture}
\caption{}
\label{Fig:Rep_Cone}
  \end{subfigure}
  \caption{A representative fan, which is spanned by the 
$2$-dim.\ cones $C_{p}^{(2)}$ for $p=1,\ldots,L$, is displayed 
in~\ref{Fig:Rep_Fan}. 
In addition, \ref{Fig:Rep_Cone} contains a $2$-dim.\ cone with a Hilbert basis 
of the two ray generators (black) and two additional minimal generators (blue). 
The ray generators span the fundamental parallelotope (red region).}
\label{Fig:Sketch}
\end{figure}
For simplicity, we illustrate the consequences for a rank two example. Assume 
that the Weyl chamber is divided into a fan generated the $2$-dimensional cones 
$C_p^{(2)}$ for $p=1,\ldots,L$, as sketched in 
Fig.~\ref{Fig:Rep_Cone}. For each cone, one has two $1$-dimensional cones
$C_{p-1}^{(1)}$, $C_{p}^{(1)}$ and the trivial cone $C^{(0)}=\{0\}$ as boundary, 
i.e.\ $\partial C_{p}^{(2)}= C_{p-1}^{(1)} \cup C_{p}^{(1)}$, where 
$C_{p-1}^{(1)} \cap C_{p}^{(1)} = C^{(0)}$.

The Hilbert basis $\Hcal(S_p^{(2)})$ for $S_p^{(2)} \coloneqq C_p^{(2)} \cap 
\Lambda_w^{\GNOG}$ contains the ray generators $\{x_{p-1},x_p\}$, such that 
$\Hcal(S_p^{(1)})= \{x_p\}$, and potentially other minimal generators 
$u_{\kappa}^{p}$ for $\kappa$ in some \emph{finite} index set. 
Although any element $s\in S_p^{(2)}$ can be generated by $\{x_{p-1},x_p, 
\{u_{\kappa}^{p} \}_\kappa\}$, the representation $s= a_0 x_{p-1} + a_1 x_{p} + 
\sum_\kappa b_\kappa u_{\kappa}^{p}$ is \emph{not} unique. Therefore, great 
care needs to be taken if one would like to sum over all elements in 
$S_p^{(2)}$. A possible realisation employs the \emph{fundamental 
parallelotope} 
\begin{equation}
 \mathcal{P}(C_p^{(2)})\coloneqq \{ a_0 x_{p-1} + a_1 x_{p} \ | \ 0 \geq 
a_0,a_1 \geq 1\} \; ,
\end{equation}
see also Fig.~\ref{Fig:Rep_Cone}. The number of points contained in 
$\mathcal{P}(C_p^{(2)})$ is computed by the discriminant
\begin{equation}
 d(C_p^{(2)}) \coloneqq |\det(x_{p-1},x_p)| \; .
\end{equation}
However, as known from solid state physics, the discriminant counts each of the 
four boundary lattice points by $\tfrac{1}{4}$; thus, there are $d(C_p^{(2)}) 
-1$ points in the interior. Remarkably, each point $s \in \mathrm{Int}( 
\mathcal{P}(C_p^{(2)}))$ is given by positive integer combinations of the 
$\{u_{\kappa}^{p} \}_\kappa$ alone. A translation of $ \mathcal{P}(C_p^{(2)}) $
by non-negative integer combinations of the ray-generators $\{x_{p-1},x_p\}$ 
fills the entire semi-group $S_p^{(2)}$ and each point is only realised once. 

Now, we employ this fact to evaluate the un-refined Hilbert series explicitly.
\begin{align}
 \HS_{\G}(t) &= \sum_{m \in \Lambda_w(\GNOG) \slash \mathcal{W}_{\GNOG}} 
t^{\Delta(m)} P_{\G}(t,m) \notag\\
&= P_{\G}(t,0) 
+ \sum_{p=0}^{L} P_{\G}(t,x_p) \sum_{n_p>0} t^{n_p \Delta(x_p)} 
+\sum_{p=1}^{L}  \sum_{n_{p-1},n_p > 0} P_{\G}(t,x_{p-1}+x_p) t^{\Delta(n_{p-1} 
x_{p-1} + n_{p} x_{p})} \notag \\ 
&\phantom{= P_{\G}(t,0) \ ; }
+\sum_{p=1}^{L} \sum_{s \in \mathrm{Int}( 
\mathcal{P}(C_p^{(2)}))} \sum_{n_{p-1},n_p \geq0} P_{\G}(t,s) t^{\Delta(s + 
n_{p-1} x_{p-1} + n_{p} x_{p})} \notag \\
&=  P_{\G}(t,0) +  \sum_{p=0}^{L} P_{\G}(t,x_p) \frac{t^{ \Delta(x_p)}}{1- 
t^{\Delta(x_p)}} + \sum_{p=1}^{L} \frac{ P_{\G}(t,x_{p-1}+x_p) \
t^{\Delta(x_{p-1}) 
+\Delta(x_p)}}{\left(1-t^{\Delta(x_{p-1})}\right) 
\left(1-t^{\Delta(x_{p})}\right)} \notag \\
&\phantom{= P_{\G}(t,0) \ ; }
+ \sum_{p=1}^{L} \sum_{s \in \mathrm{Int}( 
\mathcal{P}(C_p^{(2)}))} \frac{ P_{\G}(t,s) \
t^{\Delta(s)}}{\left(1-t^{\Delta(x_{p-1})}\right)
\left(1-t^{\Delta(x_{p})}\right)} \notag \\
&= \frac{P_{\G}(t,0) }{\prod_{p=0}^{L} \left(1-t^{\Delta(x_{p})}\right)}
\Bigg\{ \prod_{q=0}^{L} \left(1-t^{\Delta(x_{q})}\right)
+ \sum_{q=0}^{L} \frac{P_{\G}(t,x_q)}{P_{\G}(t,0)} t^{\Delta(x_q)} 
\prod_{r=0\atop r\neq q}^{L} \left(1-t^{\Delta(x_{r})}\right) 
\label{eqn:HS_generic_solved}\\
&\qquad \qquad \qquad
+\sum_{q=1}^{L} \frac{P_{\G}(t,C_q^{(2)})}{P_{\G}(t,0)} 
\bigg[t^{\Delta(x_{q-1}) +\Delta(x_q)}+  \sum_{s \in 
\mathrm{Int}( \mathcal{P}(C_q^{(2)}))} t^{\Delta(s)} 
\bigg] 
\prod_{r=0\atop{r\neq q-1,q}}^L \left(1-t^{\Delta(x_{r})}\right) 
\Bigg\} \notag \; .
 \end{align}
Next, we utilise that the classical dressing factors, for rank two examples, 
only have three different values: in the ($2$-dim.) interior of the Weyl 
chamber $W$, the residual gauge group is the maximal torus $\T$ and 
$P_{\G}(t,\mathrm{Int}W) \equiv P_2(t)= \prod_{i=1}^2 \tfrac{1}{(1-t)}$. Along 
the $1$-dimensional boundaries, the residual gauge group is a non-abelian 
subgroup $\Hh$ such that $\T \subset \Hh \subset \G$ and the $P_{\G}(t,\partial 
W \setminus \{0\}) \equiv 
P_1(t) = \prod_{i=1}^2 \tfrac{1}{(1-t^{b_i})} $, for the two degree $b_i$ 
Casimir invariants of $\Hh$. At the ($0$-dim.) boundary of the boundary, the 
group is 
unbroken and $P_{\G}(t, 0) \equiv P_0(t) = \prod_{i=1}^2 \tfrac{1}{(1-t^{d_i})} 
$ contains the Casimir invariants of $\G$ of degree $d_i$.
Thus, there are a few observations to be addressed.
\begin{enumerate}
 \item The numerator of~\eqref{eqn:HS_generic_solved}, which is everything in 
the curly brackets $\{ \ldots \}$, starts with a one and is a polynomial with 
integer coefficients, which is required for consistency.
\item The denominator of~\eqref{eqn:HS_generic_solved} is 
given by $P_{\G}(t,0)\slash \prod_{p=0}^{L} (1-t^{\Delta(x_{p})})$ and  
describes the poles due to the Casimir invariants of $\G$ and the bare monopole 
$(x_p,\Delta(x_p))$ which originate from ray generators $x_p$.
\item The numerator has contributions $\sim t^{\Delta(x_p)}$ for the ray 
generators with pre-factors $\tfrac{P_1(t)}{P_0(t)} -1$ for the two outermost 
rays $p=0$, $p=L$ and pre-factors $\tfrac{P_2(t)}{P_0(t)} -1$ for the remaining 
ray generators. None of the two pre-factors has a constant term as 
$P_i(t\to0)=1$ for each $i=0,1,2$. Also $\deg(1\slash P_0(t))\geq \deg(1\slash 
P_1(t))\geq \deg(1\slash P_2(t)) =2$ and 
\begin{equation}
 \frac{P_2(t)}{P_0(t)} = \frac{(1-t^{d_1})(1-t^{d_2})}{(1-t)(1-t)} = 
\sum_{i=0}^{d_1 -1} \sum_{j=0}^{d_2-1}t^{i+j}
\end{equation}
is a polynomial for any rank two group. For the examples considered here, we 
also obtain
\begin{equation}
 \frac{P_1(t)}{P_0(t)} = \frac{(1-t^{d_1})(1-t^{d_2})}{(1-t^{b_1})(1-t^{b_2})}
 = \frac{(1-t^{k_1 b_1})(1-t^{k_2 b_2})}{(1-t^{b_1})(1-t^{b_2})} 
= 
\sum_{i=0}^{b_1 -1} \sum_{j=0}^{b_2-1} t^{i \cdot k_1 + j \cdot k_2}
\end{equation}
for some $k_1,k_2 \in \NN$.
In summary, $(\tfrac{P_{\G}(t,x_p)}{P_{\G}(t,0)} -1)t^{\Delta(x_p)}$ describes 
the dressed monopole operators corresponding to the ray generators $x_p$.
 \item The finite sums $\sum_{s \in 
\mathrm{Int}( \mathcal{P}(C_p^{(2)}))} t^{\Delta(s)}$ are entirely determined 
by the conformal dimensions of the minimal generators $u_\kappa^p$.
\item The first contributions for the minimal generators $u_\kappa^p$ are 
of the form 
\begin{equation}
 \tfrac{P_{2}(t)}{P_{0}(t)} t^{\Delta(u_\kappa^p)} = \sum_{i=0}^{d_1 -1} 
\sum_{j=0}^{d_2-1}t^{i+j+ \Delta(u_\kappa^p)} \; ,
\end{equation}
which then comprise the bare and the dressed monopole operators simultaneously.
  \item If $C_p^{(2)}$ is simplicial, i.e.\ $\Hcal(S_p^{(2)})= 
\{x_{p-1},x_p\}$, 
then the sum over $s \in \mathrm{Int}( \mathcal{P}(C_p^{(2)})) $ 
in~\eqref{eqn:HS_generic_solved} is zero, as the interior is empty. Also 
indicated by $d(C_p^{(2)})=1$.
\end{enumerate}
In conclusion, the Hilbert series~\eqref{eqn:HS_generic_solved} suggests that 
\emph{ray generators} are to be expected in the denominator, while other 
minimal generators are manifest in the numerator. Moreover, the entire Hilbert 
series is determined by a \emph{finite} set of numbers: the conformal 
dimensions of the minimal generators $\{\Delta(x_p) \ | \ p=0,1,\ldots, L\} $ 
and $\{ \{ \Delta(u_\kappa^{(p)}) \ | \ \kappa=1,\ldots, d(C_p^{(2)})-1 \} \ | 
\ p=1,\ldots,L \}$ as well as the classical dressing factors.

Moreover, the dressing behaviour, i.e.\ number and degree, of a minimal 
generator $m$ is described by the quotient $P_{\G}(t,m) \slash P_{\G}(t,0)$. 
Consolidating evidence for this statement comes from the analysis of the 
plethystic logarithm, which we present in App.~\ref{app:PL}. 
Together, the Hilbert series and the plethystic logarithm allow a better 
understanding of the chiral ring.

We illustrate the formula~\eqref{eqn:HS_generic_solved} for the two simplest 
cases in order to hint on the differences that arise if $d(C_p^{(2)})>1$ for 
cones within the fan.
\paragraph{Example: one simplicial cone} Adapting the 
result~\eqref{eqn:HS_generic_solved} to one cone $C_1^{(2)}$ with cone / 
Hilbert basis $\{x_0,x_1\}$, we find
\begin{align}
\label{eqn:Example_HS_simplicial}
 \HS= \frac{1+ \left( \frac{P_1(t)}{P_0(t)}-1\right) 
\left(t^{\Delta(x_0)}+t^{\Delta(x_1)} \right) + \left(1-2\frac{P_1(t)}{P_0(t)} 
+\frac{P_2(t)}{P_0(t)} \right) t^{\Delta(x_0)+\Delta(x_1)}}{\prod_{i=1}^2 
\left( 1-t^{d_i} \right) \prod_{p=0}^{1} \left( 1-t^{\Delta(x_p)} \right)} \; .
\end{align}
Examples treated in this paper are as follows: firstly, the 
representation $[2,0]$ for the quotients $\Spin(4)$, $\sorm(3)\times \su$, $\su 
\times \sorm(3)$, $\mathrm{PSO}(4)$ of Sec.~\ref{subsec:A1xA1_Rep20}; secondly, 
$\USp$ for the case $N_3=0$ of Sec.~\ref{subsec:USp4_N3=0}; thirdly, $\Gtwo$ in 
the representations $[1,0]$, $[0,1]$ and $[2,0]$ of Sec.~\ref{subsec:G2_Cat1}.
The corresponding expression for the plethystic logarithm is provided 
in~\eqref{eqn:Example_PL_simplicial}.
\paragraph{Example: one non-simplicial cone} Adapting the 
result~\eqref{eqn:HS_generic_solved} to one cone $C_1^{(2)}$ with Hilbert basis 
$\{x_0,x_1,\{u_\kappa \}\}$, fundamental parallelotope $\mathcal{P}$, and 
discriminant $d>1$, we find
\begin{align}
\label{eqn:Example_HS_non-simplicial}
 \HS= \frac{1+ \left( \frac{P_1(t)}{P_0(t)}-1\right) 
\left(t^{\Delta(x_0)}+t^{\Delta(x_1)} \right) + \left(1-2\frac{P_1(t)}{P_0(t)} 
+\frac{P_2(t)}{P_0(t)} \right) 
t^{\Delta(x_0)+\Delta(x_1)} + \frac{P_2(t)}{P_0(t)} \sum_{s \in 
\mathrm{Int}(\mathcal{P})} t^{\Delta(s)}
}{\prod_{i=1}^2 \left( 1-t^{d_i} \right)  
\prod_{p=0}^{1} \left( 1-t^{\Delta(x_p)} \right)} \; .
\end{align}
An example for this case is $\sorm(4)$ with representation $[2,0]$ treated 
in Sec.~\ref{subsec:A1xA1_Rep20}. For the plethystic logarithm we refer 
to~\eqref{eqn:Example_PL_non-simplicial}.

The difference between~\eqref{eqn:Example_HS_simplicial} 
and~\eqref{eqn:Example_HS_non-simplicial} lies in the finite sum added in the 
numerator which accounts for the minimal generators that are not ray generators.
\subsection{Consequences for refined Hilbert series}
\label{subsec:summation_HS_refined}
If the centre $\Zcal(\GNOG)$ of the GNO-dual group $\GNOG$ is a non-trivial 
Lie-group of rank $\rank(\Zcal(\GNOG))=\rho$, one introduces additional 
fugacities $\vec{z}\equiv (z_i)$ for $i=1,\ldots,\rho$ such that the Hilbert 
series counts operators according to $\su_R$-spin $\Delta(m)$ and 
topological charges $\vec{J}(m)\equiv(J_i(m))$ for 
$i=1,\ldots,\rho$. Let us introduce the notation
\begin{equation}
 \vec{z}^{\vec{J}(m)} \coloneqq \prod_{i=1}^{\rho} z_i^{J_i(m)} \st 
 \vec{z}^{\vec{J}(m_1+m_2)} = 
\vec{z}^{\vec{J}(m_1)+\vec{J}(m_2)} = \vec{z}^{\vec{J}(m_1)} 
\cdot \vec{z}^{\vec{J}(m_2)} \; ,
\end{equation}
where we \emph{assumed} each component $J_i(m)$ to be a linear function 
in $m$.
By the very same arguments as in~\eqref{eqn:HS_generic_solved}, one can 
evaluate the refined Hilbert series explicitly and obtains
\begin{align}
  \HS_{\G}(t,\vec{z}) &= \sum_{m \in \Lambda_w^{\GNOG} \slash 
\mathcal{W}_{\GNOG}} \vec{z}^{\vec{J}(m)}
t^{\Delta(m)} P_{\G}(t,m) \notag\\
&= \frac{P_{\G}(t,0) }{\prod_{p=0}^{L} \left(1-\vec{z}^{\vec{J}(x_p)} 
t^{\Delta(x_{p})}\right)}
\Bigg\{ 
\prod_{q=0}^{L} \left( 1-\vec{z}^{\vec{J}(x_q)}t^{\Delta(x_{q})} \right) 
\label{eqn:HS_generic_refined}\\
&\phantom{= \frac{P_{\G}(t,0) }{\prod_{p=0}^{L} \left(1-\vec{z}^{\vec{J}(x_p)} 
t^{\Delta(x_{p})}\right)}}
+ \sum_{q=0}^{L} \frac{P_{\G}(t,x_q)}{P_{\G}(t,0)} 
\vec{z}^{\vec{J}(x_q)} t^{\Delta(x_q)} 
\prod_{r=0\atop r\neq q}^{L} 
\left( 1-\vec{z}^{\vec{J}(x_r)} t^{\Delta(x_{r})} \right) 
\notag \\
&\phantom{= \frac{P_{\G}(t,0) }{\prod_{p=0}^{L} \left(1-\vec{z}^{\vec{J}(x_p)} 
t^{\Delta(x_{p})}\right)}}
+\sum_{q=1}^{L} \frac{P_{\G}(t,C_q^{(2)})}{P_{\G}(t,0)} 
\bigg[\vec{z}^{\vec{J}(x_{q-1})+\vec{J}(x_q)} t^{\Delta(x_{q-1}) +\Delta(x_q)} 
\notag\\*
&\phantom{= \frac{P_{\G}(t,0) }{\prod_{p=0}^{L} \left(1-\vec{z}^{\vec{J}(x_p)} 
t^{\Delta(x_{p})}\right)}}
\qquad  \qquad
+  \sum_{s \in \mathrm{Int}( \mathcal{P}(C_q^{(2)}))} 
\vec{z}^{\vec{J}(s)} t^{\Delta(s)} \bigg] \prod_{r=0\atop{r\neq q-1,q}}^L 
\left(1- \vec{z}^{\vec{J}(x_r)} t^{\Delta(x_{r})}\right) 
\Bigg\} \notag \; .
\end{align}
The interpretation of the refined Hilbert 
series~\eqref{eqn:HS_generic_refined} remains the same as before: the minimal 
generators, i.e.\ their GNO-charge, $\su_R$-spin, topological charges 
$\vec{J}$, and their dressing factors, completely determine the Hilbert series. 
In principle, this data makes the (sometimes cumbersome) explicit 
summation of~\eqref{eqn:HS_refined} obsolete.
%%%%%%%%%%%%%%%%%%%%%%%%%%%%%%%%%%%%%%%%%%%%%%%%%%%%%%%%%%%%%%%%%%%%%%%%%%%%%%%%
   \section{Case: \texorpdfstring{$\boldsymbol{\uo \times \uo}$}{U(1)xU(1)}}
\label{sec:U1xU1}
In this section we analyse the abelian product $\uo\times \uo$. By 
construction, the Hilbert series simplifies as the dressing factors are constant 
throughout the lattice of magnetic weights. Consequently, abelian 
theories do not exhibit dressed monopole operators.
\subsection{Set-up}
The weight lattice of the GNO-dual of $\uo$ is simply $\Z$ and no Weyl-group 
exists due the abelian character; thus, $\Lambda_{w}(\widehat{\uo\times\uo}) 
= \Z^2$. Moreover, since $\uo\times \uo$ is abelian the classical dressing 
factors are the same for any magnetic weight $(m_1,m_2)$, i.e.
\begin{equation}
 P_{\uo\times \uo} (t,m_1,m_2) = \frac{1}{(1-t)^2} \; ,
\end{equation}
which reflects the two degree one Casimir invariants. 
\subsection{Two types of hypermultiplets}
\paragraph{Set-up}
To consider a rank $2$ abelian gauge group of the form $\uo\times\uo$ requires 
a delicate choice of matter content. If one considers $N_1$ hypermultiplets 
with charges $(a_1,b_1)\in \mathbb{N}^2$ under $\uo\times\uo$, then the 
conformal dimension reads
\begin{subequations}
\begin{equation}
 \Delta_{1 \text{h-plet}}(m_1,m_2)= \frac{N_1}{2} \left|a_1 m_1 + b_1 m_2 
\right| \for (m_1,m_2)\in \Z^2  \; .
\end{equation}
However, there exists an infinite number of points $\{m_1 =b_1 k, m_2=-a_1 k, 
k\in \Z \} $ with zero conformal dimension, i.e.\ the Hilbert series does not 
converge due to a decoupled $\uo$. Fixing this symmetry would reduce the 
rank to one.

Fortunately, we can circumvent this problem by introducing a second set of 
$N_2$ hypermultiplets with charges $(a_2,b_2)\in \mathbb{N}^2$, such that the 
matrix 
\begin{equation}
 \begin{pmatrix} a_1 & b_1 \\ a_2 & b_2   \end{pmatrix}  
\end{equation}
 has maximal rank. The relevant conformal dimension then reads
 \begin{equation}
 \Delta_{2 \text{h-plet}}(m_1,m_2)= \sum_{j=1}^2 \frac{N_j}{2} \left|a_j m_1 + 
b_j m_2 \right| \for (m_1,m_2)\in \Z^2 \; . \label{eqn:Delta:U1xU1_2hypers}
\end{equation}
Nevertheless, this set-up would introduce four charges and the summation of the 
Hilbert series becomes tricky. We evade the difficulties by the 
\emph{choice} $a_2=b_1$ and $b_2=-a_1$. Dealing with such a scenario leads to 
summation 
bounds such as 
\begin{align}
 a\, m_1 \geq b \, m_2  &\Leftrightarrow m_1 \geq \tfrac{b}{a} m_2 
\Leftrightarrow m_1 
\geq \lceil \tfrac{b}{a} m_2 \rceil \; ,\\
a\, m_1 < b\, m_2  &\Leftrightarrow m_1 < \tfrac{b}{a} m_2 \Leftrightarrow m_1 
< \lceil \tfrac{b}{a} m_2 \rceil -1 \; .
\end{align}
Having the summation variable within a floor or ceiling function seems to be 
an elaborate task with \texttt{Mathematica}. Therefore, we simplify the 
setting by \emph{assuming} $\exists$ $k \in \mathbb{N}$ such that $b_1 = k a_1$.
Then we arrive at
 \begin{equation}
 \Delta_{2 \text{h-plet}}(m_1,m_2)=  
 \frac{ a_1}{2} \left(  N_1 \left| m_1 + k m_2 \right| 
 +  N_2 \left|k m_1 - m_2 \right| \right) \for (m_1,m_2)\in \Z^2  \; .
 \label{eqn:Delta_U1xU1_2hplet}
\end{equation}
\end{subequations}
For this conformal dimension, there exists exactly one point $(m_1,m_2)$ with 
zero conformal dimension --- the trivial solution. Further, by a redefinition 
of $N_1$ and $N_2$ we can consider $a_1=1$.
\paragraph{Hilbert basis}
Consider the conformal dimension~\eqref{eqn:Delta_U1xU1_2hplet} for $a_1=1$.
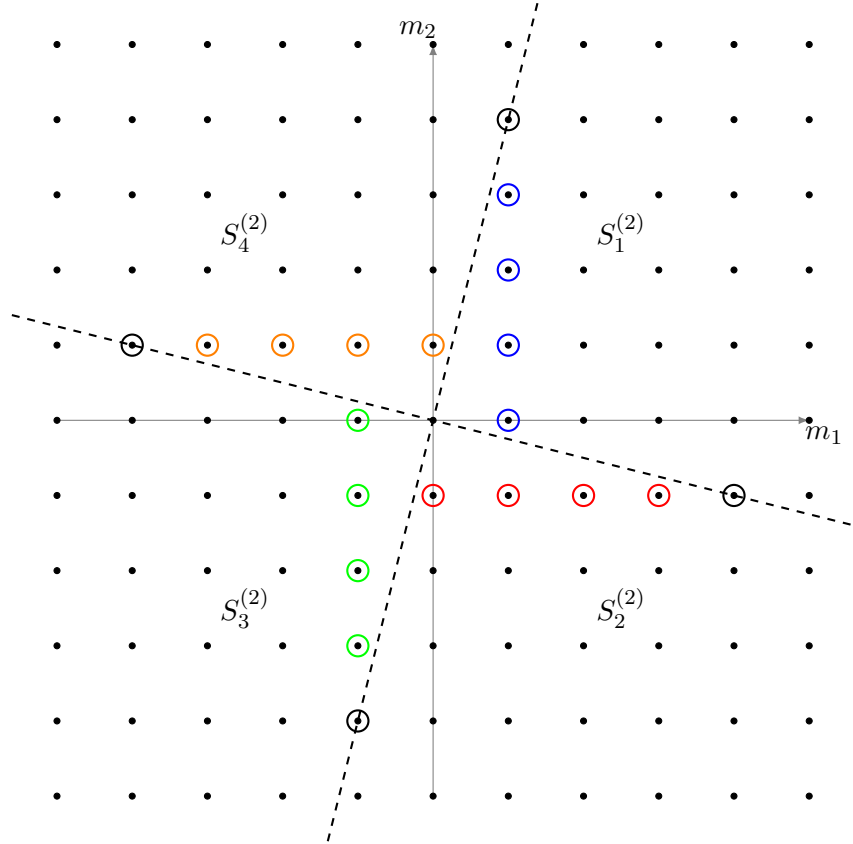
\begin{figure}[h]
\centering
\begin{tikzpicture}
  \coordinate (Origin)   at (0,0);
  \coordinate (XAxisMin) at (-5,0);
  \coordinate (XAxisMax) at (5,0);
  \coordinate (YAxisMin) at (0,-5);
  \coordinate (YAxisMax) at (0,5);
  \draw [thin, gray,-latex] (XAxisMin) -- (XAxisMax);% Draw x axis
  \draw [thin, gray,-latex] (YAxisMin) -- (YAxisMax);% Draw y axis
  \draw (5.2,-0.2) node {$m_1$};
  \draw (-0.2,5.2) node {$m_2$};
%Draw the root lattice
  \foreach \x in {-5,-4,...,5}{% Two indices running over each
      \foreach \y in {-5,-4,...,5}{% node on the grid we have drawn 
        \node[draw,circle,inner sep=0.8pt,fill,black] at (\x,\y) {};
            % Places a dot at those points
            }
            }
   \draw[black,dashed,thick]  (-1.4,-5.6) -- (1.4,5.6);  
   \draw[black,dashed,thick]  (-5.6,1.4) -- (5.6,-1.4);
 \draw[black,thick] (1,4) circle (4pt);
 \draw[black,thick] (-1,-4) circle (4pt);
 \draw[black,thick] (4,-1) circle (4pt);
 \draw[black,thick] (-4,1) circle (4pt);
 \draw[blue,thick] (1,3) circle (4pt);
 \draw[blue,thick] (1,2) circle (4pt);
 \draw[blue,thick] (1,1) circle (4pt);
 \draw[blue,thick] (1,0) circle (4pt);
 \draw[red,thick] (3,-1) circle (4pt);
 \draw[red,thick] (2,-1) circle (4pt);
 \draw[red,thick] (1,-1) circle (4pt);
 \draw[red,thick] (0,-1) circle (4pt);
 \draw[green,thick] (-1,0) circle (4pt);
 \draw[green,thick] (-1,-1) circle (4pt);
 \draw[green,thick] (-1,-2) circle (4pt);
 \draw[green,thick] (-1,-3) circle (4pt);
 \draw[orange,thick] (0,1) circle (4pt);
 \draw[orange,thick] (-1,1) circle (4pt);
 \draw[orange,thick] (-2,1) circle (4pt);
 \draw[orange,thick] (-3,1) circle (4pt);
\draw (2.5,2.5) node {$S_1^{(2)}$};
\draw (2.5,-2.5) node {$S_2^{(2)}$};
\draw (-2.5,-2.5) node {$S_3^{(2)}$};
\draw (-2.5,2.5) node {$S_4^{(2)}$};
\end{tikzpicture}
\caption{The dashed lines correspond the 
$k\, m_1=m_2$ and $m_1=-k\,m_2$ and divide the lattice $\Z^2$ into four 
semi-groups $S_j^{(2)}$ for $j=1,2,3,4$. The black 
circles denote the ray generators, while the blue circles complete the Hilbert 
basis for $S_1^{(2)}$, red circled points complete the basis for 
$S_{2}^{(2)}$. Green circles correspond to the remaining minimal generators of 
$S_{3}^{(2)}$ and orange circled points are the analogue for $S_{4}^{(2)}$. 
(Here, the example is $k=4$.)}
\label{Fig:U1xU1}
\end{figure}
By resolving the absolute 
values, we divide $\Z^2$ into four semi-groups
\begin{subequations}
\begin{align}
 S_1^{(2)} &= \left\{ (m_1,m_2)\in \Z^2 | \left( km_1 \geq m_2 \right) \wedge 
\left(  m_1\geq -k m_2 \right) \right\} \; ,\\
S_2^{(2)} &= \left\{ (m_1,m_2)\in \Z^2 | \left( km_1 \geq m_2 \right) \wedge 
\left(  m_1 \leq -k m_2 \right) \right\} \; , \\
 S_3^{(2)} &= \left\{ (m_1,m_2)\in \Z^2 | \left( km_1 \leq m_2 \right) \wedge 
\left(  m_1\geq -k m_2 \right) \right\} \; , \\
S_4^{(2)} &= \left\{ (m_1,m_2)\in \Z^2 | \left( km_1 \leq m_2 \right) \wedge 
\left(  m_1 \leq -k m_2 \right) \right\} \; ,
\end{align}
\end{subequations}
which all descend from $2$-dimensional rational polyhedral cones. The situation 
is depicted in Fig.~\ref{Fig:U1xU1}.
Next, one needs to compute the Hilbert basis $\Hcal(S)$ for each 
semi-group $S$. In this example, it follows from the drawing that
\begin{subequations}
\begin{align}
 \Hcal(S_1^{(2)}) &= \Big\{ (k,-1), \big\{ (1,l) \; \big| \; 
l=0,1, \ldots, k\big\}  \Big\} \; ,\\
 \Hcal(S_2^{(2)}) &=\Big\{ (-1,-k),\big\{ (l,-1) \; \big| \; l=0,1, 
\ldots, k\big\} \Big\} \; ,\\
 \Hcal(S_3^{(2)}) &= \Big\{ (-k,1),\{ (-1,-l) \; \big| \; l=0,1, 
\ldots, k\big\} \Big\} \; ,\\
\Hcal(S_4^{(2)}) &= \Big\{ (1,k),\{ (-l,1) \; \big| \; l=0,1, 
\ldots, k\big\} \Big\} \; .
\end{align}
\end{subequations}
% 
% % 
For a fixed $k\geq1$ we obtain $4(k+1)$ basis elements.
\paragraph{Hilbert series}
We then compute the following Hilbert series
\begin{align}
 \HS_{\uo\times \uo}^{k}(t,z_1,z_2) = \frac{1}{(1-t)^2} \sum_{m_1,m_2 \in \Z} 
z_1^{m_1} z_2^{m_2} t^{\Delta_{2 \text{h-plet}}(m_1,m_2)} \; ,
\end{align}
for which we obtain
\begin{subequations}
\label{eqn:HS_U1xU1_2hplets}
\begin{equation}
  \HS_{\uo\times \uo}^{k}(t,z_1,z_2) = \frac{R(t,z_1,z_2)}{P(t,z_1,z_2)} \; ,
\end{equation}
with denominator
\begin{align}
P(t,z_1,z_2)&= 
(1-t)^2 
\left(1-\frac{1}{z_1} t^{\frac{k N_2-N_1}{2}}\right) 
\left(1-z_1 t^{\frac{k N_2-N_1}{2}}\right) 
\left(1-\frac{1}{z_2} t^{\frac{k N_1-N_2}{2}}\right) 
\left(1-z_2 t^{\frac{k N_1-N_2}{2}}\right) \notag \\
&\quad \times
\left(1-\frac{1}{z_1} t^{\frac{k N_2+N_1}{2}}\right) 
\left(1-z_1 t^{\frac{k N_2+N_1}{2}}\right) 
\left(1-\frac{1}{z_2} t^{\frac{k N_1+N_2}{2}}\right) 
\left(1-z_2 t^{\frac{k N_1+N_2}{2}}\right)  \\
&\quad \times
\left(1-\frac{1}{z_1 z_2^{k}} t^{\frac{1}{2} \left(k^2+1\right) N_1}\right) 
\left(1-z_1 z_2^k t^{\frac{1}{2} \left(k^2+1\right) N_1}\right) \notag \\
&\quad \times
\left(1-\frac{z_1^k}{z_2} t^{\frac{1}{2} \left(k^2+1\right) N_2}\right) 
\left(1-\frac{z_2}{ z_1^{k}} t^{\frac{1}{2} \left(k^2+1\right) N_2}\right) 
\notag \; ,
\end{align}
\end{subequations}
while the numerator $R(t,z_1,z_2)$ is too long to be displayed, as it contains 
$1936$ monomials.
Nonetheless, one can explicitly verify a few properties of the Hilbert series. 
For example, the Hilbert series~\eqref{eqn:HS_U1xU1_2hplets} has a pole of 
order 
$4$ at $t\to 1$, because $R(1,z_1,z_2)=0$ and the derivatives $\tfrac{\diff^n 
}{\diff t^n}R(t,z_1,z_2) |_{t=1}=0$ for 
$n=1,2,\ldots 9$ (at least for $z_1=z_2=1$).
Moreover, the degrees of numerator and denominator depend on the relations 
between $N_1$, $N_2$, and $k$; however, one can show that the difference in 
degrees is precisely $2$, i.e.\ it matches the quaternionic dimension of the 
moduli space.
\paragraph{Discussion}
Analysing the plethystic logarithm and the Hilbert series, the monopole 
operators corresponding to the Hilbert basis can be identified as follows:
Eight poles of the Hilbert series~\eqref{eqn:HS_U1xU1_2hplets} can be 
identified with monopole generators as shown in Tab.~\ref{tab:U1xU1_Ops_ray}.
\begin{table}[h]
 \begin{subtable}{1\textwidth}
  \centering
 \begin{tabular}{c|c||c|c}
 \toprule
  $(m_1,m_2)$ & $\Delta(m_1,m_2)$ & $(m_1,m_2)$ & $\Delta(m_1,m_2)$ \\ \midrule
  $(1,0)$, $(-1,0)$ & $\frac{1}{2} \left( N_1+k N_2 \right)$ & $(0,1)$, 
$(0,-1)$ & $\frac{1}{2} \left( k N_1+ N_2 \right)$ \\
  $(1,k)$, $(-1,-k)$ & $\frac{1}{2} \left(1 +k^2 \right) N_1$ & $(-k,1)$, 
$(k,-1)$ & $\frac{1}{2} \left(1 +k^2 \right) N_2$ \\
\bottomrule
 \end{tabular}
\caption{The minimal generators which are 
ray generators or poles of the Hilbert series.}
\label{tab:U1xU1_Ops_ray}
\end{subtable}
\begin{subtable}{1\textwidth}
  \centering
 \begin{tabular}{c|c||c|c}
 \toprule
  $(m_1,m_2)$ & $\Delta(m_1,m_2)$ & $(m_1,m_2)$ & $\Delta(m_1,m_2)$ \\ \midrule
  $(1,l)$, $(-1,-l)$ & $\frac{1}{2} N_1 (k l+1)+\frac{1}{2} N_2 
(k-l)$ & $(-l,1)$, 
$(l,-1)$ & $\frac{1}{2} N_1 (k-l)+\frac{1}{2} N_2 (k l+1)$ \\
\bottomrule
 \end{tabular}
\caption{The minimal generators, labelled by $l= 1,2,\ldots , k-1$, which are 
not ray generators.}
\label{tab:U1xU1_Ops_non-ray}
\end{subtable}
\caption{The set of bare monopole operators for a $\uo\times \uo$ theory with 
conformal dimension~\eqref{eqn:Delta_U1xU1_2hplet}.}
% %
\end{table}
Studying the plethystic logarithm clearly displays the remaining set, which is 
displayed in Tab.~\ref{tab:U1xU1_Ops_non-ray}.
\paragraph{Remark}
A rather special case of~\eqref{eqn:Delta:U1xU1_2hypers} is $a_2=0=b_1$, for 
which the theory becomes the product of two $\uo$-theories with $N_1$ or $N_2$ 
electrons of charge $a$ or $b$, respectively. In detail, the 
conformal dimension is simply
\begin{equation}
  \Delta_{2 \text{h-plet}}(m_1,m_2)\stackrel{a_2=0=b_1}{=}  
\frac{N_1}{2} \left|a\, m_1 \right| + 
\frac{N_2}{2} \left| b\, m_2 \right|\for (m_1,m_2)\in \Z^2 \; ,
\end{equation}
such that the Hilbert series becomes
\begin{align}
 \HS_{\uo^2}^{a,b}(t,z_1,z_2) &= 
 \frac{1-t^{a N_1}}{ (1-t) \left(1-z_1 t^{\tfrac{a N_1}{2}}\right) 
 \left(1-\tfrac{1}{z_1} t^{\tfrac{a N_1}{2}} \right) } \times
  \frac{1-t^{b N_2}}{ (1-t) \left(1-z_2 t^{\tfrac{b N_2}{2}}\right) 
 \left(1-\tfrac{1}{z_2} t^{\tfrac{b N_2}{2}}\right) }  \notag \\
&= \HS_{\uo}^{a}(t,z_1,N_1) \times \HS_{\uo}^{b}(t,z_2,N_2) \; .
\end{align}
For the unrefined Hilbert series, that is $z_1=1=z_2$, the rational function
$\HS_{\uo}^{a}(t,N)$ equals the Hilbert series of the (abelian)  ADE-orbifold 
$\C^2 \slash \Z_{a\cdot N}$, see for instance~\cite{Cremonesi:2014xha}. Thus, 
the $\uo{\times}\uo$ Coulomb branch is the product of two A-type singularities.

Quite intuitively, taking the corresponding limit $k\to0$ 
in~\eqref{eqn:HS_U1xU1_2hplets} yields the product
\begin{equation}
 \lim_{k\to0}\HS_{\uo\times \uo}^{k}(t,z_1,z_2) = \HS_{\uo}(t,z_1,N_1) 
\times \HS_{\uo}(t,z_2,N_2)  \; ,
\end{equation}
which are $\uo$ theories with $N_1$ and $N_2$ electrons of unit charge. The 
unrefined rational functions are the Hilbert series of $\Z_{N_1}$ and 
$\Z_{N_2}$ singularities in the ADE-classification.
From Fig.~\ref{Fig:U1xU1} one observes that in the limit $k\to 0$ the relevant 
rational cones coincide with the four quadrants of $\R^2$ and the 
Hilbert basis reduces to the cone generators.
%
%%%%%%%%%%%%%%%%%%%%%%%%%%%%%%%%%%%%%%%%%%%%%%%%%%%%%%%%%%%%%%%%%%%%%%%%%%%%
%%%%%%%%%%%%%%%%%%%%%%%%%%%%%%%%%%%%%%%%%%%%%%%%%%%%%%%%%%%%%%%%%%%%%%%%%%%%
%
\subsection{Reduced moduli space of one 
\texorpdfstring{$\sorm(5)$}{SO(5)}-instanton}
Consider the Coulomb branch of the quiver gauge theory depicted 
in Fig.~\ref{Fig:Quiver_SO5_instanton} with 
conformal dimension given by 
\begin{equation}
 \Delta(m_1,m_2)= \frac{1}{2} \left( |m_1| + | m_1 -2m_2| \right) \; . 
\label{eqn:Delta_SO5_instanton}
\end{equation}
Instead of associating~\eqref{eqn:Delta_SO5_instanton} with the 
quiver of Fig.~\ref{Fig:Quiver_SO5_instanton}, one could equally well 
understand it as a special case of a $\uo^2$ theory with two different 
hypermultiplets~\eqref{eqn:Delta:U1xU1_2hypers}.
\begin{figure}[h]
\centering
\begin{tikzpicture}
 \draw[black,thick] (0,0) circle (10pt);
 \draw[black,thick] (2,0) circle (10pt);
 \draw[thick,double] (0.35,0) -- (1.65,0);
 \draw[thick,black] (1.2,0.2) -- (1,0) -- (1.2,-0.2);
 \draw[black,thick] (-0.3,1.7) rectangle (0.3,2.3);
 \draw[thick,double] (0,0.35) -- (0,1.7);
 \draw[thick,black] (0.2,1.2) -- (0,1) -- (-0.2,1.2);
 \draw (0,-0.7) node {$U(1)$};
 \draw (2,-0.7) node {$U(1)$};
 \draw (-0.8,2) node {$U(1)$};
\end{tikzpicture}
\caption{Quiver gauge theory whose Coulomb branch is the reduced 
moduli space of one $\sorm(5)$-instanton.}
\label{Fig:Quiver_SO5_instanton}
\end{figure}
\paragraph{Hilbert basis}
Similar to the previous case, the conformal dimensions induces a fan which, in 
this case, is generated by four $2$-dimensional cones
\begin{subequations}
\begin{alignat}{2}
 C_1^{(2)}&=\cone \left((2,1),(0,1) \right) \; , \qquad&
 C_2^{(2)}&=\cone \left((2,1),(0,-1) \right) \; , \\
C_3^{(2)}&=\cone \left((-2,-1),(0,-1) \right) \; ,\qquad&
 C_4^{(2)}&=\cone \left((-2,-1),(0,1) \right) \; .
\end{alignat}
\end{subequations}
The intersection with the $\Z^2$ lattice defines the semi-groups $S_p^{(2)} 
\coloneqq C_p^{(2)} \cap \Z^2$ for which we need to compute the Hilbert bases. 
Fig.~\ref{Fig:Quiver_SO5} illustrates the situation and we obtain
\begin{subequations}
 \label{eqn:Hilbert_basis_U1xU1_SO5}
\begin{alignat}{2}
 \Hcal(S_1^{(2)}) &= \left\{(2,1),(1,1),(0,1) \right\} \; ,\qquad&
 \Hcal(S_2^{(2)}) &= \left\{(2,1),(1,0),(0,-1) \right\} \; , \\
\Hcal(S_3^{(2)}) &= \left\{(-2,-1),(-1,-1),(0,-1) \right\} \; ,\qquad&
 \Hcal(S_4^{(2)}) &= \left\{(-2,-1),(-1,0),(0,1) \right\} \; .
\end{alignat}
\end{subequations}
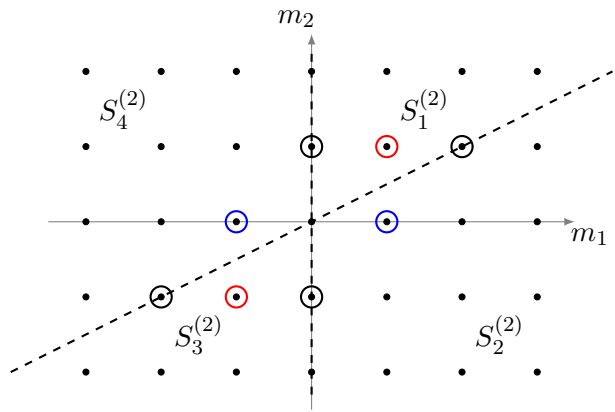
\begin{figure}[h]
\centering
\begin{tikzpicture}
  \coordinate (Origin)   at (0,0);
  \coordinate (XAxisMin) at (-3.5,0);
  \coordinate (XAxisMax) at (3.5,0);
  \coordinate (YAxisMin) at (0,-2.5);
  \coordinate (YAxisMax) at (0,2.5);
  \draw [thin, gray,-latex] (XAxisMin) -- (XAxisMax);% Draw x axis
  \draw [thin, gray,-latex] (YAxisMin) -- (YAxisMax);% Draw y axis
  \draw (3.7,-0.2) node {$m_1$};
  \draw (-0.2,2.7) node {$m_2$};
%    \clip (-3,-2) rectangle (10cm,10cm); % Clips the picture...
%Draw the root lattice
  \foreach \x in {-3,-2,...,3}{% Two indices running over each
      \foreach \y in {-2,-1,...,2}{% node on the grid we have drawn 
        \node[draw,circle,inner sep=0.8pt,fill,black] at (\x,\y) {};
            % Places a dot at those points
            }
            }
   \draw[black,dashed,thick]  (0,-2.3) -- (0,2.3);  
   \draw[black,dashed,thick]  (-2*2,-1*2) -- (2*2,2*1);
%   \filldraw[dotted,fill=yellow, fill opacity=0.2, draw=gray] (Origin) -- 
% (1.4,5.6) -- (5,5)-- (5.6,-1.4) -- cycle ;
 \draw[black,thick] (0,1) circle (4pt);
 \draw[black,thick] (0,-1) circle (4pt);
 \draw[black,thick] (2,1) circle (4pt);
 \draw[black,thick] (-2,-1) circle (4pt);
%   %  
 \draw[red,thick] (1,1) circle (4pt);
 \draw[red,thick] (-1,-1) circle (4pt);
 \draw[blue,thick] (1,0) circle (4pt);
 \draw[blue,thick] (-1,0) circle (4pt);
% %  
\draw (1.5,1.5) node {$S_1^{(2)}$};
\draw (2.5,-1.5) node {$S_2^{(2)}$};
\draw (-1.5,-1.5) node {$S_3^{(2)}$};
\draw (-2.5,1.5) node {$S_4^{(2)}$};
\end{tikzpicture}
\caption{The dashed lines correspond the 
$m_1=2m_2$ and $m_1=0$ and divide the lattice $\Z^2$ into four 
semi-groups $S_j^{(2)}$ for $j=1,2,3,4$. The black 
circles denote the ray generators, while the red circles complete the Hilbert 
bases for $S_1^{(2)}$ and $S_3^{(2)}$. Blue circled lattice points complete the 
bases for $S_2^{(2)}$ and $S_4^{(2)}$.}
\label{Fig:Quiver_SO5}
\end{figure}
\paragraph{Hilbert series}
The Hilbert series is evaluated to
\begin{subequations}
\label{eqn:HS_U1xU1_SO5}
\begin{align}
\HS_{\uo^2}^{\sorm(5)}(t,z_1,z_2)&= \frac{R(t,z_1,z_2) }{
 (1-t)^2 
 \left(1-\frac{t}{z_2}\right) 
 \left(1- z_2 t\right) 
 \left(1-\frac{t}{z_1^2 z_2}\right) 
 \left(1- z_1^2 z_2 t\right)} \; ,\\
R(t,z_1,z_2)&=1
+t \left(z_1+\frac{1}{z_1}+z_1 z_2+\frac{1}{z_1 z_2}\right) \\
&\qquad -2 t^{2} \left(1+ z_1+ \frac{1}{z_1} + z_1 z_2+\frac{1}{z_1 
z_2}\right) \notag \\
&\qquad +t^{3} \left(z_1+\frac{1}{z_1}+ z_1 z_2+\frac{1}{z_1 z_2}\right)
+t^{4} \notag \; .
\end{align}
\end{subequations}
The Hilbert series~\eqref{eqn:HS_U1xU1_SO5} has a pole of order $4$ at $t=1$, 
because one can explicitly verify that $R(t=1,z_1,z_2)=0$, $\tfrac{\diff}{\diff 
t}R(t,z_1,z_2)|_{t=1}=0$, 
but $\tfrac{\diff^2}{\diff t^2}R(t,z_1,z_2)|_{t=1}\neq0$. Thus, the complex 
dimension of the moduli space is $4$. Moreover, the difference in degrees of 
numerator and denominator is $2$, which equals the quaternionic dimension of 
the 
Coulomb branch.
\paragraph{Plethystic logarithm}
The plethystic logarithm for this scenario reads
\begin{align}
\label{eqn:PL_U1xU1_SO5}
\PL(\HS_{\uo^2}^{\sorm(5)})= & \left(2+z_1^2 z_2+\frac{1}{z_1^2 z_2}+z_1 
z_2+\frac{1}{z_1 z_2}+z_1+\frac{1}{z_1}+z_2+\frac{1}{z_2}\right)t \\
&- \bigg(4 
+z_1^2 +\frac{1}{z_1^2}
+z_2 +\frac{1}{z_2}
+z_1^2 z_2^2 +\frac{1}{z_1^2 z_2^2} 
+z_1^2 z_2 +\frac{1}{z_1^2 z_2}  \notag\\*
&\qquad \qquad \quad  +2 z_1 +\frac{2}{z_1}
+2 z_1 z_2+\frac{2}{z_1 z_2} \bigg)t^2 
+\mathcal{O}(t^{3}) \; . \notag
\end{align}
\paragraph{Symmetry enhancement}
The information conveyed by the Hilbert 
basis~\eqref{eqn:Hilbert_basis_U1xU1_SO5}, the Hilbert 
series~\eqref{eqn:HS_U1xU1_SO5}, and the plethystic 
logarithm~\eqref{eqn:PL_U1xU1_SO5} is that there are eight minimal generators 
of conformal dimension one which, together with the two Casimir invariants, 
span the adjoint representation of $\sorm(5)$. It is 
known~\cite{Cremonesi:2014xha,Hanany:2015hxa} that~\eqref{eqn:HS_U1xU1_SO5} is 
the Hilbert series 
for the reduced moduli space of one $\sorm(5)$-instanton over $\C^2$. 
% 
%%%%%%%%%%%%%%%%%%%%%%%%%%%%%%%%%%%%%%%%%%%%%%%%%%%%%%%%%%%
%%%%%%%%%%%%%%%%%%%%%%%%%%%%%%%%%%%%%%%%%%%%%%%%%%%%%%%%%%%
%
\subsection{Reduced moduli space of one 
\texorpdfstring{$\sut$}{SU(3)}-instanton}
The quiver gauge theories associated to the affine Dynkin diagram $\hat{A}_n$ 
have been studied in~\cite{Cremonesi:2013lqa}. Here, we consider the Coulomb 
branch of the $\hat{A}_2$ quiver gauge theory as depicted in 
Fig.~\eqref{Fig:Quiver_SU3_instanton} and with conformal dimension 
given by 
\begin{equation}
 \Delta(m_1,m_2)= \frac{1}{2} \left( |m_1| + |m_2| + | m_1 -m_2|  \right) \; .
\end{equation}
\begin{figure}[h]
\centering
\begin{tikzpicture}
 \draw[black,thick] (0,0) circle (10pt);
 \draw[black,thick] (2,0) circle (10pt);
 \draw[thick] (0.35,0) -- (1.65,0);
 \draw[black,thick] (-0.3,1.7) rectangle (0.3,2.3);
 \draw[thick] (0,0.35) -- (0,1.7);
 \draw[black,thick] (1.7,1.7) rectangle (2.3,2.3);
 \draw[thick] (2,0.35) -- (2,1.7);
 \draw (0,-0.7) node {$U(1)$};
 \draw (2,-0.7) node {$U(1)$};
 \draw (-0.8,2) node {$U(1)$};
 \draw (2.8,2) node {$U(1)$};
\end{tikzpicture}
\caption{Quiver gauge theory whose Coulomb branch is the reduced 
moduli space of one $\sut$-instanton.}
\label{Fig:Quiver_SU3_instanton}
\end{figure}
\paragraph{Hilbert basis}
Similar to the previous case, the conformal dimensions induces a fan which, in 
this case, is generated by six $2$-dimensional cones
\begin{subequations}
\begin{alignat}{2}
 C_1^{(2)}&=\cone \left((0,1),(1,1) \right) \; , \qquad&
 C_2^{(2)}&=\cone \left((1,1),(1,0) \right) \; , \\
C_3^{(2)}&=\cone \left((1,0),(0,-1) \right)\; , \qquad&
 C_4^{(2)}&=\cone \left((0,-1),(-1,-1) \right) \; , \\
C_5^{(2)}&=\cone \left((-1,-1),(-1,0) \right) \; ,\qquad&
 C_6^{(2)}&=\cone \left((-1,0),(0,1) \right) \; .
\end{alignat}
\end{subequations}
The intersection with the $\Z^2$ lattice defines the semi-groups $S_p^{(2)} 
\coloneqq C_p^{(2)} \cap \Z^2$ for which we need to compute the Hilbert bases. 
Fig.~\ref{Fig:Quiver_SU3} illustrates the situation. We compute the Hilbert 
bases to read
\begin{subequations}
 \label{eqn:Hilbert_basis_U1xU1_SU3}
\begin{alignat}{2}
 \Hcal(S_1^{(2)}) &= \left\{(0,1),(1,1) \right\} \qquad&
 \Hcal(S_2^{(2)}) &= \left\{(1,1),(1,0)) \right\} \; , \\
\Hcal(S_3^{(2)}) &= \left\{(1,0),(0,-1) \right\} \qquad&
 \Hcal(S_4^{(2)}) &= \left\{(0,-1),(-1,-1) \right\} \; ,\\
\Hcal(S_5^{(2)}) &= \left\{(-1,-1),(-1,0) \right\} \qquad&
 \Hcal(S_6^{(2)}) &= \left\{(-1,0),(0,1) \right\} \; .
\end{alignat}
\end{subequations}
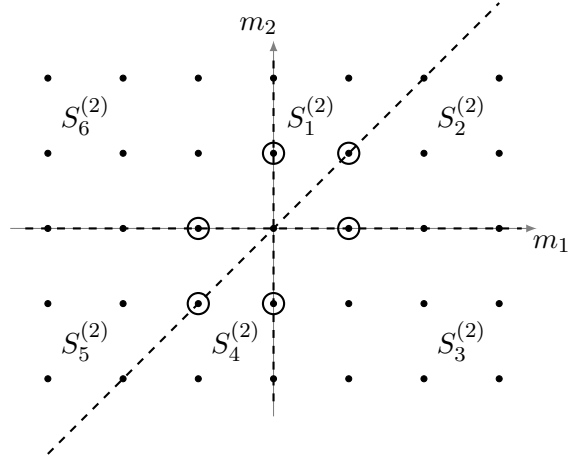
\begin{figure}[h]
\centering
\begin{tikzpicture}
  \coordinate (Origin)   at (0,0);
  \coordinate (XAxisMin) at (-3.5,0);
  \coordinate (XAxisMax) at (3.5,0);
  \coordinate (YAxisMin) at (0,-2.5);
  \coordinate (YAxisMax) at (0,2.5);
  \draw [thin, gray,-latex] (XAxisMin) -- (XAxisMax);% Draw x axis
  \draw [thin, gray,-latex] (YAxisMin) -- (YAxisMax);% Draw y axis
  \draw (3.7,-0.2) node {$m_1$};
  \draw (-0.2,2.7) node {$m_2$};
%    \clip (-3,-2) rectangle (10cm,10cm); % Clips the picture...
%Draw the root lattice
  \foreach \x in {-3,-2,...,3}{% Two indices running over each
      \foreach \y in {-2,-1,...,2}{% node on the grid we have drawn 
        \node[draw,circle,inner sep=0.8pt,fill,black] at (\x,\y) {};
            % Places a dot at those points
            }
            }
   \draw[black,dashed,thick]  (0,-2.3) -- (0,2.3);  
   \draw[black,dashed,thick]  (-3.3,0) -- (3.3,0);
   \draw[black,dashed,thick]  (-3,-3) -- (3,3);
%   \filldraw[dotted,fill=yellow, fill opacity=0.2, draw=gray] (Origin) -- 
% (1.4,5.6) -- (5,5)-- (5.6,-1.4) -- cycle ;
 \draw[black,thick] (0,1) circle (4pt);
 \draw[black,thick] (0,-1) circle (4pt);
 \draw[black,thick] (1,1) circle (4pt);
 \draw[black,thick] (-1,-1) circle (4pt);
 \draw[black,thick] (1,0) circle (4pt);
 \draw[black,thick] (-1,0) circle (4pt);
% %  
\draw (0.5,1.5) node {$S_1^{(2)}$};
\draw (2.5,1.5) node {$S_2^{(2)}$};
\draw (2.5,-1.5) node {$S_3^{(2)}$};
\draw (-0.5,-1.5) node {$S_4^{(2)}$};
\draw (-2.5,-1.5) node {$S_5^{(2)}$};
\draw (-2.5,1.5) node {$S_6^{(2)}$};
\end{tikzpicture}
\caption{The dashed lines correspond the 
$m_1=m_2$, $m_1=0$, and $m_2=0$ and divide the lattice $\Z^2$ into six 
semi-groups $S_j^{(2)}$ for $j=1,\ldots,6$. The black circled points denote the 
ray generators, which coincide with the minimal generators.}
\label{Fig:Quiver_SU3}
\end{figure}
\paragraph{Hilbert series}
\begin{subequations}
\label{eqn:HS_U1xU1_SU3}
 \begin{align}
 \HS_{\uo^2 }^{\sut}(t,z_1,z_2)&=\frac{R(t,z_1,z_2)}{
 (1-t)^2 
 \left(1-\frac{t}{z_1}\right) 
 \left(1- z_1 t \right) 
\left(1-\frac{t}{z_2}\right) 
\left(1- z_2 t \right) 
\left(1-\frac{t}{z_1 z_2}\right) 
\left(1-z_1 z_2 t \right)}\\
  R(t,z_1,z_2)&=1
  - \left(3 +z_1 +\frac{1}{z_1} +z_2 +\frac{1}{z_2} +z_1 z_2 +\frac{1}{z_1 z_2} 
\right) t^2 \\
  &\qquad +2 \left(2+ z_1+\frac{1}{z_1} + z_2+\frac{1}{z_2} + z_1 
z_2+\frac{1}{z_1 z_2}\right) t^3 \notag \\
  &\qquad -\left(3 +z_1 +\frac{1}{z_1} +z_2 +\frac{1}{z_2} + z_1 z_2 
+\frac{1}{z_1 z_2}\right)t^4  +t^6 \notag
 \end{align}
\end{subequations}
The Hilbert series~\eqref{eqn:Hilbert_basis_U1xU1_SU3} has a pole of order $4$ 
as $t\to 1$, because $R(t=1,z_1,z_2)=0$ and $\tfrac{\diff^n}{\diff t^n} 
R(t,z_1,z_2)|_{t=1,z_1=z_2=1}=0$ for $n=1,2,3$. Thus, the Coulomb branch is of 
complex dimension $4$. In addition, the difference in degrees of numerator and 
denominator is $2$, which equals the quaternionic dimension.
\paragraph{Plethystic logarithm}
\begin{align}
\label{eqn:PL_U1xU1_SU3}
 \PL(\HS_{\uo^2 }^{\sut})=  &\left(2 +z_1 +\frac{1}{z_1} +z_2 
+\frac{1}{z_2}+ z_1 z_2+\frac{1}{z_1 z_2} \right) t \\
&- \left(3 +z_1 +\frac{1}{z_1} +z_2 +\frac{1}{z_2} + z_1 z_2 +\frac{1}{z_1 z_2} 
\right) t^2 + \mathcal{O}(t^3) \notag
\end{align}
\paragraph{Symmetry enhancement}
The information conveyed by the Hilbert 
basis~\eqref{eqn:Hilbert_basis_U1xU1_SU3}, the Hilbert 
series~\eqref{eqn:HS_U1xU1_SU3}, and the plethystic 
logarithm~\eqref{eqn:PL_U1xU1_SU3} is that there are six minimal generators 
of conformal dimension one which, together with the two Casimir invariants, 
span the adjoint representation of $\sut$. As proved 
in~\cite{Cremonesi:2013lqa}, the Hilbert series~\eqref{eqn:HS_U1xU1_SU3} can be 
resumed as 
\begin{equation}
\label{eqn:HS_U1xU1_SU3_mod}
  \HS_{\uo^2}^{\sut}(t,z_1,z_2)= \sum_{k=0}^{\infty} 
\chi_{[k,k]} 
t^k
\end{equation}
with $\chi_{[k,k]}$ being the character of the $\sut$-representation $[k,k]$. 
Therefore, this theory has an explicit $\sut$-enhancement in the Coulomb 
branch. It is known~\cite{Benvenuti:2010pq} 
that~\eqref{eqn:HS_U1xU1_SU3_mod} is the reduced instanton moduli space of one 
$\sut$-instanton over $\C^2$.
%%%%%%%%%%%%%%%%%%%%%%%%%%%%%%%%%%%%%%%%%%%%%%%%%%%%%%%%%%%%%%%%%%%%%%%%%%%%%%%%
  \section{Case: \texorpdfstring{$\boldsymbol{\utwo}$}{U(2)}}
\label{sec:U2}
In this section we aim to consider two classes of $\utwo$ gauge theories 
wherein $\utwo \cong \su {\times} \uo$, i.e.\ this is effectively an $\su$ 
theory 
with varying $\uo$-charge. As a unitary group, $\utwo$ is self-dual under 
GNO-duality.
\subsection{Set-up}
To start with, let consider the two view points and elucidate the relation 
between them.
\paragraph{$\boldsymbol{\utwo}$ view point}
The GNO-dual of $\utwo$ is $\utwo$ itself; hence, the weight 
lattice is $\Lambda_w(\utwo) \cong \Z^2$. Moreover, the Weyl-group is $S_2$ 
and acts via permuting the two Cartan generators; consequently, 
$\Lambda_w(\utwo)\slash S_2 = \{ (m_1,m_2)\in \Z^2: m_1\geq m_2\}$.
\paragraph{$\boldsymbol{\uo\times\su}$ view point}
Considering $\uo\times \su$, we need to find the weight lattice of the 
GNO-dual, i.e.\ find all solutions to the Dirac quantisation condition, see for 
instance~\cite{Goddard:1976qe}. Since we consider the product, the 
exponential in~\eqref{eqn:general_Dirac} 
factorises in $\exp( 2\pi \im \ n \ T_{\uo}) $ and $ \exp(2\pi \im \ m \ 
T_{\su})$, where the $T$'s are the Cartan generators.
Besides the solution
\begin{subequations}
\begin{equation}
 (n,m)\in H_0 \coloneqq \Z^2 = \Z 
\times \Lambda_w(\sorm(3)) = \Z 
\times \Lambda_r(\su)
\end{equation}
corresponding to the weight lattice of $\uo \times \sorm(3)$, there exists also 
the solution
\begin{equation}
  (n,m)\in H_1 \coloneqq \Z^2 +(\tfrac{1}{2},\tfrac{1}{2}) = 
\left( \Z+\tfrac{1}{2} \right) \times \left( \Lambda_w(\su)\setminus 
\Lambda_r(\su) \right) \; ,
\end{equation}
\end{subequations}
for which both factors are equal to $-1$. The action of the Weyl-group $S_2$ 
restricts then to non-negative $m$ i.e.\ $H_0^+ = H_0 \cap \{m\geq0\}$ and 
$H_1^+ = H_1 \cap \{m\geq0\}$.
\paragraph{Relation between both}
To identify both views with one another, we select the $\uo$ as diagonally 
embedded, i.e.\ identify the charges as follows:
  \begin{equation}
  \label{eqn:transf_U2_to_U1xSU2}
 \begin{matrix} n\coloneqq \frac{m_1+m_2}{2} \\ m\coloneqq \frac{m_1 - m_2}{2}  
 \end{matrix} \Bigg\} \qquad \Leftrightarrow \qquad
 \Bigg\{ \begin{matrix} m_1 =n+m \\ m_2=n-m  \end{matrix} \; .
  \end{equation}
The two classes of $\utwo$-representations under consideration in this section 
are
\begin{subequations}
\begin{align}
 [1,a] &\with \chi_{[1,a]}^{\utwo} = y_1^{a+1} y_2^a + y_1^a y_2^{a+1}  \; ,\\
 [2,a] &\with \chi_{[2,a]}^{\utwo} = y_1^{a+2} y_2^a + y_1^{a+1} y_2^{a+1} + 
y_1^a y_2^{a+2} \; ,
\end{align}
\end{subequations}
for $a \in \NN_0$.
Following~\eqref{eqn:transf_U2_to_U1xSU2}, we define the fugacities 
\begin{equation}
 q \coloneqq \sqrt{y_1 \ y_2} \for \uo \und x\coloneqq \sqrt{\frac{y_1}{y_2}} 
\for \su,
\end{equation}
and consequently observe
\begin{subequations}
\begin{align}
 \chi_{[1,a]}^{\utwo} &= q^{2a+1} \left(x+ \tfrac{1}{x}\right) = 
\chi_{2a+1}^{\uo} \cdot \chi_{[1]}^{\su}  \; ,\\
  \chi_{[2,a]}^{\utwo} &= q^{2a+2} \left(x^2+ 1+ \tfrac{1}{x^2}\right) = 
\chi_{2a+2}^{\uo} \cdot \chi_{[2]}^{\su} \; ,
\end{align}
where the $\su$-characters are defined via 
\begin{equation}
\chi_{[L]}^{\su}=\sum_{r=-\tfrac{L}{2}}^{\tfrac{L}{2}} x^{2r}  \; . 
\end{equation}
\end{subequations}
Therefore, the family $[1,a]$ corresponds to the fundamental representation of 
$\su$ with \emph{odd} $\uo$-charge $2a+1$; while the family $[2,a]$ 
represents the adjoint representation of $\su$ with \emph{even} 
$\uo$-charge $2a+2$.
\paragraph{Dressing factors} 
Lastly, the calculation employs the classical dressing function
\begin{equation}
 P_{\utwo}(t^2,m)\coloneqq \begin{cases} \frac{1}{(1-t^2)^2} & \; , m\neq 0 \\
   \frac{1}{(1-t^2)(1-t^4)} &\; , m=0 \end{cases} \; ,
\label{eqn:Dressing_U2}
\end{equation}
as presented in~\cite{Cremonesi:2013lqa}. (Note that we rescaled $t$ to be 
$t^2$ for later convenience.) Following the discussion of App.~\ref{app:PL}, 
monopoles with $m\neq0$ have precisely one dressing by a $\uo$ Casimir 
invariant due 
to $P_{\utwo}(t^2,m) \slash  P_{\utwo}(t^2,0) = 1+t^2$. In contrast, there are 
no dressed monopole operators for $m=0$.
% 
%%%%%%%%%%%%%%%%%%%%%%%%%%%%%%%%%%%%%%%%%%%%%%%%%%%%%%%%%%%%%%%%%%%%%%%%%%%%%%
%%%%%%%%%%%%%%%%%%%%%%%%%%%%%%%%%%%%%%%%%%%%%%%%%%%%%%%%%%%%%%%%%%%%%%%%%%%%%%
% 
\subsection{\texorpdfstring{$N$}{N} hypermultiplets in the 
fundamental representation of \texorpdfstring{$\su$}{SU(2)}}
\label{subsec:U2_fund_SU2}
The conformal dimension for a $\utwo$ theory with $N$ hypermultiplets 
transforming in $[1,a]$ is given as
\begin{equation}
 \Delta(n,m)= \frac{N}{2} \big( \left| (2a+1) \cdot n +m \right| +  \left| 
(2a+1) \cdot n -m \right| \big) - 2 |m|\label{eqn:delta_U2_fundamental}
\end{equation}
such that the Hilbert series is computed via
\begin{equation}
 \HS_{\utwo}^{[1,a]}(t,z)= \sum_{n,m} P_{\utwo}(t^2,m) \ t^{2 \Delta(n,m)} 
z^{2n} 
\, ,
\end{equation}
where the ranges of $n,m$ have been specified above. Here we use the fugacity 
$t^2$ instead of $t$ to avoid half-integer powers.
\paragraph{Hilbert basis}
The conformal dimension~\eqref{eqn:delta_U2_fundamental} divides 
$\Lambda_w(\utwo) \slash S_2$ into semi-groups 
via the absolute values $|m|$,
$|(2a+1)n+m|$, and $|(2a+1)n-m|$. Thus, there are three semi-groups
\begin{subequations}
\begin{align}
 S_+^{(2)} &= \left\{ (m,n)\in \Lambda_w^{\utwo} \slash S_2 \ | \ (n \geq 0) \; 
 \wedge \; (0\leq m \leq (2a+1) n ) \right\} \; , \\
S_0^{(2)} &= \left\{ (m,n)\in \Lambda_w^{\utwo} \slash S_2 \ | \ -(2a+1)n \leq 
m \leq (2a+1)n  \right\} \; , \\
S_-^{(2)} &= \left\{ (m,n)\in \Lambda_w^{\utwo} \slash S_2 \ | \ ( n \leq 0) \; 
 \wedge \; ( 0\leq m \leq -(2a+1)n ) \right\} 
\end{align}
\end{subequations} 
originating from $2$-dimensional cones, see Fig.~\ref{Fig:U2_N-fund_a=odd}.
Since all these semi-groups $S_{\pm}^{(2)}$, $S_{0}^{(2)}$ are finitely 
generated, one can compute the Hilbert basis $\Hcal(S_p)$ for each $p$ and 
obtains
\begin{subequations}
\label{eqn:Hilbert_basis_U2_N-fund_a=odd}
\begin{align}
 \Hcal(S_{\pm}^{(2)}) &= \Big\{ (0,\pm1), \{(l+\tfrac{1}{2},\pm\tfrac{1}{2}) 
\  | \ l=0,1,\ldots, a \}  \Big\} \; , \\
 \Hcal(S_0^{(2)}) &= \Big\{(a+\tfrac{1}{2},\tfrac{1}{2}), (1,0), 
(a+\tfrac{1}{2}, -\tfrac{1}{2})    \Big\} \; .
\end{align}
\end{subequations}
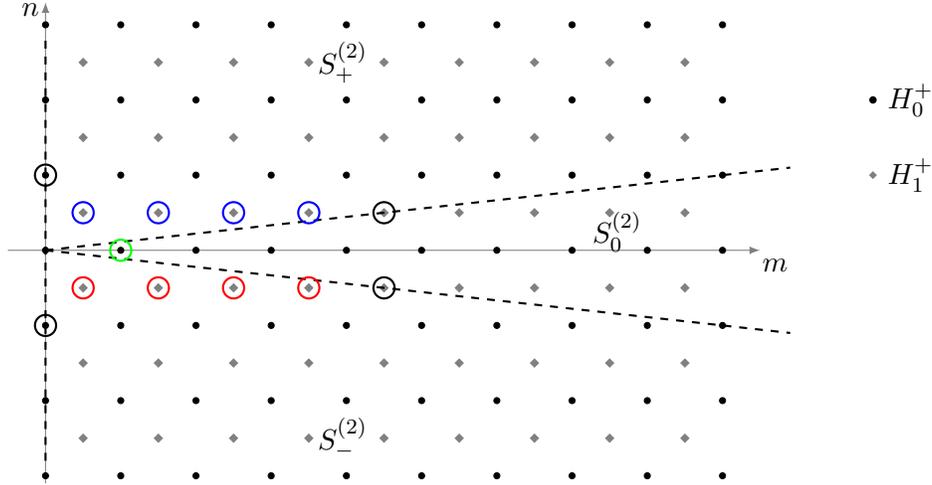
\begin{figure}[h]
\centering
\begin{tikzpicture}
  \coordinate (Origin)   at (0,0);
  \coordinate (XAxisMin) at (-0.5,0);
  \coordinate (XAxisMax) at (9.5,0);
  \coordinate (YAxisMin) at (0,-3.1);
  \coordinate (YAxisMax) at (0,3.3);
  \draw [thin, gray,-latex] (XAxisMin) -- (XAxisMax);%
  \draw [thin, gray,-latex] (YAxisMin) -- (YAxisMax);%
  \draw (9.7,-0.2) node {$m$};
  \draw (-0.2,3.2) node {$n$};
%Draw the root lattice
  \foreach \x in {0,1,...,9}{%
      \foreach \y in {-3,-2,...,3}{%
        \node[draw,circle,inner sep=0.8pt,fill,black] at (\x,\y) {};
            }
            }
  \foreach \x in {0,1,...,8}{%
      \foreach \y in {-3,-2,...,2}{%
        \node[draw,diamond,inner sep=0.8pt,fill,gray] at (\x +1/2,\y +1/2) {};
            }
            }
   \draw[black,dashed,thick]  (Origin) -- (9*1.1,1.1);  
   \draw[black,dashed,thick]  (Origin) -- (9*1.1,-1.1);
   \draw[black,dashed,thick]  (0,-2.8) -- (0,2.8);
 \draw[black,thick] (4.5,0.5) circle (4pt);
 \draw[black,thick] (4.5,-0.5) circle (4pt);
 \draw[green,thick] (1,0) circle (4pt);
%   %  
 \draw[blue,thick] (3.5,0.5) circle (4pt);
 \draw[blue,thick] (2.5,0.5) circle (4pt);
 \draw[blue,thick] (1.5,0.5) circle (4pt);
 \draw[blue,thick] (0.5,0.5) circle (4pt);
 \draw[black,thick] (0,1) circle (4pt);
% %  
 \draw[red,thick] (3.5,-0.5) circle (4pt);
 \draw[red,thick] (2.5,-0.5) circle (4pt);
 \draw[red,thick] (1.5,-0.5) circle (4pt);
 \draw[red,thick] (0.5,-0.5) circle (4pt);
  \draw[black,thick] (0,-1) circle (4pt);
  \draw (3.95,2.5) node {$S_+^{(2)}$};
  \draw (7.6,0.25) node {$S_0^{(2)}$};
  \draw (3.95,-2.5) node {$S_-^{(2)}$};
 \draw[black,thick,fill] (11,2) circle (1pt);
 \draw (11.5,2) node {$H_0^+ $};
 \draw (11,1) node[diamond,inner sep=0.8pt,fill,gray] {};
 \draw (11.5,1) node {$H_1^+ $};
\end{tikzpicture}
\caption{The Weyl-chamber for the example $a=4$.
The black circled lattice points are the ray generators. 
The blue circled lattice points complete the Hilbert basis (together with two 
ray generators) for $S_+^{(2)}$; while the red circled points analogously 
complete the Hilbert basis for $S_-^{(2)}$. The green circled point represents 
the missing minimal generator for $S_0^{(2)}$.}
\label{Fig:U2_N-fund_a=odd}
\end{figure}
% % 
% 
\paragraph{Hilbert series}
Computing the Hilbert series yields
\begin{subequations}
\label{eqn:HS_U2_fund_aOdd}
\begin{equation}
 \HS_{\utwo}^{[1,a]} (t,z,N)= \frac{R(t,z)}{P(t,z)} \; ,\\
\end{equation}
\begin{align}
P(t,z)&=
\left(1-t^2\right)^2 
\left(1-t^4\right) 
\left(1-t^{2 N-4}\right) 
\left(1-\tfrac{1}{z^2} t^{(4 a+2) N}\right) 
\left(1-z^2 t^{(4 a+2) N}\right) 
\label{eqn:HS_U2_fund_aOdd_Den}\\*
&\phantom{=\left(1-t^2\right)^2 \left(1-t^4\right)} 
\times 
\left(1-\tfrac{1}{z} t^{(2a+1)(N-2)}\right) 
\left(1-z t^{(2a+1)(N-2)}\right) \; ,\notag
 \\
R(t,z)&=
1-t^2
+t^{2 N-2}
-t^{2 N}
+2 t^{4 a N-4 a+2 N}
-t^{4 a N-8 a+2 N-4}
-t^{4 a N-8 a+2 N-2} \\
&\qquad 
-2 t^{4 a N-4 a+4 N-4}
+t^{4 a N-8 a+4 N-6}
+t^{4 a N-8 a+4 N-4}
+t^{8 a N+4 N}
+t^{8 a N+4 N+2} \notag \\
&\qquad 
-2 t^{8 a N-4 a+4 N}
-t^{8 a N+6 N-2}
-t^{8 a N+6 N}
+2 t^{8 a N-4 a+6 N-4}
-t^{12 a N-8 a+6 N-4}\notag \\
&\qquad 
+t^{12 a N-8 a+6 N-2}
-t^{12 a N-8 a+8 N-6}
+t^{12 a N-8 a+8 N-4}
\notag \\
&+\left(z + \tfrac{1}{z} \right) 
\Big(
t^{2 a N-4 a+N}
-t^{2 a N+N+2}
+t^{2 a N+3 N-2}
-t^{2 a N-4 a+3 N-4}
+t^{6 a N+3 N+2}\notag \\
&\qquad 
-t^{6 a N-8 a+3 N-2}
-t^{6 a N+5 N-2}
+t^{6 a N-8 a+5 N-6}
-t^{10 a N-4 a+5 N}
+t^{10 a N-8 a+5 N-2}\notag \\
&\qquad 
+t^{10 a N-4 a+7 N-4}
-t^{10 a N-8 a+7 N-6}
\Big) \notag \\
&+\left(z^2 + \tfrac{1}{z^2} \right) 
\big( 
t^{4 a N-4 a+2 N}
-t^{4 a N+2 N}
+t^{4 a N+4 N}
-t^{4 a N-4 a+4 N-4}
-t^{8 a N-4 a+4 N}\notag \\
&\qquad 
+t^{8 a N-8 a+4 N-4}
+t^{8 a N-4 a+6 N-4}
-t^{8 a N-8 a+6 N-4}
\big) \notag \; .
\end{align}
\end{subequations}
The Hilbert series~\eqref{eqn:HS_U2_fund_aOdd} has a pole of order $4$ at 
$t\to1$, because $R(t=1,z)=0$ and $\tfrac{\diff^n}{\diff t^n} R(t,z)|_{t=1}=0$ 
for $n=1,2,3$. Hence, the moduli space is of (complex) dimension $4$.
As a comment, the additional $(1-t^2)$-term in the denominator can be cancelled 
with a corresponding term in the numerator either explicitly for each 
$a=\mathrm{fixed}$ or for any $a$, but the resulting expressions are not 
particularly insightful.
\paragraph{Discussion}
The four poles of the Hilbert series~\eqref{eqn:HS_U2_fund_aOdd}, which are 
graded as $z^{\pm2}$ and $z^{\pm1}$, can be 
identified with the four ray generators $(0,\pm1) $ and 
$(a+\tfrac{1}{2},\pm\tfrac{1}{2})$, i.e.\ they correspond to bare monopole 
operators. In addition, the bare monopole operator for the minimal generator 
$(1,0)$ is present in the denominator~\eqref{eqn:HS_U2_fund_aOdd_Den}, too.

In contrast, the family of monopoles $\{ ( l+\tfrac{1}{2} ,\pm\tfrac{1}{2})\, 
, l=0,1,\ldots, a-1\} $ is not directly visible in 
the Hilbert series, but can be deduced unambiguously from the plethystic 
logarithm. These monopole operators correspond the \emph{minimal generators} 
of $S_{\pm}^{(2)}$ which are not ray generators. 
Tab.~\ref{tab:Ops_U2_fund_aOdd} provides as summary of the monopole 
generators and their properties.
\begin{table}[h]
\centering
 \begin{tabular}{c|c|c|c|c}
 \toprule
  $(m,n)$ & $(m_1,m_2)$ & $2\Delta(m,n)$ & $H_{(m,n)}$ & dressings \\ \midrule
 $(1,0)$ & $(1,-1)$ & $2N-4$ & $\uo^2$ & $1$ by $\uo$ \\ \midrule
 $(l+\tfrac{1}{2}, \tfrac{1}{2})$, for $l=0,1,\ldots,a$ & $(l+1,-l)$  & 
$(2a+1)N -2(2l+1)$ & $\uo^2$ & $1$ by $\uo$ \\
 $(l+\tfrac{1}{2},- \tfrac{1}{2})$, for $l=0,1,\ldots,a$ & $(l,-(l+1))$ & 
$(2a+1)N -2(2l+1)$ & $\uo^2$ & $1$ by $\uo$ \\ \midrule
$(0,\pm1)$ & $\pm(1,1)$ & $(4a+2)N$ & $\utwo$ & none \\
 \bottomrule
 \end{tabular}
\caption{Bare and dressed monopole operators for the family $[1,a]$ of 
$\utwo$-representations.}
\label{tab:Ops_U2_fund_aOdd}
\end{table}
As a remark, the family of monopole operators $(l+\tfrac{1}{2}, 
\pm\tfrac{1}{2})$ is not always completely present in the plethystic logarithm. 
We observe that $l$-th bare operator is a generator if $N\geq 2(a-l+1)$, while 
the dressing of the $l$-th object is a generator if $N>2(a-l+1)$. The reason 
for the disappearance lies in a relation at degree 
$\Delta(1,0)+\Delta(a+\tfrac{1}{2},\pm\tfrac{1}{2})+2$, which coincides with 
$\Delta(l+\tfrac{1}{2}, \pm\tfrac{1}{2})$ for $N-1=2(a-l+1)$, such that the 
terms cancel in the PL. (See also App.~\ref{app:PL}.) Thus, for 
large $N$ all above listed objects are 
generators.
%
%%%%%%%%%%%%%%%%%%%%%%%%%%%%%%%%%%%%%%%%%%%%%%%%%%%%%%%%%%%%%%%%%%%%
%%%%%%%%%%%%%%%%%%%%%%%%%%%%%%%%%%%%%%%%%%%%%%%%%%%%%%%%%%%%%%%%%%%%
%
\subsubsection{Case: \texorpdfstring{$a=0$}{a=0}, complete 
intersection}
For the choice $a=1$, we obtain the Hilbert series for the $2$-dimensional 
fundamental representation $[1,0]$ of $\utwo$ as 
\begin{equation}
 \HS_{\utwo}^{[1,0]}(t,z,N)=\frac{   \left(1-t^{2 N}\right) \left(1-t^{2 
N -2}\right)   }{ \left(1- t^2\right) \left(1-t^4\right) 
\left(1- \tfrac{1}{z} t^N\right) \left(1-z t^N\right)
\left(1-\tfrac{1}{z} t^{N-2}\right) \left(1-z t^{N-2}\right)   }
\end{equation}
which agrees with the results of~\cite{Cremonesi:2013lqa}.

Let us comment on the reduction of generators compared to the Hilbert 
basis~\eqref{eqn:Hilbert_basis_U2_N-fund_a=odd}. The minimal generators have 
conformal dimensions $2\Delta(\tfrac{1}{2},\pm\tfrac{1}{2})=N-2$, 
$2\Delta(1,0)=2N-4$, and $2\Delta(0,\pm1) =2N$. Thus, $(1,0)$ is generated by 
$(\tfrac{1}{2},\pm\tfrac{1}{2})$ and $(0,\pm1)$ are generated by utilising the 
dressed monopoles of $(\tfrac{1}{2},\pm\tfrac{1}{2})$ and suitable elements in 
their Weyl-orbits.
% 
%%%%%%%%%%%%%%%%%%%%%%%%%%%%%%%%%%%%%%%%%%%%%%%%%%%%%%%%%%%%%%%%%%%%%%%%%%%%%%
%%%%%%%%%%%%%%%%%%%%%%%%%%%%%%%%%%%%%%%%%%%%%%%%%%%%%%%%%%%%%%%%%%%%%%%%%%%%%%
% 
\subsection{\texorpdfstring{$N$}{N} hypermultiplets in the adjoint 
representation of \texorpdfstring{$\su$}{SU(2)}}
\label{subsec:U2_adj_SU2}
The conformal dimension for a $\utwo$-theory with $N$ hypermultiplets 
transforming in the adjoint representation of $\su$ and arbitrary even 
$\uo$-charge is given by
\begin{equation}
 \Delta(n,m)=\frac{N}{2} \big( 
\left|(2a+2)n+2m\right|+\left|(2a+2)n\right|+\left|(2a+2)n-2m\right| \big)  
-2|m| \; .
\label{eqn:delta_U2_adjoint}
\end{equation}
Already at this stage, one can define the four semi-groups induced by the 
conformal dimension, which originate from $2$-dimensional cones
\begin{subequations}
\label{eqn:semi-groups_U2_adjoint}
  \begin{align}
   S_{2,\pm}^{(2)} &= \Big\{ (m,n) \in \Lambda_w^{\utwo} \slash S_2 \ | \ 
(m\geq 0) \wedge (m \leq \pm (a+1) n) \wedge (\pm n \geq 0) \Big\} \; , \\
  S_{1,\pm}^{(2)} &= \Big\{ (m,n) \in \Lambda_w^{\utwo} \slash S_2 \ | \ (m\geq 
0) \wedge (m \geq \pm (a+1) n) \wedge (\pm n \geq 0) \Big\} \; .
  \end{align}
  \end{subequations}
It turns out that the precise form of the Hilbert basis depends on the 
divisibility of $a$ by $2$; thus, we split the considerations in 
two cases: $a=2k-1$ and $a=2k$.
% 
%%%%%%%%%%%%%%%%%%%%%%%%%%%%%%%%%%%%%%%%%%%%%%%%%%%%%%%%%%%%%%%%%%%%%%%%%%%%%%
%%%%%%%%%%%%%%%%%%%%%%%%%%%%%%%%%%%%%%%%%%%%%%%%%%%%%%%%%%%%%%%%%%%%%%%%%%%%%%
%
\subsubsection{Case: \texorpdfstring{$a= 1 \mod 2$}{a=1 mod 2}}
\paragraph{Hilbert basis}
The collection of semi-groups~\eqref{eqn:semi-groups_U2_adjoint} is 
depicted in Fig.~\ref{Fig:U2_N-adj_a=0mod4}. As before, we compute the Hilbert 
basis $\Hcal$ for each semi-group of the minimal generators.

 \begin{subequations}
  \begin{align}
   \Hcal(S_{2,\pm}^{(2)}) &= \Big\{ (0,\pm1),(2k,\pm1) , \{ 
(j+\tfrac{1}{2},\pm\tfrac{1}{2}) \; | \; j=0,\ldots, k-1\}  \Big\} \; , \\
  \Hcal(S_{1,\pm}^{(2)}) &= \Big\{ (2k,\pm1),  ( k+\tfrac{1}{2},\pm 
\tfrac{1}{2}) ,(1,0) \Big\} \; .
  \end{align}
\end{subequations}
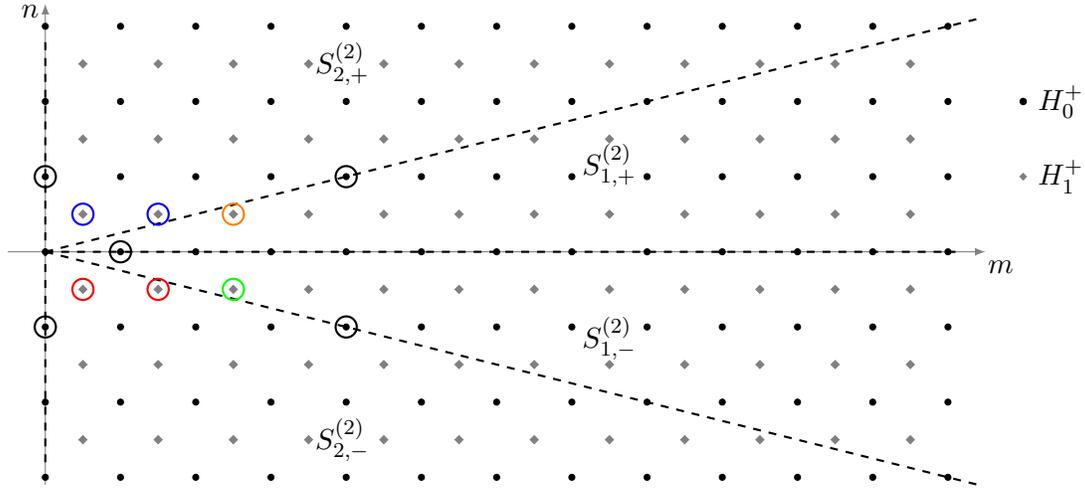
\begin{figure}[h]
\centering
\begin{tikzpicture}
  \coordinate (Origin)   at (0,0);
  \coordinate (XAxisMin) at (-0.5,0);
  \coordinate (XAxisMax) at (12.5,0);
  \coordinate (YAxisMin) at (0,-3.1);
  \coordinate (YAxisMax) at (0,3.3);
  \draw [thin, gray,-latex] (XAxisMin) -- (XAxisMax);%
  \draw [thin, gray,-latex] (YAxisMin) -- (YAxisMax);%
  \draw (12.7,-0.2) node {$m$};
  \draw (-0.2,3.2) node {$n$};
%Draw the root lattice
  \foreach \x in {0,1,...,12}{%
      \foreach \y in {-3,-2,...,3}{%
        \node[draw,circle,inner sep=0.8pt,fill,black] at (\x,\y) {};
            }
            }
  \foreach \x in {0,1,...,11}{%
      \foreach \y in {-3,-2,...,2}{%
        \node[draw,diamond,inner sep=0.8pt,fill,gray] at (\x +1/2,\y +1/2) {};
            }
            }
   \draw[black,dashed,thick]  (Origin) -- (4*3.1,3.1);  
   \draw[black,dashed,thick]  (Origin) -- (4*3.1,-3.1);
   \draw[black,dashed,thick]  (0,-2.8) -- (0,2.8);
   \draw[black,dashed,thick]  (Origin) -- (12.1,0);
 \draw[black,thick] (4,1) circle (4pt);
 \draw[black,thick] (4,-1) circle (4pt);
 \draw[black,thick] (1,0) circle (4pt);
 \draw[orange,thick] (2.5,0.5) circle (4pt);
 \draw[green,thick] (2.5,-0.5) circle (4pt);
% %
 \draw[blue,thick] (1.5,0.5) circle (4pt);
 \draw[blue,thick] (0.5,0.5) circle (4pt);
 \draw[black,thick] (0,1) circle (4pt);
% %
 \draw[red,thick] (1.5,-0.5) circle (4pt);
 \draw[red,thick] (0.5,-0.5) circle (4pt);
  \draw[black,thick] (0,-1) circle (4pt);
% %   
  \draw (3.95,2.5) node {$S_{2,+}^{(2)}$};
  \draw (7.5,1.15) node {$S_{1,+}^{(2)}$};
  \draw (7.5,-1.15) node {$S_{1,-}^{(2)}$};
  \draw (3.95,-2.5) node {$S_{2,-}^{(2)}$};
  \draw[black,thick,fill] (13,2) circle (1pt);
  \draw (13.5,2) node {$H_0^+ $};
  \draw (13,1) node[diamond,inner sep=0.8pt,fill,gray] {};
  \draw (13.5,1) node {$H_1^+ $};
\end{tikzpicture}
\caption{The Weyl-chamber for odd $a$, here with the example $a=3$. 
The black circled lattice points correspond to the ray generators originating 
from the fan. The blue/red circled points are the remaining minimal 
generators for $S_{2,\pm}^{(2)}$, respectively. Similarly, the orange/green 
circled point are the generators that complete the Hilbert basis for 
$S_{1,\pm}^{(2)}$.}
\label{Fig:U2_N-adj_a=0mod4}
\end{figure}
\paragraph{Hilbert series}
The computation of the Hilbert series yields
\begin{subequations}
\label{eqn:HS_U2_adj_0mod4}
\begin{equation}
 \HS_{\utwo}^{[2,2k-1]}(t,z,N)= \frac{R(t,z,N)}{P(t,z,N) }  \; ,
\end{equation}
\begin{align}
P(t,z,N) &= \left(1-t^2\right)^2 
 \left(1-t^4\right) 
 \left(1-t^{4 N-4}\right) 
\left(1-\tfrac{1}{z^2} t^{12 k N}\right)
\left(1-z^2 t^{12 k N}\right) \\*
&\phantom{= \left(1-t^2\right)^2  \left(1-t^4\right)}
\times 
\left(1-\tfrac{1}{z^2} t^{12 k N-8 k}\right) 
\left(1-z^2 t^{12 k N-8 k}\right) \; , \notag \\
R(t,z,N) &= 
1-t^2+t^{4 N-2}-t^{4 N}
t^{24 k N}
+t^{24 k N+2}
-t^{24 k N-16 k}
-t^{24 k N-16 k+2}\\
&\qquad
-t^{24 k N+4 N-2}
-t^{24 k N+4 N}
+t^{24 k N-16 k+4 N} 
+t^{24 k N-16 k+4 N-2}
-t^{48 k N-16 k}\notag \\
&\qquad
+t^{48 k N-16 k+2}
+t^{48 k N-16 k+4 N}
-t^{48 k N-16 k+4 N-2} \notag \\
&+ \left(z + \tfrac{1}{z} \right)
\bigg(
-t^{6 k N+2}
+t^{6 k N-4 k+2}
+t^{6 k N-4 k+2 N-2}
-t^{6 k N-4 k+2 N+2}
+t^{6 k N+4 N-2}\notag \\
&\qquad
-t^{6 k N-4 k+4 N-2}
+t^{18 k N+2} 
-t^{18 k N-4 k+2}
+t^{18 k N-8 k+2}
-t^{18 k N-12 k+2}\notag\\
&\qquad
-t^{18 k N-4 k+2 N-2}
+t^{18 k N-4 k+2 N+2}
-t^{18 k N-12 k+2 N-2}
+t^{18 k N-12 k+2 N+2}\notag\\
&\qquad
-t^{18 k N+4 N-2}
+t^{18 k N-4 k+4 N-2}
-t^{18 k N-8 k+4 N-2}
+t^{18 k N-12 k+4 N-2}
+t^{30 k N-4 k+2}\notag\\
&\qquad
-t^{30 k N-8 k+2}
+t^{30 k N-12 k+2}
-t^{30 k N-16 k+2}
+t^{30 k N-4 k+2 N-2}
-t^{30 k N-4 k+2 N+2}\notag\\
&\qquad
+t^{30 k N-12 k+2 N-2}
-t^{30 k N-12 k+2 N+2}
-t^{30 k N-4 k+4 N-2}
+t^{30 k N-8 k+4 N-2}\notag\\
&\qquad
-t^{30 k N-12 k+4 N-2}
+t^{30 k N-16 k+4 N-2}
-t^{42 k N-12 k+2}
+t^{42 k N-16 k+2}
-t^{42 k N-12 k+2 N-2}\notag\\
&\qquad
+t^{42 k N-12 k+2 N+2}
+t^{42 k N-12 k+4 N-2}
-t^{42 k N-16 k+4 N-2}
\bigg) \notag \\
&+ \left(z^2 + \tfrac{1}{z^2} \right)
\bigg(
-t^{12 k N}
+t^{12 k N-8 k+2}
+t^{12 k N+4 N}
-t^{12 k N-8 k+4 N-2}
+t^{36 k N-16 k}\notag\\
&\qquad
-t^{36 k N-8 k+2}
-t^{36 k N-16 k+4 N}
+t^{36 k N-8 k+4 N-2}
\bigg) \notag \\
&+ \left(z^3 + \tfrac{1}{z^3} \right)
\bigg(
-t^{18 k N-4 k+2}
+t^{18 k N-8 k+2}
-t^{18 k N-4 k+2 N-2}
+t^{18 k N-4 k+2 N+2} \notag\\
&\qquad
+t^{18 k N-4 k+4 N-2}
-t^{18 k N-8 k+4 N-2}
-t^{30 k N-8 k+2}
+t^{30 k N-12 k+2} \notag\\
&\qquad
+t^{30 k N-12 k+2 N-2}
-t^{30 k N-12 k+2 N+2}
+t^{30 k N-8 k+4 N-2}
-t^{30 k N-12 k+4 N-2}
\bigg) \notag \; .
\end{align}
\end{subequations}
Inspection of the Hilbert series~\eqref{eqn:HS_U2_adj_0mod4} reveals that it 
has 
a pole of order $4$ as 
$t\to1$ because one explicitly verifies $R(t=1,z,N)=0$, $\tfrac{\diff}{\diff t} 
R(t,z,N) 
|_{t=1}=0$, and $\tfrac{\diff^n}{\diff t^n} R(t,z,N) 
|_{t=1,z=1}=0$ for $n=2,3$. 
\paragraph{Discussion}
The denominator of the Hilbert series~\eqref{eqn:HS_U2_adj_0mod4} displays 
poles for the five bare monopole operators $(0,\pm1)$, $(2k,\pm1)$, and 
$(1,0)$, which are ray generators and charged under $\uo_J$ as $\pm2$, $\pm2$, 
and $0$, respectively. The remaining operators, corresponding to the minimal 
generators which are not ray generators, are apparent in the analysis of the 
plethystic logarithm.
The relevant bare and dressed monopole operators are summarised in 
Tab.~\ref{tab:Ops_U2_adj_0mod4}.
\begin{table}[h]
\centering
 \begin{tabular}{c|c|c|c|c}
 \toprule
  $(m,n)$ & $(m_1,m_2)$ & $2\Delta(m,n)$ & $H_{(m,n)} $ & dressings \\ 
\midrule
$(1,0)$ & $(1,-1)$  & $4N-4$ & $\uo^2$ & $1$ by $\uo$ \\\midrule
 $(j+\tfrac{1}{2},\tfrac{1}{2})$, for $j=0,\ldots,k-1$ & $(j+1,-j)$  & 
$6kN-4j-2$ & $\uo^2$ & $1$ by $\uo$ \\
  $(j+\tfrac{1}{2},-\tfrac{1}{2})$, for $j=0,\ldots,k-1$ & $(j,-(j+1))$  & 
$6kN-4j-2$ & $\uo^2$ & $1$ by $\uo$ \\
$(k+\tfrac{1}{2},\tfrac{1}{2})$ & $(k+1,-k)$  & 
$6kN+2N-4k-2$ & $\uo^2$ & $1$ by $\uo$ \\
  $(k+\tfrac{1}{2},-\tfrac{1}{2})$ & $(k,-(k+1))$  & 
$6kN+2N-4k-2$ & $\uo^2$ & $1$ by $\uo$ \\ \midrule
 $(0,\pm1)$ & $\pm(1,1)$  & $12kN$ & $\utwo$ & none \\
 $(2k,1)$ & $(2k+1,1-2k)$  & $12kN-8k$ & $\uo^2$ & $1$ by $\uo$ \\
$(2k,-1)$ & $(2k-1,-(2k+1))$  & $12kN-8k$ & $\uo^2$ & $1$ by $\uo$ \\
\bottomrule
 \end{tabular}
\caption{Summary of the monopole operators for odd $a$.}
\label{tab:Ops_U2_adj_0mod4}
\end{table}

The plethystic logarithm, moreover, displays that not always all monopoles of 
the family  $(j+\tfrac{1}{2},\pm\tfrac{1}{2})$ are generators (in the sense of 
the PL). The observation is: if $k-j < N$ then the $j$-th operator (bare as 
well as dressed) is truely a generator in the PL. The reason behind lies in a 
relation at degree $\Delta(k-\tfrac{1}{2},\pm \tfrac{1}{2}) + \Delta(1,0)$, 
which coincides with $\Delta(j+\tfrac{1}{2},\pm\tfrac{1}{2})$ for $k-j=N$. 
(See also App.~\ref{app:PL}.)
Hence, for large enough $N$ all above listed operators are generators.
% 
%%%%%%%%%%%%%%%%%%%%%%%%%%%%%%%%%%%%%%%%%%%%%%%%%%%%%%%%%%%%
%%%%%%%%%%%%%%%%%%%%%%%%%%%%%%%%%%%%%%%%%%%%%%%%%%%%%%%%%%%%
%
\subsubsection{Case: \texorpdfstring{$a= 0 \mod 2$}{a=0 mod 2}}
\begin{figure}[h]
\centering
\begin{tikzpicture}
  \coordinate (Origin)   at (0,0);
  \coordinate (XAxisMin) at (-0.5,0);
  \coordinate (XAxisMax) at (12.5,0);
  \coordinate (YAxisMin) at (0,-3.1);
  \coordinate (YAxisMax) at (0,3.3);
  \draw [thin, gray,-latex] (XAxisMin) -- (XAxisMax);%
  \draw [thin, gray,-latex] (YAxisMin) -- (YAxisMax);%
  \draw (12.7,-0.2) node {$m$};
  \draw (-0.2,3.2) node {$n$};
%Draw the root lattice
  \foreach \x in {0,1,...,12}{%
      \foreach \y in {-3,-2,...,3}{%
        \node[draw,circle,inner sep=0.8pt,fill,black] at (\x,\y) {};
            }
            }
  \foreach \x in {0,1,...,11}{%
      \foreach \y in {-3,-2,...,2}{%
        \node[draw,diamond,inner sep=0.8pt,fill,gray] at (\x +1/2,\y +1/2) {};
            }
            }
   \draw[black,dashed,thick]  (Origin) -- (5*2.5,2.5);  
   \draw[black,dashed,thick]  (Origin) -- (5*2.5,-2.5);
   \draw[black,dashed,thick]  (0,-2.8) -- (0,2.8);
   \draw[black,dashed,thick]  (Origin) -- (12.1,0);
 \draw[black,thick] (5/2,1/2) circle (4pt);
 \draw[black,thick] (5/2,-1/2) circle (4pt);
 \draw[black,thick] (1,0) circle (4pt);
% %
 \draw[blue,thick] (1.5,0.5) circle (4pt);
 \draw[blue,thick] (0.5,0.5) circle (4pt);
 \draw[black,thick] (0,1) circle (4pt);
% %
 \draw[red,thick] (1.5,-0.5) circle (4pt);
 \draw[red,thick] (0.5,-0.5) circle (4pt);
  \draw[black,thick] (0,-1) circle (4pt);
% %   
  \draw (3.95,2.5) node {$S_{2,+}^{(2)}$};
  \draw (8.5,1.15) node {$S_{1,+}^{(2)}$};
  \draw (8.5,-1.15) node {$S_{1,-}^{(2)}$};
  \draw (3.95,-2.5) node {$S_{2,-}^{(2)}$};
  \draw[black,thick,fill] (13,2) circle (1pt);
  \draw (13.5,2) node {$H_0^+ $};
  \draw (13,1) node[diamond,inner sep=0.8pt,fill,gray] {};
  \draw (13.5,1) node {$H_1^+ $};
\end{tikzpicture}
\caption{The Weyl-chamber for $a=0 \mod 2$, here with the example $a=4$. 
The black circled lattice points correspond to the ray generators originating 
from the fan. The blue/red circled points are the remaining minimal 
generators for $S_{2,\pm}^{(2)}$, respectively.}
\label{Fig:U2_N-adj_a=2mod4}
\end{figure}
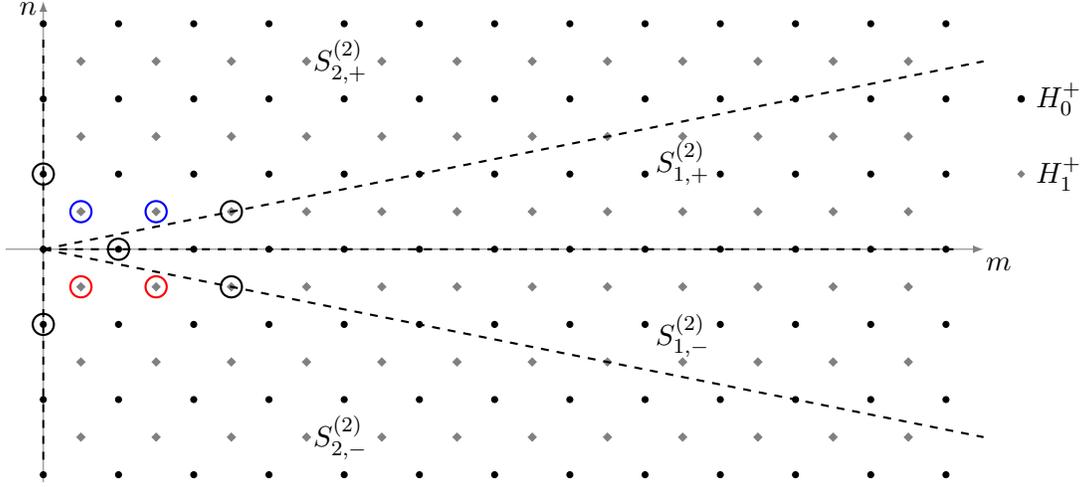

\paragraph{Hilbert basis}
The diagram for the minimal generators is provided in 
Fig.~\ref{Fig:U2_N-adj_a=2mod4}. Again, the appearing (bare) monopoles 
correspond to the Hilbert basis of the semi-groups.
 \begin{subequations}
  \begin{align}
   \Hcal(S_{2,\pm}^{(2)}) &= \Big\{ (0,\pm1), \{ (j+\tfrac{1}{2},\pm 
\tfrac{1}{2}) \; , \; j=0,1,\ldots, k\}  \Big\} \; , 
\\
  \Hcal(S_{1,\pm}^{(2)}) &= \Big\{ ( k+\tfrac{1}{2},\pm 
\tfrac{1}{2}),(1,0) \Big\} \; .
    \end{align} 
\end{subequations}
\paragraph{Hilbert series}
The computation of the Hilbert series for this case yields
\begin{subequations}
\label{eqn:HS_U2_adj_2mod4}
\begin{equation}
 \HS_{\utwo}^{[2,2k]}(t,z,N) = \frac{R(t,z,N)}{P(t,z,N)} \; ,
\end{equation}
\begin{align}
P(t,z,N)&=\left(1-t^2\right)^2 
 \left(1-t^4\right) 
 \left(1-t^{4 N-4}\right)
 \left(1-\tfrac{1}{z} t^{6 k N-4 k+3 N-2}\right) 
 \left(1-z t^{6 k N-4 k+3 N-2}\right) \\*
 &\qquad \times
 \left(1-\tfrac{1}{z^2} t^{12 k N+6 N}\right) 
 \left(1-z^2 t^{12 k N+6 N}\right) \; , \notag\\
R(t,z,N) &= 
1-t^2+t^{4 N-2}-t^{4 N}
+2 t^{12 k N-4 k+6 N}
-t^{12 k N-8 k+6 N-4}
-t^{12 k N-8 k+6 N-2} \\
&\qquad 
-2 t^{12 k N-4 k+10 N-4}
+t^{12 k N-8 k+10 N-6} 
+t^{12 k N-8 k+10 N-4}
+t^{24 k N+12 N}
+t^{24 k N+12 N+2}\notag \\
&\qquad 
-2 t^{24 k N-4 k+12 N}
-t^{24 k N+16 N-2}
-t^{24 k N+16 N}
+2 t^{24 k N-4 k+16 N-4}\notag \\
&\qquad 
-t^{36 k N-8 k+18 N-4}
+t^{36 k N-8 k+18 N-2}
-t^{36 k N-8 k+22 N-6}
+t^{36 k N-8 k+22 N-4}
\notag \\
&+\left(z +\tfrac{1}{z} \right) \bigg(
-t^{6 k N+3 N+2}
+t^{6 k N-4 k+3 N}
+t^{6 k N+7 N-2}
-t^{6 k N-4 k+7 N-4}
+t^{18 k N+9 N+2} \notag \\
&\qquad 
-t^{18 k N-8 k+9 N-2}
-t^{18 k N+13 N-2}
+t^{18 k N-8 k+13 N-6}
-t^{30 k N-4 k+15 N}\notag \\
&\qquad 
+t^{30 k N-8 k+15 N-2} 
+t^{30 k N-4 k+19 N-4}
-t^{30 k N-8 k+19 N-6}
\bigg) \notag \\
&+\left(z^2 +\tfrac{1}{z^2} \right) \bigg(
-t^{12 k N+6 N}
+t^{12 k N-4 k+6 N}
+t^{12 k N+10 N}
-t^{12 k N-4 k+10 N-4}\notag \\
&\qquad 
-t^{24 k N-4 k+12 N} 
+t^{24 k N-8 k+12 N-4} 
+t^{24 k N-4 k+16 N-4}
-t^{24 k N-8 k+16 N-4}
\bigg) \notag \; .
\end{align}
\end{subequations}
The Hilbert series~\eqref{eqn:HS_U2_adj_2mod4} has a pole of order $4$ as 
$t\to1$ because one can explicitly verify that $R(t=1,z,N)=0$, 
$\tfrac{\diff}{\diff t} R(t,z,N) |_{t=1}=0$, and 
$\tfrac{\diff^n}{\diff t^n} R(t,z,N) |_{t=1,z=1}=0$  for $n=2,3$.
\paragraph{Discussion}
The five monopoles corresponding to the ray generators, i.e.\ $(0,\pm1)$, 
$(k+\tfrac{1}{2},\pm \tfrac{1}{2})$, and $(1,0)$, appear as poles in 
the Hilbert series~\eqref{eqn:HS_U2_adj_2mod4} and are charged under $\uo_J$ as 
$\pm2$, $\pm1$, and $0$, respectively. The remaining minimal generator can be 
deduced by inspecting the plethystic logarithm. We summarise the monopole 
generators in Tab.~\ref{tab:Ops_U2_adj_2mod4}.
\begin{table}[h]
\centering
 \begin{tabular}{c|c|c|c|c}
 \toprule
  $(m,n)$ & $(m_1,m_2)$ & $2\Delta(m,n) $ & $H_{(m,n)}$ & dressings \\ \midrule
 $(1,0)$ & $(1,-1) $ & $4N-4$ & $\uo^2$ & $1$ by $\uo$ \\ \midrule
 $(j+\tfrac{1}{2},\tfrac{1}{2})$, for $j=0,1,\ldots,k$ & $(j+1,-j)$ & 
$6kN+3N-4j-2 $ &  $\uo^2$ & $1$ by $\uo$ \\
$(j+\tfrac{1}{2},- \tfrac{1}{2})$, for $j=0,1,\ldots,k$ & $(j,-(j+1)) $  & 
$6kN+3N-4j-2 $ &  
$\uo^2$ & $1$ by $\uo$ \\ \midrule
 $(0,\pm 1)$ & $\pm(1,1)$ & $12kN+6N $ &  $\utwo$ & none \\ \bottomrule
 \end{tabular}
\caption{Summary of the monopole operators for even $a$.}
\label{tab:Ops_U2_adj_2mod4}
\end{table}
Similarly to the case of odd $a$, the plethystic logarithm displays 
that not always all monopoles of 
the family  $(j+\tfrac{1}{2},\pm\tfrac{1}{2})$ are generators. The observation 
is: if $k-j +1 \geq N$ then the $j$-th bare operator is a generator in 
the PL, while for $k-j+2\geq N$ then also the dressing of the $j$-th monopole 
is a generator. The reason behind lies, again, in a 
relation at degree $\Delta(k-\tfrac{1}{2},\pm \tfrac{1}{2}) + \Delta(1,0)+2$, 
which coincides with $\Delta(j+\tfrac{1}{2},\pm\tfrac{1}{2})$ for $k-j=N$. (See 
also App.~\ref{app:PL}.)
Hence, for large enough $N$ all above listed operators are generators.
% 
%%%%%%%%%%%%%%%%%%%%%%%%%%%%%%%%%%%%%%%%%%%%%%%%%%%%%%%%%%%%%%%%%%%%%%%%%%%%%%
%%%%%%%%%%%%%%%%%%%%%%%%%%%%%%%%%%%%%%%%%%%%%%%%%%%%%%%%%%%%%%%%%%%%%%%%%%%%%%
%
\subsection{Direct product of \texorpdfstring{$\su$}{SU(2)} and 
\texorpdfstring{$\uo$}{U(1)}}
A rather simple example is obtained by considering the non-interacting product 
of an $\su$ and a $\uo$ theory. Nonetheless, it illustrates how the rank two 
Coulomb branches contain the product of rank one Coulomb branches as subclasses.

As first example, take $N_1$ fundamentals of $\su$ and $N_2$ hypermultiplets 
charged under $\uo$ with charges $a \in \NN$. The conformal dimension is given 
by
\begin{equation}
 \Delta(m,n)= (N_1 -2) |m| + \frac{N_2 \cdot a}{2}|n| \for m\in \NN \und n \in\Z
\end{equation}
and the dressing factor splits as
\begin{equation}
 P_{\su}(t,m,n) = P_{\su}(t,m)\times P_{\uo}(t,n) \; ,
\end{equation}
such that the Hilbert series factorises 
\begin{equation}
 \HS_{\su \times \uo}^{[1],a}(t,N_1,N_2) = \HS_{\su}^{[1]}(t,N_1) \times 
\HS_{\uo}^{a}(t,N_2) \; .
\end{equation}
The rank one Hilbert series have been presented in~\cite{Cremonesi:2013lqa}. 
Moreover, $\HS_{\uo}^{a}(t,N_2)$ equals the $A_{a\cdot N_2 -1}$ singularity 
$\C^2\slash \Z_{a\cdot N_2}$; whereas  $\HS_{\su}^{[1]}(t,N_1)$ is precisely 
the $D_{N_1}$ singularity.

The second, follow-up example is simply  a theory comprise of $N_1$ 
hypermultiplets in the adjoint representation of $\su$ and $N_2$ 
hypermultiplets charged under $\uo$ as above. The conformal dimension is 
modified to 
\begin{equation}
 \Delta(m,n)= 2(N_1 -1) |m| + \frac{N_2 \cdot a}{2}|n| \for m\in \NN \und n 
\in\Z
\end{equation}
and Hilbert series is obtained as
\begin{equation}
 \HS_{\su \times \uo}^{[2],a}(t,N_1,N_2) = \HS_{\su}^{[2]}(t,N_1) \times 
\HS_{\uo}^{a}(t,N_2) \; .
\end{equation}
Applying the results of~\cite{Cremonesi:2013lqa}, $\HS_{\su}^{[2]}(t,N_1)$ is 
the Hilbert series of the $D_{2N_1}$-singularity on $\C^2$. 

Summarising, the direct product of these $\su$-theories with $\uo$-theories 
results in moduli spaces that are products of A and D type singularities, 
which are complete intersections. Moreover, any non-trivial interactions 
between these two gauge groups, as discussed in 
Subsec.~\ref{subsec:U2_fund_SU2} and~\ref{subsec:U2_adj_SU2}, leads to a very
elaborate expression for the Hilbert series as rational functions. Also, the 
Hilbert basis becomes an important concept for understanding the moduli space.
%%%%%%%%%%%%%%%%%%%%%%%%%%%%%%%%%%%%%%%%%%%%%%%%%%%%%%%%%%%%%%%%%%%%%%%%%%%%%%%%
  \section{Case: \texorpdfstring{$\boldsymbol{A_1 \times A_1}$}{A1xA1}}
\label{sec:A1xA1}
This section concerns all Lie groups with Lie algebra $D_2$, which allows to 
study products of the rank one gauge groups $\sorm(3)$ and $\su$, but 
also the proper rank two group $\sorm(4)$.
\subsection{Set-up}
\begin{figure}
\centering
\begin{tikzpicture}
  \coordinate (Origin)   at (0,0);
  \coordinate (XAxisMin) at (-0.5,0);
  \coordinate (XAxisMax) at (5,0);
  \coordinate (YAxisMin) at (0,-3);
  \coordinate (YAxisMax) at (0,3);
  \draw [thin, gray,-latex] (XAxisMin) -- (XAxisMax);%
  \draw [thin, gray,-latex] (YAxisMin) -- (YAxisMax);%
  \draw (5.2,-0.2) node {$m_1$};
  \draw (-0.2,3.2) node {$m_2$};
%Draw the root lattice
  \foreach \x in {0,1,...,2}{%
      \foreach \y in {-1,0,...,1}{%
        \node[draw,circle,inner sep=0.8pt,fill,black] at (2*\x,2*\y) {};
            }
            }
    \foreach \x in {0,1,...,2}{%
      \foreach \y in {0,1}{%
        \node[draw,circle,inner sep=0.8pt,fill,black] at (2*\x-1,2*\y-1) {};
            }
            }            
%Draw K² lattice
    \foreach \x in {0,1,...,2}{%
      \foreach \y in {-1,0,...,1}{%
        \node[draw,regular polygon,regular polygon sides=3,black,inner 
sep=0.7pt] at (2*\x-1,2*\y) {};
            }
            }
    \foreach \x in {0,1,...,2}{%
      \foreach \y in {0,1}{%
        \node[draw,regular polygon,regular polygon sides=3,black,inner 
sep=0.7pt] at (2*\x,2*\y-1) {};
            }
            }            
% Draw K¹ lattice            
    \foreach \x in {0,1,...,2}{%
      \foreach \y in {-1,0,...,1}{%
        \node[draw,diamond,black,inner sep=0.8pt] at (2*\x +1/2,2*\y +1/2) 
{};
            }
            }  
      \foreach \x in {0,1,...,2}{%
      \foreach \y in {-1,0,...,1}{%
        \node[draw,diamond,black,inner sep=0.8pt] at (2*\x-1 
+1/2,2*\y-1+1/2) {};
            }
            } 
%Draw K³ lattice
    \foreach \x in {0,1,...,2}{%
      \foreach \y in {-1,0,...,1}{%
        \node[draw,cross out,thick,black,inner 
sep=0.8pt] at (2*\x-1+1/2,2*\y+1/2) {};
            }
            }
     \foreach \x in {0,1,...,2}{%
      \foreach \y in {-1,0,...,1}{%
        \node[draw,cross out,thick,black,inner 
sep=0.8pt] at (2*\x +1/2,2*\y-1 +1/2) {};
            }
            }
  \draw[black,dashed,thick]  (0,0) -- (3.2,3.2);  
  \draw[black,dashed,thick]  (0,0) -- (3.2,-3.2);
  \filldraw[dotted,fill=yellow, fill opacity=0.3, draw=gray] (Origin) -- 
(3.2,3.2) -- (4.6,3.2) -- (4.6,-3.2) -- (3.2,-3.2) -- cycle ;
% set up the explanation
\draw (6,2) node[circle,inner sep=0.8pt,fill,black] {};
\draw (7.2,2.05) node { $K^{[0]}$ lattice};
\draw (6,1) node[draw,diamond,inner sep=0.8pt,black] {};
\draw (7.2,1.05) node { $K^{[1]}$ lattice};
\draw (6,0) node[draw,regular polygon,regular polygon sides=3,inner 
sep=0.8pt,black] {};
\draw (7.2,0.05) node { $K^{[2]}$ lattice};
\draw (6,-1 ) node[draw,cross out,thick,inner sep=0.8pt,black] {};
\draw (7.2,-0.95) node { $K^{[3]}$ lattice};
\draw (6,-2) node[rectangle,inner sep=8pt,fill,yellow,fill opacity=0.35] 
{};
\draw (9,-2) node { Weyl chamber $m_1\geq |m_2| $};
\end{tikzpicture}%
\caption{The four different sublattices of the covering group of $\sorm(4)$. 
One recognises the root lattice $\Lambda_r^{\widetilde{\sorm}(4)}=K^{[0]}$ and 
the weight lattice $ \Lambda_w^{\widetilde{\sorm}(4)} =K^{[0]}\cup K^{[1]}\cup 
K^{[2]}\cup K^{[3]}$.}%
\label{fig:A1xA1_sublattices}%
\end{figure}
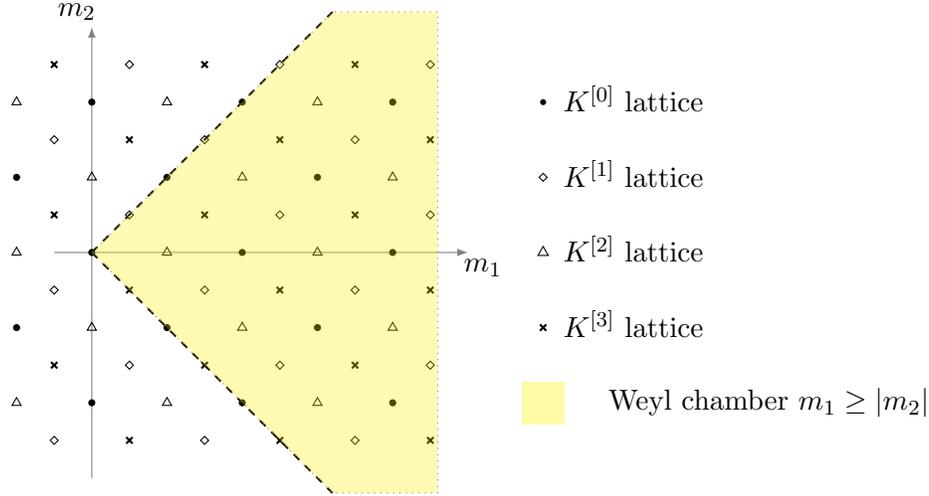

Let us consider the Lie algebra $D_2 \cong A_1\times A_1$. 
Following~\cite{Goddard:1976qe}, there are five different groups with this 
Lie algebra. The reason is that the universal covering 
group $\widetilde{\sorm}(4)$ of $\sorm(4)$ has a 
non-trivial centre $\Zcal(\widetilde{\sorm}(4)) =\Z_2\times\Z_2$ of order 
4. The quotient of $\widetilde{\sorm}(4)$ by any of the five different 
subgroups $\Zcal(\widetilde{\sorm}(4))$ yields a Lie group with the same 
Lie algebra. Fortunately, working with $\sorm(4)$ allows to use the isomorphism 
 $\widetilde{\sorm}(4) = \Spin(4)\cong \su\times \su$. We can 
summarise the setting as displayed in Tab.~\ref{tab:A1xA1_quotients}.
\begin{table}[h]
\centering
\begin{doublespacing}
\begin{tabular}{c|c||c|c|c}
\toprule
 Quotient & isomorphic group $\G$ & GNO-dual $\GNOG$ & 
$\Zcal(\GNOG) $ & GNO-charges $(m_1,m_2)$ \\ \midrule
$\frac{\widetilde{\sorm}(4)}{\{1\}}$ & $\su\times \su$ & 
$\sorm(3)\times\sorm(3)$ & $\{1\}$ & $  K^{[0]}$ \\
$\frac{\widetilde{\sorm}(4)}{\Z_2\times\{1\}} $ & $\sorm(3) \times \su$ & $\su 
\times \sorm(3)$ & $\Z_2 \times \{1\} $ & $ K^{[0]}\cup K^{[1]}$ 
\\
$\frac{\widetilde{\sorm}(4)}{\diag(\Z_2)} $ & $\sorm(4) $ & $\sorm(4)$ & 
$\Z_2$ & $ K^{[0]}\cup K^{[2]}$ \\
$\frac{\widetilde{\sorm}(4)}{\{1\} \times \Z_2} $ & $ \su 
\times \sorm(3)$ & $\sorm(3)\times \su $ & $\{1\} \times \Z_2 $ & $ K^{[0]}\cup 
K^{[3]}$ \\
$\frac{\widetilde{\sorm}(4)}{\Z_2 \times \Z_2} $ & $ \sorm(3) 
\times \sorm(3)$ & $\su \times \su $ & $\Z_2 \times \Z_2  $ & $ K^{[0]}\cup 
K^{[1]}\cup K^{[2]}\cup K^{[3]}$ \\
\bottomrule
\end{tabular}
\end{doublespacing}
\caption{All the Lie groups that arise taking the quotient of 
$\widetilde{\sorm}(4)$ by a subgroup of its centre; hence, their Lie algebra is 
$D_2$. }
\label{tab:A1xA1_quotients}
\end{table}
Here, we employed $\widehat{\su}=\sorm(3)$ and that for semi-simple 
groups $\G_1$, $\G_2$ 
\begin{equation}
 \widehat{\G_1\times  \G_2}=\GNOG_1 \times \GNOG_2
\end{equation}
holds~\cite{Goddard:1976qe}. Moreover, the GNO-charges are defined via the 
following sublattices of the weight lattice of $\Spin(4)$ (see also 
Fig.~\ref{fig:A1xA1_sublattices})
\begin{subequations}
\begin{align}
 K^{[0]} &= \Big\{(m_1,m_2) \; | \;  m_i = p_i \in \Z \, , \; p_1+p_2 
=\text{even} \Big\} \; ,\\
K^{[1]} &= \Big\{(m_1,m_2) \; | \;  m_i = p_i + \tfrac{1}{2} \, , \; p_i \in \Z 
\, , \; p_1+p_2 =\text{even} \Big\}  \; , \\
 K^{[2]} &= \Big\{(m_1,m_2) \; | \;  m_i = p_i \in \Z \, , \; p_1+p_2 
=\text{odd} \Big\} \; , \\
K^{[3]} &= \Big\{(m_1,m_2) \; | \;  m_i = p_i + \tfrac{1}{2} \, , \; p_i \in \Z 
\, , \; p_1+p_2 =\text{odd} \Big\} \; .
\end{align}
\end{subequations}
The important consequence of this set-up is that the fan defined by the 
conformal dimension will be the same for a given representation in each of the 
five quotients, but the semi-groups will differ due to the different lattices 
$\Lambda_w(\GNOG)$ used in the intersection. Hence, we will find different 
Hilbert basis in each quotient group. Nevertheless, we are forced to consider 
representations on the root lattice as we otherwise cannot compare all 
quotients.
\paragraph{Dressings}
In addition, we have chosen to parametrise the principal Weyl chamber via 
$m_1\geq |m_2|$ such that the classical dressing factors are given 
by~\cite{Cremonesi:2013lqa} 
\begin{equation}
 P_{A_1\times A_1}(t,m_1,m_2)= \begin{cases} \frac{1}{(1-t^2)^2} \; ,& \for 
m_1=m_2=0 \; , \\
		  \frac{1}{(1-t)(1-t^2)}\; , & \for m_1=|m_2|>0 \; , \\
				  \frac{1}{(1-t)^2} \; , & \for m_1>|m_2| \geq 
0 
\; .
                    \end{cases}
\end{equation}
Regardless of the quotient $\widetilde{\sorm(4)} \slash \Gamma$, the space of 
Casimir invariance is $2$-dimensional. We choose a basis such that the 
two degree $2$ Casimir invariants stem either from $\su$ or $\sorm(3)$, 
i.e.~\footnote{In a different basis, the Casimir invariants for $\sorm(4)$ are 
the quadratic Casimir and the Pfaffian.} 
\begin{equation}
 \diag (\Phi) = (\phi_1,\phi_2) \quad \longrightarrow \quad \Casi{2}^{(i)}= 
(\phi_i)^2 \; .
\end{equation}
Next, we can clarify all relevant bare and dressed monopole 
operators for an $(m_1,m_2)$ that is a minimal generator. There are two cases: 
On the one hand, for $m_2 = \pm m_1$, i.e. at the boundary of the Weyl 
chamber, the residual gauge group is either $\uo_i\times \su_j$ or $\uo_i 
\times \sorm(3)_j$ (for $i,j=1,2$ and $i\neq j$), depending on the quotient 
under consideration. Thus, only the degree $1$ Casimir invariant of the $\uo_i$ 
can be employed for a dressing, as the Casimir invariant of $\su_j$ or 
$\sorm(3)_j$ belongs to the quotient $\widetilde{\sorm(4)} \slash \Gamma$ 
itself. Hence, we get
\begin{subequations}
\label{eqn:A1xA1_dress}
\begin{equation}
 V_{(m_1,\pm m_1)}^{\mathrm{dress},0} = (m_1,\pm m_1) \und 
 V_{(m_1,\pm m_1)}^{\mathrm{dress},1} = \phi_i \ (m_1,\pm m_1) \; .
\end{equation}
Alternatively, we can apply the results of App.~\ref{app:PL} and deduce the 
dressing behaviour at the boundary of the Weyl chamber to be $P_{A_1\times 
A_1}(t,m_1,\pm m_1) \slash P_{A_1\times A_1}(t,0,0) = 1+t$, i.e.\ only one 
dressed monopole arises.

On the other hand, for $m_1>|m_2|\geq0$, i.e. in the interior of the Weyl 
chamber, the residual gauge group is $\uo^2$. From the resulting two degree $1$ 
Casimir invariants one constructs the following monopole operators:
\begin{equation}
 V_{(m_1, m_2)}^{\mathrm{dress},0} = (m_1,m_2) 
 \quad \longrightarrow \quad 
 \left\{ \begin{matrix}
  V_{(m_1,m_2)}^{\mathrm{dress},1,i} = \phi_i \ (m_1,m_2) \; ,& \for i=1,2 
\\ V_{(m_1,m_2)}^{\mathrm{dress},2} = \phi_1 \phi_2 \ (m_1, m_2)\; .
 \end{matrix} \right.
\end{equation}
\end{subequations}
Using App.~\ref{app:PL}, we obtain that monopole operator with GNO-charge 
in the interior of the Weyl chamber exhibit the following 
dressings $P_{A_1\times 
A_1}(t,m_1,m_2) \slash P_{A_1\times A_1}(t,0,0) = 1+2t+t^2$, which agrees with 
our discussion above.
% 
%%%%%%%%%%%%%%%%%%%%%%%%%%%%%%%%%%%%%%%%%%%%%%%%%
%%%%%%%%%%%%%%%%%%%%%%%%%%%%%%%%%%%%%%%%%%%%%%%%%
%
\subsection{Representation \texorpdfstring{$[2,0]$}{[2,0]}}
\label{subsec:A1xA1_Rep20}
The conformal dimension for this case reads
\begin{equation}
\label{eqn:delta_A1xA1_Rep20}
 \Delta(m_1,m_2)=(N-1)\left(|m_1 + m_2| + |m_1 - m_2|\right) \; .
\end{equation}
Following the ideas outlined earlier, the conformal 
dimension~\eqref{eqn:delta_A1xA1_Rep20} defines a fan in the dominant Weyl 
chamber. In this example, $\Delta$ is already a linear function on the entire 
dominant Weyl chamber; thus, we generate a fan which just consists of one 
$2$-dimensional rational cone
 \begin{equation}
 \label{eqn:fan_A1xA1_Rep20}
  C^{(2)} = \Big\{ (m_1 \geq  m_2) \wedge (m_1 \geq -m_2) \Big\} \; .
 \end{equation} 
\subsubsection{Quotient \texorpdfstring{$\Spin(4)$}{Spin(4)}}
\begin{figure}[h]
\centering
\begin{tikzpicture}
  \coordinate (Origin)   at (0,0);
  \coordinate (XAxisMin) at (-0.5,0);
  \coordinate (XAxisMax) at (5,0);
  \coordinate (YAxisMin) at (0,-3);
  \coordinate (YAxisMax) at (0,3);
  \draw [thin, gray,-latex] (XAxisMin) -- (XAxisMax);%
  \draw [thin, gray,-latex] (YAxisMin) -- (YAxisMax);%
  \draw (5.2,-0.2) node {$m_1$};
  \draw (-0.2,3.2) node {$m_2$};
%Draw the root lattice
  \foreach \x in {0,1,...,2}{%
      \foreach \y in {-1,0,...,1}{%
        \node[draw,circle,inner sep=0.8pt,fill,black] at (2*\x,2*\y) {};
            }
            }
    \foreach \x in {0,1,...,2}{%
      \foreach \y in {0,1}{%
        \node[draw,circle,inner sep=0.8pt,fill,black] at (2*\x-1,2*\y-1) {};
            }
            }           
  \draw[black,dashed,thick]  (0,0) -- (3.2,3.2);  
  \draw[black,dashed,thick]  (0,0) -- (3.2,-3.2);
% set up the explanation
\draw (6,2) node[circle,inner sep=0.8pt,fill,black] {};
\draw (7.2,2.05) node { $K^{[0]}$ lattice};
\draw[black,thick] (1,1) circle (4pt);
\draw[black,thick] (1,-1) circle (4pt);
\draw (4,1.3) node {$S^{(2)}$};
\end{tikzpicture}
\caption{The semi-group $S^{(2)}$ and its ray-generators (black circled points) 
for the quotient $\Spin(4)$ and the representation $[2,0]$.}
\label{Fig:Hilbert_basis_Spin4_Rep20}
\end{figure}
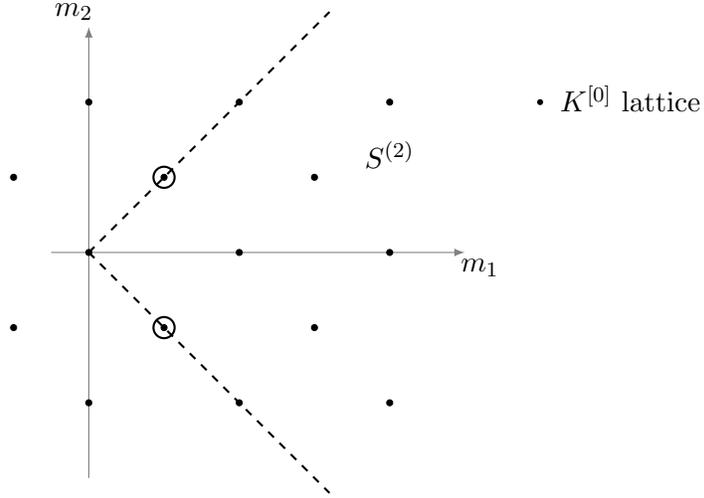

\paragraph{Hilbert basis}
Starting from the fan~\eqref{eqn:fan_A1xA1_Rep20} with the cone $C^{(2)}$, the 
Hilbert basis for the semi-group $S^{(2)} \coloneqq C^{(2)} \cap K^{[0]}$ is 
simply given by the ray generators 
 \begin{equation}
  \Hcal(S^{(2)}) = \Big\{ (1,1), (1,-1) \Big\}  \;,
 \end{equation}
see for instance Fig.~\ref{Fig:Hilbert_basis_Spin4_Rep20}.
Both minimal generators exhibit a bare monopole operator and one dressed 
operators, as explained in~\eqref{eqn:A1xA1_dress}.
\paragraph{Hilbert series}
We compute the Hilbert series to
\begin{equation}
\HS_{\Spin(4)}^{[2,0]}(t,N)= \frac{\left(1-t^{4 
N-2}\right)^2}{\left(1-t^2\right)^2 \left(1-t^{2 N-2}\right)^2 \left(1-t^{2 
N-1}\right)^2} \; ,
\label{eqn:HS_Spin4_Rep20}
\end{equation}
which is a complete intersection with $6$ generators and $2$ relations. The 
generators are given in Tab.~\ref{tab:Ops_Spin4_Rep20}.
\begin{table}[h]
\centering
 \begin{tabular}{c|c|c|c|c|c}
 \toprule
  object & $(m_1,m_2)$ & lattice & $\Delta(m_1,m_2)$ & $\Hh_{(m_1,m_2)}$ & 
dressings \\ \midrule
 Casimirs & --- &  --- & $2$ & --- & --- \\
 monopole & $(1,\pm1)$ & $K^{[0]}$ & $2N-2 $ & $\uo \times \su$ & $1$ by $\uo$\\
 \bottomrule
 \end{tabular}
\caption{Bare and dressed monopole generators for a $\Spin(4)$ gauge theory 
with matter transforming in $[2,0]$.}
\label{tab:Ops_Spin4_Rep20}
\end{table}
\paragraph{Remark} The Hilbert series~\eqref{eqn:HS_Spin4_Rep20} can be 
compared to the case of $\su$ with $n$ fundamentals and 
$n_a$ adjoints such that $2N=n+2n_a$, c.f.~\cite{Cremonesi:2013lqa}. One 
derives 
at
\begin{equation}
 \HS_{\Spin(4)}^{[2,0]}(t,N) = \HS_{\su}^{[1]+[2]}(t,n,n_a) \times  
\HS_{\su}^{[1]+[2]}(t,n,n_a) \; ,
\end{equation}
which equals the product of two $D_{2N}$ singularities. As a consequence, 
the minimal generator $(1,1)$ belongs to one $\su$ Hilbert series with adjoint 
matter content, while $(1,-1)$ belongs to the other.
\subsubsection{Quotient \texorpdfstring{$\sorm(4)$}{SO(4)}}
\begin{figure}[h] 
\centering
\begin{tikzpicture}
  \coordinate (Origin)   at (0,0);
  \coordinate (XAxisMin) at (-0.5,0);
  \coordinate (XAxisMax) at (5,0);
  \coordinate (YAxisMin) at (0,-3);
  \coordinate (YAxisMax) at (0,3);
  \draw [thin, gray,-latex] (XAxisMin) -- (XAxisMax);%
  \draw [thin, gray,-latex] (YAxisMin) -- (YAxisMax);%
  \draw (5.2,-0.2) node {$m_1$};
  \draw (-0.2,3.2) node {$m_2$};
%Draw the root lattice
  \foreach \x in {0,1,...,2}{%
      \foreach \y in {-1,0,...,1}{%
        \node[draw,circle,inner sep=0.8pt,fill,black] at (2*\x,2*\y) {};
            }
            }
    \foreach \x in {0,1,...,2}{%
      \foreach \y in {0,1}{%
        \node[draw,circle,inner sep=0.8pt,fill,black] at (2*\x-1,2*\y-1) {};
            }
            }            
%Draw K² lattice
    \foreach \x in {0,1,...,2}{%
      \foreach \y in {-1,0,...,1}{%
        \node[draw,cross out,inner sep=0.8pt,thick,black] at (2*\x-1,2*\y) {};
            }
            }
    \foreach \x in {0,1,...,2}{%
      \foreach \y in {0,1}{%
        \node[draw,cross out,inner sep=0.8pt,thick,black] at (2*\x,2*\y-1) {};
            }
            }              
  \draw[black,dashed,thick]  (0,0) -- (3.2,3.2);  
  \draw[black,dashed,thick]  (0,0) -- (3.2,-3.2);
% set up the explanation
\draw (6,2) node[circle,inner sep=0.8pt,fill,black] {};
\draw (7.2,2.05) node { $K^{[0]}$ lattice};
\draw (6,1) node[draw, cross out,inner sep=0.8pt,thick,black] {};
\draw (7.2,1.05) node { $K^{[2]}$ lattice};
\draw[black,thick] (1,1) circle (4pt);
\draw[black,thick] (1,-1) circle (4pt);
\draw[red,thick] (1,0) circle (4pt);
\draw (4,1.3) node {$S^{(2)}$};
\end{tikzpicture}
\caption{The semi-group $S^{(2)}$ and its ray-generators (black circled 
points) for the quotient $\sorm(4)$ and the representation $[2,0]$. The red 
circled lattice point completes the Hilbert basis for $S^{(2)}$.}
\label{Fig:Hilbert_basis_SO4_Rep20}
\end{figure}
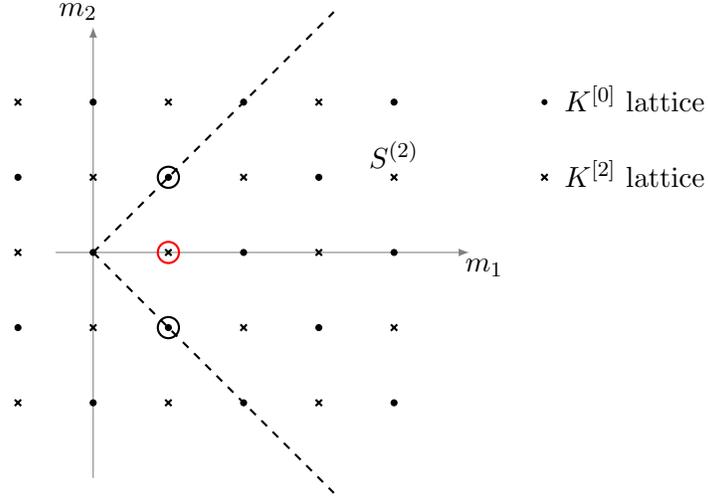

The centre of the GNO-dual $\sorm(4)$ is a $\Z_2$, which we choose to count if 
$(m_1,m_2)$ belongs to $K^{[0]}$ or $K^{[2]}$. A realisation is given by
\begin{equation}
 z^{m_1+m_2} = \begin{cases} z^{\text{even}}=1 & \for (m_1,m_2)\in K^{[0]} \; 
,\\
              z^{\text{odd}}=z & \for (m_1,m_2)\in K^{[2]} \; .
             \end{cases}
\end{equation}
In other words, $z$ is a $\Z_2$-fugacity.
\paragraph{Hilbert basis}
The semi-group $S^{(2)}\coloneqq C^{(2)} \cap \left(K^{[0]} \cup K^{[2]} 
\right)$ has a Hilbert basis as displayed in 
Fig.~\ref{Fig:Hilbert_basis_SO4_Rep20} or explicitly
\begin{equation}
   \Hcal(S^{(2)}) = \Big\{ (1,\pm1), (1,0) \Big\} \; .
\end{equation}
\paragraph{Hilbert series}
The Hilbert series for $\sorm(4)$ is given by
\begin{equation}
\label{eqn:HS_SO4_Rep20}
 \HS_{\sorm(4)}^{[2,0]}(t,z,N)=\frac{1
 +t^{2 N-2}
 +2 t^{2 N-1}
 +z t^{2 N}
 +2 z t^{2 N-1}
 +z t^{4 N-2}
 }{\left(1-t^2\right)^2 
\left(1-t^{2 N-2}\right) \left(1-z t^{2 N-2}\right)}  \; ,
\end{equation}
which is a rational function with a palindromic polynomial 
of degree $4N-2$ as numerator, while the denominator is of degree $4N$. Hence, 
the difference in degrees is $2$, i.e.\ the quaternionic dimension of the 
moduli 
space.
In addition, the denominator~\eqref{eqn:HS_SO4_Rep20} has a pole of order $4$ 
at $t\to1$, which equals the complex dimension of the moduli space.
\paragraph{Plethystic logarithm}
Analysing the PL yields for $N\geq3$
\begin{align}
 \PL(\HS_{\sorm(4)}^{[2,0]}) = 2t^2 &+ z t^{\Delta(1,0)}(1+2t^2+t^2)
 +2t^{\Delta(1,\pm1)} (1+t) \\
 &-t^{2\Delta(1,0)} (1+2(1+z)t + (6+4z)t^2 + 2(1+z)t^3 +t^4 ) +\ldots \notag
\end{align}
and for $N=2$
\begin{align}
 \PL(\HS_{\sorm(4)}^{[2,0]})= 2 t^2 + zt^2(1+2t+t^2) + 2t^2(1+t)
 -  t^4(1+ 2(1  + z) t + (6 + 
    4 z) t^2 ) + \ldots
\end{align}
such that we have generators as summarised in Tab.~\ref{tab:Ops_SO4_Rep20}.
\begin{table}[h]
\centering
 \begin{tabular}{c|c|c|c|c|c}
 \toprule
  object & $(m_1,m_2)$ & lattice & $\Delta(m_1,m_2)$ & $\Hh_{(m_1,m_2)}$ & 
dressings 
\\ \midrule
 Casimirs & --- &  --- & $2$ & --- & --- \\
 monopole & $(1,0)$ & $K^{[2]}$ & $2N-2 $ & $\uo \times \uo$ & $3$ by $\uo^2$\\
 monopole & $(1,\pm1)$ & $K^{[0]}$ & $2N-2 $ & $\uo \times \su$ & $1$ by $\uo$
\\
 \bottomrule
 \end{tabular}
\caption{Bare and dressed monopole generators for a $\sorm(4)$ gauge theory 
with matter transforming in $[2,0]$.}
\label{tab:Ops_SO4_Rep20}
\end{table}
\paragraph{Gauging a $\boldsymbol{\Z_2}$}
Although the Hilbert series~\eqref{eqn:HS_SO4_Rep20} is not a complete 
intersection, the gauging of the topological $\Z_2$ reproduces the $\Spin(4)$ 
result~\eqref{eqn:HS_Spin4_Rep20}, that is
\begin{equation}
  \HS_{\Spin(4)}^{[2,0]}(t,N) = \frac{1}{2} \left(
 \HS_{\sorm(4)}^{[2,0]}(t,z{=}1,N) + 
 \HS_{\sorm(4)}^{[2,0]}(t,z{=}-1,N)\right) \; .
 \label{eqn:SO4_gauging_Rep20}
\end{equation}
\subsubsection{Quotient \texorpdfstring{$\sorm(3) \times \su$}{SO(3)xSU(2)}}
\begin{figure}[h]
\centering
\begin{tikzpicture}
  \coordinate (Origin)   at (0,0);
  \coordinate (XAxisMin) at (-0.5,0);
  \coordinate (XAxisMax) at (5,0);
  \coordinate (YAxisMin) at (0,-3);
  \coordinate (YAxisMax) at (0,3);
  \draw [thin, gray,-latex] (XAxisMin) -- (XAxisMax);%
  \draw [thin, gray,-latex] (YAxisMin) -- (YAxisMax);%
  \draw (5.2,-0.2) node {$m_1$};
  \draw (-0.2,3.2) node {$m_2$};
%Draw the root lattice
  \foreach \x in {0,1,...,2}{%
      \foreach \y in {-1,0,...,1}{%
        \node[draw,circle,inner sep=0.8pt,fill,black] at (2*\x,2*\y) {};
            }
            }
    \foreach \x in {0,1,...,2}{%
      \foreach \y in {0,1}{%
        \node[draw,circle,inner sep=0.8pt,fill,black] at (2*\x-1,2*\y-1) {};
            }
            }                      
% Draw K¹ lattice            
    \foreach \x in {0,1,...,2}{%
      \foreach \y in {-1,0,...,1}{%
        \node[draw,diamond,inner sep=0.8pt,black] at (2*\x +1/2,2*\y +1/2) 
{};
            }
            }  
      \foreach \x in {0,1,...,2}{%
      \foreach \y in {-1,0,...,1}{%
        \node[draw,diamond,inner sep=0.8pt,black] at (2*\x-1 +1/2,2*\y-1+1/2) 
{};
            }
            }
  \draw[black,dashed,thick]  (0,0) -- (3.2,3.2);  
  \draw[black,dashed,thick]  (0,0) -- (3.2,-3.2);
% set up the explanation
\draw (6,2) node[circle,inner sep=0.8pt,fill,black] {};
\draw (7.2,2.05) node { $K^{[0]}$ lattice};
\draw (6,1) node[draw,diamond,inner sep=0.8pt,black] {};
\draw (7.2,1.05) node { $K^{[1]}$ lattice};
\draw[black,thick] (1/2,1/2) circle (4pt);
\draw[black,thick] (1,-1) circle (4pt);
\draw (4,1.3) node {$S^{(2)}$};
\end{tikzpicture}
\caption{The semi-group $S^{(2)}$ for the quotient $\sorm(3)\times \su$ and the 
representation $[2,0]$. The black circled points are the ray generators.}
\label{Fig:Hilbert_basis_SO3xSU2_Rep20}
\end{figure}
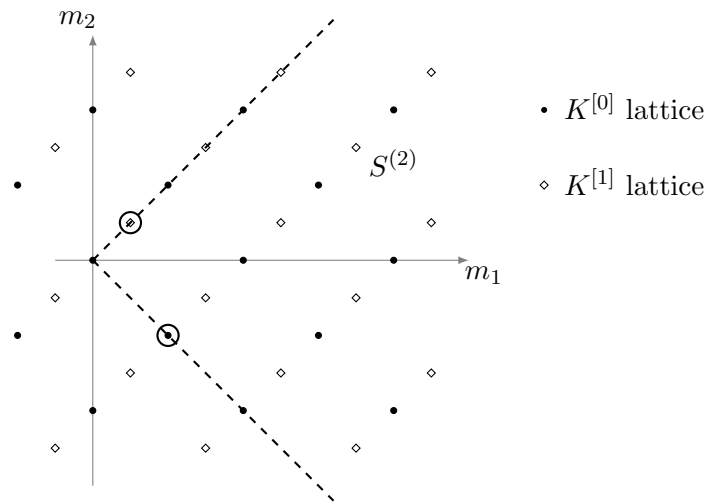

The dual group is $\su \times \sorm(3)$ and the summation extends over 
$(m_1,m_2) \in K^{[0]} \cup K^{[1]}$. The non-trivial centre $ \Z_2 \times \{1 
\}$ gives rise to a $\Z_2$-action, which we choose to distinguish the two 
lattices $K^{[0]}$ and $K^{[1]}$ as follows:
\begin{equation}
 z_1^{m_1+m_2} =\begin{cases} z_1^{p_1+p_2} = z_1^{\text{even}} =1 & \for 
(m_1,m_2) \in K^{[0]} \; , \\
   z_1^{p_1+ \tfrac{1}{2}+p_2 + \tfrac{1}{2}} = z_1^{\text{even}+1} =z_1 & \for 
(m_1,m_2) \in K^{[1]}   \; .           
                \end{cases}
\end{equation}
\paragraph{Hilbert basis}
The semi-group $S^{(2)}\coloneqq C^{(2)} \cap \left(K^{[0]} \cup K^{[1]} 
\right)$ has a Hilbert basis comprised of the ray generators. We refer to 
Fig.~\ref{Fig:Hilbert_basis_SO3xSU2_Rep20} and provide 
the minimal generators for completeness:
\begin{equation}
   \Hcal(S^{(2)}) = \Big\{ (\tfrac{1}{2},\tfrac{1}{2}), (1,-1) \Big\} \; .
  \label{eqn:Hilbert_basis_SO3xSU2_Rep20}
\end{equation}
\paragraph{Hilbert series}
Computing the Hilbert series and using explicitly the $\Z_2$-properties of 
$z_1$ yields
\begin{equation}
\HS_{\sorm(3)\times \su}^{[2,0]}(t,z_1,N)= \frac{\left(1-t^{2 N}\right) 
\left(1-t^{4 N-2}\right)}{\left(1-t^2\right)^2 \left(1-t^{2 N-2}\right) 
\left(1-t^{2 N-1}\right) \left(1-z_1 t^{N-1}\right) 
\left(1-z_1 t^{N}\right)} \; ,
\label{eqn:HS_SO3xSU2_Rep20}
\end{equation}
which is a complete intersection with $6$ generators and $2$ relations. The 
generators are displayed in Tab.~\ref{tab:Ops_SO3xSU2_Rep20}.
\begin{table}[h]
\centering
 \begin{tabular}{c|c|c|c|c|c}
 \toprule
  object & $(m_1,m_2)$ & lattice & $\Delta(m_1,m_2)$ & $\Hh_{(m_1,m_2)}$ & 
dressings 
\\ \midrule
 Casimirs & --- &  --- & $2$ & --- & --- \\
 monopole & $(\tfrac{1}{2},\tfrac{1}{2})$ & $K^{[1]}$ & $N-1 $ & $\uo \times 
\su$ & $1$ by $\uo$\\
 monopole & $(1,-1)$ & $K^{[0]}$ & $2N-2 $ & $\uo \times \su$ & $1$ by $\uo$ \\
 \bottomrule
 \end{tabular}
\caption{Bare and dressed monopole generators for a $\sorm(3)\times \su$ gauge 
theory with matter transforming in $[2,0]$.}
\label{tab:Ops_SO3xSU2_Rep20}
\end{table}
\paragraph{Remark}
Comparing to the case of $\su$ with $n_a$ adjoints and $\sorm(3)$ with $n$ 
fundamentals presented in~\cite{Cremonesi:2013lqa}, we can re-express the 
Hilbert series~\eqref{eqn:HS_SO3xSU2_Rep20} as the product
\begin{equation}
 \HS_{\sorm(3)\times \su}^{[2,0]}(t,z_1,N) = \HS_{\sorm(3)}^{[1]}(t,z_1,n=N) 
\times \HS_{\su}^{[2]}(t,n_a=N) \; ,
\end{equation}
where the $z_1$-grading belongs to $\sorm(3)$ with $N$ fundamentals. The 
minimal generator $(\tfrac{1}{2},\tfrac{1}{2})$ is the minimal generator for 
$\sorm(3)$ with $N$ fundamentals, while $(1,-1)$ is the minimal generator for 
$\su$ with $N$ adjoints.
\subsubsection{Quotient \texorpdfstring{$\su\times \sorm(3)$}{SU(2)xSO(3)}}
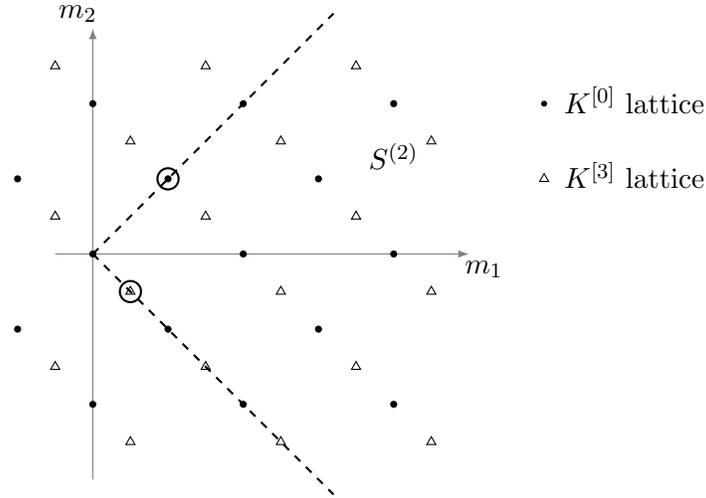
\begin{figure}[h]
\centering
\begin{tikzpicture}
  \coordinate (Origin)   at (0,0);
  \coordinate (XAxisMin) at (-0.5,0);
  \coordinate (XAxisMax) at (5,0);
  \coordinate (YAxisMin) at (0,-3);
  \coordinate (YAxisMax) at (0,3);
  \draw [thin, gray,-latex] (XAxisMin) -- (XAxisMax);%
  \draw [thin, gray,-latex] (YAxisMin) -- (YAxisMax);%
  \draw (5.2,-0.2) node {$m_1$};
  \draw (-0.2,3.2) node {$m_2$};
%Draw the root lattice
  \foreach \x in {0,1,...,2}{%
      \foreach \y in {-1,0,...,1}{%
        \node[draw,circle,inner sep=0.8pt,fill,black] at (2*\x,2*\y) {};
            }
            }
    \foreach \x in {0,1,...,2}{%
      \foreach \y in {0,1}{%
        \node[draw,circle,inner sep=0.8pt,fill,black] at (2*\x-1,2*\y-1) {};
            }
            }            
%Draw K³ lattice
    \foreach \x in {0,1,...,2}{%
      \foreach \y in {-1,0,...,1}{%
        \node[draw,regular polygon,regular polygon sides=3,inner 
sep=0.7pt,black] at (2*\x-1+1/2,2*\y+1/2) {};
            }
            }
     \foreach \x in {0,1,...,2}{%
      \foreach \y in {-1,0,...,1}{%
        \node[draw,regular polygon,regular polygon sides=3,inner 
sep=0.7pt,black] at 
(2*\x +1/2,2*\y-1 +1/2) {};
            }
            }
  \draw[black,dashed,thick]  (0,0) -- (3.2,3.2);  
  \draw[black,dashed,thick]  (0,0) -- (3.2,-3.2);
\draw (6,2) node[circle,inner sep=0.8pt,fill,black] {};
\draw (7.2,2.05) node { $K^{[0]}$ lattice};
\draw (6,1) node[draw,regular polygon,regular polygon sides=3,inner 
sep=0.7pt,black] {};
\draw (7.2,1.05) node { $K^{[3]}$ lattice};
\draw[black,thick] (1/2,-1/2) circle (4pt);
\draw[black,thick] (1,1) circle (4pt);
\draw (4,1.3) node {$S^{(2)}$};
\end{tikzpicture}
\caption{The semi-group $S^{(2)}$ for the quotient $\su \times \sorm(3)$ and 
the representation $[2,2]$. The black circled points are the ray generators.}
\label{Fig:Hilbert_basis_SU2xSO3_Rep20}
\end{figure}

The dual group is $\sorm(3) \times \su$ and the summation extends over 
$(m_1,m_2) \in K^{[0]} \cup K^{[3]}$. The non-trivial centre $ \{1\} \times 
\Z_2 $ gives rise to a $\Z_2$-action, which we choose to distinguish the two 
lattices $K^{[0]}$ and $K^{[3]}$ as follows:
\begin{equation}
 z_2^{p_1+p_2} =\begin{cases} z_2^{\text{even}} =1 & \for 
(m_1,m_2) \in K^{[0]} \; , \\
    z_2^{\text{odd}} =z_2 & \for 
(m_1,m_2) \in K^{[3]}   \; .  \end{cases}
\end{equation}
\paragraph{Hilbert basis}
The semi-group $S^{(2)}\coloneqq C^{(2)} \cap \left(K^{[0]} \cup K^{[3]} 
\right)$  has as Hilbert basis the set of 
ray generators
\begin{equation}
   \Hcal(S^{(2)}) = \Big\{ (1,1), (\tfrac{1}{2},-\tfrac{1}{2}) \Big\} \; .
  \label{eqn:Hilbert_basis_SU2xSO3_Rep20}
\end{equation}
Fig.~\ref{Fig:Hilbert_basis_SU2xSO3_Rep20} depicts the situation.
We observe that bases~\eqref{eqn:Hilbert_basis_SO3xSU2_Rep20} 
and~\eqref{eqn:Hilbert_basis_SU2xSO3_Rep20} are related by reflection along the 
$m_2=0$ axis, which in turn corresponds to the interchange of $K^{[1]}$ and 
$K^{[3]}$.
\paragraph{Hilbert series}
Similar to the previous case, employing the $\Z_2$-properties of $z_2$ we 
obtain the following Hilbert series:
\begin{equation}
\HS_{\su \times \sorm(3)}^{[2,0]}(t,z_2,N)= \frac{\left(1-t^{2 N}\right) 
\left(1-t^{4 N-2}\right)}{\left(1-t^2\right)^2 \left(1-t^{2 N-2}\right) 
\left(1-t^{2 N-1}\right) \left(1-z_2 t^{N-1}\right) 
\left(1-z_2 t^{N}\right)} \; ,
\end{equation}
which is a complete intersection with $6$ generators and $2$ relations. We 
summarise the generators in Tab.~\ref{tab:Ops_SU2xSO3_Rep20}.
\begin{table}[h]
\centering
 \begin{tabular}{c|c|c|c|c|c}
 \toprule
  object & $(m_1,m_2)$ & lattice & $\Delta(m_1,m_2)$ & $\Hh_{(m_1,m_2)}$ & 
dressings \\ \midrule
 Casimirs & --- &  --- & $2$ & --- & --- \\
 monopole & $(\tfrac{1}{2},-\tfrac{1}{2})$ & $K^{[3]}$ & $N-1 $ & $\uo \times 
\su$ & $1$ by $\uo$\\
 monopole & $(1,1)$ & $K^{[0]}$ & $2N-2 $ & $\uo \times \su$ & $1$ by $\uo$ \\
 \bottomrule
 \end{tabular}
\caption{Bare and dressed monopole generators for a $\su\times\sorm(3)$ gauge 
theory with matter transforming in $[2,0]$.}
\label{tab:Ops_SU2xSO3_Rep20}
\end{table}
\paragraph{Remark}
Also, the equivalence
\begin{equation}
  \HS_{\sorm(3)\times \su}^{[2,0]}(t,z_1,N) \xleftrightarrow{\quad z_1 
\leftrightarrow z_2 \quad }
\HS_{\su \times \sorm(3)}^{[2,0]}(t,z_2,N) 
\label{eqn:Exchange_z1-z2_Rep20}
\end{equation}
holds, which then also implies
 \begin{equation}
 \HS_{\su \times \sorm(3)}^{[2,0]}(t,z_2,N) = \HS_{\sorm(3)}^{[1]}(t,z_2,n=N) 
\times \HS_{\su}^{[2]}(t,n_a=N) \; .
\end{equation}
Thus, the moduli space is a product of two complete intersections.
\subsubsection{Quotient \texorpdfstring{$\mathrm{PSO}(4)$}{PSO(4)}}
Taking the quotient with respect to the entire centre of $\widetilde{\sorm(4)}$ 
yields the projective group $\mathrm{PSO}(4)$, which has as GNO-dual 
$\Spin(4)\cong \su\times \su$. Consequently, the summation extends over the 
whole weight lattice $K^{[0]}\cup K^{[1]}\cup K^{[2]}\cup K^{[3]}$ and there 
is an 
action of $\Z_2 \times \Z_2$ on this lattice, which is chosen as displayed in 
Tab.~\ref{tab:PSO4_Z2-grading}.
\begin{table}[h]
\centering
\begin{tabular}{c|c||c}
\toprule
 lattice & $\Z_2\times \Z_2$ &  $\widetilde{\Z}_2\times \widetilde{\Z}_2$ \\ 
\midrule
 $K^{[0]}$ & $(z_1)^0,(z_2)^0 $ & $(w_1)^0,(w_2)^0 $ \\
 $K^{[1]}$ & $(z_1)^1,(z_2)^0 $ & $(w_1)^1,(w_2)^1 $ \\
 $K^{[2]}$ & $(z_1)^0,(z_2)^1 $ & $(w_1)^0,(w_2)^1 $ \\
 $K^{[3]}$ & $(z_1)^1,(z_2)^1 $ & $(w_1)^1,(w_2)^0 $ \\
 \bottomrule
\end{tabular}
\caption{The $\Z_2\times \Z_2$ distinguishes the four different lattice 
$K^{[j]}$, $j=0,1,2,3$. The choice of fugacities $z_1$, $z_2$ is used in the 
computation, while the second choice $w_1$, $w_2$ is convenient for gauging 
$\mathrm{PSO}(4)$ to $\su\times \sorm(3)$.}
\label{tab:PSO4_Z2-grading}
\end{table}
\paragraph{Hilbert basis}
The semi-group $S^{(2)}\coloneqq C^{(2)} \cap \left(K^{[0]} \cup K^{[1]} \cup 
K^{[2]} \cup K^{[3]} \right)$ has a Hilbert basis that is determined by the 
ray generators. Fig.~\ref{Fig:Hilbert_basis_PSO4_Rep20} depicts the 
situation and the Hilbert basis reads
\begin{equation}
   \Hcal(S^{(2)}) = \Big\{ (\tfrac{1}{2},\tfrac{1}{2}) ,  
(\tfrac{1}{2},-\tfrac{1}{2}) \Big\} \; .
  \label{eqn:Hilbert_basis_PSO4_Rep20}
\end{equation}
\begin{figure}[h]
 \centering
\begin{tikzpicture}
  \coordinate (Origin)   at (0,0);
  \coordinate (XAxisMin) at (-0.5,0);
  \coordinate (XAxisMax) at (5,0);
  \coordinate (YAxisMin) at (0,-3);
  \coordinate (YAxisMax) at (0,3);
  \draw [thin, gray,-latex] (XAxisMin) -- (XAxisMax);%
  \draw [thin, gray,-latex] (YAxisMin) -- (YAxisMax);%
  \draw (5.2,-0.2) node {$m_1$};
  \draw (-0.2,3.2) node {$m_2$};
%Draw the root lattice
  \foreach \x in {0,1,...,2}{%
      \foreach \y in {-1,0,...,1}{%
        \node[draw,circle,inner sep=0.8pt,fill,black] at (2*\x,2*\y) {};
            }
            }
    \foreach \x in {0,1,...,2}{%
      \foreach \y in {0,1}{%
        \node[draw,circle,inner sep=0.8pt,fill,black] at (2*\x-1,2*\y-1) {};
            }
            }            
%Draw K² lattice
    \foreach \x in {0,1,...,2}{%
      \foreach \y in {-1,0,...,1}{%
        \node[draw,cross out,inner sep=0.8pt,thick,black] at (2*\x-1,2*\y) {};
            }
            }
    \foreach \x in {0,1,...,2}{%
      \foreach \y in {0,1}{%
        \node[draw,cross out,inner sep=0.8pt,thick,black] at (2*\x,2*\y-1) {};
            }
            }            
% Draw K¹ lattice            
    \foreach \x in {0,1,...,2}{%
      \foreach \y in {-1,0,...,1}{%
        \node[draw,diamond,inner sep=0.8pt,black] at (2*\x +1/2,2*\y +1/2) 
{};
            }
            }  
      \foreach \x in {0,1,...,2}{%
      \foreach \y in {-1,0,...,1}{%
        \node[draw,diamond,inner sep=0.8pt,black] at (2*\x-1 +1/2,2*\y-1+1/2) 
{};
            }
            } 
%Draw K³ lattice
    \foreach \x in {0,1,...,2}{%
      \foreach \y in {-1,0,...,1}{%
        \node[draw,regular polygon, regular polygon sides=3,inner 
sep=0.7pt,black] at (2*\x-1+1/2,2*\y+1/2) {};
            }
            }
     \foreach \x in {0,1,...,2}{%
      \foreach \y in {-1,0,...,1}{%
        \node[draw,regular polygon, regular polygon sides=3,inner 
sep=0.7pt,black] at (2*\x +1/2,2*\y-1 +1/2) {};
            }
            }
  \draw[black,dashed,thick]  (0,0) -- (3.2,3.2);  
  \draw[black,dashed,thick]  (0,0) -- (3.2,-3.2);
\draw (6,3) node[circle,inner sep=0.8pt,fill,black] {};
\draw (7.2,3.05) node { $K^{[0]}$ lattice};
\draw (6,2) node[draw,diamond,inner sep=0.8pt,black] {};
\draw (7.2,2.05) node { $K^{[1]}$ lattice};
\draw (6,1) node[draw,cross out,inner sep=0.8pt,thick,black] {};
\draw (7.2,1.05) node { $K^{[2]}$ lattice};
\draw (6,0 ) node[draw,regular polygon,regular polygon sides=3,inner 
sep=0.7pt,black] {};
\draw (7.2,0.05) node { $K^{[3]}$ lattice};
\draw[black,thick] (1/2,1/2) circle (4pt);
\draw[black,thick] (1/2,-1/2) circle (4pt);
\draw (4,1.3) node {$S^{(2)}$};
\end{tikzpicture}
\caption{The semi-group $S^{(2)}$ and its ray-generators (black circled 
points) for the quotient $\mathrm{PSO}(4)$ and the representation $[2,0]$.}
\label{Fig:Hilbert_basis_PSO4_Rep20}
\end{figure}
\paragraph{Hilbert series}
An evaluation of the Hilbert series yields
\begin{equation}
\HS_{\mathrm{PSO}(4)}^{[2,0]}(t,z_1,z_2,N)= \frac{\left(1-t^{2 
N}\right)^2}{\left(1-t^2\right)^2 \left(1-z_1 t^{N-1}\right) \left(1-z_1 
t^{N}\right) \left(1-z_1 z_2 t^{N-1}\right) \left(1-z_1 z_2 
t^{N}\right)} \; ,
\label{eqn:HS_PSO4_Rep20}
\end{equation}
which is a complete intersection with $6$ generators and $2$ 
relations. Tab.~\ref{tab:Ops_PSO4_Rep20} summarises the generators with their 
properties.
\begin{table}[h]
\centering
 \begin{tabular}{c|c|c|c|c|c}
 \toprule
  object & $(m_1,m_2)$ & lattice & $\Delta(m_1,m_2)$ & $\Hh_{(m_1,m_2)}$ & 
dressings 
\\ \midrule
 Casimirs & --- &  --- & $2$ & --- & --- \\
 monopole & $(\tfrac{1}{2},\tfrac{1}{2})$ & $K^{[1]}$ & $N-1 $ & $\uo \times 
\su$ & $1$ by $\uo$\\
 monopole & $(\tfrac{1}{2},-\tfrac{1}{2})$ & $K^{[3]}$ & $N-1 $ & $\uo \times 
\su$ & $1$ by $\uo$ \\
\bottomrule
 \end{tabular}
\caption{Bare and dressed monopole generators for a $\mathrm{PSO}(4)$ gauge 
theory with matter transforming in $[2,0]$.}
\label{tab:Ops_PSO4_Rep20}
\end{table}
\paragraph{Gauging a $\boldsymbol{\Z_2}$}
Now, we utilise the $\Z_2 \times \Z_2$ global symmetry to recover the Hilbert 
series for \emph{all} five quotients solely from the $\mathrm{PSO}(4)$ result.
Firstly, to obtain the $\sorm(4)$ result, we need to average out the 
contributions of $K^{[1]}$ and $K^{[3]}$, which is achieved for $z_1 \to \pm 1$ 
(we also relabel $z_2$ for consistence), see also 
Tab.~\ref{tab:PSO4_Z2-grading}. This yields
\begin{subequations}
\label{eqn:PSO4_gauging_Rep20}
\begin{equation}
 \HS_{\sorm(4)}^{[2,0]}(t,z,N) = \frac{1}{2} \left(
 \HS_{\mathrm{PSO}(4)}^{[2,0]}(t,z_1{=}1,z_2{=}z,N) + 
 \HS_{\mathrm{PSO}(4)}^{[2,0]}(t,z_1{=}-1,z_2{=}z,N)\right) \; .
\end{equation}
Secondly, a subsequent gauging leads to the $\Spin(4)$ result as demonstrated 
in~\eqref{eqn:SO4_gauging_Rep20}, because one averages the $K^{[2]}$ 
contributions out.
Thirdly, one can gauge the other $\Z_2$-factor corresponding to $z_2 \to \pm1$, 
which then eliminates the contributions of $K^{[2]}$ and $K^{[3]}$ due to the 
choices of Tab.~\ref{tab:PSO4_Z2-grading}. The result then reads
\begin{equation}
  \HS_{\sorm(3)\times\su}^{[2,0]}(t,z_1,N) = \frac{1}{2} \left(
 \HS_{\mathrm{PSO}(4)}^{[2,0]}(t,z_1,z_2{=}1,N) + 
 \HS_{\mathrm{PSO}(4)}^{[2,0]}(t,z_1,z_2{=}-1,N)\right) \; .
\end{equation}
Lastly, for obtaining the $\su\times \sorm(3)$ Hilbert series one needs to 
eliminate the $K^{[1]}$ and $K^{[2]}$ contributions. For that, we have to 
redefine the $\Z_2$-fugacities conveniently. One choice is
\begin{equation}
 z_1 \cdot z_2 \mapsto w_1 \; , \qquad z_1 \mapsto w_1 \cdot w_2 \; , \und z_2 
\mapsto w_2 \; ,
\label{eqn:redefine_Z2-grading_A1xA1}
\end{equation}
which is consistent in $\Z_2 \times \Z_2$. The effect on the lattices is 
summarised in Tab.~\ref{tab:PSO4_Z2-grading}. Hence, $w_2\to \pm1$ has the 
desired effect and leads to 
\begin{equation}
  \HS_{\su\times\sorm(3)}^{[2,0]}(t,z_2{=} w_1,N) = \frac{1}{2} \left(
 \HS_{\mathrm{PSO}(4)}^{[2,0]}(t,w_1,w_2{=}1,N) + 
 \HS_{\mathrm{PSO}(4)}^{[2,0]}(t,w_1,w_2{=}-1,N)\right) \; .
\end{equation}
\end{subequations}
Consequently, the Hilbert series for \emph{all} five quotients can be computed 
from the $\mathrm{PSO}(4)$-result by gauging $\Z_2$-factors.
\paragraph{Remark}
As for most of the cases in this section, the Hilbert 
series~\eqref{eqn:HS_PSO4_Rep20} can be written as a product of two complete 
intersections. Employing the results of~\cite{Cremonesi:2013lqa} for $\sorm(3)$ 
with $n$ fundamentals, we obtain
 \begin{equation}
 \HS_{\mathrm{PSO}(4)}^{[2,0]}(t,z_1,z_2,N) = \HS_{\sorm(3)}^{[1]}(t,z_1,n=N) 
\times  \HS_{\sorm(3)}^{[1]}(t,z_1 z_2,n=N) \; .
\end{equation}
% % 
%
%%%%%%%%%%%%%%%%%%%%%%%%%%%%%%%%%%%%%%%%%%%%%%%%%%%%%%%%%%%%%%%%%%%%%%%%%%
%%%%%%%%%%%%%%%%%%%%%%%%%%%%%%%%%%%%%%%%%%%%%%%%%%%%%%%%%%%%%%%%%%%%%%%%%%
%
\subsection{Representation \texorpdfstring{$[2,2]$}{[2,2]}}
Let us use the representation $[2,2]$ to further compare the results for the 
five different quotient groups. The conformal dimension reads
\begin{equation}
\label{eqn:delta_A1xA1_Rep22}
\Delta(m_1,m_2)= N (\left| m_1-m_2\right| +\left| m_1+m_2\right| 
+2 \left| m_1\right| +2 \left| m_2\right| )-\left| 
m_1-m_2\right| -\left| m_1+m_2\right| \; .
\end{equation}
As described in the introduction, the conformal 
dimension~\eqref{eqn:delta_A1xA1_Rep22} defines a fan in the dominant Weyl 
chamber, which is spanned by two $2$-dimensional rational cones
 \begin{equation}
 \label{eqn:fan_A1xA1_Rep22}
  C_{\pm}^{(2)} = \Big\{ (m_1 \geq \pm m_2) \wedge (m_2 \geq \pm 0) \Big\} \; .
 \end{equation}
%
%%%%%%%%%%%%%%%%%%%%%%%%%%%%%%%%%%%%%%%%%%%%%%%%%%%%%%%%%%%%%%%%%%%%%%%%%%%% 
% 
\subsubsection{Quotient \texorpdfstring{$\Spin(4)$}{Spin(4)}}
\begin{figure}[h]
\centering
\begin{tikzpicture}
  \coordinate (Origin)   at (0,0);
  \coordinate (XAxisMin) at (-0.5,0);
  \coordinate (XAxisMax) at (5,0);
  \coordinate (YAxisMin) at (0,-3);
  \coordinate (YAxisMax) at (0,3);
  \draw [thin, gray,-latex] (XAxisMin) -- (XAxisMax);%
  \draw [thin, gray,-latex] (YAxisMin) -- (YAxisMax);%
  \draw (5.2,-0.2) node {$m_1$};
  \draw (-0.2,3.2) node {$m_2$};
%Draw the root lattice
  \foreach \x in {0,1,...,2}{%
      \foreach \y in {-1,0,...,1}{%
        \node[draw,circle,inner sep=0.8pt,fill,black] at (2*\x,2*\y) {};
            }
            }
    \foreach \x in {0,1,...,2}{%
      \foreach \y in {0,1}{%
        \node[draw,circle,inner sep=0.8pt,fill,black] at (2*\x-1,2*\y-1) {};
            }
            }           
  \draw[black,dashed,thick]  (0,0) -- (3.2,3.2);  
  \draw[black,dashed,thick]  (0,0) -- (3.2,-3.2);
  \draw[black,dashed,thick]  (0,0) -- (4.5,0);  
\draw (6,2) node[circle,inner sep=0.8pt,fill,black] {};
\draw (7.2,2.05) node { $K^{[0]}$ lattice};
\draw (2,0) circle (4pt);
\draw (1,1) circle (4pt);
\draw (1,-1) circle (4pt);
\draw (4,1.3) node {$S_+^{(2)}$};
\draw (4,-1.3) node {$S_-^{(2)}$};
\end{tikzpicture}
\caption{The semi-groups and their ray-generators (black circled points) for 
the quotient $\Spin(4)$ and the representation $[2,2]$.}
\label{Fig:Hilbert_basis_Spin4_Rep22}
\end{figure}
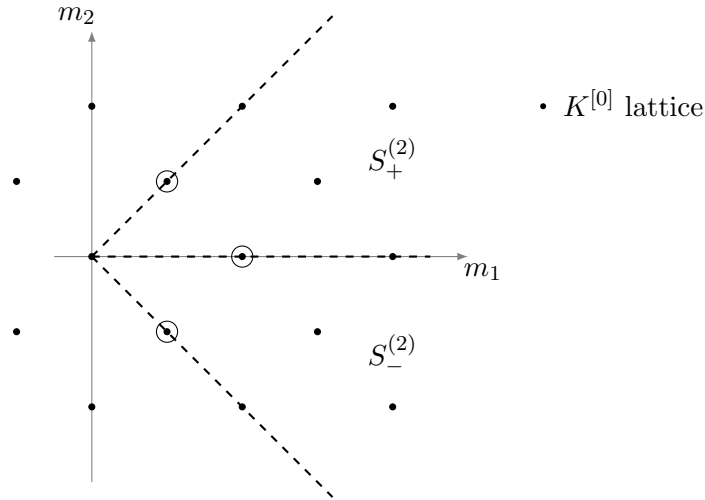

\paragraph{Hilbert basis}
Starting from the fan~\eqref{eqn:fan_A1xA1_Rep22} with cones $C_{\pm}^{(2)}$, 
the Hilbert bases for the semi-groups $S_{\pm}^{(2)} \coloneqq C_{\pm}^{(2)} 
\cap K^{[0]}$ are simply given by the ray generators, see for instance 
Fig.~\ref{Fig:Hilbert_basis_Spin4_Rep22}. 
 \begin{equation}
  \Hcal(S_{\pm}^{(2)}) = \Big\{ (1,\pm1), (2,0) \Big\} \; .
 \end{equation}
\paragraph{Hilbert series}
The GNO-dual $\sorm(3)\times \sorm(3)$ has a trivial centre and the  
Hilbert series reads
\begin{equation}
\label{eqn:HS_Spin4_Rep22}
 \HS_{\Spin(4)}^{[2,2]}(t,N)=\frac{1+t^{6 N-2}+2 t^{6 N-1}+2 t^{8 
N-3}+t^{8 N-2}+t^{14 N-4}}{\left(1-t^2\right)^2 \left(1-t^{6 
N-2}\right) \left(1-t^{8 N-4}\right)} \; .
\end{equation}
The numerator of~\eqref{eqn:HS_Spin4_Rep22} is a palindromic polynomial of 
degree $14N-4$; while the denominator is a polynomial of degree $14N-2$. Hence, 
the difference in degree is two, which equals the quaternionic dimension of the 
moduli space.
In addition, denominator of~\eqref{eqn:HS_Spin4_Rep22} has a pole of order four 
at $t=1$, which equals the complex dimension of the moduli space.
\paragraph{Plethystic logarithm}
The plethystic logarithm takes the form
\begin{align}
 \PL(\HS_{\Spin(4)}^{[2,2]})= 2t^2 &+ 2t^{\Delta(1,\pm1)} (1+t) 
 + t^{\Delta(2,0)}(1+2t+t^2) \\
 &-t^{2\Delta(1,\pm1)} (1 + 2t + 3t^2 + 2t^3 + 4t^4 + 2t^5 + 3t^6 + 2t^7 
+t^8)+\ldots \notag
\end{align}
The appearing terms agree with the minimal generators of the Hilbert 
bases~\eqref{Fig:Hilbert_basis_Spin4_Rep22}.
One has two independent degree two Casimir invariants. Further, there are 
monopole operators of GNO-charge $(1,1)$ and $(1,-1)$ at conformal 
dimension $6N-2$ with an independent dressed monopole generator of conformal 
dimension $6N-1$ for both charges.
Moreover, there is a monopole operator of GNO-charge $(2,0)$ at dimension 
$8N-4$ with two dressing operators at dimension $8N-3$ and 
one at $8N-2$. 
% 
%%%%%%%%%%%%%%%%%%%%%%%%%%%%%%%%%%%%%%%%
%%%%%%%%%%%%%%%%%%%%%%%%%%%%%%%%%%%%%%%%%
% 
\subsubsection{Quotient \texorpdfstring{$\sorm(4)$}{SO(4)}}
\paragraph{Hilbert basis}
The semi-groups $S_{\pm}^{(2)}\coloneqq C_{\pm}^{(2)} \cap \left(K^{[0]} \cup 
K^{[2]} \right)$ have Hilbert bases which again equal (the now different) ray 
generators. The situation is depicted in Fig.~\ref{Fig:Hilbert_basis_SO4_Rep22} 
and the Hilbert bases are as follows:
\begin{equation}
   \Hcal(S_{\pm}^{(2)}) = \Big\{ (1,\pm1), (1,0) \Big\} \; .
\end{equation}
\begin{figure}[h]
\centering
\begin{tikzpicture}
  \coordinate (Origin)   at (0,0);
  \coordinate (XAxisMin) at (-0.5,0);
  \coordinate (XAxisMax) at (5,0);
  \coordinate (YAxisMin) at (0,-3);
  \coordinate (YAxisMax) at (0,3);
  \draw [thin, gray,-latex] (XAxisMin) -- (XAxisMax);%
  \draw [thin, gray,-latex] (YAxisMin) -- (YAxisMax);%
  \draw (5.2,-0.2) node {$m_1$};
  \draw (-0.2,3.2) node {$m_2$};
%Draw the root lattice
  \foreach \x in {0,1,...,2}{%
      \foreach \y in {-1,0,...,1}{%
        \node[draw,circle,inner sep=0.8pt,fill,black] at (2*\x,2*\y) {};
            }
            }
    \foreach \x in {0,1,...,2}{%
      \foreach \y in {0,1}{%
        \node[draw,circle,inner sep=0.8pt,fill,black] at (2*\x-1,2*\y-1) {};
            }
            }            
%Draw K² lattice
    \foreach \x in {0,1,...,2}{%
      \foreach \y in {-1,0,...,1}{%
        \node[draw,cross out,inner sep=0.8pt,thick,black] at (2*\x-1,2*\y) {};
            }
            }
    \foreach \x in {0,1,...,2}{%
      \foreach \y in {0,1}{%
        \node[draw,cross out,inner sep=0.8pt,thick,black] at (2*\x,2*\y-1) {};
            }
            }           
  \draw[black,dashed,thick]  (0,0) -- (3.2,3.2);  
  \draw[black,dashed,thick]  (0,0) -- (3.2,-3.2);
  \draw[black,dashed,thick]  (0,0) -- (4.5,0);  
\draw (6,2) node[circle,inner sep=0.8pt,fill,black] {};
\draw (7.2,2.05) node { $K^{[0]}$ lattice};
\draw (6,1) node[draw,cross out,inner sep=0.8pt,thick,black] {};
\draw (7.2,1.05) node { $K^{[2]}$ lattice};
\draw (1,0) circle (4pt);
\draw (1,1) circle (4pt);
\draw (1,-1) circle (4pt);
\draw (4,1.5) node {$S_+^{(2)}$};
\draw (4,-1.5) node {$S_-^{(2)}$};
\end{tikzpicture}
\caption{The semi-groups and their ray-generators (black circled points) for 
the quotient $\sorm(4)$ and the representation $[2,2]$.}
\label{Fig:Hilbert_basis_SO4_Rep22}
\end{figure}
\paragraph{Hilbert series}
The Hilbert series reads
\begin{equation}
\label{eqn:HS_SO4_Rep22}
 \HS_{\sorm(4)}^{[2,2]}(t,z,N)= \frac{1+ z t^{4 N}+2 z t^{4 N-1}+t^{6 
N-2}+2 t^{6 N-1}+z t^{10 N-2}}{\left(1-t^2\right)^2 \left(1-z t^{4 
N-2}\right)\left(1-t^{6 N-2}\right) } \; .
\end{equation}
The numerator of~\eqref{eqn:HS_SO4_Rep22} is a palindromic polynomial of degree 
$10N-2$ (neglecting the dependence on $z$); while the 
denominator is a polynomial of degree $10N$. Hence, the difference in degree 
is two equals the quaternionic dimension of the moduli space.
Moreover, the denominator has a pole of order four at $t=1$, which equals the 
complex dimension of the moduli space.
\paragraph{Plethystic logarithm} 
Studying the PL, we observe
\begin{align}
 \PL(\HS_{\sorm(4)}^{[2,2]})=2t^2 &+ z t^{\Delta(1,0)} (1+2t^2+t) + 2 
t^{\Delta(1,\pm1)} (1+t) \\
 &-t^{2\Delta(1,0)+2}(3+2t^2 + t^2 + 2t^3 + 4t^4 + 2t^5 + t^6 + 2 t^7 + 3 t^8) 
+ \ldots \notag
\end{align}
such that we can associate the generators as follows: two degree two Casimir 
invariants of $\sorm(4)$, i.e.\ the quadratic Casimir and the Pfaffian; 
A monopole of GNO-charge $(1,0)\in K^{[2]}$ at conformal dimension 
$4N-2$ with two dressings at dimension 
$4N-1$ and another dressing at $4N$; and two monopole operators of GNO-charges 
$(1,1)$, $(1,-1) \in K^{[0]}$ at dimension $6N-2$ one dressed monopoles at 
dimension $6N-1 $ each.
\paragraph{Gauging the $\Z_2$} In addition, one can gauge the topological 
$\Z_2$ in~\eqref{eqn:HS_SO4_Rep22} and obtains
\begin{equation}
\label{eqn:SO4_gauging_Rep22}
\HS_{\Spin(4)}^{[2,2]}(t,N)= \frac{1}{2} 
\left(\HS_{\sorm(4)}^{[2,2]}(t,z{=}1,N)+\HS_{\sorm(4)}^{[2,2]}(t,z{=}-1,N) 
\right) \; .
\end{equation}
% 
%%%%%%%%%%%%%%%%%%%%%%%%%%%%%%%%%%%%%%%%%%%%%%%%%%%%%%%%%%%%%%%%%%%%%%
%%%%%%%%%%%%%%%%%%%%%%%%%%%%%%%%%%%%%%%%%%%%%%%%%%%%%%%%%%%%%%%%%%%%%%
%
\subsubsection{Quotient \texorpdfstring{$\sorm(3)\times \su $}{SO(3)xSU(2)}}
\begin{figure}[h]
\centering
\begin{tikzpicture}
  \coordinate (Origin)   at (0,0);
  \coordinate (XAxisMin) at (-0.5,0);
  \coordinate (XAxisMax) at (5,0);
  \coordinate (YAxisMin) at (0,-3);
  \coordinate (YAxisMax) at (0,3);
  \draw [thin, gray,-latex] (XAxisMin) -- (XAxisMax);%
  \draw [thin, gray,-latex] (YAxisMin) -- (YAxisMax);%
  \draw (5.2,-0.2) node {$m_1$};
  \draw (-0.2,3.2) node {$m_2$};
%Draw the root lattice
  \foreach \x in {0,1,...,2}{%
      \foreach \y in {-1,0,...,1}{%
        \node[draw,circle,inner sep=0.8pt,fill,black] at (2*\x,2*\y) {};
            }
            }
    \foreach \x in {0,1,...,2}{%
      \foreach \y in {0,1}{%
        \node[draw,circle,inner sep=0.8pt,fill,black] at (2*\x-1,2*\y-1) {};
            }
            }                      
% Draw K¹ lattice            
    \foreach \x in {0,1,...,2}{%
      \foreach \y in {-1,0,...,1}{%
        \node[draw,diamond,inner sep=0.8pt,black] at (2*\x +1/2,2*\y +1/2) {};
            }
            }  
      \foreach \x in {0,1,...,2}{%
      \foreach \y in {-1,0,...,1}{%
        \node[draw,diamond,inner sep=0.8pt,black] at (2*\x-1 
+1/2,2*\y-1+1/2) {};
            }
            }
  \draw[black,dashed,thick]  (0,0) -- (3.2,3.2);  
  \draw[black,dashed,thick]  (0,0) -- (3.2,-3.2);
  \draw[black,dashed,thick]  (0,0) -- (4.5,0);  
\draw (6,2) node[circle,inner sep=0.8pt,fill,black] {};
\draw (7.2,2.05) node { $K^{[0]}$ lattice};
\draw (6,1) node[draw,diamond,inner sep=0.8pt,black] {};
\draw (7.2,1.05) node { $K^{[1]}$ lattice};
\draw[black,thick] (1/2,1/2) circle (4pt);
\draw[black,thick] (1,-1) circle (4pt);
\draw[red,thick] (1+1/2,-1+1/2) circle (4pt);
\draw[black,thick] (2,0) circle (4pt);
\draw (4,1.5) node {$S_+^{(2)}$};
\draw (4,-1.5) node {$S_-^{(2)}$};
\end{tikzpicture}
\caption{The semi-groups for the quotient $\sorm(3)\times \su$ and the 
representation $[2,2]$. The black circled points are the ray generators and the 
red circled point completes the Hilbert basis for $S_{-}^{(2)}$.}
\label{Fig:Hilbert_basis_SO3xSU2_Rep22}
\end{figure}
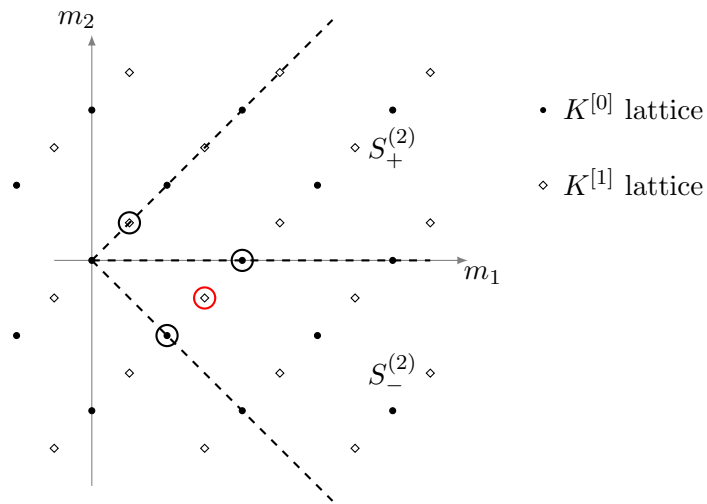

\paragraph{Hilbert basis}
The semi-groups $S_{\pm}^{(2)}\coloneqq C_{\pm}^{(2)} \cap \left(K^{[0]} \cup 
K^{[1]} \right)$ have Hilbert bases that go beyond the set of 
ray generators. We refer to Fig.~\ref{Fig:Hilbert_basis_SO3xSU2_Rep22} and 
the Hilbert bases are obtained as follows:
\begin{equation}
   \Hcal(S_{+}^{(2)}) = \Big\{ (\tfrac{1}{2},\tfrac{1}{2}), (2,0) \Big\} 
\und
   \Hcal(S_{-}^{(2)}) = \Big\{ (1,-1), (\tfrac{3}{2},-\tfrac{1}{2}) , (2,0) 
\Big\} \; .
\label{eqn:Hilbert_basis_SO3xSU2_Rep22}
\end{equation}
\paragraph{Hilbert series}
The Hilbert series is computed to be
\begin{subequations}
\label{eqn:HS_SO3xSU2_Rep22}
\begin{align}
\HS_{\sorm(3)\times \su}^{[2,2]}(t,z_1,N)&= 
\frac{R(t,z_1,N)}{\left(1-t^2\right)^2 \left(1-t^{6 N-2}\right) \left(1-t^{8 
N-4}\right)} \; , \\
R(t,z_1,N)&=1 
+z_1 t^{3 N}
+z_1 t^{3 N-1}
+t^{6 N-2}
+2 t^{6 N-1}
+z_1 t^{7 N-3}
+2 z_1 t^{7 N-2}
+z_1 t^{7 N-1} \notag\\
&\qquad 
+2 t^{8 N-3}
+t^{8 N-2}
+z_1 t^{11 N-4}
+z_1 t^{11 N-3}
+t^{14 N-4} \; .
\end{align}
\end{subequations}
Again, the numerator of~\eqref{eqn:HS_SO3xSU2_Rep22} is a palindromic 
polynomial of degree $14N-4$; while the denominator is a polynomial of degree 
$14N-2$. Hence, the difference in degree is two, which matches the quaternionic 
dimension of the moduli space.
Also, the denominator has a pole of order four at $t=1$, which equals the 
complex dimension of the moduli space.
\paragraph{Plethystic logarithm}
The inspection of the PL for $N\geq 2$ reveals 
\begin{align}
 \PL(\HS_{\sorm(3)\times \su}^{[2,2]}) = 2 t^2 &+ z_1 
t^{\Delta(\frac{1}{2},\frac{1}{2})}(1 + t) + 
t^{\Delta(1,\pm1)}(1+t-t^2) \\
&+z_1 t^{\Delta(1+\frac{1}{2},-1+\frac{1}{2})} 
(1+2t+t^2) 
+ t^{\Delta(2,0)}(1+2t+t^2)  \notag \\
&-z_1 t^{3 \Delta(\frac{1}{2},\frac{1}{2})} (1+2t+t^2)+ \ldots  \;. \notag
\end{align}
We summarise the generators in Tab.~\ref{tab:Ops_SO3xSU2_Rep22}.
\begin{table}[h]
\centering
 \begin{tabular}{c|c|c|c|c}
 \toprule
  $(m_1,m_2)$ & lattice  & $\Delta(m_1,m_2)$ & $\Hh_{(m_1,m_2)}$ & dressings \\ 
\midrule
  $(\tfrac{1}{2},\tfrac{1}{2})$ & $K^{[1]}$ & $3N-1$ & $\uo\times \su$ & $1$ by 
$\uo$\\
  $(1,- 1)$ & $K^{[0]}$ & $6N-2$ & $\uo\times \su$ & $1$ by $\uo$ \\
  $(\tfrac{3}{2},-\tfrac{1}{2})$ & $K^{[1]}$ & $7N-3$ & $\uo\times \uo$ & 
$3$ by $\uo^2$ \\
  $(2,0)$ & $K^{[0]}$ & $8N-4$ & $\uo\times \uo$ & $3$ by $\uo^2$ \\
  \bottomrule
 \end{tabular}
\caption{The generators for the chiral ring of a $\sorm(3)\times \su$ gauge 
theory with matter in $[2,2]$.}
\label{tab:Ops_SO3xSU2_Rep22}
\end{table}
%
%
%%%%%%%%%%%%%%%%%%%%%%%%%%%%%%%%%%%%%%%%%%%%%%%%%%%%%%%%%%%%%%%%%%%%%%%%%
%%%%%%%%%%%%%%%%%%%%%%%%%%%%%%%%%%%%%%%%%%%%%%%%%%%%%%%%%%%%%%%%%%%%%%%%%
% 
\subsubsection{Quotient \texorpdfstring{$\su\times\sorm(3) $}{SU(2)xSO(3)}}
\begin{figure}[h]
\centering
\begin{tikzpicture}
  \coordinate (Origin)   at (0,0);
  \coordinate (XAxisMin) at (-0.5,0);
  \coordinate (XAxisMax) at (5,0);
  \coordinate (YAxisMin) at (0,-3);
  \coordinate (YAxisMax) at (0,3);
  \draw [thin, gray,-latex] (XAxisMin) -- (XAxisMax);%
  \draw [thin, gray,-latex] (YAxisMin) -- (YAxisMax);%
  \draw (5.2,-0.2) node {$m_1$};
  \draw (-0.2,3.2) node {$m_2$};
%Draw the root lattice
  \foreach \x in {0,1,...,2}{%
      \foreach \y in {-1,0,...,1}{%
        \node[draw,circle,inner sep=0.8pt,fill,black] at (2*\x,2*\y) {};
            }
            }
    \foreach \x in {0,1,...,2}{%
      \foreach \y in {0,1}{%
        \node[draw,circle,inner sep=0.8pt,fill,black] at (2*\x-1,2*\y-1) {};
            }
            }            
%Draw K³ lattice
    \foreach \x in {0,1,...,2}{%
      \foreach \y in {-1,0,...,1}{%
        \node[draw,regular polygon,regular polygon sides=3,inner 
sep=0.7pt,black] at (2*\x-1+1/2,2*\y+1/2) {};
            }
            }
     \foreach \x in {0,1,...,2}{%
      \foreach \y in {-1,0,...,1}{%
        \node[draw,regular polygon,regular polygon sides=3,inner 
sep=0.7pt,black] at (2*\x +1/2,2*\y-1 +1/2) {};
            }
            }
  \draw[black,dashed,thick]  (0,0) -- (3.2,3.2);  
  \draw[black,dashed,thick]  (0,0) -- (3.2,-3.2);
  \draw[black,dashed,thick]  (0,0) -- (4.5,0);  
\draw (6,2) node[circle,inner sep=0.8pt,fill,black] {};
\draw (7.2,2.05) node { $K^{[0]}$ lattice};
\draw (6,1) node[draw,regular polygon,regular polygon sides=3,inner 
sep=0.7pt,black] {};
\draw (7.2,1.05) node { $K^{[3]}$ lattice};
\draw[black,thick] (1/2,-1/2) circle (4pt);
\draw[black,thick] (1,1) circle (4pt);
\draw[red,thick] (3/2,1/2) circle (4pt);
\draw[black,thick] (2,0) circle (4pt);
\draw (4,1.5) node {$S_+^{(2)}$};
\draw (4,-1.5) node {$S_-^{(2)}$};
\end{tikzpicture}
\caption{The semi-groups for the quotient $\su \times \sorm(3)$ and the 
representation $[2,2]$. The black circled points are the ray generators and the 
red circled point completes the Hilbert basis for $S_{+}^{(2)}$.}
\label{Fig:Hilbert_basis_SU2xSO3_Rep22}
\end{figure}
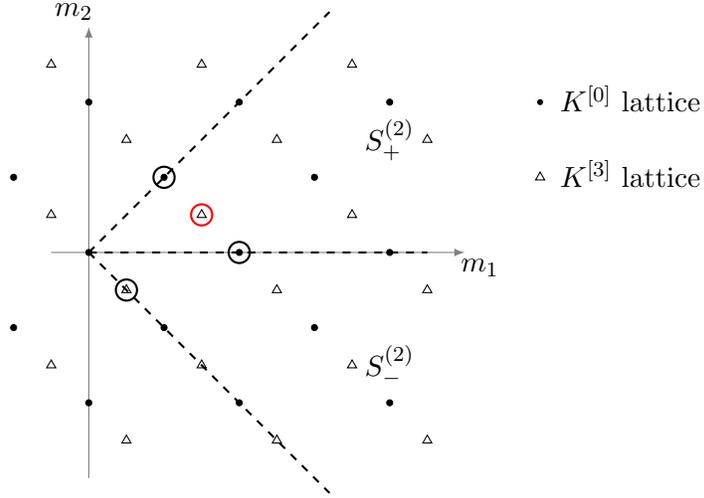

\paragraph{Hilbert basis}
The semi-groups $S_{\pm}^{(2)}\coloneqq C_{\pm}^{(2)} \cap \left(K^{[0]} \cup 
K^{[3]} \right)$ have Hilbert bases that go beyond the set 
of ray generators. Fig.~\ref{Fig:Hilbert_basis_SU2xSO3_Rep22} depicts the 
situation and the Hilbert bases are computed to be
\begin{equation}
   \Hcal(S_{+}^{(2)}) = \Big\{ (1,1), (\tfrac{3}{2},\tfrac{1}{2}) , (2,0) 
\Big\} \und
   \Hcal(S_{-}^{(2)}) = \Big\{ (\tfrac{1}{2},-\tfrac{1}{2}), (2,0) \Big\} \; . 
\label{eqn:Hilbert_basis_SU2xSO3_Rep22}
\end{equation}
We observe that the bases~\eqref{eqn:Hilbert_basis_SO3xSU2_Rep22} 
and~\eqref{eqn:Hilbert_basis_SU2xSO3_Rep22} are related by reflection along the 
$m_2=0$ axis, which in turn corresponds to the interchange of $K^{[1]}$ and 
$K^{[3]}$.
\paragraph{Hilbert series}
The Hilbert series reads
\begin{subequations}
\label{eqn:HS_SU2xSO3_Rep22}
\begin{align}
\HS_{\su \times \sorm(3)}^{[2,2]}(t,z_2,N)&= 
\frac{R(t,z_2,N)}{\left(1-t^2\right)^2 \left(1-t^{6 N-2}\right) \left(1-t^{8 
N-4}\right)} \; , \\
R(t,z_2,N)&=1
+z_2 t^{3 N}
+z_2 t^{3 N-1}
+t^{6 N-2}
+2 t^{6 N-1}
+z_2 t^{7 N-3}
+2 z_2 t^{7 N-2}
+z_2 t^{7 N-1} \notag\\
&\qquad
+2 t^{8 N-3}
+t^{8 N-2}
+z_2 t^{11 N-4}
+z_2 t^{11 N-3}
+t^{14 N-4} \; .
\end{align}
\end{subequations}
The numerator of~\eqref{eqn:HS_SU2xSO3_Rep22} is palindromic polynomial of 
degree $14N-4$; while the denominator is a polynomial of degree $14N-2$. Hence, 
the difference in degree is two, which equals the quaternionic dimension of the 
moduli space.
In addition, the denominator has a pole of order four at $t=1$, which matches 
the complex dimension of the moduli space.

As before, comparing the quotients $\sorm(3)\times \su$ and $\su \times 
\sorm(3)$ as well as the symmetry of~\eqref{eqn:delta_A1xA1_Rep22}, it is 
natural to expect the relationship
\begin{equation}
 \HS_{\sorm(3)\times \su}^{[2,2]}(t,z_1,N) 
 \xleftrightarrow{\quad z_1\leftrightarrow z_2 \quad }
 \HS_{\su \times \sorm(3)}^{[2,2]}(t,z_2,N) \; ,
 \label{eqn:Exchange_z1-z2_Rep22}
\end{equation}
which is verified explicitly for~\eqref{eqn:HS_SO3xSU2_Rep22} 
and~\eqref{eqn:HS_SU2xSO3_Rep22}.
\paragraph{Plethystic logarithm}
The equivalence to $\sorm(3)\times \su$ is further confirmed by 
the inspection of the PL for $N\geq 2$  
\begin{align}
 \PL(\HS_{\su \times \sorm(3)}^{[2,2]})= 2 t^2 &+ z_2 
t^{\Delta(\tfrac{1}{2},-\tfrac{1}{2})}(1 + t) + t^{\Delta(1,1)}(1+t-t^2) \\
&+z_2 t^{\Delta(\tfrac{3}{2},\tfrac{1}{2})} (1+2t+t^2) 
+ t^{\Delta(2,0)}(1+2t+t^2)+ \ldots \notag
\end{align}
where we can summarise the monopole generators as in 
Tab.~\ref{tab:Ops_SU2xSO3_Rep22}.
\begin{table}[h]
\centering
 \begin{tabular}{c|c|c|c|c}
 \toprule
  $(m_1,m_2)$ & lattice  & $\Delta(m_1,m_2)$ & $\Hh_{(m_1,m_2)}$ & dressings \\ 
\midrule
  $(\tfrac{1}{2},-\tfrac{1}{2})$ & $K^{[3]}$ & $3N-1$ & $\uo\times \su$ & $1$ 
by $\uo$ \\
  $(1, 1)$ & $K^{[0]}$ & $6N-2$ & $\uo\times \su$ & $1$ by $\uo$ \\
  $(\tfrac{3}{2},\tfrac{1}{2})$ & $K^{[3]}$ & $7N-3$ & $\uo\times \uo$ & 
$3$ by $\uo^2$ \\
  $(2,0)$ & $K^{[0]}$ & $8N-4$ & $\uo\times \uo$ & $3 $ by $\uo^2$ \\
  \bottomrule
 \end{tabular}
\caption{The generators for the chiral ring of a $\su\times \sorm(3)$ gauge 
theory with matter in $[2,2]$.}
\label{tab:Ops_SU2xSO3_Rep22}
\end{table}
Note the change in GNO-charges in accordance with the use of $K^{[3]}$ instead 
of $K^{[1]}$.
% 
% 
%%%%%%%%%%%%%%%%%%%%%%%%%%%%%%%%%%%%%%%%%%%%%%%%%%%%%%%%%%%%%%%%%%%%%%%%%
%%%%%%%%%%%%%%%%%%%%%%%%%%%%%%%%%%%%%%%%%%%%%%%%%%%%%%%%%%%%%%%%%%%%%%%%%
% 
\subsubsection{Quotient \texorpdfstring{$\mathrm{PSO}(4) $}{PSO(4)}}
\paragraph{Hilbert basis}
The semi-groups $S_{\pm}^{(2)}\coloneqq C_{\pm}^{(2)} \cap \left(K^{[0]} \cup 
K^{[1]} \cup K^{[2]} 
\cup K^{[3]} \right)$ have Hilbert bases that are determined by the 
ray generators. Fig.~\ref{Fig:Hilbert_basis_PSO4_Rep22} depicts the 
situation and the Hilbert bases read
\begin{equation}
\label{eqn:Hilbert_basis_PSO4_Rep22}
   \Hcal(S_{\pm}^{(2)}) = \Big\{ (\tfrac{1}{2},\pm\tfrac{1}{2}) , (1,0) 
\Big\} \; .
\end{equation}
\begin{figure}[h]
 \centering
\begin{tikzpicture}
  \coordinate (Origin)   at (0,0);
  \coordinate (XAxisMin) at (-0.5,0);
  \coordinate (XAxisMax) at (5,0);
  \coordinate (YAxisMin) at (0,-3);
  \coordinate (YAxisMax) at (0,3);
  \draw [thin, gray,-latex] (XAxisMin) -- (XAxisMax);%
  \draw [thin, gray,-latex] (YAxisMin) -- (YAxisMax);%
  \draw (5.2,-0.2) node {$m_1$};
  \draw (-0.2,3.2) node {$m_2$};
%Draw the root lattice
  \foreach \x in {0,1,...,2}{%
      \foreach \y in {-1,0,...,1}{%
        \node[draw,circle,inner sep=0.8pt,fill,black] at (2*\x,2*\y) {};
            }
            }
    \foreach \x in {0,1,...,2}{%
      \foreach \y in {0,1}{%
        \node[draw,circle,inner sep=0.8pt,fill,black] at (2*\x-1,2*\y-1) {};
            }
            }            
%Draw K² lattice
    \foreach \x in {0,1,...,2}{%
      \foreach \y in {-1,0,...,1}{%
        \node[draw,cross out,inner sep=0.8pt,thick,black] at (2*\x-1,2*\y) {};
            }
            }
    \foreach \x in {0,1,...,2}{%
      \foreach \y in {0,1}{%
        \node[draw,cross out,inner sep=0.8pt,thick,black] at (2*\x,2*\y-1) {};
            }
            }            
% Draw K¹ lattice            
    \foreach \x in {0,1,...,2}{%
      \foreach \y in {-1,0,...,1}{%
        \node[draw,diamond,inner sep=0.8pt,black] at (2*\x +1/2,2*\y +1/2) {};
            }
            }  
      \foreach \x in {0,1,...,2}{%
      \foreach \y in {-1,0,...,1}{%
        \node[draw,diamond,inner sep=0.8pt,black] at (2*\x-1 
+1/2,2*\y-1+1/2) {};
            }
            } 
%Draw K³ lattice
    \foreach \x in {0,1,...,2}{%
      \foreach \y in {-1,0,...,1}{%
        \node[draw,regular polygon,regular polygon sides=3,inner 
sep=0.7pt,black] at (2*\x-1+1/2,2*\y+1/2) {};
            }
            }
     \foreach \x in {0,1,...,2}{%
      \foreach \y in {-1,0,...,1}{%
        \node[draw,regular polygon,regular polygon sides=3,inner 
sep=0.7pt,black] at (2*\x +1/2,2*\y-1 +1/2) {};
            }
            }
  \draw[black,dashed,thick]  (0,0) -- (3.2,3.2);  
  \draw[black,dashed,thick]  (0,0) -- (3.2,-3.2);
  \draw[black,dashed,thick]  (0,0) -- (4.5,0);
\draw (6,3) node[circle,inner sep=0.8pt,fill,black] {};
\draw (7.2,3.05) node { $K^{[0]}$ lattice};
\draw (6,2) node[draw,diamond,inner sep=0.8pt,black] {};
\draw (7.2,2.05) node { $K^{[1]}$ lattice};
\draw (6,1) node[draw,cross out,inner sep=0.8pt,thick,black] {};
\draw (7.2,1.05) node { $K^{[2]}$ lattice};
\draw (6,0) node[draw,regular polygon,regular polygon sides=3,inner 
sep=0.7pt,black] {};
\draw (7.2,0.05) node { $K^{[3]}$ lattice};
\draw[black,thick] (1/2,1/2) circle (4pt);
\draw[black,thick] (1/2,-1/2) circle (4pt);
\draw[black,thick] (1,0) circle (4pt);
\draw (4,1.5) node {$S_+^{(2)}$};
\draw (4,-1.5) node {$S_-^{(2)}$};
\end{tikzpicture}
\caption{The semi-groups and their ray-generators (black circled points) for 
the quotient $\mathrm{PSO}(4)$ and the representation $[2,2]$.}
\label{Fig:Hilbert_basis_PSO4_Rep22}
\end{figure}
\paragraph{Hilbert series}
The Hilbert series reads
\begin{subequations}
\label{eqn:HS_PSO4_Rep22}
 \begin{align}
 \HS_{\mathrm{PSO}(4)}^{[2,2]}(t,z_1,z_2,N)&= 
\frac{R(t,z_1,z_2,N)}{\left(1-t^2\right)^2 \left(1-t^{6 N-2}\right) \left(1-z_2 
t^{4 N-2}\right)} \; ,\\
R(t,z_1,z_2,N)&=1
+z_1 t^{3 N}
+z_1 t^{3 N-1}
+z_1 z_2 t^{3 N}
+z_1 z_2 t^{3 N-1}
+z_2 t^{4 N}
+2 z_2 t^{4 N-1} \\
&\qquad
+t^{6 N-2}
+2 t^{6 N-1}
+z_1 z_2 t^{7 N-2}
+z_1 z_2 t^{7 N-1} \notag \\
&\qquad
+z_1 t^{7 N-2}
+z_1 t^{7 N-1}
+z_2 t^{10 N-2}\notag \; .
 \end{align}
\end{subequations}
The numerator of~\eqref{eqn:HS_PSO4_Rep22} is palindromic polynomial of degree 
$10N-2$; while the denominator is a polynomial of degree $10N$. Hence, the 
difference in degree is two, which corresponds to the quaternionic dimension 
of the moduli space.
Similarly to the previous cases, the denominator of~\eqref{eqn:HS_PSO4_Rep22} 
has a pole of order four at $t=1$, which equals the complex dimension of the 
moduli space.
\paragraph{Gauging a $\Z_2$}
As before, by gauging the $\Z_2$-factor corresponding to $z_1$ we recover the 
$\sorm(4)$-result
\begin{subequations}
\label{eqn:PSO4_gauging_Rep22}
\begin{equation}
 \HS_{\sorm(4)}^{[2,2]}(t,z,N) = \frac{1}{2} 
\left(\HS_{\mathrm{PSO}(4)}^{[2,2]}(t,z_1{=}1,z_2{=}z,N) +
\HS_{\mathrm{PSO}(4)}^{[2,2]}(t,z_1{=}-1,z_2{=}z,N)\right) \; ,
\end{equation}
while gauging the $\Z_2$-factor with fugacity $z_2$ provides the 
$\sorm(3)\times\su$-result
\begin{equation}
 \HS_{\sorm(3)\times\su}^{[2,2]}(t,z_1,N) = \frac{1}{2} 
\left(\HS_{\mathrm{PSO}(4)}^{[2,2]}(t,z_1,z_2{=}1,N) +
\HS_{\mathrm{PSO}(4)}^{[2,2]}(t,z_1,z_2{=}{-1},N)\right) \; .
\end{equation}
Furthermore, employing the redefined fugacities $w_1$, $w_2$ 
of~\eqref{eqn:redefine_Z2-grading_A1xA1} one reproduces the $\su\times 
\sorm(3)$ 
Hilbert series as follows:
\begin{equation}
 \HS_{\su\times\sorm(3)}^{[2,2]}(t,z_2{=}w_1,N) = \frac{1}{2} 
\left(\HS_{\mathrm{PSO}(4)}^{[2,2]}(t,w_1,w_2{=}1,N) +
\HS_{\mathrm{PSO}(4)}^{[2,2]}(t,w_1,w_2{=}{-1},N)\right) \; .
\end{equation}
\end{subequations}
Therefore, one can obtain the Hilbert series for \emph{all} five quotients from 
the $\mathrm{PSO}(4)$-result~\eqref{eqn:HS_PSO4_Rep22} by employing the 
$\Z_2$-gaugings~\eqref{eqn:SO4_gauging_Rep22} 
and~\eqref{eqn:PSO4_gauging_Rep22}.
\paragraph{Plethystic logarithm}
Inspecting the PL leads to
\begin{align}
 \PL(\HS_{\mathrm{PSO}(4)}^{[2,2]})= 2 t^2 + z_1 
t^{\Delta(\tfrac{1}{2},\tfrac{1}{2})}(1+t) 
 + z_1 z_2 t^{\Delta(\tfrac{1}{2},-\tfrac{1}{2})}(1+t)
 +z_2 t^{\Delta(1,0)} (1+2t+t^2) +\ldots
\end{align}
such that we can summarise the monopole generators as in 
Tab.~\ref{tab:Ops_PSO4_Rep22}.
\begin{table}[h]
\centering
 \begin{tabular}{c|c|c|c|c}
 \toprule
  $(m_1,m_2)$ & lattice  & $\Delta(m_1,m_2)$ & $\Hh_{(m_1,m_2)}$ & dressings \\ 
\midrule
  $(\tfrac{1}{2},\tfrac{1}{2})$ & $K^{[1]}$ & $3N-1$ & $\uo\times \su$ & $1$ 
by $\uo$\\
  $(\tfrac{1}{2},-\tfrac{1}{2})$ & $K^{[3]}$ & $3N-1$ & $\uo\times \su$ & 
$1$ by $\uo$ \\
  $(1, 0)$ & $K^{[2]}$ & $4N-2$ & $\uo\times \uo$ & $3$ by $\uo^2$ \\
  \bottomrule
 \end{tabular}
\caption{The generators for the chiral ring of a $\mathrm{PSO}(4)$ gauge 
theory with matter in $[2,2]$.}
\label{tab:Ops_PSO4_Rep22}
\end{table}
%
%%%%%%%%%%%%%%%%%%%%%%%%%%%%%%%%%%%%%%%%%%%%%%%%%
%%%%%%%%%%%%%%%%%%%%%%%%%%%%%%%%%%%%%%%%%%%%%%%%%
%
\subsection{Representation \texorpdfstring{$[4,2]$}{[4,2]}}
The conformal dimension for this case reads
\begin{align}
 \Delta(m_1,m_2)&=N\big( 
 \left| 3 m_1-m_2\right|
 +\left| m_1-3 m_2\right| 
 +\left| m_1+m_2\right| 
 +3 \left| m_1-m_2\right|  
 +2 \left| m_1\right| 
 +2 \left| m_2\right|\big)  
\notag \\*
&\qquad -|m_1 + m_2| - |m_1 - m_2|\; .
\label{eqn:delta_A1xA1_Rep42}
\end{align}
The interesting feature of this representation is its asymmetric behaviour 
under exchange of $m_1$ and $m_2$.

As before, the conformal dimension~\eqref{eqn:delta_A1xA1_Rep42} defines a fan 
in the dominant Weyl chamber of, which is spanned by three $2$-dimensional cones
\begin{subequations}
\label{eqn:fan_A1xA1_Rep42}
 \begin{align}
  C_{1}^{(2)} &= \Big\{ (m_1 \geq - m_2) \wedge (m_2 \leq 0) \Big\} \; , \\
  C_{2}^{(2)} &= \Big\{ (m_1 \geq 3 m_2) \wedge (m_2 \geq 0) \Big\} \; , \\
  C_{3}^{(2)} &= \Big\{ (m_1 \geq  m_2) \wedge  (m_1 \leq 3 m_2)  \Big\} \; .
 \end{align}
\end{subequations}
\subsubsection{Quotient \texorpdfstring{$\Spin(4)$}{Spin(4)}}
\begin{figure}[h]
\centering
\begin{tikzpicture}
  \coordinate (Origin)   at (0,0);
  \coordinate (XAxisMin) at (-0.5,0);
  \coordinate (XAxisMax) at (5,0);
  \coordinate (YAxisMin) at (0,-3);
  \coordinate (YAxisMax) at (0,3);
  \draw [thin, gray,-latex] (XAxisMin) -- (XAxisMax);%
  \draw [thin, gray,-latex] (YAxisMin) -- (YAxisMax);%
  \draw (5.2,-0.2) node {$m_1$};
  \draw (-0.2,3.2) node {$m_2$};
%Draw the root lattice
  \foreach \x in {0,1,...,2}{%
      \foreach \y in {-1,0,...,1}{%
        \node[draw,circle,inner sep=0.8pt,fill,black] at (2*\x,2*\y) {};
            }
            }
    \foreach \x in {0,1,...,2}{%
      \foreach \y in {0,1}{%
        \node[draw,circle,inner sep=0.8pt,fill,black] at (2*\x-1,2*\y-1) {};
            }
            }           
  \draw[black,dashed,thick]  (0,0) -- (3.2,3.2);  
  \draw[black,dashed,thick]  (0,0) -- (3*1.5,1*1.5);  
  \draw[black,dashed,thick]  (0,0) -- (3.2,-3.2);
  \draw[black,dashed,thick]  (0,0) -- (4.5,0);  
\draw (6,2) node[circle,inner sep=0.8pt,fill,black] {};
\draw (7.2,2.05) node { $K^{[0]}$ lattice};
\draw[thick,black] (2,0) circle (4pt);
\draw[thick,black] (1,1) circle (4pt);
\draw[thick,black] (1,-1) circle (4pt);
\draw[thick,black] (3,1) circle (4pt);
\draw (3.8,2.3) node {$S_3^{(2)}$};
\draw (3.8,0.7) node {$S_2^{(2)}$};
\draw (4,-1.3) node {$S_1^{(2)}$};
\end{tikzpicture}
\caption{The semi-groups and their ray-generators (black circled points) for 
the quotient $\Spin(4)$ and the representation $[4,2]$.}
\label{Fig:Hilbert_basis_Spin4_Rep42}
\end{figure}
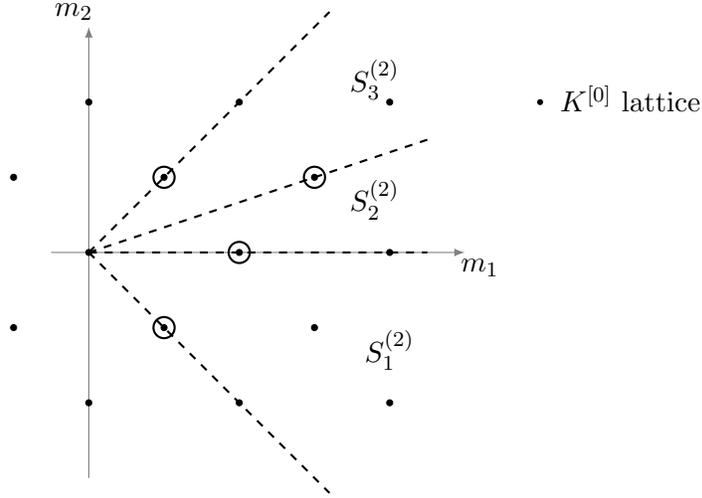

\paragraph{Hilbert basis}
Starting from the fan~\eqref{eqn:fan_A1xA1_Rep42} with cones $C_p^{(2)}$ 
($p=1,2,3$), the Hilbert bases for the semi-groups $S_p^{(2)} \coloneqq 
C_p^{(2)} \cap K^{[0]}$ are simply given by the ray generators, see for 
instance Fig.~\ref{Fig:Hilbert_basis_Spin4_Rep42}. 
 \begin{equation}
  \Hcal(S_{1}^{(2)}) = \Big\{ (2,0), (1,-1) \Big\}   , \quad 
  \Hcal(S_{2}^{(2)}) = \Big\{ (3,1), (2,0) \Big\}  , \quad 
  \Hcal(S_{3}^{(2)}) = \Big\{ (1,1), (3,1) \Big\}    .
 \end{equation}
\paragraph{Hilbert series}
The Hilbert series reads
\begin{subequations}
\label{eqn:HS_Spin4_Rep42}
\begin{align}
\HS_{\Spin(4)}^{[4,2]}(t,N)&= \frac{R(t,N)}{\left(1-t^2\right)^2 \left(1-t^{18 
N-2}\right) \left(1-t^{20 N-4}\right) \left(1-t^{26 
N-6}\right)} \; , \\
R(t,N)&= 1
+t^{10 N-2}(1+t)
+t^{18 N-1}
+t^{20 N-4}(1+3t+t^2)
\\
&\qquad
+ t^{26 N-5}(2+t)
-t^{28 N-4}(1+t)
+t^{36 N-7}(1+t)
\notag \\
&\qquad
-t^{38 N-6}(1+2t)
-t^{44 N-8}(1+3t+t^2)
-t^{46 N-9} \notag \\
&\qquad
-t^{54 N-9}(1+t)
-t^{64 N-10} \notag \; .
\end{align}
\end{subequations}
The numerator of~\eqref{eqn:HS_Spin4_Rep42} is an anti-palindromic polynomial 
of degree $64N-10$, while the denominator is of degree $64N-8$. Consequently, 
the difference in degree is two.
Moreover, the rational function~\eqref{eqn:HS_Spin4_Rep42} has a pole of order 
four as $t\to1$ because $R(t{=}1,N)=0$, but $\tfrac{\diff}{\diff t} R(t,N) 
|_{t=1}\neq0$.
\paragraph{Plethystic logarithm}
Inspecting the PL yields for $N\geq 3$
\begin{align}
 \PL(\HS_{\Spin(4)}^{[4,2]}) = 2t^2 &+ t^{\Delta(1,1)} (1+t)  + 
t^{\Delta(1,-1)} 
(1+t) 
  + t^{\Delta(2,0)} (1+2t)  + t^{\Delta(3,1)} (1+2t+t^2) \\
 &-t^{\Delta(1,1)+\Delta(1,-1)} (1+2t+t^2)
 -t^{\Delta(1,1)+\Delta(2,0)}(1+3t+3t^2+t^3) + \ldots \notag 
\end{align}
leads to an identification of generators as in Tab.~\ref{tab:Ops_Spin4_Rep42}.
\begin{table}[h]
\centering
 \begin{tabular}{c|c|c|c|c|c}
 \toprule
  object & $(m_1,m_2)$ & lattice & $\Delta(m_1,m_2)$ & $\Hh_{(m_1,m_2)}$ & \#
dressings \\ \midrule
 Casimirs & --- &  --- & $2$ & --- & --- \\
 monopole & $(1,1)$ & $K^{[0]}$ & $10N-2 $ & $\uo \times 
\su$ & $1$ by $\uo$ \\
 monopole & $(1,-1)$ & $K^{[0]}$ & $18N-2 $ & $\uo \times 
\su$ & $1$ by $\uo$ \\
 monopole & $(2,0)$ & $K^{[0]}$ & $20N-4 $ & $\uo \times 
\uo$ & $2$ by $\uo^2$\\
 monopole & $(3,1)$ & $K^{[0]}$ & $26N-6 $ & $\uo \times 
\uo$ & $3$ by $\uo^2$ \\
\bottomrule
 \end{tabular}
\caption{The chiral ring generators for a $\Spin(4)$ gauge 
theory with matter transforming in $[4,2]$.}
\label{tab:Ops_Spin4_Rep42}
\end{table}
We observe that $(2,0)$ has only $2$ dressings, although we would 
expect $3$. We know from other examples that there should be a relation at $2 
\Delta(1,1) +2 =20N-2$ which is precisely the dimension of the second dressing 
of $(2,0)$.
\subsubsection{Quotient \texorpdfstring{$\sorm(4)$}{SO(4)}}
\begin{figure}[h]
\centering
\begin{tikzpicture}
  \coordinate (Origin)   at (0,0);
  \coordinate (XAxisMin) at (-0.5,0);
  \coordinate (XAxisMax) at (5,0);
  \coordinate (YAxisMin) at (0,-3);
  \coordinate (YAxisMax) at (0,3);
  \draw [thin, gray,-latex] (XAxisMin) -- (XAxisMax);%
  \draw [thin, gray,-latex] (YAxisMin) -- (YAxisMax);%
  \draw (5.2,-0.2) node {$m_1$};
  \draw (-0.2,3.2) node {$m_2$};
%Draw the root lattice
  \foreach \x in {0,1,...,2}{%
      \foreach \y in {-1,0,...,1}{%
        \node[draw,circle,inner sep=0.8pt,fill,black] at (2*\x,2*\y) {};
            }
            }
    \foreach \x in {0,1,...,2}{%
      \foreach \y in {0,1}{%
        \node[draw,circle,inner sep=0.8pt,fill,black] at (2*\x-1,2*\y-1) {};
            }
            }            
%Draw K² lattice
    \foreach \x in {0,1,...,2}{%
      \foreach \y in {-1,0,...,1}{%
        \node[draw,cross out,inner sep=0.8pt,thick,black] at (2*\x-1,2*\y) {};
            }
            }
    \foreach \x in {0,1,...,2}{%
      \foreach \y in {0,1}{%
        \node[draw,cross out,inner sep=0.8pt,thick,black] at (2*\x,2*\y-1) {};
            }
            }           
    \draw[black,dashed,thick]  (0,0) -- (3.2,3.2);  
    \draw[black,dashed,thick]  (0,0) -- (3*1.5,1*1.5);  
    \draw[black,dashed,thick]  (0,0) -- (3.2,-3.2);
    \draw[black,dashed,thick]  (0,0) -- (4.5,0); 
\draw (6,2) node[circle,inner sep=0.8pt,fill,black] {};
\draw (7.2,2.05) node { $K^{[0]}$ lattice};
\draw (6,1) node[draw,cross out,inner sep=0.8pt,thick,black] {};
\draw (7.2,1.05) node { $K^{[2]}$ lattice};
\draw[black,thick] (1,0) circle (4pt);
\draw[black,thick] (1,1) circle (4pt);
\draw[black,thick] (1,-1) circle (4pt);
\draw[red,thick] (2,1) circle (4pt);
\draw[black,thick] (3,1) circle (4pt);
\draw (3.8,2.3) node {$S_3^{(2)}$};
\draw (3.6,0.6) node {$S_2^{(2)}$};
\draw (4.2,-1.4) node {$S_1^{(2)}$};
\end{tikzpicture}
\caption{The semi-groups for the quotient $\sorm(4)$ and the 
representation $[4,2]$. The black circled points are the ray generators and the 
red circled point completes the Hilbert basis for $S_{3}^{(2)}$.}
\label{Fig:Hilbert_basis_SO4_Rep42}
\end{figure}
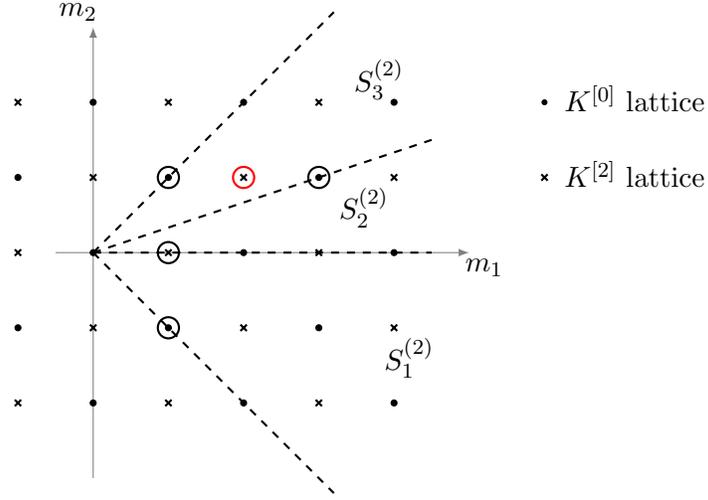

\paragraph{Hilbert basis}
The semi-groups $S_p^{(2)}\coloneqq C_p^{(2)} \cap \left(K^{[0]} \cup K^{[2]} 
\right)$ have Hilbert bases as shown in Fig.~\ref{Fig:Hilbert_basis_SO4_Rep42} 
or explicitly:
\begin{subequations}
\begin{alignat}{2}
   \Hcal(S_{1}^{(2)}) &= \Big\{ (1,0), (1,-1) \Big\} \; , \quad &
   \Hcal(S_{2}^{(2)}) &= \Big\{ (3,1), (1,0)  \Big\} \; , \\
    \Hcal(S_{3}^{(2)}) &= \Big\{ (1,1), (2,1) , (3,0) \Big\} \; .
\end{alignat}
\end{subequations}
\paragraph{Hilbert series}
\begin{subequations}
\label{eqn:HS_SO4_Rep42}
\begin{align}
\HS_{\sorm(4)}^{[4,2]}(t,z,N) &= \frac{R(t,z,N)}{
 \left(1-t^2\right)^2 
\left(1-t^{10 N-2}\right) 
\left(1-t^{18 N-2}\right) 
\left(1-t^{26 N-6}\right) 
\left(1-z t^{10 N-2}\right)} \; ,\\
R(t,z,N)&=1
+t^{10 N-1}
+ z t^{10 N-1} (2+t)
+z t^{18 N-4}(1+2t+t^3)
+t^{18 N-1} \notag \\
&\qquad 
-z t^{20 N-4}(1+3t+t^2)
+2 t^{26 N-5}(2+t)
\label{eqn:HS_SO4_Rep42_Num}\\
&\qquad
-t^{28 N-6}(1+2t+2t^2+2t^3)
-z t^{28 N-3}\notag\\
&\qquad
-t^{36 N-7} 
- z t^{36 N-7}(2+2t+2t^2+t^3)
+z t^{38 N-6}(1+2t)
\notag\\
&\qquad
-t^{44 N-8} (1+3t+t^2)
+z t^{46 N-9}
+t^{46 N-8} (1+2t+t^2)
\notag\\
&\qquad
+t^{54 N-10}(1+2t)
+z t^{54 N-9}
+z t^{64 N-10} \; . \notag
\end{align}
\end{subequations}
The numerator~\eqref{eqn:HS_SO4_Rep42_Num} is a palindromic polynomial of 
degree 
$64N-10 $, while the denominator is of degree $64N-8$. Consequently, the 
difference of the degree is two.
Also, the Hilbert series~\eqref{eqn:HS_SO4_Rep42} has a pole of order four as 
$t\to 1$, because $R(t{=}1,z,N)=0$ and $\tfrac{\diff}{\diff t} R(t,z,N) |_{t=1} 
=0$, but $\tfrac{\diff^2}{\diff t^2} R(t,z,N) |_{t=1} \neq0$.
\paragraph{Plethystic logarithm}
Inspecting the PL reveals
\begin{align}
 \PL(\HS_{\sorm(4)}^{[4,2]})= 2 t^2 &+ z t^{\Delta(1,0)} (1+2t +t^2) 
  +t^{\Delta(1,1)} (1+t)
  +z t^{\Delta(2,1)} (1+ 2t + t^2) \\
  &+ t^{\Delta(1,-1)} (1+t)
  -z t^{2 \Delta(1,0)} (1+ 3t + 3t^2 + t^3)
  -t^{2 \Delta(1,1) +2} (4+2t +t^2) \notag \\
  &+t^{\Delta(3,1)} (1+2t+t^2) +\ldots \; , \notag
\end{align}
such that the monopole generators can be summarised as in 
Tab.~\ref{tab:Ops_SO4_Rep42}.
\begin{table}[h]
\centering
 \begin{tabular}{c|c|c|c|c|c}
 \toprule
  object & $(m_1,m_2)$ & lattice & $\Delta(m_1,m_2)$ & $\Hh_{(m_1,m_2)}$ & 
dressings 
\\ \midrule
 Casimirs & --- &  --- & $2$ & --- & --- \\
 monopole & $(1,0)$ & $K^{[2]}$ & $10N-2 $ & $\uo \times 
\uo$ & $3$ by $\uo^2$\\
 monopole & $(1,1)$ & $K^{[0]}$ & $10N-2 $ & $\uo \times 
\su$ & $1$ by $\uo$ \\
 monopole & $(2,1)$ & $K^{[2]}$ & $18N-4 $ & $\uo \times 
\uo$ & $3$ by $\uo^2$ \\
 monopole & $(1,-1)$ & $K^{[0]}$ & $18N-2 $ & $\uo \times 
\su$ & $1$ by $\uo$ \\
 monopole & $(3,1)$ & $K^{[0]}$ & $26N-6 $ & $\uo \times 
\uo$ & $3$ by $\uo^2$ \\
\bottomrule
 \end{tabular}
\caption{The chiral ring generators for a $\sorm(4)$ gauge 
theory with matter transforming in $[4,2]$.}
\label{tab:Ops_SO4_Rep42}
\end{table}
\paragraph{Gauging the $\boldsymbol{\Z_2}$}
Again, one can gauge the finite symmetry to recover the $\Spin(4)$ Hilbert 
series
\begin{equation}
  \HS_{\Spin(4)}^{[4,2]}(t,N) = \frac{1}{2} \left(
 \HS_{\sorm(4)}^{[4,2]}(t,z{=}1,N) + 
 \HS_{\sorm(4)}^{[4,2]}(t,z{=}-1,N)\right) \; .
 \label{eqn:SO4_gauging_Rep42}
\end{equation}
% 
% 
%%%%%%%%%%%%%%%%%%%%%%%%%%%%%%%%%%%%%%%%%%%%%%%%%%%
%%%%%%%%%%%%%%%%%%%%%%%%%%%%%%%%%%%%%%%%%%%%%%%%%%%
% 
\subsubsection{Quotient \texorpdfstring{$\sorm(3) \times \su$}{SO(3)xSU(2)}}
\paragraph{Hilbert basis}
The semi-groups $S_p^{(2)}\coloneqq C_p^{(2)} \cap \left(K^{[0]} \cup K^{[1]} 
\right)$ ($p=1,2,3$) have Hilbert bases that go beyond the set of ray 
generators. We refer to Fig.~\ref{Fig:Hilbert_basis_SO3xSU2_Rep42} and the 
Hilbert bases are obtained as follows:
\begin{subequations}
\label{eqn:Hilbert_basis_SO3xSU2_Rep42}
\begin{alignat}{2}
   \Hcal(S_{1}^{(2)}) &= \Big\{ (2,1),  (\tfrac{3}{2},-\tfrac{1}{2}), (1,-1) 
\Big\} \; , 
\qquad &
   \Hcal(S_{2}^{(2)}) &= \Big\{ (3,1), (\tfrac{5}{2},\tfrac{1}{2}) , (2,0) 
\Big\} \; , \\
   \Hcal(S_{3}^{(2)}) &= \Big\{ (1,1), (3,1) \Big\} \; .
\end{alignat}
\end{subequations}
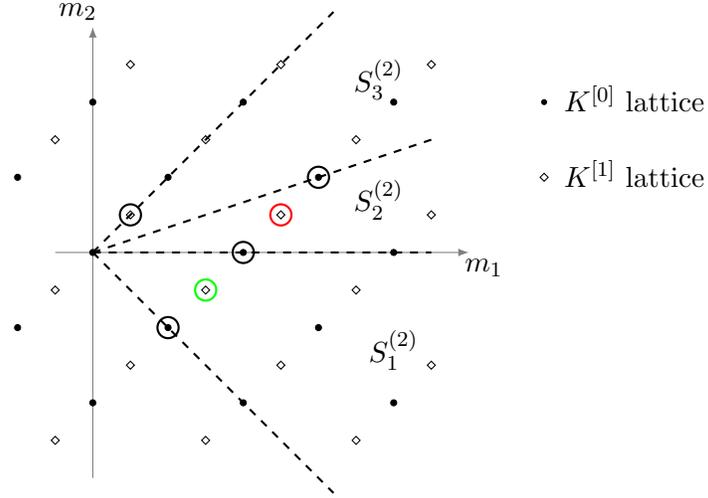
\begin{figure}[h]
\centering
\begin{tikzpicture}
  \coordinate (Origin)   at (0,0);
  \coordinate (XAxisMin) at (-0.5,0);
  \coordinate (XAxisMax) at (5,0);
  \coordinate (YAxisMin) at (0,-3);
  \coordinate (YAxisMax) at (0,3);
  \draw [thin, gray,-latex] (XAxisMin) -- (XAxisMax);%
  \draw [thin, gray,-latex] (YAxisMin) -- (YAxisMax);%
  \draw (5.2,-0.2) node {$m_1$};
  \draw (-0.2,3.2) node {$m_2$};
%Draw the root lattice
  \foreach \x in {0,1,...,2}{%
      \foreach \y in {-1,0,...,1}{%
        \node[draw,circle,inner sep=0.8pt,fill,black] at (2*\x,2*\y) {};
            }
            }
    \foreach \x in {0,1,...,2}{%
      \foreach \y in {0,1}{%
        \node[draw,circle,inner sep=0.8pt,fill,black] at (2*\x-1,2*\y-1) {};
            }
            }                      
% Draw K¹ lattice            
    \foreach \x in {0,1,...,2}{%
      \foreach \y in {-1,0,...,1}{%
        \node[draw,diamond,inner sep=0.8pt,black] at (2*\x +1/2,2*\y +1/2) {};
            }
            }  
      \foreach \x in {0,1,...,2}{%
      \foreach \y in {-1,0,...,1}{%
        \node[draw,diamond,inner sep=0.8pt,black] at (2*\x-1 +1/2,2*\y-1+1/2) 
{};
            }
            }
    \draw[black,dashed,thick]  (0,0) -- (3.2,3.2);  
    \draw[black,dashed,thick]  (0,0) -- (3*1.5,1*1.5);  
    \draw[black,dashed,thick]  (0,0) -- (3.2,-3.2);
    \draw[black,dashed,thick]  (0,0) -- (4.5,0); 
\draw (6,2) node[circle,inner sep=0.8pt,fill,black] {};
\draw (7.2,2.05) node { $K^{[0]}$ lattice};
\draw (6,1) node[draw,diamond,inner sep=0.8pt,black] {};
\draw (7.2,1.05) node { $K^{[1]}$ lattice};
\draw[thick,black] (1/2,1/2) circle (4pt);
\draw[thick,black] (1,-1) circle (4pt);
\draw[thick,green] (3/2,-1/2) circle (4pt);
\draw[thick,black] (2,0) circle (4pt);
\draw[thick,red] (5/2,1/2) circle (4pt);
\draw[thick,black] (3,1) circle (4pt);
\draw (3.8,2.3) node {$S_3^{(2)}$};
\draw (3.8,0.7) node {$S_2^{(2)}$};
\draw (4,-1.3) node {$S_1^{(2)}$};
\end{tikzpicture}
\caption{The semi-groups for the quotient $\sorm(3)\times \su$ and the 
representation $[4,2]$. The black circled points are the ray generators, the 
red circled point completes the Hilbert basis for $S_{2}^{(2)}$, while the 
green circled point completes the Hilbert basis of $S_{1}^{(2)}$.}
\label{Fig:Hilbert_basis_SO3xSU2_Rep42}
\end{figure}
\paragraph{Hilbert series}
We compute the Hilbert series to
\begin{subequations}
\label{eqn:HS_SO3xSU2_Rep42}
\begin{align}
\HS_{\sorm(3)\times \su}^{[4,2]}(t,z_1,N)&= 
\frac{R(t,z_1,N)}{\left(1-t^2\right)^2 
\left(1-t^{18 N-2}\right) 
\left(1-t^{20 N-4}\right) 
\left(1-t^{26 N-6}\right)} \; ,\\
R(t,z_1,N)&= 1
+z_1 t^{5 N-1}(1+t)
+t^{10 N-2}(1+t)
+z_1 t^{15 N-3}(1+t)
\\
&\qquad
+t^{18 N-1} 
+z_1 t^{19 N-3}(1+2t+t^3)
\notag \\
&\qquad
+t^{20 N-4}(1+3t+t^2)
+z_1 t^{23 N-5}(1+2t-t^3)
\notag \\
&\qquad
+ t^{26 N-5}(2+t)
-t^{28 N-4}(1+t)
+z_1 t^{31 N-6}(1+t)
\notag \\
&\qquad
-z_1 t^{33 N-5}(1+t)
+t^{36 N-7}(1+t)
-t^{38 N-6}(1+2t)
\notag \\
&\qquad
+z_1 t^{41 N-8} (1-2t^2-t^3)
-t^{44 N-8}(1+3t+t^2)
\notag \\
&\qquad
-z_1 t^{45 N-9}(1+2t+t^2)
-t^{46 N-9}
-z_1 t^{49 N-8}(1+t)
\notag \\
&\qquad
-t^{54 N-9}(1+t)
-z_1 t^{59 N-10}(1+t)
-t^{64 N-10} \; . \notag
\end{align}
\end{subequations}
The numerator of~\eqref{eqn:HS_SO3xSU2_Rep42} is an anti-palindromic polynomial 
of degree $64N-10$, 
while the denominator is of degree $64N-8$. Thus, the difference in degrees is 
again 2.
In addition, the Hilbert series~\eqref{eqn:HS_SO3xSU2_Rep42} has a pole of 
order 
4 as $t\to1$, because $R(t{=}1,z_1,N)=0$, but $\tfrac{\diff}{\diff t} 
R(t,z_1,N)|_{t=1}\neq0$.
\paragraph{Plethystic logarithm}
Analysing the PL yields
\begin{align}
 PL = 2t^2 &+ z_1 t^{\Delta(\tfrac{1}{2},\tfrac{1}{2})} (1+t) - 
t^{\Delta(\tfrac{1}{2},\tfrac{1}{2})+2}
+t^{\Delta(1,-1)} (1+t) \\
&+z_1 t^{\Delta(\tfrac{3}{2},-\tfrac{1}{2})} (1+2t+t^2)
+t^{\Delta(2,0)} (1+2t+t^2) \notag \\
&+z_1 t^{\Delta(\tfrac{5}{2},\tfrac{1}{2})}(1+2t +\textcolor{red}{1})
-z_1 t^{\Delta(\tfrac{1}{2},\tfrac{1}{2})+\Delta(1,-1)} (\textcolor{red}{1} + 
2t +t^2) \notag \\
&-t^{\Delta(\tfrac{1}{2},\tfrac{1}{2})+\Delta(\tfrac{3}{2},-\tfrac{1}{2})} 
(1+3t+3t^2+t^3) \notag\\
&-z_1 t^{\Delta(\tfrac{1}{2},\tfrac{1}{2})+\Delta(2,0)} (1+3t+3t^2+t^3)\notag \\
&+t^{\Delta(3,1)} (1+2t+t^2) +\ldots \notag \; ,
\end{align}
verfies the set of generators as presented in 
Tab.~\ref{tab:Ops_SO3xSU2_Rep42}.
\begin{table}[h]
\centering
 \begin{tabular}{c|c|c|c|c|c}
 \toprule
  object & $(m_1,m_2)$ & lattice & $\Delta(m_1,m_2)$ & $\Hh_{(m_1,m_2)}$ & 
dressings 
\\ \midrule
 Casimirs & --- &  --- & $2$ & --- & --- \\
 monopole & $(\tfrac{1}{2},\tfrac{1}{2})$ & $K^{[1]}$ & $5N-1 $ & $\uo \times 
\su$ & $1$ by $\uo$\\
 monopole & $(1,-1)$ & $K^{[0]}$ & $18N-2 $ & $\uo \times 
\su$ & $1$ by $\uo$ \\
 monopole & $(\tfrac{3}{2},-\tfrac{1}{2})$ & $K^{[1]}$ & $19N-3 $ & $\uo \times 
\uo$ & $3$ by $\uo^2$ \\
 monopole & $(2,0)$ & $K^{[0]}$ & $20N-4 $ & $\uo \times 
\uo$ & $3$ by $\uo$ \\
 monopole & $(\tfrac{5}{2},\tfrac{1}{2})$ & $K^{[1]}$ & $23N-5 $ & $\uo \times 
\uo$ & $\textcolor{red}{3} (2)$ by $\uo^2$ \\
 monopole & $(3,1)$ & $K^{[0]}$ & $26N-6 $ & $\uo \times 
\uo$ & $3$ by $\uo^2$ \\
\bottomrule
 \end{tabular}
\caption{The chiral ring generators for a $\sorm(3)\times \su$ gauge 
theory with matter transforming in $[4,2]$.}
\label{tab:Ops_SO3xSU2_Rep42}
\end{table}
The coloured term indicates that we suspect a cancellation 
between one dressing of $(\tfrac{5}{2},\tfrac{1}{2})$ and one relation because 
$\Delta(\tfrac{5}{2},\tfrac{5}{2})+2= 23N-3 = 
\Delta(\tfrac{1}{2},\tfrac{1}{2})+\Delta(1,-1)=5N-1+18N-2$.
% 
% 
%%%%%%%%%%%%%%%%%%%%%%%%%%%%%%%%%%%%%%%%%
%%%%%%%%%%%%%%%%%%%%%%%%%%%%%%%%%%%%%%%%%
% 
\subsubsection{Quotient \texorpdfstring{$\su \times \sorm(3)$}{SU(2)xSO(3)}}
\paragraph{Hilbert basis}
The semi-groups $S_p^{(2)}\coloneqq C_p^{(2)} \cap \left(K^{[0]} \cup K^{[3]} 
\right)$ (for $p=1,2,3$) have Hilbert bases consist of the ray generators as 
shown 
in Fig.~\ref{Fig:Hilbert_basis_SU2xSO3_Rep42} and we obtain explicitly
\begin{equation}
\label{eqn:Hilbert_basis_SU2xSO3_Rep42}
 \Hcal(S_{1}^{(2)}) = \Big\{ (2,0),  (\tfrac{1}{2},-\tfrac{1}{2}) \Big\} , 
\quad
   \Hcal(S_{2}^{(2)}) = \Big\{ (\tfrac{3}{2},\tfrac{1}{2}),(2,0) 
\Big\} , \quad 
   \Hcal(S_{3}^{(2)}) = \Big\{ (1,1), (\tfrac{3}{2},\tfrac{1}{2}) \Big\} .
\end{equation}
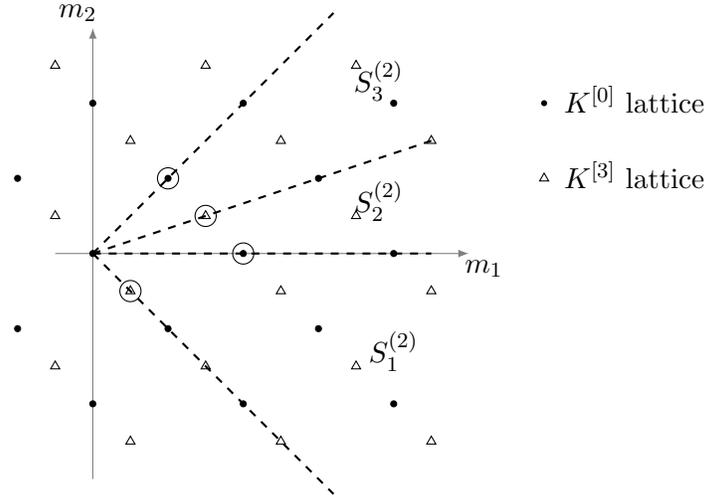
\begin{figure}[h]
\centering
\begin{tikzpicture}
  \coordinate (Origin)   at (0,0);
  \coordinate (XAxisMin) at (-0.5,0);
  \coordinate (XAxisMax) at (5,0);
  \coordinate (YAxisMin) at (0,-3);
  \coordinate (YAxisMax) at (0,3);
  \draw [thin, gray,-latex] (XAxisMin) -- (XAxisMax);%
  \draw [thin, gray,-latex] (YAxisMin) -- (YAxisMax);%
  \draw (5.2,-0.2) node {$m_1$};
  \draw (-0.2,3.2) node {$m_2$};
%Draw the root lattice
  \foreach \x in {0,1,...,2}{%
      \foreach \y in {-1,0,...,1}{%
        \node[draw,circle,inner sep=0.8pt,fill,black] at (2*\x,2*\y) {};
            }
            }
    \foreach \x in {0,1,...,2}{%
      \foreach \y in {0,1}{%
        \node[draw,circle,inner sep=0.8pt,fill,black] at (2*\x-1,2*\y-1) {};
            }
            }            
%Draw K³ lattice
    \foreach \x in {0,1,...,2}{%
      \foreach \y in {-1,0,...,1}{%
        \node[draw,regular polygon,regular polygon sides=3,inner 
sep=0.7pt,black] at (2*\x-1+1/2,2*\y+1/2) {};
            }
            }
     \foreach \x in {0,1,...,2}{%
      \foreach \y in {-1,0,...,1}{%
        \node[draw,regular polygon,regular polygon sides=3,inner 
sep=0.7pt,black] at (2*\x +1/2,2*\y-1 +1/2) {};
            }
            }
    \draw[black,dashed,thick]  (0,0) -- (3.2,3.2);  
    \draw[black,dashed,thick]  (0,0) -- (3*1.5,1*1.5);  
    \draw[black,dashed,thick]  (0,0) -- (3.2,-3.2);
    \draw[black,dashed,thick]  (0,0) -- (4.5,0); 
\draw (6,2) node[circle,inner sep=0.8pt,fill,black] {};
\draw (7.2,2.05) node { $K^{[0]}$ lattice};
\draw (6,1) node[draw,regular polygon,regular polygon sides=3,inner 
sep=0.7pt,black] {};
\draw (7.2,1.05) node { $K^{[3]}$ lattice};
\draw (1/2,-1/2) circle (4pt);
\draw (1,1) circle (4pt);
\draw (3/2,1/2) circle (4pt);
\draw (2,0) circle (4pt);
\draw (3.8,2.3) node {$S_3^{(2)}$};
\draw (3.8,0.7) node {$S_2^{(2)}$};
\draw (4,-1.3) node {$S_1^{(2)}$};
\end{tikzpicture}
\caption{The semi-groups for the quotient $\su \times \sorm(3)$ and the 
representation $[4,2]$. The black circled points are the ray generators.}
\label{Fig:Hilbert_basis_SU2xSO3_Rep42}
\end{figure}
\paragraph{Hilbert series}
We compute the Hilbert series to
\begin{subequations}
\label{eqn:HS_SU2xSO3_Rep42}
\begin{align}
\HS_{\su \times \sorm(3)}^{[4,2]}(t,z_2,N) &= 
\frac{R(t,z_2,N)}{\left(1-t^2\right)^2 
\left(1-t^{18 N-2}\right) 
\left(1-t^{20 N-4}\right) 
\left(1-t^{26 N-6}\right)} \; , \\
R(t,z_2,N)&=1
+z_2 t^{9 N-1} (1+t)
+t^{10 N-2}(1+t)
+z_2 t^{13 N-3} (1+2t+t^2)
\notag \\
&\qquad
+t^{18 N-1}
+t^{20 N-4}(1+3t+t^2)
+z_2 t^{23 N-5}(1+2t+t^2)
\notag \\
&\qquad
+ t^{26 N-5}(2+t)
-t^{28 N-4}(1+t)
+z_2 t^{29 N-4}(1+t)
\\
&\qquad
-z_2 t^{31 N-5}(1+2t+t^2)
+z_2 t^{33 N-7}(1+2t+t^2)
\notag \\
&\qquad
-z_2 t^{35 N-7}(1+t)
+t^{36 N-7}(1+t)
-t^{38 N-6}(1+2t)
\notag \\
&\qquad
-z_2 t^{41 N-7}(1+2t+t^2)
-t^{44 N-8}(1+3t+t^2)
\notag \\
&\qquad
-t^{46 N-9}
-z_2 t^{51 N-9}(1+2t+t^2)
-t^{54 N-9}(1+t)
\notag \\
&\qquad
-z_2 t^{55 N-10}(1+t)
-t^{64 N-10} \; . \notag
\end{align}
\end{subequations}
As before, we can try to compare the quotients $\sorm(3)\times \su$ 
and $\su \times \sorm(3)$. However, due to the asymmetry in $m_1$, $m_2$ or 
the asymmetry of the fan in the Weyl chamber, the Hilbert series for the two 
quotients are \emph{not} related by an exchange of $z_1$ and $z_2$.
\paragraph{Plethystic logarithm}
Upon analysing the PL we find
\begin{align}
 \PL(\HS_{\su \times \sorm(3)}^{[4,2]}) = 2t^2 &+ z_2 
t^{\Delta(\tfrac{1}{2},-\tfrac{1}{2})} (1+t)
 + t^{\Delta(1,1)} (1+t)
 + z_2 t^{\Delta(\tfrac{3}{2},\tfrac{1}{2})} (1+2t +t^2) \\
 &-t^{2\Delta(\tfrac{1}{2},-\tfrac{1}{2}) +2}  - 
z_2 t^{\Delta(\tfrac{1}{2},-\tfrac{1}{2})+ \Delta(1,1)} (1+2t+t^2) 
\notag \\
&+t^{\Delta(2,0)} (1+2t) - 
t^{\Delta(\tfrac{1}{2},-\tfrac{1}{2})+\Delta(\tfrac{3}{2},\tfrac{1}{2})} 
(1+3t+3t^2+t^3) +\dots \notag \; ,
 \end{align}
through which one identifies the generators as given in 
Tab.~\ref{tab:Ops_SU2xSO3_Rep42}.
\begin{table}[h]
\centering
 \begin{tabular}{c|c|c|c|c|c}
 \toprule
  object & $(m_1,m_2)$ & lattice & $\Delta(m_1,m_2)$ & $\Hh_{(m_1,m_2)}$ & 
dressings 
\\ \midrule
 Casimirs & --- &  --- & $2$ & --- & --- \\
 monopole & $(\tfrac{1}{2},-\tfrac{1}{2})$ & $K^{[3]}$ & $9N-1 $ & $\uo \times 
\su$ & $1$ by $\uo$\\
 monopole & $(1,1)$ & $K^{[0]}$ & $10N-2 $ & $\uo \times 
\su$ & $1$ by $\uo$ \\
 monopole & $(\tfrac{3}{2},\tfrac{1}{2})$ & $K^{[3]}$ & $13N-3 $ & $\uo \times 
\uo$ & $3$ by $\uo^2$ \\
 monopole & $(2,0)$ & $K^{[0]}$ & $20N-4 $ & $\uo \times 
\uo$ & $3$ by $\uo^2$ \\
\bottomrule
 \end{tabular}
\caption{The chiral ring generators for a $\su\times \sorm(3)$ gauge 
theory with matter transforming in $[4,2]$.}
\label{tab:Ops_SU2xSO3_Rep42}
\end{table}
The terms in the denominator of the Hilbert series can be seen to reproduce 
these generators
\begin{subequations}
\begin{align}
 (1-t^{18N-2}) &= (1-z_2 t^{9N-1}) (1+z_2 t^{9N-1}) \; , \\
 (1-t^{26N-6}) &= (1-z_2 t^{13N-3}) (1+z_2 t^{13N-3}) \; .
\end{align}
\end{subequations}
Unfortunately, we are unable to reduce the numerator accordingly.
%
%%%%%%%%%%%%%%%%%%%%%%%%%%%%%%%%%%%%%%%%%%%%%%%%%%%%%%%%%%
%%%%%%%%%%%%%%%%%%%%%%%%%%%%%%%%%%%%%%%%%%%%%%%%%%%%%%%%%%
% 
\subsubsection{Quotient \texorpdfstring{$\mathrm{PSO}(4)$}{PSO(4)}}
\paragraph{Hilbert basis}
The semi-groups $S_p^{(2)}\coloneqq C_p^{(2)} \cap \left(K^{[0]} \cup K^{[1]} 
\cup K^{[2]} \cup K^{[3]} \right)$ (for $p=1,2,3$) have Hilbert bases that are 
determined by the ray generators. Fig.~\ref{Fig:Hilbert_basis_PSO4_Rep42} 
depicts the situation and the Hilbert bases read:
\begin{equation}
\label{eqn:Hilbert_basis_PSO4_Rep42}
\Hcal(S_{1}^{(2)}) = \Big\{ (1,0), (\tfrac{1}{2},-\tfrac{1}{2}) \Big\} , \quad 
   \Hcal(S_{2}^{(2)}) = \Big\{ (\tfrac{3}{2},\tfrac{1}{2}), (1,0)  \Big\}, \quad
   \Hcal(S_{3}^{(2)}) = \Big\{ (\tfrac{1}{2},\tfrac{1}{2}), 
(\tfrac{3}{2},\tfrac{1}{2})  \Big\} .
\end{equation}
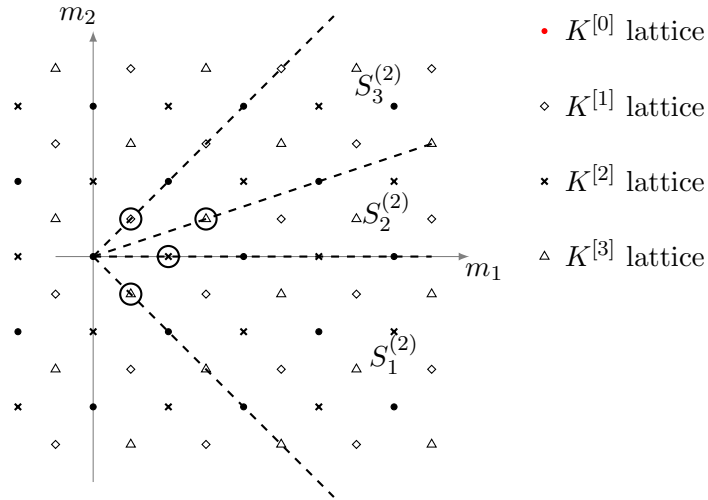
\begin{figure}[h]
\centering
\begin{tikzpicture}
  \coordinate (Origin)   at (0,0);
  \coordinate (XAxisMin) at (-0.5,0);
  \coordinate (XAxisMax) at (5,0);
  \coordinate (YAxisMin) at (0,-3);
  \coordinate (YAxisMax) at (0,3);
  \draw [thin, gray,-latex] (XAxisMin) -- (XAxisMax);%
  \draw [thin, gray,-latex] (YAxisMin) -- (YAxisMax);%
  \draw (5.2,-0.2) node {$m_1$};
  \draw (-0.2,3.2) node {$m_2$};
%Draw the root lattice
  \foreach \x in {0,1,...,2}{%
      \foreach \y in {-1,0,...,1}{%
        \node[draw,circle,inner sep=0.8pt,fill,black] at (2*\x,2*\y) {};
            }
            }
    \foreach \x in {0,1,...,2}{%
      \foreach \y in {0,1}{%
        \node[draw,circle,inner sep=0.8pt,fill,black] at (2*\x-1,2*\y-1) {};
            }
            }            
%Draw K² lattice
    \foreach \x in {0,1,...,2}{%
      \foreach \y in {-1,0,...,1}{%
        \node[draw,cross out,inner sep=0.8pt,thick,black] at (2*\x-1,2*\y) {};
            }
            }
    \foreach \x in {0,1,...,2}{%
      \foreach \y in {0,1}{%
        \node[draw,cross out,inner sep=0.8pt,thick,black] at (2*\x,2*\y-1) {};
            }
            }            
% Draw K¹ lattice            
    \foreach \x in {0,1,...,2}{%
      \foreach \y in {-1,0,...,1}{%
        \node[draw,diamond,inner sep=0.8pt,black] at (2*\x +1/2,2*\y +1/2) {};
            }
            }  
      \foreach \x in {0,1,...,2}{%
      \foreach \y in {-1,0,...,1}{%
        \node[draw,diamond,inner sep=0.8pt,black] at (2*\x-1 +1/2,2*\y-1+1/2) 
{};
            }
            } 
%Draw K³ lattice
    \foreach \x in {0,1,...,2}{%
      \foreach \y in {-1,0,...,1}{%
        \node[draw,regular polygon,regular polygon sides=3,inner 
sep=0.7pt,black] at (2*\x-1+1/2,2*\y+1/2) {};
            }
            }
     \foreach \x in {0,1,...,2}{%
      \foreach \y in {-1,0,...,1}{%
        \node[draw,regular polygon,regular polygon sides=3,inner 
sep=0.7pt,black] at (2*\x +1/2,2*\y-1 +1/2) {};
            }
            }
    \draw[black,dashed,thick]  (0,0) -- (3.2,3.2);  
    \draw[black,dashed,thick]  (0,0) -- (3*1.5,1*1.5);  
    \draw[black,dashed,thick]  (0,0) -- (3.2,-3.2);
    \draw[black,dashed,thick]  (0,0) -- (4.5,0); 
\draw (6,3) node[circle,inner sep=0.8pt,fill,red] {};
\draw (7.2,3.05) node { $K^{[0]}$ lattice};
\draw (6,2) node[draw,diamond,inner sep=0.8pt,black] {};
\draw (7.2,2.05) node { $K^{[1]}$ lattice};
\draw (6,1) node[draw,cross out,inner sep=0.8pt,thick,black] {};
\draw (7.2,1.05) node { $K^{[2]}$ lattice};
\draw (6,0 ) node[draw,regular polygon,regular polygon sides=3,inner 
sep=0.7pt,black] {};
\draw (7.2,0.05) node { $K^{[3]}$ lattice};
\draw[black,thick] (1/2,1/2) circle (4pt);
\draw[black,thick] (1/2,-1/2) circle (4pt);
\draw[black,thick] (1,0) circle (4pt);
\draw[black,thick] (3/2,1/2) circle (4pt);
\draw (3.8,2.3) node {$S_3^{(2)}$};
\draw (3.9,0.6) node {$S_2^{(2)}$};
\draw (4,-1.3) node {$S_1^{(2)}$};
\end{tikzpicture}
\caption{The semi-groups and their ray-generators (black circled points) for 
the quotient $\mathrm{PSO}(4)$ and the representation $[4,2]$.}
\label{Fig:Hilbert_basis_PSO4_Rep42}
\end{figure}
\paragraph{Hilbert series}
We obtain the following Hilbert series
\begin{subequations}
\label{eqn:HS_PSO4_Rep42}
\begin{align}
\HS_{\mathrm{PSO}(4)}^{[4,2]}(t,z_1,z_2,N)&= 
\frac{R(t,z_1,z_2,N)}{\left(1-t^2\right)^2 \left(1-t^{10 N-2}\right) 
\left(1-t^{18 N-2}\right) \left(1-t^{26 N-6}\right) \left(1-t^{10 N-2} 
z_2\right)} \; , \\
R(t,z_1,z_2,N)&= 1
+z_1 t^{5 N-1}(1+t)
+z_1 z_2 t^{9 N-1}(1+t)
+z_1 z_2 t^{9 N}
+t^{10 N-1} \\
&\qquad
+z_2 t^{10 N-1} (2+t)
+z_1 z_2 t^{13 N-3} (1+2t+t^2)
-z_1 z_2 t^{15 N-3} (1+t)
\notag \\
&\qquad
+z_2 t^{18 N-4} (1+2t+t^2)
+t^{18 N-1}
-z_1 z_2 t^{19 N-3} (1+t)
\notag \\
&\qquad
+z_1 t^{19 N-2}(1+t)
-z_2 t^{20 N-4}(1+3t+t^2) 
-z_1 t^{23 N-3}(1+t)
\notag \\
&\qquad
+ t^{26 N-5}(2+t) 
-t^{28 N-6}(1+2t+2t^2+2t^3)
-z_2 t^{28 N-3}
\notag \\
&\qquad
-z_1 t^{29 N-4}(1+t)
+z_1 t^{31 N-6} (1+t)
-z_1 z_2 t^{31 N-5}(1+2t+t)
\notag \\
&\qquad
-z_1 t^{33 N-7}(1+2t+t^2)
+z_1 z_2 t^{33 N-5}(1+t)
\notag \\
&\qquad
-z_1 z_2 t^{35 N-7}(1+t)
- z_2 t^{36 N-7}(2+2t+2t^2+t^3)
-t^{36 N-7}
\notag \\
&\qquad
+z_2 t^{38 N-6}(1+2t)
-z_1 z_2 t^{41 N-8} (1+t)
-t^{44 N-8} (1+3t+t^2)\notag \\
&\qquad
+z_1 z_2 t^{45 N-9} (1+t)
-z_1 t^{45 N-8}(1+t)
\notag \\
&\qquad
+z_2 t^{46 N-9}
+t^{46 N-8} (1+2t+t^2)
-z_1 t^{49 N-8}(1+t)
\notag \\
&\qquad
+z_1 t^{51 N-9}(1+2t+t^2)
+t^{54 N-10}(1+2t)
+z_2 t^{54 N-9}
\notag \\
&\qquad
+z_1 t^{55 N-10}(1+t)
+z_1 z_2 t^{59 N-10}(1+t)
+z_2 t^{64 N-10} \notag \; .
\end{align}
\end{subequations}
The numerator of~\eqref{eqn:HS_PSO4_Rep42} is a palindromic polynomial of 
degree 
$64N-10$, while the denominator is of degree $64N-8 $. Hence, the difference in 
degrees is again 2.
Moreover, the Hilbert series~\eqref{eqn:HS_PSO4_Rep42} has a pole of order 4 as 
$t\to 1$ because $R(1,z_1,z_2,N)=0$ and $\tfrac{\diff}{\diff t} 
R(t,z_1,z_2,N)|_{t\to1}=0$, while $\tfrac{\diff^2}{\diff t^2} 
R(t,z_1,z_2,N)|_{t\to1}\neq0$.
\paragraph{Plethystic logarithm}
Working with the PL instead reveals further insights
\begin{align}
 \PL(\HS_{\mathrm{PSO}(4)}^{[4,2]})= 2t^2 &+ z_1 
t^{\Delta(\tfrac{1}{2},\tfrac{1}{2})} (1+t) 
 + z_1 z_2 t^{\Delta(\tfrac{1}{2},-\tfrac{1}{2})} (1+t)
 +z_2 t^{\Delta(1,0)} (1+2t +t^2) \\
 &-t^{2\Delta(\tfrac{1}{2},\tfrac{1}{2}) +2}
 +z_1 z_2 t^{\Delta(\tfrac{3}{2},\tfrac{1}{2})} (1+2t+t^2) \notag \\
 &-z_2 t^{\Delta(\tfrac{1}{2},\tfrac{1}{2})+\Delta(\tfrac{1}{2},\tfrac{1}{2})} 
(1+2t +t^2) 
 -z_1 z_2 
t^{\Delta(\tfrac{1}{2},\tfrac{1}{2})+\Delta(1,0)} 
(1+3t +3t^2+t^3)  + \ldots \notag \; .
 \end{align}
The list of generators, together with their properties, is provided in 
Tab.~\ref{tab:Ops_PSO4_Rep42}.
\begin{table}[h]
\centering
 \begin{tabular}{c|c|c|c|c|c}
 \toprule
  object & $(m_1,m_2)$ & lattice & $\Delta(m_1,m_2)$ & $\Hh_{(m_1,m_2)}$ & 
dressings 
\\ \midrule
 Casimirs & --- &  --- & $2$ & --- & --- \\
 monopole & $(\tfrac{1}{2},\tfrac{1}{2})$ & $K^{[1]}$ & $5N-1 $ & $\uo \times 
\su$ & $1$ by $\uo$\\
 monopole & $(\tfrac{1}{2},-\tfrac{1}{2})$ & $K^{[3]}$ & $9N-1 $ & $\uo \times 
\su$ & $1$ by $\uo$ \\
monopole & $(1,0)$ & $K^{[2]}$ & $10N-2 $ & $\uo \times 
\uo$ & $3$ by $\uo^2$\\
 monopole & $(\tfrac{3}{2},\tfrac{1}{2})$ & $K^{[3]}$ & $13N-3 $ & $\uo \times 
\uo$ & $3$ by $\uo^2$ \\
\bottomrule
 \end{tabular}
\caption{The chiral ring generators for a $\mathrm{PSO}(4)$ gauge 
theory with matter transforming in $[4,2]$.}
\label{tab:Ops_PSO4_Rep42}
\end{table}
\paragraph{Gauging a $\Z_2$}
The global $\Z_2\times \Z_2$ symmetry allows us to compute the Hilbert series 
for all five quotients from the $\mathrm{PSO}(4)$ result. 
We start by gauging the $\Z_2$-factor with fugacity $z_1$ (and relabel $z_2$ as 
$z$) and recover the $\sorm(4)$-result
\begin{subequations}
\label{eqn:PSO4_gauging_Rep42}
\begin{equation}
 \HS_{\sorm(4)}^{[4,2]}(t,z,N) = \frac{1}{2} 
\left(\HS_{\mathrm{PSO}(4)}^{[4,2]}(t,z_1{=}1,z_2{=}z,N) +
\HS_{\mathrm{PSO}(4)}^{[4,2]}(t,z_1{=}-1,z_2{=}z,N)\right) \; .
\end{equation}
In contrast, gauging the other $\Z_2$-factor with fugacity $z_1$ provides the 
$\sorm(3)\times\su$-result
\begin{equation}
 \HS_{\sorm(3)\times\su}^{[4,2]}(t,z_1,N) = \frac{1}{2} 
\left(\HS_{\mathrm{PSO}(4)}^{[4,2]}(t,z_1,z_2{=}1,N) +
\HS_{\mathrm{PSO}(4)}^{[4,2]}(t,z_1,z_2{=}{-1},N)\right) \; .
\end{equation}
Lastly, switching to $w_1$, $w_2$ fugacities as 
in~\eqref{eqn:redefine_Z2-grading_A1xA1} allows to recover the Hilbert series 
for $\su\times\sorm(3)$ as follows:
\begin{equation}
 \HS_{\su\times\sorm(3)}^{[4,2]}(t,z_2{=}w_1,N) = \frac{1}{2} 
\left(\HS_{\mathrm{PSO}(4)}^{[4,2]}(t,w_1,w_2{=}1,N) +
\HS_{\mathrm{PSO}(4)}^{[4,2]}(t,w_1,w_2{=}{-1},N)\right) \; .
\end{equation}
\end{subequations}
In conclusion, the $\mathrm{PSO}(4)$ result is sufficient to obtain the 
remaining four quotients by gauging of various $\Z_2$ global symmetries as 
in~\eqref{eqn:PSO4_gauging_Rep42} and~\eqref{eqn:SO4_gauging_Rep42}.
% 
%%%%%%%%%%%%%%%%%%%%%%%%%%%%%%%%%%%%%%%%%%%%%%%%%%%%%%%%%%%%%%%%%%%%%%%%%%%%%%%%
%%%%%%%%%%%%%%%%%%%%%%%%%%%%%%%%%%%%%%%%%%%%%%%%%%%%%%%%%%%%%%%%%%%%%%%%%%%%%%%%
%
\subsection{Comparison to \texorpdfstring{$\orm(4)$}{O(4)}}
In this subsection we explore the orthogonal group $\orm(4)$, 
related to $\sorm(4)$ by $\Z_2$. To begin with, we summarise the set-up as 
presented in~\cite[App.~A]{Cremonesi:2014uva}. The dressing factor 
$P_{\orm(4)}(t)$ and the GNO lattice of $\orm(4)$ equal those of $\sorm(5)$. 
Moreover, the dominant Weyl chamber is parametrised by $(m_1,m_2)$ subject to 
$m_1 \geq m_2 \geq 0$. Graphically, the Weyl chamber is the upper half of the 
yellow-shaded region in Fig.~\ref{fig:A1xA1_sublattices} with the lattices 
$K^{[0]}\cup K^{[2]}$ present.
Consequently, the dressing function is given as
\begin{equation}
 P_{\orm(4)}(t,m_1,m_2) = \begin{cases} \frac{1}{\left(1-t^2 \right) 
\left(1-t^4\right)} \; ,& m_1=m_2=0 \; ,\\
\frac{1}{\left(1-t \right) \left(1-t^2 \right)}\; , & m_1=m_2 >0 \; , \\
  \frac{1}{\left(1-t \right) \left(1-t^2 \right)} \; , & m_1>0, \ m_2=0 \; ,\\
\frac{1}{\left(1-t \right)^2}  \; , & m_1>m_2>0 \; .
                          \end{cases}
\end{equation}
It is apparent that $\orm(4)$ has a different Casimir invariant as $\sorm(4)$, 
which comes about as the Levi-Civita tensor $\varepsilon$ is not an invariant 
tensor under $\orm(4)$. In other words, the Pfaffian of $\sorm(4)$ is not an 
invariant of $\orm(4)$.

Now, we provide the Hilbert series for the three different representations 
studied above.
\subsubsection{Representation \texorpdfstring{$[2,0]$}{[2,0]}}
The conformal dimension is the same as in~\eqref{eqn:delta_A1xA1_Rep20} and the 
rational cone of the Weyl chamber is simply
\begin{equation}
 C^{(2)} = \cone\left((1,0),(1,1)\right) \; ,
\end{equation}
such that the cone generators and the Hilbert basis for $S^{(2)}\coloneqq 
C^{(2)} \cap \left( K^{[0]}\cup K^{[2]} \right)$ coincide. The upper half-space 
of Fig.~\ref{Fig:Hilbert_basis_SO4_Rep20} depicts the situation.

The Hilbert series is then computed to read
\begin{equation}
 \HS_{\orm(4)}^{[2,0]}(t,N)= \frac{1 +2 t^{2 N-1} + 2 t^{2 N} +2 t^{2 N+1} + 
t^{4N}}{\left(1-t^2\right) \left(1-t^4\right) \left(1-t^{2 N-2}\right)^2}\; ,
\label{eqn:HS_O4_Rep20}
\end{equation}
which clearly displays the palindromic numerator, the order four pole for $t\to 
1$, 
and the order two pole for $t\to \infty$, i.e.\ the difference in degrees of 
denominator and numerator is two.
By inspection of~\eqref{eqn:HS_O4_Rep20} and use of the plethystic logarithm 
\begin{equation}
 \PL(\HS_{\orm(4)}^{[2,0]}) = t^2 +t^4 + t^{\Delta(1,0)}(1+t+t^2+t^3) + 
t^{\Delta(1,1)}(1+t+t^2+t^3) - \mathcal{O}(t^{2\Delta(1,0) +2}) \; ,
\end{equation}
for $N\geq2$, we can summarise the generators as in Tab.~\ref{tab:Ops_O4_Rep20}.
\begin{table}[h]
\centering
 \begin{tabular}{c|c|c|c|c|c}
 \toprule
  object & $(m_1,m_2)$ & lattice & $\Delta(m_1,m_2)$ & $\Hh_{(m_1,m_2)}$ & 
dressings 
\\ \midrule
 Casimirs & --- &  --- & $2$, $4$ & --- & --- \\
 monopole & $(1,0)$ & $K^{[2]}$ & $2N-2 $ & $\utwo$ & $3$ \\
 monopole & $(1,1)$ & $K^{[0]}$ & $2N-2 $ & $\uo \times \orm(2)$ & $3$ 
\\
 \bottomrule
 \end{tabular}
\caption{Bare and dressed monopole generators for a $\orm(4)$ gauge theory with 
matter transforming in $[2,0]$.}
\label{tab:Ops_O4_Rep20}
\end{table}
The different dressing behaviour of the $\orm(4)$ monopole generators $(1,0)$ 
and $(1,1)$ compared to their $\sorm(4)$ counterparts can be deduced from 
dividing the relevant dressing factor by the trivial one. In detail
\begin{equation}
 \frac{P_{\orm(4)}(t,\{(1,0) \text{ or } (1,1)\})}{P_{\orm(4)}(t,0,0)} = 
\frac{(1-t^2)(1-t^4)}{(1-t)(1-t^2)}= 1+t+t^2+t^3 \; .
\end{equation}
% 
% 
%%%%%%%%%%%%%%%%%%%%%%%%%%%%%%%%%%%%%%%%
% 
\subsubsection{Representation \texorpdfstring{$[2,2]$}{[2,2]}}
The conformal dimension is the same as in~\eqref{eqn:delta_A1xA1_Rep22} and 
the 
rational cone of the Weyl chamber is still
\begin{equation}
 C^{(2)} = \cone\left((1,0),(1,1)\right) \; ,
\end{equation}
such that the cone generators and the Hilbert basis for $S^{(2)}\coloneqq 
C^{(2)} \cap \left( K^{[0]}\cup K^{[2]} \right)$ coincide. The upper half-space 
of Fig.~\ref{Fig:Hilbert_basis_SO4_Rep22} depicts the situation. We note that 
the Weyl chamber for $\sorm(4)$ is already divided into a fan by two rational 
cones, while the Weyl chamber for $\orm(4)$ is not.

The computation of the Hilbert series then yields
\begin{equation}
 \HS_{\orm(4)}^{[2,2]}(t,N)= 
 \frac{1+t^{4 N-1}+ t^{4 N} +t^{4 N+1}
 +t^{6 N-1} +t^{6 N} +t^{6 N+1} +t^{10 N}}{\left(1-t^2\right) 
\left(1-t^4\right) \left(1-t^{4 N-2}\right) \left(1-t^{6 N-2}\right)}
 \; .
\label{eqn:HS_O4_Rep22}
\end{equation}
Again, the rational function clearly displays a palindromic numerator, an order 
four pole for $t\to 1$, 
and an order two pole for $t\to \infty$, i.e.\ the difference in degrees of 
denominator and numerator is two.
By inspection of~\eqref{eqn:HS_O4_Rep22} and use of the plethystic logarithm 
\begin{equation}
 \PL(\HS_{\orm(4)}^{[2,2]}) = t^2 +t^4 + t^{\Delta(1,0)}(1+t+t^2+t^3) + 
t^{\Delta(1,1)}(1+t+t^2+t^3) - \mathcal{O}(t^{2\Delta(1,0) +2}) \; ,
\end{equation}
for $N\geq2$, we can summarise the generators as in Tab.~\ref{tab:Ops_O4_Rep22}.
\begin{table}[h]
\centering
 \begin{tabular}{c|c|c|c|c|c}
 \toprule
  object & $(m_1,m_2)$ & lattice & $\Delta(m_1,m_2)$ & $\Hh_{(m_1,m_2)}$ & 
dressings 
\\ \midrule
 Casimirs & --- &  --- & $2$, $4$ & --- & --- \\
 monopole & $(1,0)$ & $K^{[2]}$ & $4N-2 $ & $\utwo$ & $3$ \\
 monopole & $(1,1)$ & $K^{[0]}$ & $6N-2 $ & $\uo \times \orm(2)$ & $3$ 
\\
 \bottomrule
 \end{tabular}
\caption{Bare and dressed monopole generators for a $\orm(4)$ gauge theory with 
matter transforming in $[2,2]$.}
\label{tab:Ops_O4_Rep22}
\end{table}
The dressings behave as discussed earlier.
% 
% 
%%%%%%%%%%%%%%%%%%%%%%%%%%%%%%%%%%%%%%%%
% 
\subsubsection{Representation \texorpdfstring{$[4,2]$}{[4,2]}}
The conformal dimension is given in~\eqref{eqn:delta_A1xA1_Rep42} and 
the Weyl chamber is split into a fan generated by two rational cones
\begin{equation}
 C_2^{(2)} = \cone\left((1,0),(3,1)\right)  \und 
 C_3^{(2)} = \cone\left((3,1),(1,1)\right) \; ,
\end{equation}
where we use the notation of the $\sorm(4)$ setting, see the upper half plan of
Fig.~\ref{Fig:Hilbert_basis_SO4_Rep42}.
The Hilbert bases for $S_p^{(2)}\coloneqq 
C_p^{(2)} \cap \left( K^{[0]}\cup K^{[2]} \right)$ differ from the cone 
generators and are obtained as
\begin{equation}
 \Hcal(S_2^{(2)}) = \left\{(1,0),(3,1) \right\} \und
 \Hcal(S_3^{(2)}) = \left\{(3,1),(2,1),(1,1) \right\} \; .
\end{equation}
The computation of the Hilbert series then yields
\begin{subequations}
\label{eqn:HS_O4_Rep42}
\begin{align}
 \HS_{\orm(4)}^{[4,2]}(t,N)&= 
\frac{R(t,N)}{\left(1-t^2\right) \left(1-t^4\right) \left(1-t^{10 N-2}\right) 
\left(1-t^{26 N-6}\right)}
 \; , \\
R(t,N)&= 1 +t^{10 N-2}+2 t^{10 N-1} + 2 t^{10 N} +2 t^{10 N+1} \\
&\qquad +t^{18 N-4}+2 t^{18 N-3} +2 t^{18 N-2} + 2 t^{18 N-1} +t^{18 N} \notag 
\\
&\qquad +2 t^{26 N-5}+2 t^{26 N-4}+2 t^{26 N-3}+t^{26 N-2}+t^{36 N-4} \notag
\end{align}
\end{subequations}
As before, the rational function~\eqref{eqn:HS_O4_Rep42} clearly displays a 
palindromic numerator, an order 
four pole for $t\to 1$, 
and an order two pole for $t\to \infty$, i.e.\ the difference in degrees of 
denominator and numerator is two.
By inspection of~\eqref{eqn:HS_O4_Rep42} and use of the plethystic logarithm 
\begin{align}
 \PL(\HS_{\orm(4)}^{[4,2]}) = t^2 +t^4 &+ t^{\Delta(1,0)}(1+t+t^2+t^3) + 
t^{\Delta(1,1)}(1+t+t^2+t^3) \label{eqn:PL_O4_Rep42} \\
&+t^{\Delta(2,1)}(1+2(t+t^2+t^3)+t^4)\notag \\
&-t^{\Delta(1,0)+\Delta(1,1)}(1+2t+5t^2+6t^3+7t^4+4t^5+3t^6) \notag \\
&+t^{\Delta(3,1)}(1+2(t+t^2+t^3)+t^4)
- \mathcal{O}(t^{\Delta(1,0) +\Delta(2,1)}) \; , \notag
\end{align}
for $N\geq2$, we can summarise the generators as in Tab.~\ref{tab:Ops_O4_Rep42}.
\begin{table}[h]
\centering
 \begin{tabular}{c|c|c|c|c|c}
 \toprule
  object & $(m_1,m_2)$ & lattice & $\Delta(m_1,m_2)$ & $\Hh_{(m_1,m_2)}$ & 
dressings 
\\ \midrule
 Casimirs & --- &  --- & $2$, $4$ & --- & --- \\
 monopole & $(1,0)$ & $K^{[2]}$ & $10N-2 $ & $\utwo$ & $3$ \\
 monopole & $(1,1)$ & $K^{[0]}$ & $10N-2 $ & $\uo \times \orm(2)$ & $3$ 
\\
 monopole & $(2,1)$ & $K^{[2]}$ & $18N-4 $ & $\uo^2 $ & $7$ 
\\
 monopole & $(3,1)$ & $K^{[0]}$ & $26N-6 $ & $\uo^2$ & $7$ 
\\
 \bottomrule
 \end{tabular}
\caption{Bare and dressed monopole generators for a $\orm(4)$ gauge theory with 
matter transforming in $[4,2]$.}
\label{tab:Ops_O4_Rep42}
\end{table}
The dressing behaviour of $(1,0)$, $(1,1)$ is as discussed earlier; however, we 
need to describe the dressings of $(2,1)$ and $(3,1)$ as it differs from the 
$\sorm(4)$ counterparts. Again, we compute the quotient of the dressing factor 
of the maximal torus divided by the trivial one, i.e.\
\begin{equation}
  \frac{P_{\orm(4)}(t,m_1>m_2>0)}{P_{\orm(4)}(t,0,0)}= 
\frac{(1-t^2)(1-t^4)}{(1-t)^2}= 1+2(t+t^2+t^3)+t^4 \; .
\end{equation}
Consequently, each bare monopole $(2,1)$, $(3,1)$ is accompanied by seven 
dressings, which is in agreement with~\eqref{eqn:PL_O4_Rep42}.
%%%%%%%%%%%%%%%%%%%%%%%%%%%%%%%%%%%%%%%%%%%%%%%%%%%%%%%%%%%%%%%%%%%%%%%%%%%%%%%%
  \section{Case: \texorpdfstring{$\boldsymbol{\USp}$}{USp(4)}}
\label{sec:USp4}
This section is devoted to the study of the compact symplectic group $\USp$ 
with corresponding Lie algebra $C_2$. GNO-duality relates them with the special 
orthogonal group $\sorm(5)$ and the Lie algebra $B_2$.

\subsection{Set-up}
For studying the non-abelian group $\USp$, we start by providing the 
contributions of $N_{a,b}$ hypermultiplets in various 
representations $[a,b]$ of $\USp$ to the conformal dimensions
\begin{subequations}
\begin{align}
 \Delta_{\mathrm{h-plet}}^{[1,0]} &= N_{1,0} \sum_i |m_i| \; ,\\
 \Delta_{\mathrm{h-plet}}^{[0,1]} &= N_{0,1} \Big( \sum_{i<j} |m_i-m_j| +  
\sum_{i<j} |m_i+m_j| \Big) \; ,\\
 \Delta_{\mathrm{h-plet}}^{[2,0]} &=2N_{2,0} \sum_i |m_i| 
 +  N_{2,0} \Big(  \sum_{i<j} |m_i-m_j| +  \sum_{i<j} |m_i+m_j| \Big) \; ,\\
 \Delta_{\mathrm{h-plet}}^{[0,2]} &=2N_{0,2} \sum_i |m_i| 
 +  3 N_{0,2} \Big(  \sum_{i<j} |m_i-m_j| +  \sum_{i<j} |m_i+m_j| \Big) \; 
,\\
 \Delta_{\mathrm{h-plet}}^{[1,1]} &=2N_{1,1} \sum_i |m_i| 
 +  N_{1,1} \Big(  \sum_{i<j} \left(  |2m_i-m_j| +|m_i-2m_j| \right) \\*
 &\phantom{ =2N_{1,1} \sum_i |m_i|  +  N_{1,1} \Big( \sum_{i<j} }
 +  \sum_{i<j} \left( |2m_i+m_j| + |m_i+2m_j| \right) \Big) \; 
,\notag \\
\Delta_{\mathrm{h-plet}}^{[3,0]} &=5N_{3,0} \sum_i |m_i| 
 +  N_{3,0} \Big(  \sum_{i<j} \left(  |2m_i-m_j| +|m_i-2m_j| \right) \\*
 &\phantom{ =5N_{3,0} \sum_i |m_i|  +  N_{3,0} \Big(  \sum_{i<j} }
 +  \sum_{i<j} \left( |2m_i+m_j| + |m_i+2m_j| \right) \Big) \; , \notag
\end{align}
wherein $i,j =1,2$, and the contribution of the vector multiplet is given by
\begin{equation}
 \Delta_{\mathrm{V-plet}}=- 2 \sum_i |m_i| 
 -  \Big(  \sum_{i<j} |m_i-m_j| +  \sum_{i<j} |m_i+m_j| \Big) \; .
\end{equation}
\end{subequations}
Such that we will consider the following conformal dimension
\begin{subequations}
\begin{align}
\label{eqn:delta_USp4_generic}
 \Delta(m_1,m_2) &= (N_1 -2) (|m_1| +|m_2|)  + (N_2-1) \left(  |m_1-m_2| +  
 |m_1+m_2| \right) \\
&\qquad +N_3\left( |2m_1-m_2| +|m_1-2m_2|  +|2m_1+m_2| + |m_1+2m_2| \right) 
\notag
\end{align}
and we can vary the representation content via
\begin{align}
 N_1 &= N_{1,0}+2N_{2,0}+2N_{0,2}+2N_{1,1}+5N_{3,0} \; , \\
N_2 &= N_{0,1} + N_{2,0}+3N_{0,2}  \; ,\\
N_3 &= N_{1,1} + N_{3,0} \; .
\end{align}
\end{subequations}
The Hilbert series is computed as usual
\begin{equation}
 \HS_{\USp}(t,N) = \sum_{m_1\geq m_2 \geq 0} t^{\Delta(m_1,m_2)} 
P_{\USp}(t,m_1,m_2) \; ,
\end{equation}
where the summation for $m_1,m_2$ has been restricted to the principal Weyl 
chamber of the GNO-dual group $\sorm(5)$, whose Weyl group is $S_2\ltimes 
(\Z_2)^2$. Thus, we use the reflections to restrict to non-negative 
$m_i\geq0$ and the permutations to restrict to a ordering $m_1\geq m_2$.
The classical dressing factor takes the following form~\cite{Cremonesi:2013lqa}:
\begin{equation}
 P_{\USp}(t,m_1,m_2)= \begin{cases} \frac{1}{(1-t)^2}\; , & m_1 >m_2 >0  \; ,\\
              \frac{1}{(1-t)(1-t^2)} \;,  & (m_1>m_2=0) \vee (m_1=m_2>0)  \; ,\\
                       \frac{1}{(1-t^2)(1-t^4)} \; , & m_1=m_2=0 \; .           
   
    \end{cases}
\end{equation}
%
%%%%%%%%%%%%%%%%%%%%%%%%%%%%%%%%%%%%%%%%%%%%%%%%%%%%%%%%%%%%%%%%%%%%%%%%
%%%%%%%%%%%%%%%%%%%%%%%%%%%%%%%%%%%%%%%%%%%%%%%%%%%%%%%%%%%%%%%%%%%%%%%%
%
\subsection{Hilbert basis}
\begin{figure}
 \begin{center}
  \begin{subfigure}{0.485\textwidth}
  \centering
   \begin{tikzpicture}
  \coordinate (Origin)   at (0,0);
  \coordinate (XAxisMin) at (-0.5,0);
  \coordinate (XAxisMax) at (4.5,0);
  \coordinate (YAxisMin) at (0,-0.5);
  \coordinate (YAxisMax) at (0,4.5);
  \draw [thin, gray,-latex] (XAxisMin) -- (XAxisMax);% Draw x axis
  \draw [thin, gray,-latex] (YAxisMin) -- (YAxisMax);% Draw y axis
  \draw (4.7,-0.2) node {$m_1$};
  \draw (-0.3,4.3) node {$m_2$};
%Draw the root lattice
  \foreach \x in {0,1,...,4}{% Two indices running over each
      \foreach \y in {0,1,...,4}{% node on the grid we have drawn 
        \node[draw,circle,inner sep=0.8pt,fill,black] at (\x,\y) {};
            % Places a dot at those points
            }
            }
   \draw[black,dashed,thick]  (Origin) -- (4.2,4.2);  
   \draw[black,dashed,thick]  (Origin) -- (4.2,0);
   \draw[black,dashed,thick]  (Origin) -- (2.2*2,2.2*1);
% % 
 \draw[black,thick] (1,1) circle (4pt);
 \draw[black,thick] (1,0) circle (4pt);
 \draw[black,thick] (2,1) circle (4pt);
  \draw (3.5,2.5) node {$S_+^{(2)}$};
  \draw (3.5,1.3) node {$S_-^{(2)}$};
\end{tikzpicture}
\caption{$N_3\neq 0$}
\label{Fig:Hilbert_basis_USp4_N3>0}
  \end{subfigure}
    \begin{subfigure}{0.485\textwidth}
    \centering
   \begin{tikzpicture}
  \coordinate (Origin)   at (0,0);
  \coordinate (XAxisMin) at (-0.5,0);
  \coordinate (XAxisMax) at (4.5,0);
  \coordinate (YAxisMin) at (0,-0.5);
  \coordinate (YAxisMax) at (0,4.5);
  \draw [thin, gray,-latex] (XAxisMin) -- (XAxisMax);% Draw x axis
  \draw [thin, gray,-latex] (YAxisMin) -- (YAxisMax);% Draw y axis
  \draw (4.7,-0.2) node {$m_1$};
  \draw (-0.3,4.3) node {$m_2$};
%Draw the root lattice
  \foreach \x in {0,1,...,4}{% Two indices running over each
      \foreach \y in {0,1,...,4}{% node on the grid we have drawn 
        \node[draw,circle,inner sep=0.8pt,fill,black] at (\x,\y) {};
            % Places a dot at those points
            }
            }
    \draw[black,dashed,thick]  (Origin) -- (4.2,4.2);  
    \draw[black,dashed,thick]  (Origin) -- (4.2,0);
% % 
 \draw[black,thick] (1,1) circle (4pt);
 \draw[black,thick] (1,0) circle (4pt);
  \draw (3.5,1.5) node {$S^{(2)}$};
\end{tikzpicture}
\caption{$N_3=0$}
\label{Fig:Hilbert_basis_USp4_N3=0}
  \end{subfigure}
 \end{center}
\caption{The various semi-groups for $\USp$ depending on whether $N_3\neq0$ 
or $N_3=0$. For both cases the black circled points are the ray generators.}
\label{Fig:Hilbert_basis_USp4_generic}
\end{figure}
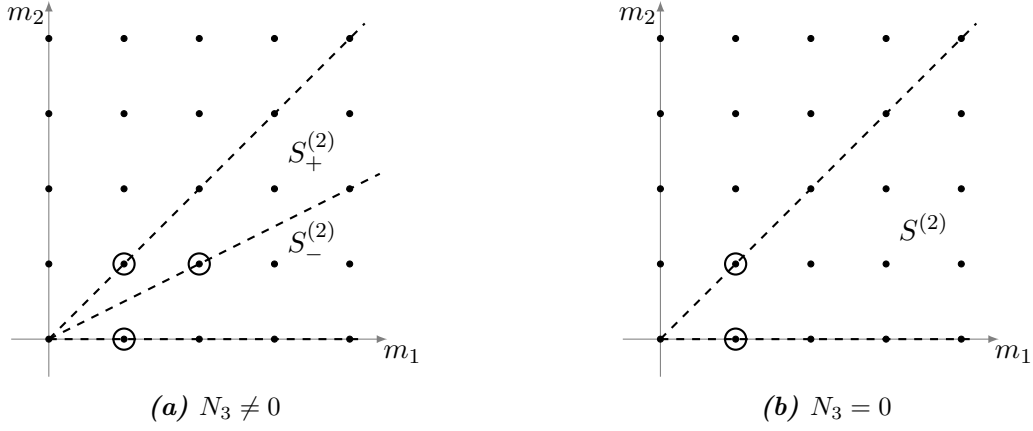

The conformal dimension~\eqref{eqn:delta_USp4_generic} divides the dominant Weyl 
chamber of $\sorm(5)$ into a fan. The intersection with the corresponding weight 
lattice $\Lambda_w(\sorm(5))$ introduces semi-groups $S_p$, which are sketched 
in Fig.~\ref{Fig:Hilbert_basis_USp4_generic}.
As displayed, the set of semi-groups (and rational cones that constitute the 
fan) differ if $N_3\neq0$. The Hilbert bases for both case are readily computed, 
because they coincide with the set of ray generators.
\begin{itemize}
 \item For $N_3\neq 0$, which is displayed in 
Fig.~\ref{Fig:Hilbert_basis_USp4_N3>0}, there exists one hyperplane 
$|m_1-2m_2|=0 $ which intersects the Weyl chamber non-trivially. Therefore, 
$\Lambda_w(\sorm(5)) \slash \mathcal{W}_{\sorm(5)}$ becomes a fan generated by 
two $2$-dimensional cones. The Hilbert bases of the corresponding semi-groups 
are computed to
\begin{equation}
  \Hcal(S_{+}^{(2)}) = \Big\{ (1,1),(2,1) \Big\}   \; , \qquad 
  \Hcal(S_{-}^{(2)}) = \Big\{ (2,1),(1,0)\Big\} \; .
  \label{eqn:Hilbert_basis_USp4_N3>0}
 \end{equation}
 \item For $N_3 =0$, as shown in Fig.~\ref{Fig:Hilbert_basis_USp4_N3=0}, there 
exists no hyperplane that intersects the dominant Weyl chamber non-trivially. As 
a consequence, the $\Lambda_w(\sorm(5)) \slash \mathcal{W}_{\sorm(5)}$ is 
described by one rational polyhedral cone of dimension $2$. The Hilbert basis 
for the semi-group is given by
 \begin{equation}
  \Hcal(S^{(2)}) = \Big\{ (1,1),(1,0) \Big\} \; .
  \label{eqn:Hilbert_basis_USp4_N3=0}
 \end{equation}
\end{itemize}
% 
%%%%%%%%%%%%%%%%%%%%%%%%%%%%%%%%%%%%%%%%%%%%%%%%%%%%%%%%%%%%%%%%%%%%%%%%%
%%%%%%%%%%%%%%%%%%%%%%%%%%%%%%%%%%%%%%%%%%%%%%%%%%%%%%%%%%%%%%%%%%%%%%%%%
%
\subsection{Dressings}
Before evaluating the Hilbert series, let us analyse the classical dressing 
factors for the minimal generators~\eqref{eqn:Hilbert_basis_USp4_N3>0} 
or~\eqref{eqn:Hilbert_basis_USp4_N3=0}.
Firstly, the classical Lie group $\USp$ has two Casimir invariants of 
degree $2$ and $4$ and can they can be written as $\tr(\Phi^2) = \sum_{i=1}^2 
(\phi_i)^2$ and $\tr(\Phi^4)= \sum_{i=1}^2 (\phi_i)^4$, respectively. 
Again, we employ the diagonal form of the adjoint valued scalar field $\Phi$.

Secondly, the bare monopole operator corresponding to GNO-charge $(1,0)$ has 
conformal dimension $N_1+ 2N_2 +6N_3 -4$ and the residual gauge group is 
$\Hh_{(1,0)}=\uo\times\su$, i.e.\ allowing for dressings by degree $1$ and $2$ 
Casimirs. The resulting set of bare and dressed monopoles is
\begin{subequations}
\begin{align}
 V_{(1,0)}^{\mathrm{dress},0} &= (1,0) + (-1,0) + (0,1) + (0,-1) \; , \\
V_{(1,0)}^{\mathrm{dress},2} &= 
\left( (1,0) + (-1,0) \right) (\phi_2)^2 
+ \left((0,1) + (0,-1) \right) (\phi_1)^2 \; ,\\
V_{(1,0)}^{\mathrm{dress},1} &= 
\left( (1,0) - (-1,0) \right) \phi_1 
+ \left((0,1) - (0,-1) \right) \phi_2 \; ,\\
V_{(1,0)}^{\mathrm{dress},3} &= 
\left( (1,0) - (-1,0) \right) (\phi_1)^3 
+ \left((0,1) - (0,-1) \right) (\phi_2)^3 \; .
\end{align}
\end{subequations}
 
Thirdly, the bare monopole operators of GNO-charge $(1,1)$ has conformal 
dimension 
$2N_1+2N_2+8N_3-6$ and residual gauge group $\Hh_{(1,1)} = \uo \times \su$. The 
bare and dressed monopole operators can be written as 
\begin{subequations}
 \begin{align}
 V_{(1,1)}^{\mathrm{dress},0} &= (1,1) + (1,-1)+  (-1,1) +(-1,-1) \; , \\
 V_{(1,1)}^{\mathrm{dress},2} &= ( (1,1) + (-1,-1)) ( (\phi_1)^2 + (\phi_2)^2 ) 
+  (1,-1) (\phi_2)^2 +  (-1,1) (\phi_1)^2 \; ,\\
 V_{(1,1)}^{\mathrm{dress},1} &= (1,1) (\phi_1+ \phi_2) + (-1,-1) 
(-\phi_1-\phi_2) + (1,-1)(-\phi_2) + (-1,1)(-\phi_1) \; ,\\
V_{(1,1)}^{\mathrm{dress},3} &= (1,1) ((\phi_1)^3+ (\phi_2)^3) + (-1,-1) 
(-(\phi_1)^3-(\phi_2)^3) \\
&\phantom{= (1,1) ((\phi_1)^3+ (\phi_2)^3) \; } 
+ (1,-1)(-(\phi_2)^3) + (-1,1)(-(\phi_1)^3) \; .\notag
 \end{align}
\end{subequations}
The two magnetic weights $(1,0)$, $(1,1)$ lie at the boundary of the dominant 
Weyl chamber such that the dressing behaviour can be predicted by 
$P_{\USp}(t,m_1,m_2) \slash P_{\USp}(t,0,0) = 1+t+t^2+t^3 $, following 
App.~\ref{app:PL}. The above description of the bare and dressed monopole 
operators is therefore a valid choice of generating elements for the chiral 
ring.

Lastly, the bare monopole for $(2,1)$ has conformal dimension 
$3N_1+4N_2+12N_3-10$ and residual gauge group $H_{(2,1)}=\uo^2$. Thus, the 
dressing proceeds by two independent degree $1$ Casimir invariants.
\begin{subequations}
 \begin{align}
  V_{(2,1)}^{\mathrm{dress},0} &= (2,1) + (2,-1) + (-2,1) + (1,2) + (1,-2) 
+ (-1,2)+ (-1,-2) + (-2,-1) \notag \\
&\equiv (2,1) + (2,-1) + (-2,1) + (-2,-1)+\text{permutations} \; ,\\
 V_{(2,1)}^{\mathrm{dress},2j-1,1}&= 
 (2,1) (\phi_1)^{2j-1} 
 + (2,-1)(\phi_1)^{2j-1} 
 + (-2,1)(-\phi_1)^{2j-1} \\
 &\qquad + (-2,-1)(-\phi_1)^{2j-1}+\text{permutations} \for j=1,2 \; , \notag\\
 V_{(2,1)}^{\mathrm{dress},2j-1,2}&= 
 (2,1) (\phi_2)^{2j-1} 
 + (2,-1)(-\phi_2)^{2j-1} 
 + (-2,1)(\phi_2)^{2j-1} \\
 &\qquad + (-2,-1)(-\phi_2)^{2j-1}+\text{permutations} \for j=1,2 \; , \notag\\
 V_{(2,1)}^{\mathrm{dress},2,1}&= 
 (2,1) (\phi_1)^2 
 + (2,-1)(-(\phi_1)^2)
 + (-2,1)(-(\phi_1)^2) \\
 &\qquad + (-2,-1)(\phi_1)^2+\text{permutations} \; , \notag \\
 V_{(2,1)}^{\mathrm{dress},2,2}&= 
 (2,1) (\phi_1 \phi_2) 
 + (2,-1)(-\phi_1 \phi_2)
 + (-2,1)(-\phi_1 \phi_2) \\
 &\qquad 
 + (-2,-1)(\phi_1 \phi_2)+\text{permutations}  \; , \notag\\
 V_{(2,1)}^{\mathrm{dress},4}&= 
 (2,1) (\phi_1^3 \phi_2) 
 + (2,-1)(-(\phi_1)^3 \phi_2)
 + (-2,1)(-(\phi_1)^3 \phi_2) \\
&\qquad + (-2,-1)((\phi_1)^3 \phi_2)+\text{permutations} \; . \notag
 \end{align}
\end{subequations}
The number and the degrees of dressed monopole operators of charge $(2,1)$ are 
consistent with the quotient $P_{\USp}(t,m_1>m_2>0) \slash P_{\USp}(t,0,0) = 
1+2t+2t^2+2t^3+t^4 $ of the dressing factors.

For ``generic'' values of $N_1$, $N_2$ and $N_3$ the Coulomb branch will be 
generated by the two Casimir invariants together with the bare and dressed 
monopole operators corresponding to the minimal generators of the Hilbert 
bases. However, we will  encounter choices of the three parameters such that 
the set of monopole generators can be further reduced; for example, in the case 
of complete intersections.
% 
%%%%%%%%%%%%%%%%%%%%%%%%%%%%%%%%%%%%%%%%%%%%%%%%%%%%%%%%%%%%%%% 
% 
\subsection{Generic case}
The computation for arbitrary $N_1$, $N_2$, and $N_3$ yields
\begin{subequations}
\label{eqn:HS_USp4_generic}
\begin{equation}
 \HS_{\USp}(t,N_1,N_2,N_3)=\frac{R(t,N_1,N_2,N_3)}{P(t,N_1,N_2,N_3)} \; ,
\end{equation}
with
\begin{align}
 P(t,N_1,N_2,N_3)&=\left(1-t^2\right) 
\left(1-t^4\right) 
\left(1-t^{N_1+2 N_2+6 N_3-4}\right) 
\left(1-t^{2 N_1+2 N_2+8 N_3-6}\right) 
\label{eqn:HS_USp4_Den}\\*
&\phantom{\left(1-t^2\right) \left(1-t^4\right)} 
\times
\left(1-t^{3 N_1+4 N_2+12 N_3-10}\right) \; , \notag\\
R(t,N_1,N_2,N_3) &=1
+t^{N_1+2 N_2+6 N_3-3} ( 1+t+t^2)
+t^{2 N_1+2 N_2+8 N_3-5} (1+t+t^2)
\label{eqn:HS_USp4_Num} \\
&\qquad
+t^{3 N_1+4 N_2+12 N_3-9} (2+2t+2t^2+t^3)
-t^{3 N_1+4 N_2+14 N_3-10} (1+2t+2t^2+2t^3)
\notag \\
&\qquad
-t^{4 N_1+6 N_2+18 N_3-13}(1+t+t^2)
-t^{5 N_1+6 N_2+20 N_3-15}(1+t+t^2)
\notag \\
&\qquad
-t^{6 N_1+8 N_2+26 N_3-16} \; .
\notag
\end{align}
\end{subequations}
The numerator~\eqref{eqn:HS_USp4_Num} is an anti-palindromic polynomial of 
degree $6N_1+8N_2+26 N_3-16$; while the denominator is of degree $6 N_1+8N_2+26 
N_3-14$. The difference in degrees is $2$, which equals the quaternionic 
dimension of the moduli space.
In addition, the pole of~\eqref{eqn:HS_USp4_generic} at $t\to1$ is of order 
$4$, which matches the complex dimension of the 
moduli space. For that, one verifies explicitly $R(t=1,N_1,N_2,N_3)=0$, but 
$\tfrac{\diff}{\diff t} R(t,N_1,N_2,N_3) |_{t=1}\neq0$.

Consequently, the above interpretation of bare and dressed monopoles from the 
Hilbert series~\eqref{eqn:HS_USp4_generic} is correct for ``generic'' choices 
of $N_1$, $N_2$, and $N_3$. In particular, $N_3\neq 0$ for this arguments to 
hold. Moreover, we will now exemplify the effects of the Casimir invariance in 
various special case of~\eqref{eqn:HS_USp4_generic} explicitly. There are 
cases for which the inclusion of the Casimir invariance, i.e.\ dressed 
monopole operators, leads to a reduction of basis of monopole 
generators.
% 
% 
%%%%%%%%%%%%%%%%%%%%%%%%%%%%%%%%%%%%%%%%%%%%%%%%
%%%%%%%%%%%%%%%%%%%%%%%%%%%%%%%%%%%%%%%%%%%%%%%%
%
\subsection{Category \texorpdfstring{$N_3=0$}{N3=0}}
\label{subsec:USp4_N3=0}
\subsubsection{Representation \texorpdfstring{$[1,0]$}{[1,0]}}
\paragraph{Hilbert series}
This choice is realised for $N_1=N$, $N_2=N_3=0$ and the Hilbert series 
simplifies drastically to a complete intersection
\begin{equation}
 \HS_{\USp}^{[1,0]}(t,N)=
 \frac{(1-t^{2N-4}) (1-t^{2N-2})}{(1-t^2) (1-t^4) 
(1-t^{N-4}) (1-t^{N-3}) (1-t^{N-2}) (1-t^{N-1})} \; ,
\label{eqn:HS_USp4_Rep10}
\end{equation}
which was first obtained in~\cite{Cremonesi:2013lqa}.
Due to the complete intersection property, the plethystic logarithm terminates 
and for $N>4$ we obtain
\begin{align}
 \mathrm{PL}(\HS_{\USp}^{[1,0]})=t^2+t^4&+t^{N-4}(1+t+t^2+t^3) -t^{2N-4} - 
t^{2N-2} \; .
\end{align}
\paragraph{Hilbert basis}
Naively, the Hilbert series~\eqref{eqn:HS_USp4_Rep10} should be generated by 
the Hilbert basis~\eqref{eqn:Hilbert_basis_USp4_N3=0} plus their dressings. 
However, due to the particular form~\eqref{eqn:delta_USp4_generic} in 
representation $[1,0]$ and the Casimir invariance, the bare monopole operator 
of 
GNO-charge $(1,1)$ can be generated by the dressings of $(1,0)$. To see this, 
consider the Weyl-orbit $ \mathcal{O}_{\mathcal{W}} (1,0) = \big\{ (1,0), 
(0,1), 
(-1,0), (0,-1) \big\}$ and note the conformal dimensions align suitably, i.e.\ 
$\Delta(V_{(1,0)}^{\mathrm{dress},1})= N-3$, while 
$\Delta(V_{(1,1)}^{\mathrm{dress},0})=2N-6$. Thus, we can symbolically write
\begin{equation} 
V_{(1,1)}^{\mathrm{dress},0}=V_{(1,0)}^{\mathrm{dress},1}
+V_{(0,1)}^{\mathrm{dress},1} \; .
\end{equation}
The moduli space is then generated by the Casimir invariants and the bare and 
dressed monopole operators corresponding to $(1,0)$, but this is to be 
understood as a rather ``non-generic'' situation.
% 
%%%%%%%%%%%%%%%%%%%%%%%%%%%%%%%%%%%%%%%%%%%%%%%%
%%%%%%%%%%%%%%%%%%%%%%%%%%%%%%%%%%%%%%%%%%%%%%%%
%
\subsubsection{Representation \texorpdfstring{$[0,1]$}{[0,1]}}
This choice is realised for $N_2=N$, and $N_1=N_3=0$ and the Hilbert series 
simplifies to
\begin{equation}
\label{eqn:HS_USp4_Rep01}
 \HS_{\USp}^{[0,1]}(t,N)=
 \frac{1+t^{2 N-5}+t^{2 N-4}+2 t^{2 N-3}+t^{2 
N-2}+t^{2 N-1}+t^{4 N-6}}{\left(1-t^2\right) 
\left(1-t^4\right) \left(1-t^{2 N-6}\right) \left(1-t^{2 N-4}\right)} \; .
\end{equation}
The Hilbert series~\eqref{eqn:HS_USp4_Rep01} has a pole of order $4$ at $t=1$ 
as well as a palindromic polynomial as numerator. Moreover, the 
result~\eqref{eqn:HS_USp4_Rep01} reflects the expected basis of monopole 
operators as given in the Hilbert basis~\eqref{eqn:Hilbert_basis_USp4_N3=0}. 
% 
%%%%%%%%%%%%%%%%%%%%%%%%%%%%%%%%%%%%%%%%%%%%%%%%
%%%%%%%%%%%%%%%%%%%%%%%%%%%%%%%%%%%%%%%%%%%%%%%%
%
\subsubsection{Representation \texorpdfstring{$[2,0]$}{[2,0]}}
This choice is realised for $N_1=2N$, $N_2=N$, and $N_3=0$ and the Hilbert 
series reduces to
\begin{equation}
\label{eqn:HS_USp4_Rep20}
 \HS_{\USp}^{[2,0]}(t,N)=
\frac{1+t^{4 N-3}+t^{4 N-2}+t^{4 N-1}+t^{6 
N-5}+t^{6 N-4}+t^{6 N-3}+t^{10 
N-6}}{\left(1-t^2\right) \left(1-t^4\right) \left(1-t^{4 
N-4}\right) \left(1-t^{6 N-6}\right)} \; .
\end{equation}
Also, the rational function~\eqref{eqn:HS_USp4_Rep20} has a pole of order $4$ 
for $t\to1$ and a palindromic numerator.
Evaluating the plethystic logarithm yields for all $N>1$
\begin{align}
 \mathrm{PL}(\HS_{\USp}^{[2,0]})=t^2+t^4&+t^{4N-4}(1+t+t^2+t^3) \\
 &+ t^{6N-6}(1+t+t^2+t^3) - t^{8N-6} +\mathcal{O}(t^{8N-5}) \notag \; .
\end{align}
This proves that bare monopole operators, corresponding to the
the minimal generators of~\eqref{eqn:Hilbert_basis_USp4_N3=0}, together with 
their dressing generate all other monopole operators.
%
%%%%%%%%%%%%%%%%%%%%%%%%%%%%%%%%%%%%%%%%%%%%%%%%
%%%%%%%%%%%%%%%%%%%%%%%%%%%%%%%%%%%%%%%%%%%%%%%%
%
\subsubsection{Representation \texorpdfstring{$[0,2]$}{[0,2]}}
For $N_1=2N$, $N_2=3N$, and $N_3=0$ and the Hilbert 
series is given by
\begin{equation}
\label{eqn:HS_USp4_Rep02}
 \HS_{\USp}^{[0,2]}(t,N)=
\frac{1+t^{8 N-3}+t^{8 N-2}+t^{8 N-1}+t^{10 
N-5}+t^{10 N-4}+t^{10 N-3}+t^{18 
N-6}}{\left(1-t^2\right) \left(1-t^4\right) \left(1-t^{8N-4}\right) 
\left(1-t^{10 N-6}\right)} \; .
\end{equation}
Evaluating the plethystic logarithm yields for all $N>1$
\begin{subequations}
\begin{align}
 \PL(\HS_{\USp}^{[0,2]})=t^2+t^4&+t^{8N-4}(1+t+t^2+t^3) \\
 &+ t^{10N-6}(1+t+t^2+t^3) - t^{16N-6} +\mathcal{O}(t^{16N-5}) \notag \; ,
\end{align}
and for $N=1$
\begin{align}
 \PL(\HS_{\USp}^{[0,2]})=t^2+t^4&+t^{4}(1+t+t^2+t^3) \\
 &+ t^{4}(1+t+t^2+t^3) - 3t^{10} +\mathcal{O}(t^{11}) \notag \; .
\end{align}
\end{subequations}
The inspection of the Hilbert series~\eqref{eqn:HS_USp4_Rep02}, together with 
the PL, proves that Hilbert basis~\eqref{eqn:Hilbert_basis_USp4_N3=0}, 
alongside 
all their dressings, are a sufficient set for all monopole operators.
% 
%%%%%%%%%%%%%%%%%%%%%%%%%%%%%%%%%%%%%%%%%%%%%%%%
%%%%%%%%%%%%%%%%%%%%%%%%%%%%%%%%%%%%%%%%%%%%%%%%
%
\subsection{Category \texorpdfstring{$N_3 \neq0$}{N3>0}}
\subsubsection{Representation \texorpdfstring{$[1,1]$}{[1,1]}}
This choice corresponds to $N_1=2N$, $N_2=0$, and $N_3=N$ and we obtain the  
Hilbert series to be
\begin{subequations}
\label{eqn:HS_USp4_Rep11}
\begin{align}
 \HS_{\USp}^{[1,1]}(t,N)&=
\frac{R(t,N)}{\left(1-t^2\right) \left(1-t^4\right) 
\left(1-t^{8 N-4}\right) \left(1-t^{12 N-6}\right) \left(1-t^{18 N-10}\right)} 
\; ,
\\
R(t,N)&=1+t^{8 N-3}(1+t+t^2) +t^{12 N-5}(1+t+t^2) \\
&\qquad+ t^{18 N-9}(2+2t+2t^2+t^3) -t^{20 N-10}(1+2t+2t^2+2t^3)\notag \\
&\qquad-t^{26 N-13}(1+t+t^2) -t^{30 N-15}(1+t+t^2)-t^{38 N-16} \notag \; .
\end{align}
\end{subequations}
Considering the plethystic logarithm, we observe the following behaviour:
\begin{subequations}
\begin{itemize}
 \item For $N\geq5$ 
 \begin{align}
  \mathrm{PL}(\HS_{\USp}^{[1,1]})= t^2+t^4 &+ t^{8 N-4}(1+t+t^2+t^3) 
  + t^{12 N-6}(1+t+t^2+t^3) \\
  &-t^{2(8 N-4)+2}(1+t+2t^2+t^3+t^4) \notag \\
  &+ t^{18 N-10}(1+2t+2t^2+2t^3+t^4) \notag \\
&-t^{20N-10}(1+2t+3t^2+4t^3+3t^4+2t^5+t^6)+\ldots \notag
 \end{align}
  \item For $N=4$
  \begin{align}
  \mathrm{PL}(\HS_{\USp}^{[1,1]})= t^2+t^4 
  &+ t^{28}(1+t+t^2+t^3) 
  + t^{42}(1+t+t^2+t^3) \\
  &-t^{58}(1+t+2t^2+t^3+\textcolor{red}{t^4}) \notag \\
  &+ t^{62}(\textcolor{red}{1}+2t+2t^2+2t^3+t^4) \notag \\
&-t^{70}(1+2t+3t^2+4t^3+3t^4+2t^5+t^6)+\ldots \notag
 \end{align}
 We see, employing the previous results for $N>4$, that the bare monopole 
$(2,1)$ and the last relation at $t^{62}$ coincide. Hence, the term $\sim 
t^{62}$ disappears from the PL.
  \item For $N=3$
  \begin{align}
  \mathrm{PL}(\HS_{\USp}^{[1,1]})= t^2+t^4 
  &+ t^{20}(1+t+t^2+t^3) 
  + t^{30}(1+t+t^2+t^3) \\
&-t^{42}(1+t
+\textcolor{red}{2t^2}
+\textcolor{ForestGreen}{t^3}
+\textcolor{blue}{t^4}
) \notag \\
  &+ 
t^{44}(\textcolor{red}{1}
+\textcolor{ForestGreen}{2t}+\textcolor{blue}{2t^2}
+2t^3+t^4) \notag \\
&-t^{70}(1+2t+3t^2+4t^3+3t^4+2t^5+t^6)+\ldots \notag
 \end{align}
 We see, employing again the previous results for $N>4$, that the some monopole 
contributions of $(2,1)$ and the some of the relations coincide, c.f. the 
coloured terms. Hence, there are, presumably, cancellations between generators 
and relations.
 \item For $N=2$
  \begin{align}
  \mathrm{PL}(\HS_{\USp}^{[1,1]})= t^2+t^4 
  &+ t^{12}(1+t+t^2+t^3) 
  + t^{18}(1+t+t^2+t^3) \\
&-t^{26}(\textcolor{red}{1}
+\textcolor{ForestGreen}{t}
+\textcolor{blue}{2t^2}
+\textcolor{violet}{t^3}
+\textcolor{brown}{t^4}
) \notag \\
  &+ 
t^{26}(\textcolor{red}{1}
+\textcolor{ForestGreen}{2t}
+\textcolor{blue}{2t^2}
+\textcolor{violet}{2t^3}
+\textcolor{brown}{t^4}) \notag \\
&-t^{30}(\textcolor{brown}{1}+2t+3t^2+4t^3+3t^4+2t^5+t^6)+\ldots \notag \\
= t^2+t^4 
  &+ t^{12}(1+t+t^2+t^3) 
  + t^{18}(1+t+t^2+t^3) \\
  &+ 
t^{26}(\textcolor{red}{0}
+\textcolor{ForestGreen}{t}
+\textcolor{blue}{0}
+\textcolor{violet}{t^3}
+\textcolor{brown}{0}) \notag \\
&-t^{30}(\textcolor{brown}{1}+2t+3t^2+4t^3+3t^4+2t^5+t^6)+\ldots \notag
 \end{align}
\item For $N=1$
\begin{align}
   \mathrm{PL}(\HS_{\USp}^{[1,1]})= 
t^2 + 2t^4 + t^5 + 2t^6+2t^7+2t^8+3t^9-t^{11} + \ldots
\end{align}
\end{itemize}
\end{subequations}
Summarising, the Hilbert series~\eqref{eqn:HS_USp4_Rep11} and its plethystic 
logarithm display that the minimal generators 
of~\eqref{eqn:Hilbert_basis_USp4_N3>0} are indeed the basis for the bare 
monopole operators, and the corresponding dressings generate the remaining 
operators.
% 
%%%%%%%%%%%%%%%%%%%%%%%%%%%%%%%%%%%%%%%%%%%%%%%%
%%%%%%%%%%%%%%%%%%%%%%%%%%%%%%%%%%%%%%%%%%%%%%%%
%
\subsubsection{Representation \texorpdfstring{$[3,0]$}{[3,0]}}
For the choice $N_1=5N$, $N_2=0$, and $N_3=N$ the Hilbert 
series is given by
\begin{subequations}
\label{eqn:HS_USp4_Rep30}
\begin{align}
 \HS_{\USp}^{[3,0]}(t,N)&=
\frac{R(t,N)}{\left(1-t^2\right) \left(1-t^4\right) 
\left(1-t^{11 N-4}\right) \left(1-t^{18 N-6}\right) 
\left(1-t^{27 N-10}\right)} \; ,
\\
R(t,N)&=1+t^{11 N-3}(1+t+t^2)
+t^{18 N-5}(1+t+t^2)  \\
&\qquad + t^{27 N-9}(2+2t+2t^2+t^3) 
-t^{29 N-10}(1+2t+2t^2+2t^3) \notag \\
&\qquad
-t^{38 N-13}(1+t+t^2)
-t^{45 N-15}(1+t+t^2)
-t^{56N-16}\notag \; .
\end{align}
\end{subequations}
The inspection of the plethystic logarithm provides further insights:
\begin{subequations}
\begin{itemize}
 \item For $N\geq3$
  \begin{align}
  \PL(\HS_{\USp}^{[3,0]})= t^2+t^4 &+ t^{11 N-4}(1+t+t^2+t^3) 
  + t^{18 N-6}(1+t+t^2+t^3) \\
  &-t^{2(11 N-4)+2}(1+t+2t^2+t^3+t^4) \notag \\
  &+ t^{27 N-10}(1+2t+2t^2+2t^3+t^4) \notag \\
&-t^{(11 N-4)+(18N-6)}(1+2t+3t^2+4t^3+3t^4+2t^5+t^6)+\ldots \notag
 \end{align}
\item For $N=2$
  \begin{align}
  \PL(\HS_{\USp}^{[3,0]})= t^2+t^4 &+ t^{18}(1+t+t^2+t^3) 
  + t^{30}(1+t+t^2+t^3) \\
  &-t^{38}(1+t+2t^2+t^3+t^4) \notag \\
  &+ t^{44}(1+2t+2t^2+2t^3+\textcolor{red}{t^4}) \notag \\
&-t^{48}(\textcolor{red}{1} +2t+3t^2+4t^3+3t^4+2t^5+t^6)+\ldots \notag
 \end{align}
 We see that, presumably, one generator and one relation cancel at $~t^{48}$.
\item For $N=1$
  \begin{align}
  \mathrm{PL}(\HS_{\USp}^{[3,0]})= 
t^2+t^4+t^7(1+t+t^2+t^3)+t^{12}(1+t+t^2+t^3)-t^{16}-t^{20}+\ldots
 \end{align}
\end{itemize}
\end{subequations}
Again, we confirm that the minimal generators of the Hilbert 
basis~\eqref{eqn:Hilbert_basis_USp4_N3>0} are the relevant generators (together 
with their dressings) for the moduli space.
%
%%%%%%%%%%%%%%%%%%%%%%%%%%%%%%%%%%%%%%%%%%%%%%%%%%%%%%%%%%%%%%%%%%%%%%%%%%%%%%%%
  \section{Case: \texorpdfstring{$\boldsymbol{\Gtwo}$}{G2}}
\label{sec:G2}
Here, we study the Coulomb branch for the only exceptional simple Lie group of 
rank two. 
\subsection{Set-up}
The group $\Gtwo$ has irreducible representations labelled by two Dynkin labels 
and the dimension formula reads
\begin{equation}
\dim[a,b]= \frac{1}{120} (a+1) (b+1) (a+b+2) (a+2 b+3) (a+3 b+4) (2 a+3 b+5) \;.
\end{equation}
In the following, we study the representations given in 
Tab.~\ref{tab:G2_overview_reps}. The three categories defined are due to the 
similar form of the conformal dimensions.
\begin{table}[h]
\centering
 \begin{tabular}{c|c|c|c|c|c|c|c|c}
 \toprule
  Dynkin label & $[1,0]$ & $[0,1]$ & $[2,0]$ & $[1,1]$ & $[0,2]$ & $[3,0]$ & 
$[4,0]$ & $[2,1]$ \\ 
\midrule
  Dim. & $7$ & $14$ & $27$ & $64$ & $77$ & $77$ & $182$ & $189$ \\
   & \multicolumn{3}{c|}{category 1} & \multicolumn{3}{c|}{category 2} & 
\multicolumn{2}{c}{category 3} \\
\bottomrule
 \end{tabular}
\caption{An overview of the $\Gtwo$-representations considered in this paper.}
\label{tab:G2_overview_reps}
\end{table}

The Weyl group of $\Gtwo$ is $D_6$ and the GNO-dual group is another $\Gtwo$. 
Any element in the Cartan subalgebra $\hfrak=\mathrm{span}(H_1,H_2)$ can 
be written as $H=n_1 H_1 + n_2 H_2$. Restriction to the principal Weyl chamber 
is realised via $n_1,n_2\geq0$.

The group $\Gtwo$ has two Casimir invariants of degree $2$ and $6$. Therefore, 
the classical dressing function is~\cite{Cremonesi:2013lqa}
\begin{align}
 P_{\Gtwo}(t,n_1,n_2)= \left\{\begin{matrix} \frac{1}{(1-t^2)(1-t^6)} \; ,& 
n_1=n_2=0 \; ,\\
\frac{1}{(1-t)(1-t^2)} \; , & {n_1>0,n_2=0 }\text{ or }{n_1=0,n_2>0} \; , \\
\frac{1}{(1-t)^2} \; ,& n_1,n_2>0 \; .
\end{matrix} \right.
\end{align}
%
%%%%%%%%%%%%%%%%%%%%%%%%%%%%%%%%%%%%%%%%%%%%%%%%%%%%%%%%%%%%%%%%%%%%%%%%%
%
\subsection{Category 1}
\label{subsec:G2_Cat1}
\paragraph{Hilbert basis}
The representations $[1,0]$, $[0,1]$, and $[2,0]$ have schematically conformal 
dimensions of the form
\begin{equation}
 \Delta(n_1,n_2)= \sum_j A_j | a_j n_1 + b_j n_2| + B_1 |n_1|+ B_2 |n_2| 
 \label{eqn:delta_G2_Cat1}
\end{equation}
for $a_j,b_j \in \NN$ and $ A_j,B_1,B_2 \in \Z$.
As a consequence, the usual fan within the Weyl chamber is simply one 
$2$-dimensional rational polyhedral cone
 \begin{equation}
  C^{(2)}= \cone((1,0),(0,1)) \; .
 \end{equation}
The intersection with the weight lattice $\Lambda_w(\Gtwo)$ yields the 
relevant semi-group $S^{(2)}$, as depicted in 
Fig.~\ref{Fig:Hilbert_basis_G2_Cat1}.
The Hilbert bases are trivially given by the ray generators
\begin{equation}
 \Hcal(S^{(2)}) = \Big\{(1,0),(0,1) \Big\} \; .
 \label{eqn:Hilbert_basis_G2_Cat1}
\end{equation}
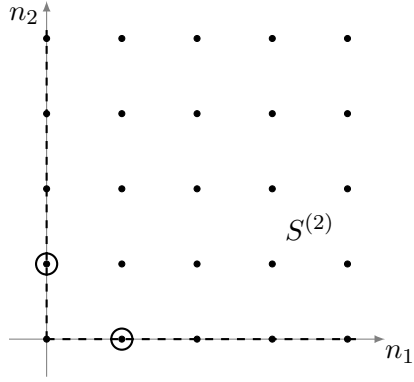
\begin{figure}[h]
\centering
   \begin{tikzpicture}
  \coordinate (Origin)   at (0,0);
  \coordinate (XAxisMin) at (-0.5,0);
  \coordinate (XAxisMax) at (4.5,0);
  \coordinate (YAxisMin) at (0,-0.5);
  \coordinate (YAxisMax) at (0,4.5);
  \draw [thin, gray,-latex] (XAxisMin) -- (XAxisMax);%
  \draw [thin, gray,-latex] (YAxisMin) -- (YAxisMax);%
  \draw (4.7,-0.2) node {$n_1$};
  \draw (-0.3,4.3) node {$n_2$};
  \foreach \x in {0,1,...,4}{%
      \foreach \y in {0,1,...,4}{%
        \node[draw,circle,inner sep=0.8pt,fill,black] at (\x,\y) {};
            }
            }
    \draw[black,dashed,thick]  (Origin) -- (0,4.2);  
    \draw[black,dashed,thick]  (Origin) -- (4.2,0);
% % 
 \draw[black,thick] (1,0) circle (4pt);
 \draw[black,thick] (0,1) circle (4pt);
  \draw (3.5,1.5) node {$S^{(2)}$};
\end{tikzpicture}
\caption{The semi-group $S^{(2)}$ for the representations $[1,0]$, $[0,1]$, and 
$[2,0]$ obtained from the $\Gtwo$ Weyl chamber (considered as rational cone) 
and its ray generators (black circled points).}
\label{Fig:Hilbert_basis_G2_Cat1}
\end{figure}
\paragraph{Dressings}
The two minimal generators lie at the boundary of the Weyl chamber and, 
therefore, have residual gauge group $\Hh_{(1,0)}=\Hh_{(0,1)}=\utwo$. Recalling 
that $\Gtwo$ has two Casimir invariants $\mathcal{C}_2$, $\mathcal{C}_6$ at 
degree $2$ and $6$, one can analyse the dressed monopole operators 
associated to $(1,0)$ and $(0,1)$.

The residual gauge group $\utwo \subset \Gtwo$ has a degree one Casimir 
$C_1\coloneqq \phi_1+\phi_2$ and a degree two Casimir 
$C_2\coloneqq\phi_1^2+\phi_2^2$. Again, we employed the diagonal form of the 
adjoint-valued scalar $\Phi$.
Consequently, the bare monopole $V_{(0,1)}^{\mathrm{dress},0}$ exhibits five 
dressed monopoles  $V_{(0,1)}^{\mathrm{dress},i}$ ($i=1,\ldots,5$) 
of degrees $\Delta(0,1)+1, \ldots,\Delta(0,1)+5$. Since the 
highest degree Casimir invariant is of order $6$ and the degree $2$ Casimir 
invariant of $\Gtwo$ differs from the pure sum of squares~\cite{Okubo}, one can 
build all dressings as follows:
\begin{align}
 C_1 (0,1) \; , \quad C_2 (0,1)\; , \quad C_1 C_2 (0,1) \; , \quad C_1^2 C_2 
(0,1)\; , \quad (C_1 C_2^2 + C_1^2 C_2) (0,1) \; .
\end{align}
The very same arguments applies for the bare and dressed monopole generators 
associated to $(1,0)$. Thus, we expect six monopole operators: one bare  
$V_{(1,0)}^{\mathrm{dress},0}$ and five dressed  $V_{(1,0)}^{\mathrm{dress},i}$ 
($i=1,\ldots,5$).

Comparing with App.~\ref{app:PL}, we find that a magnetic weight at the 
boundary of the dominant Weyl chamber has dressings given by 
$P_{\Gtwo}(t,\{n_1=0 \text{ or }n_2=0\}) \slash 
P_{\Gtwo}(t,0,0)=1+t+t^2+t^3+t^4+t^5 $, which is then consistent with the 
exposition above.

We will now exemplify the three different representations.
\subsubsection{Representation \texorpdfstring{$[1,0]$}{[1,0]}}
The relevant computation has been presented in~\cite{Cremonesi:2013lqa} and the 
conformal dimension reads
\begin{align}
\label{eqn:delta_G2_Rep10}
\Delta(n_1,n_2)= &N (\left| n_1+n_2\right| +\left| 2 
n_1+n_2\right| +\left| n_1\right| ) \\
&-(\left| 
n_1+n_2\right| +\left| 2 n_1+n_2\right| +\left| 3 
n_1+n_2\right| +\left| 3 n_1+2 n_2\right| +\left| 
n_1\right| +\left| n_2\right| ) \notag \; .
\end{align}
Evaluating the Hilbert series for $N>3$ yields
\begin{equation}
\label{eqn:HS_G2_Rep10}
 \HS_{\Gtwo}^{[1,0]}(t,N) =
 \frac{1+t^{2 N-4}+t^{2 N-3}+t^{2 
N-2}+t^{2 N-1}+t^{4 N-5}}{\left(1-t^2\right) \left(1-t^6\right) 
\left(1-t^{2 N-6}\right) \left(1-t^{2 N-5}\right)} \; .
\end{equation}
We observe that the numerator of~\eqref{eqn:HS_G2_Rep10} is a palindromic 
polynomial of degree $4N-5$; while, the denominator has degree 
$4N-3$. Hence, the difference in degree between denominator and numerator is 
$2$, which equals the quaternionic dimension of moduli space.
 In addition, the Hilbert series~\eqref{eqn:HS_G2_Rep10} has a pole of order 
$4$ 
as $t\to 1$, which matches the complex dimension of the moduli space.

As discussed in~\cite{Cremonesi:2013lqa}, the plethystic logarithm has the 
following behaviour:
\begin{equation}
\mathrm{PL}(\HS_{\Gtwo}^{[1,0]}(t,N)) = t^2 + t^6+ 
t^{2N-6}(1+t+t^2+t^3+t^4+t^5) -t^{4N-8} + \ldots \; .
\label{eqn:PL_G2_7dim}
\end{equation}
\paragraph{Hilbert basis}
According to~\cite{Cremonesi:2013lqa}, the monopole corresponding to GNO-charge 
$(1,0)$, which has $\Delta(1,0)=4N-10$, can be generated. Again, this is due to 
the specific form~\eqref{eqn:delta_G2_Rep10}.
% 
%%%%%%%%%%%%%%%%%%%%%%%%%%%%%%%%%%%%%%%%%%%%%%%%%%%%%%%%%%%%%%%%%%%%%%%%
%%%%%%%%%%%%%%%%%%%%%%%%%%%%%%%%%%%%%%%%%%%%%%%%%%%%%%%%%%%%%%%%%%%%%%%%
%
\subsubsection{Representation \texorpdfstring{$[0,1]$}{[0,1]}}
\paragraph{Hilbert series}
For this representation, the conformal dimension is given as
\begin{align}
 \Delta(n_1,n_2)=(N-1) (\left| n_1+n_2\right| +\left| 2 
n_1+n_2\right| +\left| 3 n_1+n_2\right| +\left| 3 
n_1+2 n_2\right| +\left| n_1\right| +\left| n_2\right| ) \; ,
\end{align}
and the computation of the Hilbert series for $N>1$ yields
\begin{equation}
\HS_{\Gtwo}^{[0,1]}(t,N)= \frac{1
+t^{6 N-5}(1+t+t^2+t^3+t^4)
+t^{10 N-9}(1+t+t^2+t^3+t^4)
+t^{16 N-10}}{\left(1-t^2\right) \left(1-t^6\right) \left(1-t^{6 
(N-1)}\right) \left(1-t^{10 (N-1)}\right)} \; .
\label{eqn:HS_G2_Rep01}
\end{equation}
The numerator of~\eqref{eqn:HS_G2_Rep01}  is a palindromic polynomial of 
degree $16N-10$; while, the denominator is of degree $16N-8$. Hence, the 
difference in degree between denominator and numerator is $2$, which matches
the quaternionic dimension of moduli space.
Moreover, the Hilbert series has a pole of order $4$ as $t\to 1$, i.e.\ it 
equals complex dimension of the moduli space.
Employing the knowledge of the Hilbert basis~\eqref{eqn:Hilbert_basis_G2_Cat1}, 
the appearing objects in~\eqref{eqn:HS_G2_Rep01} can be interpreted as 
in Tab.~\ref{tab:Ops_G2_Rep01}.
\begin{table}[h]
\centering
 \begin{tabular}{c|c|c|c}
 \toprule
  object & & $\Delta(n_1,n_2)$ & $\Hh_{(n_1,n_2)}$\\ \midrule
  Casimir & $\mathcal{C}_2$ & $2$ & ---\\
	  & $ \mathcal{C}_6$ & $6$ & ---\\ \midrule
  bare monopole & $V_{(0,1)}^{\mathrm{dress},0}$ & $6(N-1)$ & $\utwo$\\
 dressings $(i=1,\ldots,5)$ & $V_{(0,1)}^{\mathrm{dress},i}$ & $6(N-1)+i$ & --- 
\\ \midrule
  bare monopole & $V_{(1,0)}^{\mathrm{dress},0}$ & $10(N-1)$& $\utwo$\\
 dressings $(i=1,\ldots,5)$ & $V_{(1,0)}^{\mathrm{dress},i}$ & $10(N-1)+i$ 
&---\\
\bottomrule
 \end{tabular}
\caption{The chiral ring generators for a $\Gtwo$ gauge theory and matter 
transforming in $[0,1]$.}
\label{tab:Ops_G2_Rep01}
\end{table}
\paragraph{Plethystic logarithm}
For $N\geq3$ the PL takes the form
\begin{align}
\PL(\HS_{\Gtwo}^{[0,1]}(t,N)) = t^2 + t^6&+ 
t^{6(N-1)}(1+t+t^2+t^3+t^4+t^5)\\
&+ t^{10(N-1)}(1+t+t^2+t^3+t^4+t^5) -t^{12N-10} + \ldots \notag
\end{align}
while for $N=2$ the PL is 
\begin{align}
\PL(\HS_{\Gtwo}^{[0,1]}(t,2)) = 
t^2 + t^6
+ 
t^{6}(1+t+t^2+t^3+t^4+t^5)
+ t^{10}(1+t+t^2+t^3) -2t^{16} + \ldots 
\end{align}
In other words, the 4th and 5th dressing of $(1,0)$ are absent, because they 
can be generated.
% 
%%%%%%%%%%%%%%%%%%%%%%%%%%%%%%%%%%%%%%%%%%%%%%%%%
%%%%%%%%%%%%%%%%%%%%%%%%%%%%%%%%%%%%%%%%%%%%%%%%%
%
\subsubsection{Representation \texorpdfstring{$[2,0]$}{[2,0]}}
\paragraph{Hilbert series}
For this representation, the conformal dimensions is given by 
\begin{align}
 \Delta(n_1,n_2)=&N \Big( 2 \left| 
n_1+n_2\right| +2 \left| 2 n_1+n_2\right| +\left| 3 
n_1+n_2\right| +\left| 2 n_1+2 n_2\right| +\left| 3 
n_1+2 n_2\right| \\
&\qquad +\left| 4 n_1+2 n_2\right| +2 \left| 
n_1\right| +\left| 2 n_1\right| +\left| n_2\right| \Big) \notag \\
&-\left(\left| 
n_1+n_2\right| +\left| 2 n_1+n_2\right| +\left| 3 
n_1+n_2\right| +\left| 3 n_1+2 n_2\right| +\left| 
n_1\right| +\left| n_2\right| \right) \notag \; .
\end{align}
The calculation for the Hilbert series is analogous to the previous cases and 
we obtain
\begin{equation}
\HS_{\Gtwo}^{[2,0]}(t,N)= \frac{1
 +t^{12 N-5}(1+t+t^2+t^3+t^4)
 +t^{22 N-9}(1+t+t^2+t^3+t^4)
 +t^{34 N-10}}{\left(1-t^2\right) 
 \left(1-t^6\right) 
 \left(1-t^{12 N-6}\right) 
 \left(1-t^{22 N-10}\right)} \; .
 \label{eqn:HS_G2_Rep20}
\end{equation}
One readily observes, the numerator of~\eqref{eqn:HS_G2_Rep20} is a 
palindromic polynomial of degree $34N-10$ and the denominator is of 
degree $34N-8$. Hence, the difference in degree between denominator and 
numerator is $2$, which is precisely the quaternionic dimension of moduli space.
Also, the Hilbert series has a pole of order $4$ as $t\to 1$, which equals the 
complex dimension of the moduli space.
Having in mind the minimal generators~\eqref{eqn:Hilbert_basis_G2_Cat1}, the 
appearing objects in~\eqref{eqn:HS_G2_Rep20} can be summarised as in 
Tab.~\ref{tab:Ops_G2_Rep20}.
\begin{table}[h]
\centering
 \begin{tabular}{c|c|c|c}
 \toprule
  object & & $\Delta(n_1,n_2)$ & $\Hh_{(n_1,n_2)}$\\ \midrule
  Casimir & $\mathcal{C}_2$ & $2$ & ---\\
	  & $ \mathcal{C}_6$ & $6$ & ---\\ \midrule
  bare monopole & $V_{(0,1)}^{\mathrm{dress},0}$ & $12N-6$ & $\utwo$\\
 dressings $(i=1,\ldots,5)$ & $V_{(0,1)}^{\mathrm{dress},i}$ & $12N-6+i$ & --- 
\\ \midrule
  bare monopole & $V_{(1,0)}^{\mathrm{dress},0}$ & $22N-10$& $\utwo$\\
 dressings $(i=1,\ldots,5)$ & $V_{(1,0)}^{\mathrm{dress},i}$ & $22N-10+i$ &--- 
\\
\bottomrule
 \end{tabular}
\caption{The chiral ring generators for a $\Gtwo$ gauge theory and matter 
transforming in $[2,0]$.}
\label{tab:Ops_G2_Rep20}
\end{table}
\paragraph{Plethystic Logarithm}
\begin{itemize}
 \item For $N\geq3$ the PL takes the form
\begin{align}
\PL(\HS_{\Gtwo}^{[2,0]}(t,N)) = t^2 + t^6&+ 
t^{12N-6}(1+t+t^2+t^3+t^4+t^5)\\
&+ t^{22N-10}(1+t+t^2+t^3+t^4+t^5) -t^{12N-10} + \ldots \notag
\end{align}
\item While for $N=2$ the PL is 
\begin{align}
\PL(\HS_{\Gtwo}^{[2,0]}(t,2)) = t^2 + t^6&+ 
t^{18}(1+t+t^2+t^3+t^4+t^5) \\
&+ t^{34}(1+t+t^2+t^3) -2t^{40} + \ldots \notag 
\end{align}
By the very same reasoning as before, $V_{(1,0)}^{\mathrm{dress},4}$ and 
$V_{(1,0)}^{\mathrm{dress},5}$ can be generated by monopoles associated 
to $(0,1)$.
\item Moreover, for $N=1$ the PL looks as follows
\begin{equation}
\PL(\HS_{\Gtwo}^{[2,0]}(t,1)) = t^2 + t^6+ 
t^{6}(1+t+t^2+t^3+t^4+t^5)
+ t^{12}(1+t) -t^{16} + \ldots 
\end{equation}
Looking at the conformal dimensions reveals that the missing dressed monopoles 
of GNO-charge $(1,0)$ can be generated.
\end{itemize}
% 
%%%%%%%%%%%%%%%%%%%%%%%%%%%%%%%%%%%%%%%%%%%%%%%%
%%%%%%%%%%%%%%%%%%%%%%%%%%%%%%%%%%%%%%%%%%%%%%%%
%
\subsection{Category 2}
\label{subsec:G2_Cat2}
\begin{figure}[h]
\centering
   \begin{tikzpicture}
  \coordinate (Origin)   at (0,0);
  \coordinate (XAxisMin) at (-0.5,0);
  \coordinate (XAxisMax) at (4.5,0);
  \coordinate (YAxisMin) at (0,-0.5);
  \coordinate (YAxisMax) at (0,4.5);
  \draw [thin, gray,-latex] (XAxisMin) -- (XAxisMax);%
  \draw [thin, gray,-latex] (YAxisMin) -- (YAxisMax);%
  \draw (4.7,-0.2) node {$n_1$};
  \draw (-0.3,4.3) node {$n_2$};
%Draw the root lattice
  \foreach \x in {0,1,...,4}{%
      \foreach \y in {0,1,...,4}{%
        \node[draw,circle,inner sep=0.8pt,fill,black] at (\x,\y) {};
            }
            }
    \draw[black,dashed,thick]  (Origin) -- (0,4.2);  
    \draw[black,dashed,thick]  (Origin) -- (4.2,0);
    \draw[black,dashed,thick]  (Origin) -- (4.2,4.2);
% % 
 \draw[black,thick] (1,0) circle (4pt);
 \draw[black,thick] (0,1) circle (4pt);
 \draw[black,thick] (1,1) circle (4pt);
  \draw (1.5,3.5) node {$S_2^{(2)}$};
  \draw (3.5,1.5) node {$S_1^{(2)}$};
\end{tikzpicture}
\caption{The semi-groups $S_p^{(2)}$ ($p=1,2$) for the representations $[1,1]$, 
$[0,2]$, and $[3,0]$ obtained from the $\Gtwo$ Weyl chamber (considered as 
rational cone) and their ray generators (black circled points).}
\label{Fig:Hilbert_basis_G2_Cat2}
\end{figure}
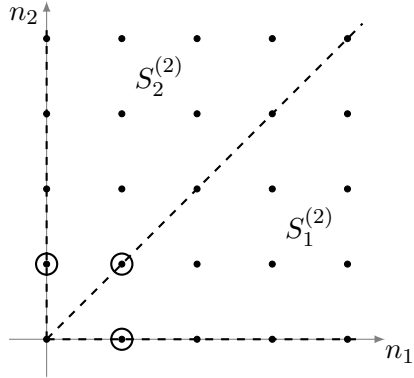

\paragraph{Hilbert basis}
The representations $[1,1]$, $[0,2]$, and $[3,0]$ have schematically conformal 
dimensions of the form
\begin{equation}
\label{eqn:delta_G2_Cat2}
 \Delta(n_1,n_2)= \sum_j A_j | a_j n_1 + b_j n_2| + B_1 |n_1|+ B_2 |n_2| + 
C|n_1-n_2| 
\end{equation}
for $a_j,b_j \in \NN$ and $A_j,B_1,B_2,C\in \Z$.
The novelty of this conformal dimension, compared to~\eqref{eqn:delta_G2_Cat1}, 
is the difference $|n_1-n_2|$, i.e.\ a hyperplane that intersects the 
Weyl chamber non-trivially.
As a consequence, there is a fan generated by two $2$-dimensional rational 
polyhedral cones 
\begin{equation}
  C_1^{(2)}= \cone((1,0),(1,1)) \und C_2^{(2)}= \cone((1,1),(0,1)) \; .
 \end{equation}
The intersection with the weight lattice $\Lambda_w(\Gtwo)$ yields the 
relevant semi-groups $S_p$ ($p=1,2$), as depicted in 
Fig.~\ref{Fig:Hilbert_basis_G2_Cat2}.
The Hilbert bases are again given by the ray generators
\begin{equation}
  \label{eqn:Hilbert_basis_G2_Cat2}
 \Hcal(S_1^{(2)}) = \Big\{(1,0),(1,1) \Big\} \und  \Hcal(S_2^{(2)}) = 
\Big\{(1,1),(0,1) \Big\} \; .
\end{equation}
\paragraph{Dressings}
The three minimal generators have different residual gauge groups, as two lie 
on the boundary and one in the interior of the Weyl chamber. The GNO-charges 
$(1,0)$ and $(0,1)$ are to be treated as in Subsec.~\ref{subsec:G2_Cat1}.

The novelty is the magnetic weight $(1,1)$ with $\Hh_{(1,1)}=\uo^2$.
Thus, the dressing can be constructed with two independent $\uo$-Casimir 
invariants, 
proportional to $\phi_1$ and $\phi_2$. We choose a basis of dressed monopoles
\begin{subequations}
\begin{align}
 V_{(1,1)}^{\mathrm{dress},j,\alpha} &= (1,1) (\phi_\alpha)^j \; , \for 
j=1,\ldots 5 \; , \; \alpha=1,2 \; , \\
V_{(1,1)}^{\mathrm{dress},6} &=  (1,1) \left( (\phi_1)^6 + (\phi_2)^6\right) \; 
.
\end{align}
\end{subequations}
The reason behind the large number of dressings of the bare monopole $(1,1)$ 
lies in the delicate $\Gtwo$ structure~\cite{Okubo}, i.e.\ the degree two 
Casimir $\mathcal{C}_2$ is not just the sum of the squares of $\phi_i$ and the  
next $\Gtwo$-Casimir $\mathcal{C}_6$ is by four higher in degree and has a 
complicated structure as well.

The number and degrees of the dressed monopole operators associated to $(1,1)$ 
can be confirmed by $P_{\Gtwo}(t,n_1>0,n_2>0)\slash 
P_{\Gtwo}(t,0,0)=1+2t+2t^2+2t^3+2t^4+2t^5+t^6$, following App.~\ref{app:PL}.

We will now exemplify the three different representations.
\subsubsection{Representation \texorpdfstring{$[1,1]$}{[1,1]}}
\paragraph{Hilbert series}
The conformal dimension of the $64$-dimensional representation is given by
\begin{align}
 \Delta(n_1,n_2) 
&=N \Big(\left| n_1-n_2\right| 
+8 \left| n_1+n_2\right| 
+8 \left| 2 n_1+n_2\right| 
+2 \left| 3 n_1+n_2\right| 
+\left| 4 n_1+n_2\right| \\
&\qquad +\left| n_1+2 n_2\right| 
+2 \left| 3 n_1+2 n_2\right| 
+\left| 5 n_1+2 n_2\right| 
+\left| 4 n_1+3 n_2\right| 
+\left| 5 n_1+3 n_2\right|  \notag \\
&\qquad 
+8 \left| n_1\right| 
+2 \left| n_2\right| \Big) \notag\\
&\qquad -\Big( \left| n_1+n_2\right| 
+\left| 2 n_1+n_2\right| 
+\left| 3 n_1+n_2\right| 
+\left| 3 n_1+2 n_2\right| 
+\left| n_1\right| 
+\left| n_2\right| \Big) \notag \; .
\end{align}
Computing the Hilbert series provides the following expression
\begin{subequations}
\label{eqn:HS_G2_Rep11}
\begin{align}
\HS_{\Gtwo}^{[1,1]}(t,N)&= \frac{R(t,N)}{
\left(1-t^2\right) \left(1-t^6\right) 
\left(1-t^{36 N-6}\right) 
\left(1-t^{64 N-10}\right) 
\left(1-t^{98 N-16}\right)} \; ,\\
R(t,N)&=1+
t^{36 N-5}(1+t+t^2+t^3+t^4)
+t^{64 N-9}(1+t+t^2+t^3+t^4) 
\label{eqn:HS_G2_Rep11_Num}\\
&\qquad
+ t^{98 N-15}(2+2t+2t^2+2t^3+2t^4+t^5)
-t^{100 N-16} (1+2t+2t^2+2t^3+2t^4+2t^5)
\notag \\
&\qquad
-t^{134 N-21}(1+t+t^2+t^3+t^4)
-t^{162 N-25}(1+t+t^2+t^3+t^4)
-t^{198 N-26} \notag \; .
\end{align}
\end{subequations}
The numerator~\eqref{eqn:HS_G2_Rep11_Num} is a anti-palindromic 
polynomial of degree $198N-26$; whereas the denominator is of degree 
$198N-24$. Hence, the difference in degree 
between denominator and numerator is $2$, which coincides 
with the quaternionic dimension of moduli space.
The Hilbert series~\eqref{eqn:HS_G2_Rep11} has a pole of order $4$ as $t\to 1$, 
which agrees with the complex dimension of the moduli space. (One can 
explicitly show that $R(t=1,N)=0$, but $\tfrac{\diff}{\diff 
t}R(t,N)|_{t=1}\neq0$.)
The appearing operators agree with the general setting outlined above and we 
summarise them in Tab.~\ref{tab:Ops_G2_Rep11}.
\begin{table}[h]
\centering
 \begin{tabular}{c|c|c|c}
 \toprule
  object & & $\Delta(n_1,n_2)$ & $\Hh_{(n_1,n_2)}$\\ \midrule
  Casimir & $\mathcal{C}_2$ & $2$ & ---\\
	  & $ \mathcal{C}_6$ & $6$ & ---\\ \midrule
  bare monopole & $V_{(0,1)}^{\mathrm{dress},0}$ & $36N-6$ & $\utwo$\\
 dressings $(i=1,\ldots,5)$ & $V_{(0,1)}^{\mathrm{dress},i}$ & $36N-6+i$ & --- 
\\ \midrule
  bare monopole & $V_{(1,0)}^{\mathrm{dress},0}$ & $64N-10$& $\utwo$\\
 dressings $(i=1,\ldots,5)$ & $V_{(1,0)}^{\mathrm{dress},i}$ & $64N-10+i$ 
&---\\ \midrule
  bare monopole & $V_{(1,1)}^{\mathrm{dress},0}$ & $98N-16$& $\uo\times \uo$\\
 dressings $(i=1,\ldots,5 ; \alpha=1,2)$ & 
$V_{(1,1)}^{\mathrm{dress},i,\alpha}$ & $98N-16+i$ 
&--- \\
dressing  & 
$V_{(1,1)}^{\mathrm{dress},6}$ & $98N-16+6$ 
&--- \\ 
\bottomrule
 \end{tabular}
\caption{The chiral ring generators for a $\Gtwo$ gauge theory and matter 
transforming in $[1,1]$.}
\label{tab:Ops_G2_Rep11}
\end{table}
The new monopole corresponds to GNO-charge $(1,1)$ and displays a different 
dressing behaviour than $(1,0)$ and $(0,1)$. The reason behind lies in the 
residual gauge group being $\uo^2$. 
\paragraph{Plethystic Logarithm}
Although the bare monopole $V_{(1,1)}^{\mathrm{dress},0}$ is generically a 
necessary generator due to its origin as an ray generators 
of~\eqref{eqn:Hilbert_basis_G2_Cat2}, not all dressings 
$V_{(1,1)}^{\mathrm{dress}}$ might be independent.
\begin{itemize}
 \item For $N\geq4$ the PL takes the form
\begin{align}
\PL(\HS_{\Gtwo}^{[1,1]}(t,N)) = t^2 + t^6&+ 
t^{36N-6}(1+t+t^2+t^3+t^4+t^5)\\
&+ t^{64N-10}(1+t+t^2+t^3+t^4+t^5) \notag\\
&-t^{2(36N-6)+2}(1+t+2t^2+2t^3+3t^4+2t^5+2t^6+t^7+t^8) \notag \\
&+t^{98N-16}(1+2t+2t^2+2t^3+2t^4+2t^5+t^6) -t^{100N-16} +\ldots \notag
\end{align}
\item For $N=3$ the PL is 
\begin{align}
\PL(\HS_{\Gtwo}^{[1,1]}(t,N=3)) = t^2 + t^6&+ 
t^{102}(1+t+t^2+t^3+t^4+t^5)\\
&+ t^{182}(1+t+t^2+t^3+t^4+t^5) \notag\\
&-t^{206}(1+t+2t^2+2t^3+3t^4+2t^5+2t^6+t^7+t^8) \notag \\
&+t^{278}(1+2t+2t^2+2t^3+2t^4+2t^5) - 2t^{285} +\ldots \notag 
\end{align}
Here, $\Delta(1,0)+\Delta(0,1)=284$ is precisely the conformal dimension of 
$V_{(1,1)}^{\mathrm{dress},6}$; i.e.\ it is generated and absent from the PL.
\item For $N=2$ the PL is 
\begin{align}
\PL(\HS_{\Gtwo}^{[1,1]}(t,N=2)) = t^2 + t^6&+ 
t^{66}(1+t+t^2+t^3+t^4+t^5)\\
&+ t^{118}(1+t+t^2+t^3+t^4+t^5) \notag\\
&-t^{134}(1+t+2t^2+2t^3+3t^4+2t^5+2t^6+t^7+t^8) \notag \\
&+t^{180}(1+2t+2t^2+2t^3+t^4) - 2t^{186} +\ldots \notag 
\end{align}
Here, $\Delta(1,0)+\Delta(0,1)=184$ is precisely the conformal dimension of 
$V_{(1,1)}^{\mathrm{dress},4,\alpha}$; i.e.\ only one of the dressings by the 
4th power of $\uo$-Casimir is a generator. Consequently, the other one is 
absent from the PL.
 \item For $N=1$ the PL is 
\begin{align}
\PL(\HS_{\Gtwo}^{[1,1]}(t,N=1)) = t^2 + t^6&+ 
t^{30}(1+t+t^2+t^3+t^4+t^5)\\
&+ t^{54}(1+t+t^2+t^3+t^4+t^5) \notag\\
&-t^{62}(1+t+2t^2+2t^3+3t^4+2t^5+2t^6+t^7+t^8) \notag \\
&+t^{82}(1+2t+t^2) - t^{62} +\ldots \notag 
\end{align}
Here, $\Delta(1,0)+\Delta(0,1)=64$ is precisely the conformal dimension of 
$V_{(1,1)}^{\mathrm{dress},2,\alpha}$; i.e.\ only one of the dressings by the 
2th power of $\uo$-Casimir is a generator. Consequently, the other one is 
absent from the PL.
\end{itemize}
% 
%%%%%%%%%%%%%%%%%%%%%%%%%%%%%%%%%%%%%%%%%%%%%%%%%%%%%%%%%%%%%
%%%%%%%%%%%%%%%%%%%%%%%%%%%%%%%%%%%%%%%%%%%%%%%%%%%%%%%%%%%%%
%
\subsubsection{Representation \texorpdfstring{$[3,0]$}{[3,0]}}
\paragraph{Hilbert series}
The conformal dimension in this representation is given by
\begin{align}
\Delta(n_1,n_2) 
&= N\Big( \left|5n_1+3n_2 \right| 
+\left|5n_1+2n_2 \right|
+\left|4n_1 +3n_2 \right|
+\left|4n_1+n_2 \right| 
+\left|n_1+2n_2 \right|
+\left|n_1-n_2 \right|\notag\\
&\qquad
+10\big(\left|2n_1+n_2 \right|
+\left|n_1+n_2 \right|
+\left|n_1 \right|\big)  
+3\big(
\left|3n_1+2n_2 \right|
+\left|3n_1+n_2 \right|
+\left| n_2\right|\big)
\Big) \\
&\qquad -\Big(\left| n_1+n_2\right| +\left| 2 n_1+n_2\right| +\left| 3 
n_1+n_2\right| +\left| 3 n_1+2 n_2\right| +\left| 
n_1\right| +\left| n_2\right| \Big) \notag \; ,
\end{align}
such that we obtain for the Hilbert series
\begin{subequations}
\label{eqn:HS_G2_Rep30}
\begin{align}
\HS_{\Gtwo}{[3,0]}(t,N)&= \frac{R(t,N)}{\left(1-t^2\right) \left(1-t^6\right) 
\left(1-t^{46 N-6}\right) \left(1-t^{82 N-10}\right) 
\left(1-t^{126 N-16}\right)} \; ,\\
R(t,N)&=1
+t^{46 N-5}(1+t+t^2+t^3+t^4)
+t^{82 N-9}(1+t+t^2+t^3+t^4)
\label{eqn:HS_G2_Rep30_Num} \\
&\qquad
+ t^{126 N-15}(2+2t+2t^2+2t^3+2t^4+t^5)
\notag \\
&\qquad
-t^{128 N-16}(1+2t+2t^2+2t^3+2t^4+2t^5)
\notag \\
&\qquad
-t^{172 N-21}(1+t+t^2+t^3+t^4)
-t^{208 N-25}(1+t+t^2+t^3+t^4)
-t^{254 N-26} \notag \; .
\end{align}
\end{subequations}
The numerator~\eqref{eqn:HS_G2_Rep30_Num} is a anti-palindromic 
polynomial of degree $254N-26$; while the denominator is of degree $254N-24$. 
Hence, the difference in degree 
between denominator and numerator is $2$, which coincides with the 
quaternionic dimension of moduli space.
The Hilbert series~\eqref{eqn:HS_G2_Rep30} has a pole of order $4$ as $t\to 1$, 
which equals the complex dimension of the moduli space. (One can 
explicitly show that $R(t=1,N)=0$, but $\tfrac{\diff}{\diff 
t}R(t,N)|_{t=1}\neq0$.)
Interpreting the appearing operators leads to a list of chiral ring generators 
as presented in Tab.~\ref{tab:Ops_G2_Rep30}.
\begin{table}[h]
\centering
 \begin{tabular}{c|c|c|c}
 \toprule
  object & & $\Delta(n_1,n_2)$ & $\Hh_{(n_1,n_2)}$\\ \midrule
  Casimir & $\mathcal{C}_2$ & $2$ & ---\\
	  & $ \mathcal{C}_6$ & $6$ & ---\\ \midrule
  bare monopole & $V_{(0,1)}^{\mathrm{dress},0}$ & $46N-6$ & $\utwo$\\
 dressings $(i=1,\ldots,5)$ & $V_{(0,1)}^{\mathrm{dress},i}$ & $46N-6+i$ & --- 
\\ \midrule
  bare monopole & $V_{(1,0)}^{\mathrm{dress},0}$ & $82N-10$& $\utwo$\\
 dressings $(i=1,\ldots,5)$ & $V_{(1,0)}^{\mathrm{dress},i}$ & $82N-10+i$ 
&---\\ \midrule
  bare monopole & $V_{(1,1)}^{\mathrm{dress},0}$ & $126N-16$& $\uo\times \uo$\\
 dressings $(i=1,\ldots,5 ; \alpha=1,2)$ & 
$V_{(1,1)}^{\mathrm{dress},i,\alpha}$ & $126N-16+i$ 
&--- \\
dressing  & 
$V_{(1,1)}^{\mathrm{dress},6}$ & $126N-16+6$ 
&--- \\
\bottomrule
 \end{tabular}
\caption{The chiral ring generators for a $\Gtwo$ gauge theory and matter 
transforming in $[3,0]$.}
\label{tab:Ops_G2_Rep30}
\end{table}
The behaviour of the Hilbert series is absolutely identical to the case 
$[1,1]$, because the conformal dimension is structurally identical. Therefore, 
we do not provide further details.
% 
%%%%%%%%%%%%%%%%%%%%%%%%%%%%%%%%%%%%%%%%%%%%%%%%%%%%%%%%%%%%%
%%%%%%%%%%%%%%%%%%%%%%%%%%%%%%%%%%%%%%%%%%%%%%%%%%%%%%%%%%%%%
%
\subsubsection{Representation \texorpdfstring{$[0,2]$}{[0,2]}}
\paragraph{Hilbert series}
The following conformal dimension reads 
\begin{align}
 \Delta(n_1,n_2)&=N\Big( \left|5n_1+3n_2 \right| 
+ \left|5n_1+2n_2 \right|
+\left|4n_1+3n_2 \right| 
+ \left|4n_1+n_2 \right| 
+\left|n_1+2n_2 \right| 
+\left|n_1-n_2 \right| \notag\\
&\qquad
+10\big( \left|2n_1+n_2 \right| 
+\left|n_1+n_2 \right|
+\left|n_1 \right| \big)
+ 5\big( \left|3n_1+2n_2 \right|
+ \left|3n_1+n_2 \right|
+ \left|n_2 \right|\big)
\Big)  \\
&\qquad -\Big(\left| n_1+n_2\right| 
+\left| 2 n_1+n_2\right| 
+\left| 3 n_1+n_2\right| 
+\left| 3 n_1+2 n_2\right| 
+\left| n_1\right| 
+\left| n_2\right| \Big)  \notag 
\end{align}
The computation of the Hilbert series results in
\begin{subequations}
\label{eqn:HS_G2_Rep02}
\begin{align}
\HS_{\Gtwo}^{[0,2]}(t,N) &= \frac{R(t,N)}{
\left(1-t^2\right) \left(1-t^6\right) 
\left(1-t^{52 N-6}\right) 
\left(1-t^{90 N-10}\right) 
\left(1-t^{140 N-16}\right)} \; ,\\
R(t,N)&=1
+t^{52 N-5}(1+t+t^2+t^3+t^4)
+t^{90 N-9}(1+t+t^2+t^3+t^4)
\label{eqn:HS_G2_Rep02_Num} \\
&\qquad
+t^{140 N-15}(2+2t+2t^2+2t^3+2t^4+t^5)
\notag \\
&\qquad
-t^{142 N-16}(1+2t+2t^2+2t^3+2t^4+2t^5)
\notag \\
&\qquad
-t^{192 N-21}(1+t+t^2+t^3+t^4)
-t^{230 N-25}(1+t+t^2+t^3+t^4)
-t^{282 N-26} \; . \notag
\end{align}
\end{subequations}
The numerator~\eqref{eqn:HS_G2_Rep02_Num} is a anti-palindromic 
polynomial of degree $282N-26$; while, the denominator is of degree $282N-24$. 
Hence, the difference in degree 
between denominator and numerator is $2$, which agrees with the 
quaternionic dimension of moduli space.
The Hilbert series~\eqref{eqn:HS_G2_Rep02} has a pole of order $4$ as $t\to 1$, 
which equals complex dimension of the moduli space. (One can explicitly 
show that $R(t=1,N)=0$, but $\tfrac{\diff}{\diff t}R(t,N)|_{t=1}\neq0$.)
Tab.~\ref{tab:Ops_G2_Rep02} summarises the appearing operators.
\begin{table}[h]
\centering
 \begin{tabular}{c|c|c|c}
 \toprule
  object & & $\Delta(n_1,n_2)$ & $\Hh_{(n_1,n_2)}$\\ \midrule
  Casimir & $\mathcal{C}_2$ & $2$ & ---\\
	  & $ \mathcal{C}_6$ & $6$ & ---\\ \midrule
  bare monopole & $V_{(0,1)}^{\mathrm{dress},0}$ & $52N-6$ & $\utwo$\\
 dressings $(i=1,\ldots,5)$ & $V_{(0,1)}^{\mathrm{dress},i}$ & $52N-6+i$ & --- 
\\ \midrule
  bare monopole & $V_{(1,0)}^{\mathrm{dress},0}$ & $90N-10$& $\utwo$\\
 dressings $(i=1,\ldots,5)$ & $V_{(1,0)}^{\mathrm{dress},i}$ & $90N-10+i$ 
&---\\ \midrule
  bare monopole & $V_{(1,1)}^{\mathrm{dress},0}$ & $140N-16$& $\uo\times \uo$\\
 dressings $(i=1,\ldots,5 ; \alpha=1,2)$ & 
$V_{(1,1)}^{\mathrm{dress},i,\alpha}$ & $140N-16+i$ 
&--- \\
dressing  & 
$V_{(1,1)}^{\mathrm{dress},6}$ & $140N-16+6$ 
&--- \\
\bottomrule
 \end{tabular}
\caption{The chiral ring generators for a $\Gtwo$ gauge theory and matter 
transforming in $[0,2]$.}
\label{tab:Ops_G2_Rep02}
\end{table}
The behaviour of the Hilbert series is identical to the cases $[1,1]$ and 
$[3,0]$, because the conformal dimension is structurally identical.
Again, we do not provide further details.
% 
%%%%%%%%%%%%%%%%%%%%%%%%%%%%%%%%%%%%%%%%%%%%%%%%%%%%%%%%%%%%%
%%%%%%%%%%%%%%%%%%%%%%%%%%%%%%%%%%%%%%%%%%%%%%%%%%%%%%%%%%%%%
%
\subsection{Category 3}
\paragraph{Hilbert basis}
Investigating the representations $[2,1]$ and $[4,0]$, one recognises the 
common structural form of the conformal dimensions 
\begin{equation}
 \Delta(n_1,n_2)= \sum_j A_j | a_j n_1 + b_j n_2| + B_1 |n_1|+ B_2 |n_2| + 
C|n_1-n_2| + D|2n_1 -n_2|
\end{equation}
for $a_j,b_j \in \NN$ and $A_j,B_1,B_2,C,D\in \Z$.
The novelty of this conformal dimension, compared to~\eqref{eqn:delta_G2_Cat1} 
and~\eqref{eqn:delta_G2_Cat2}, 
is the difference $|2n_1-n_2|$, i.e.\ a second hyperplane that intersects
the Weyl chamber non-trivially.
As a consequence, the Weyl chamber is decomposed into a fan 
generated by three rational polyhedral cones of dimension 2. These are
 \begin{align}
  C_1^{(2)}= \cone((1,0),(1,1))\; , \quad 
  C_2^{(2)}= \cone((1,1),(1,2)) 
  \und   C_3^{(2)}= \cone((1,2),(0,1)) \; .
 \end{align}
The intersection with the weight lattice $\Lambda_w(\Gtwo)$ yields the 
relevant semi-groups $S_p$ (for $p=1,2,3$), as depicted in 
Fig.~\ref{Fig:Hilbert_basis_G2_Cat3}.
The Hilbert bases are again given by the ray generators
\begin{equation}
  \label{eqn:Hilbert_basis_G2_Cat3}
 \Hcal(S_1^{(2)}) = \Big\{(1,0),(1,1) \Big\} \; , \quad  
 \Hcal(S_2^{(2)}) = \Big\{(1,1),(1,2) \Big\} \und 
\Hcal(S_3^{(2)}) = \Big\{(1,2),(0,1) \Big\} \; .
\end{equation}
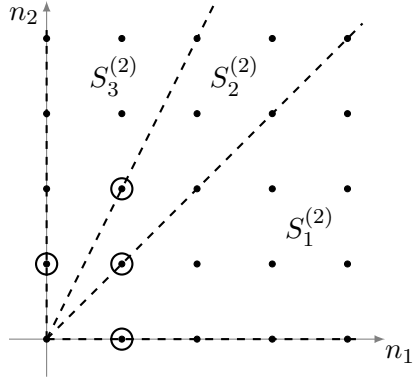
\begin{figure}[h]
\centering
   \begin{tikzpicture}
  \coordinate (Origin)   at (0,0);
  \coordinate (XAxisMin) at (-0.5,0);
  \coordinate (XAxisMax) at (4.5,0);
  \coordinate (YAxisMin) at (0,-0.5);
  \coordinate (YAxisMax) at (0,4.5);
  \draw [thin, gray,-latex] (XAxisMin) -- (XAxisMax);%
  \draw [thin, gray,-latex] (YAxisMin) -- (YAxisMax);%
  \draw (4.7,-0.2) node {$n_1$};
  \draw (-0.3,4.3) node {$n_2$};
%Draw the root lattice
  \foreach \x in {0,1,...,4}{%
      \foreach \y in {0,1,...,4}{%
        \node[draw,circle,inner sep=0.8pt,fill,black] at (\x,\y) {};
            }
            }
    \draw[black,dashed,thick]  (Origin) -- (0,4.2);  
    \draw[black,dashed,thick]  (Origin) -- (4.2,0);
    \draw[black,dashed,thick]  (Origin) -- (4.2,4.2);
    \draw[black,dashed,thick]  (Origin) -- (2.25*1,2.25*2);
 \draw[black,thick] (1,0) circle (4pt);
 \draw[black,thick] (0,1) circle (4pt);
 \draw[black,thick] (1,1) circle (4pt);
 \draw[black,thick] (1,2) circle (4pt);
  \draw (0.9,3.5) node {$S_3^{(2)}$};
  \draw (2.5,3.5) node {$S_2^{(2)}$};
  \draw (3.5,1.5) node {$S_1^{(2)}$};
\end{tikzpicture}
\caption{The semi-groups $S_p^{(2)}$ (p=1,2,3) for the representations $[2,1]$ 
and $[4,0]$ obtained from the $\Gtwo$ Weyl chamber (considered as rational 
cone) and their ray generators (black circled points).}
\label{Fig:Hilbert_basis_G2_Cat3}
\end{figure}
\paragraph{Dressings}
Compared to Subsec.~\ref{subsec:G2_Cat1} and~\ref{subsec:G2_Cat2}, the 
additional magnetic weight $(1,2)$ has the same dressing behaviour as $(1,1)$, 
because the residual gauge groups is $\uo^2$, too. Thus, the additional 
necessary monopole operators are the bare operator 
$V_{(1,2)}^{\mathrm{dress},0}$ and the dressed monopoles   
$V_{(1,2)}^{\mathrm{dress},i,\alpha}$ for $i=1,\ldots,5$, $\alpha=1,2$ as well 
as $V_{(1,2)}^{\mathrm{dress},6}$.

We will now exemplify the three different representations.
\subsubsection{Representation \texorpdfstring{$[4,0]$}{[4,0]}}
\paragraph{Hilbert series}
The conformal dimension reads
\begin{align}
\Delta(n_1,n_2)&= N \Big(
3 \left| n_1-n_2\right| 
+\left| 2 n_1-n_2\right| 
+27 \left| n_1+n_2\right| 
+30 \left| 2 n_1+n_2\right| 
+7 \left| 3 n_1+n_2\right|\\
&\qquad 
+3 \left| 4 n_1+n_2\right| 
+\left| 5 n_1+n_2\right| 
+3 \left| n_1+2 n_2\right|
+7 \left| 3 n_1+2 n_2\right| 
+3 \left| 5 n_1+2 n_2\right| \notag\\
&\qquad 
+\left| 2 n_1+3 n_2\right|
+3 \left| 4 n_1+3 n_2\right| 
+3 \left| 5 n_1+3 n_2\right|
+\left| 7 n_1+3 n_2\right|
+\left| 5 n_1+4 n_2\right| \notag\\
&\qquad
+\left| 7 n_1+4 n_2\right|
+27 \left| n_1\right| 
+7 \left| n_2\right|  
\Big) \notag \\
&\qquad -\Big(\left| n_1+n_2\right| +\left| 2 
n_1+n_2\right| +\left| 3 n_1+n_2\right| +\left| 3 n_1+2 n_2\right| 
+\left| n_1\right| +\left| n_2\right| \Big) \notag \; ,
\end{align}
from which we compute the Hilbert series to be
\begin{subequations}
\label{eqn:HS_G2_Rep40}
\begin{align}
\HS_{\Gtwo}^{[4,0]}(t,N)&=
 \frac{R(t,N)}{\left(1-t^2\right) 
\left(1-t^6\right) \left(1-t^{134 N-6}\right) \left(1-t^{238 
N-10}\right) \left(1-t^{364 N-16}\right) \left(1-t^{496 
N-22}\right)}  , \\
R(t,N)&=1
+t^{134 N-5}(1+t+t^2+t^3+t^4)
+t^{238 N-9}(1+t+t^2+t^3+t^4) 
\label{eqn:HS_G2_Rep40_Num}\\
&\qquad
+ t^{364 N-15}(2+2t+2t^2+2t^3+2t^4+t^5)
\notag\\
&\qquad
-t^{372 N-16}(1+2t+2t^2+2t^3+2t^4+2t^5)
\notag\\
&\qquad
+ t^{496 N-21}(2+2t+2t^2+2t^3+2t^4+t^5)
\notag\\
&\qquad
-t^{498 N-22}(1+3t+3t^2+3t^3+3t^4+3t^5+t^6)
\notag\\
&\qquad
-t^{602 N-25}(1+t+t^2+t^3+t^4)
-t^{630 N-27}(1+t+t^2+t^3+t^4)
\notag\\
&\qquad
-t^{734 N-32}(1+3t+3t^2+3t^3+3t^4+3t^5+t^6)
\notag\\
&\qquad
+t^{736 N-32}(1+2t+2t^2+2t^3+2t^4+2t^5)
\notag\\
&\qquad
- t^{860 N-37}(2+2t+2t^2+2t^3+2t^4+t^5)
\notag\\
&\qquad
+t^{868 N-38}(1+2t+2t^2+2t^3+2t^4+2t^5)
\notag\\
&\qquad
+t^{994 N-43}(1+t+t^2+t^3+t^4)
+t^{1098 N-47}(1+t+t^2+t^3+t^4)
+t^{1232 N-48} \notag \; .
\end{align}
\end{subequations}
The numerator~\eqref{eqn:HS_G2_Rep40_Num} is a palindromic polynomial of 
degree $1232N-48$; while, the denominator is of degree $1232N-46$. Hence, 
the difference in degree between denominator and numerator is $2$, which equals 
the quaternionic dimension of moduli space.
The Hilbert series~\eqref{eqn:HS_G2_Rep40} has a pole of order $4$ as $t\to 1$, 
which coincides with the complex dimension of the moduli space. (One can 
explicitly show that $R(t=1,N)=0$ and $\tfrac{\diff}{\diff t}R(t,N)|_{t=1}=0$, 
but $\tfrac{\diff^2}{\diff t^2}R(t,N)|_{t=1}\neq0$.)
The appearing operators can be summarised as in Tab.~\ref{tab:Ops_G2_Rep40}.
\begin{table}[h]
\centering
 \begin{tabular}{c|c|c|c}
 \toprule
  object & & $\Delta(n_1,n_2)$ & $\Hh_{(n_1,n_2)}$\\ \midrule
  Casimir & $\mathcal{C}_2$ & $2$ & ---\\
	  & $ \mathcal{C}_6$ & $6$ & ---\\ \midrule
  bare monopole & $V_{(0,1)}^{\mathrm{dress},0}$ & $134N-6$ & $\utwo$\\
 dressings $(i=1,\ldots,5)$ & $V_{(0,1)}^{\mathrm{dress},i}$ & $134N-6+i$ & --- 
\\ \midrule
  bare monopole & $V_{(1,0)}^{\mathrm{dress},0}$ & $238N-10$& $\utwo$\\
 dressings $(i=1,\ldots,5)$ & $V_{(1,0)}^{\mathrm{dress},i}$ & $238N-10+i$ 
&---\\ \midrule
  bare monopole & $V_{(1,1)}^{\mathrm{dress},0}$ & $364N-16$& $\uo\times \uo$\\
 dressings $(i=1,\ldots,5 ; \alpha=1,2)$ & 
$V_{(1,1)}^{\mathrm{dress},i,\alpha}$ & $364N-16+i$ 
&--- \\
dressing  & 
$V_{(1,1)}^{\mathrm{dress},6}$ & $364N-16+6$ 
&---\\ \midrule
  bare monopole & $V_{(1,2)}^{\mathrm{dress},0}$ & $496N-22$& $\uo\times \uo$\\
 dressings $(i=1,\ldots,5 ; \alpha=1,2)$ & 
$V_{(1,2)}^{\mathrm{dress},i,\alpha}$ & $496N-22+i$ 
&--- \\
dressing  & 
$V_{(1,2)}^{\mathrm{dress},6}$ & $496N-22+6$ 
&--- \\
\bottomrule
 \end{tabular}
\caption{The chiral ring generators for a $\Gtwo$ gauge theory and matter 
transforming in $[4,0]$.}
\label{tab:Ops_G2_Rep40}
\end{table}
The new monopole corresponds to GNO-charge $(1,2)$ and displays the same 
dressing behaviour as $(1,1)$. 
Contrary to the cases $[1,1]$, $[3,0]$, and $[0,2]$, the bare and dressed 
monopoles of GNO-charge $(1,1)$ are always independent generators as 
\begin{equation}
 \Delta(1,1)=364N-16 <372N-16=134N-6 + 238N-10= \Delta(0,1)+ \Delta(1,0)
\end{equation}
holds for all $N\geq1$.
\paragraph{Plethystic Logarithm}
By means of the minimal generators~\eqref{eqn:Hilbert_basis_G2_Cat3}, the bare 
monopole $V_{(1,2)}^{\mathrm{dress},0}$ is a necessary generator. 
Nevertheless, not all dressings $V_{(1,2)}^{\mathrm{dress}}$ need to be 
independent.
For $N\geq1$ the PL takes the form
\begin{align}
\mathrm{PL}(\HS_{\Gtwo}^{[0,2]}(t,N)) = t^2 + t^6&+ 
t^{134N-6}(1+t+t^2+t^3+t^4+t^5)\\
&+ t^{238N-10}(1+t+t^2+t^3+t^4+t^5) \notag\\
&-t^{2(134N-6)+2}(1+t+2t^2+2t^3+3t^4+2t^5+2t^6+t^7+t^8) \notag \\
&+t^{364N-16}(1+2t+2t^2+2t^3+2t^4+2t^5+t^6) -t^{372N-16} +\ldots \notag 
\end{align}
Based purely in conformal dimension and GNO-charge, we can argue the following:
\begin{itemize}
\item For $N=3$, $\Delta(1,1)+\Delta(0,1)=1472$ is precisely the conformal 
dimension of 
$V_{(1,2)}^{\mathrm{dress},6}$, i.e.\ it is generated.
\item  For $N=2$, $\Delta(1,1)+\Delta(0,1)=974$ is precisely the conformal 
dimension of 
$V_{(1,2)}^{\mathrm{dress},4,\alpha}$, i.e.\ only one of the dressings by the 
4th power of $\uo$-Casimir is a generator.
 \item For $N=1$, $\Delta(1,1)+\Delta(0,1)=476$ is precisely the conformal 
dimension of 
$V_{(1,2)}^{\mathrm{dress},2,\alpha}$, i.e.\ only one of the dressings by the 
2th power of $\uo$-Casimir is a generator. 
\end{itemize}
% 
% 
%%%%%%%%%%%%%%%%%%%%%%%%%%%%%%%%%%%%%%%%%%%%%%%%%%%%%%%%%%%%%
%%%%%%%%%%%%%%%%%%%%%%%%%%%%%%%%%%%%%%%%%%%%%%%%%%%%%%%%%%%%%
%
\subsubsection{Representation \texorpdfstring{$[2,1]$}{[2,1]}}
\paragraph{Hilbert series}
The conformal dimension reads
\begin{align}
\Delta(n_1,n_2)&=
N \Big(
3 \left| n_1-n_2\right| 
+\left| 2 n_1-n_2\right| 
+24 \left| n_1+n_2\right| 
+24 \left| 2 n_1+n_2\right| 
+8 \left| 3 n_1+n_2\right| \\
&\qquad
+3 \left| 4 n_1+n_2\right| 
+\left| 5 n_1+n_2\right| 
+3 \left| n_1+2 n_2\right|
+8 \left| 3 n_1+2 n_2\right| 
+3 \left| 5 n_1+2 n_2\right|\notag \\
&\qquad
+\left| 2 n_1+3 n_2\right|
+3 \left| 4 n_1+3 n_2\right| 
+3 \left| 5 n_1+3 n_2\right|
+\left| 7 n_1+3 n_2\right| 
+\left| 5 n_1+4 n_2\right| \notag \\
&\qquad 
+\left| 7 n_1+4 n_2\right| 
+24 \left| n_1\right| 
+8 \left| n_2\right|  
\Big) \notag \\
&\qquad
-\Big(\left| 
n_1+n_2\right| +\left| 2 n_1+n_2\right| +\left| 3 
n_1+n_2\right| +\left| 3 n_1+2 n_2\right| +\left| 
n_1\right| +\left| n_2\right| \Big) \; , \notag
\end{align}
from which we compute the Hilbert series to be
\begin{subequations}
\label{eqn:HS_G2_Rep21}
\begin{align}
\HS_{\Gtwo}^{[2,1]}(t,N)&=
 \frac{R(t,N)}{ \left(1-t^2\right) 
 \left(1-t^6\right) 
 \left(1-t^{132 N-6}\right) 
 \left(1-t^{232 N-10}\right) 
 \left(1-t^{356 N-16}\right) 
 \left(1-t^{486 N-22}\right)} , \\
R(t,N)&=1
+t^{132 N-5}(1+t+t^2+t^3+t^4)
+t^{232 N-9}(1+t+t^2+t^3+t^4)
\label{eqn:HS_G2_Rep21_Num}\\
&\qquad
+ t^{356 N-15}(2+2t+2t^2+2t^3+2t^4+t^5)
\notag\\
&\qquad
-t^{364 N-16}(1+2t+2t^2+2t^3+2t^4+2t^5)
\notag\\
&\qquad
+ t^{486 N-21}(2+2t+2t^2+2t^3+2t^4+t^5)
\notag\\
&\qquad
-t^{488 N-22}(1+3t+3t^2+3t^3+3t^4+3t^5+t^6)
\notag\\
&\qquad
-t^{588 N-25}(1+t+t^2+t^3+t^4)
-t^{618 N-27}(1+t+t^2+t^3+t^4)
\notag\\
&\qquad
-t^{718 N-32}(1+3t+3t^2+3t^3+3t^4+3t^5+t^6)
\notag\\
&\qquad
+t^{720 N-32}(1+2t+2t^2+2t^3+2t^4+2t^5)
\notag\\
&\qquad
- t^{842 N-37}(2+2t+2t^2+2t^3+2t^4+t^5)
\notag\\
&\qquad
+t^{850 N-38}(1+2t+2t^2+2t^3+2t^4+2t^5)
\notag\\
&\qquad
+t^{974 N-43}(1+t+t^2+t^3+t^4)
+t^{1074 N-47}(1+t+t^2+t^3+t^4)
+t^{1206 N-48} \notag .
\end{align}
\end{subequations}
The numerator~\eqref{eqn:HS_G2_Rep21_Num} is a palindromic polynomial of 
degree $1206N-48$; whereas, the denominator is of degree $1206N-46$. Hence, the 
difference in degree between denominator and numerator is $2$, which agrees 
with the quaternionic dimension of moduli space.
The Hilbert series~\eqref{eqn:HS_G2_Rep21} has a pole of order $4$ as $t\to 1$, 
which equals the complex dimension of the moduli space. (One can 
explicitly show that $R(t=1,N)=0$ and $\tfrac{\diff}{\diff t}R(t,N)|_{t=1}=0$, 
but $\tfrac{\diff^2}{\diff t^2}R(t,N)|_{t=1}\neq0$.)
The list of appearing operators is presented in Tab.~\ref{tab:Ops_G2_Rep21}.
\begin{table}[h]
\centering
 \begin{tabular}{c|c|c|c}
 \toprule
  object & & $\Delta(n_1,n_2)$ & $\Hh_{(n_1,n_2)}$\\ \midrule
  Casimir & $\mathcal{C}_2$ & $2$ & ---\\
	  & $ \mathcal{C}_6$ & $6$ & ---\\ \midrule
  bare monopole & $V_{(0,1)}^{\mathrm{dress},0}$ & $132N-6$ & $\utwo$\\
 dressings $(i=1,\ldots,5)$ & $V_{(0,1)}^{\mathrm{dress},i}$ & $132N-6+i$ & --- 
\\ \midrule
  bare monopole & $V_{(1,0)}^{\mathrm{dress},0}$ & $232N-10$& $\utwo$\\
 dressings $(i=1,\ldots,5)$ & $V_{(1,0)}^{\mathrm{dress},i}$ & $232N-10+i$ 
&---\\ \midrule
  bare monopole & $V_{(1,1)}^{\mathrm{dress},0}$ & $356N-16$& $\uo\times \uo$\\
 dressings $(i=1,\ldots,5 ; \alpha=1,2)$ & 
$V_{(1,1)}^{\mathrm{dress},i,\alpha}$ & $356N-16+i$ 
&--- \\
dressing  & 
$V_{(1,1)}^{\mathrm{dress},6}$ & $356N-16+6$ 
&---\\ \midrule
  bare monopole & $V_{(1,2)}^{\mathrm{dress},0}$ & $486N-22$& $\uo\times \uo$\\
 dressings $(i=1,\ldots,5 ; \alpha=1,2)$ & 
$V_{(1,2)}^{\mathrm{dress},i,\alpha}$ & $486N-22+i$ 
&--- \\
dressing  & 
$V_{(1,2)}^{\mathrm{dress},6}$ & $486N-22+6$ 
&--- \\
\bottomrule
 \end{tabular}
\caption{The chiral ring generators for a $\Gtwo$ gauge theory and matter 
transforming in $[2,1]$.}
\label{tab:Ops_G2_Rep21}
\end{table}
Due to the structure of the conformal dimension the behaviour of the $[2,1]$ 
representation is identical to that of $[4,0]$. Consequently, we do not discuss 
further details.
%%%%%%%%%%%%%%%%%%%%%%%%%%%%%%%%%%%%%%%%%%%%%%%%%%%%%%%%%%%%%%%%%%%%%%%%%%%%%%%%
  \section{Case: \texorpdfstring{$\boldsymbol{\sut}$}{SU(3)}}
\label{sec:SU3}
The last rank two example we would like to cover is $\sut$, for which the 
computation takes a detour over the corresponding $\ut$ theory, similar 
to~\cite{Cremonesi:2013lqa}. The advantage is that we can simultaneously 
investigate the rank three example $\ut$ and demonstrate that the method of 
Hilbert bases for semi-groups works equally well in higher rank cases.
\subsection{Set-up}
In the following, we systematically study a number of $\sut$ representation, 
where we understand a $\sut$-representation $[a,b]$ as an $\ut$-representation 
with a fixed $\uo$-charge. 

\paragraph{Preliminaries for $\boldsymbol{\ut}$} 
The GNO-dual group of $\ut$, which is again a $\ut$, has a weight lattice 
characterised by $m_1,m_2,m_3\in \Z$ 
and the dominant Weyl chamber is given by the restriction $m_1 \geq m_2 \geq 
m_3$, c.f.~\cite{Cremonesi:2013lqa}. The classical dressing factors associated 
to the interior and boundaries of the dominant Weyl chamber are the following:
\begin{equation}
 P_{\ut}(t^2,m_1,m_2,m_3) = \left\{
 \begin{matrix} \frac{1}{(1-t^2)^3} \; , & m_1>m_2>m_3 \; , \\ 
                \frac{1}{(1-t^2)^2(1-t^4)} \; , & (m_1=m_2 > m_3) \vee       
(m_1>m_2 = m_3)  \; ,\\
\frac{1}{(1-t^2)(1-t^4)(1-t^6)} \; , & m_1=m_2=m_3 \; .
                          \end{matrix} \right.
\label{eqn:dress_fct_U3}
\end{equation}
Note that we already introduced the fugacity $t^2$ instead of $t$.
Moreover, the GNO-dual $\ut$ has a non-trivial centre, i.e.\ $\mathcal{Z}(\ut) 
= 
\uo_J$; thus, the topological symmetry is a $\uo_J$ counted by 
$z^{m_1+m_2+m_3}$.

The contributions of $N_{(a,b)}$ 
hypermultiplets transforming in $[a,b]$ to the conformal dimension are as 
follows:
\begin{subequations}
\label{eqn:delta_h-plet_SU3}
\begin{align}
 \Delta^{[1,0]}_{\mathrm{h-plet}} &= \frac{N_{(1,0)}}{2} \sum_i | m_i |  \;,\\
\Delta^{[2,0]}_{\mathrm{h-plet}} &= \frac{3N_{(2,0)}}{2} \sum_i | m_i |  \;,\\
\Delta^{[1,1]}_{\mathrm{h-plet}} &= N_{(1,1)} \sum_{i<j} | m_i -m_j|  \;,\\
\Delta^{[3,0]}_{\mathrm{h-plet}} &= \frac{3N_{(3,0)}}{2} \sum_i | m_i | 
+N_{[3,0]} \sum_{i<j} | m_i -m_j| \;,\\
\Delta^{[2,2]}_{\mathrm{h-plet}} &= 3N_{(2,2)} \sum_i | m_i | 
+4N_{(2,2)} \sum_{i<j} | m_i -m_j| \; ,\\
\Delta^{[2,1]}_{\mathrm{h-plet}} &= 4N_{(2,1)} \sum_i | m_i | 
+\frac{N_{(2,1)}}{2} \sum_{i<j} \left( |2 m_i -m_j| + | m_i - 2m_j| \right) \; ,
\end{align}
\end{subequations}
where $i,j=1,2,3$. In addition, the contribution of the vector-multiplets reads 
as
\begin{equation}
 \Delta_{\mathrm{v-plet}} = - \sum_{i<j} |m_i-m_j| \; .
\end{equation}
Consequently, one can study a pretty wild matter content if one considers the 
conformal dimension to be of the form
\begin{equation}
\label{eqn:delta_U3_generic}
 \Delta(m_1,m_2,m_3) = \frac{N_F}{2} \sum_i |m_i| + \left( N_A-1 \right) 
\sum_{i<j} |m_i - m_j| + \frac{N_R}{2}\sum_{i<j} \left( |2m_i - m_j| + |m_i - 2
m_j|\right)
\end{equation}
and the relation to the various representations~\eqref{eqn:delta_h-plet_SU3} is 
established via
\begin{subequations}
 \begin{align}
  N_F &= N_{(1,0)} + 3N_{(2,0)} + 3 N_{(3,0)} + 6 N_{(2,2)} + 4 N_{(2,1)} \; , 
\\
 N_A &= N_{(1,1)} + N_{(3,0)} + 4 N_{(2,2)} \; , \\
 N_R &= N_{(2,1)} \; .
 \end{align}
\end{subequations}
\paragraph{Preliminaries for $\boldsymbol{\sut}$}
As noted in~\cite{Cremonesi:2013lqa}, the reduction from $\ut$ to $\sut$ (with 
the same matter content) is realised by averaging over $\uo_J$, for the purpose 
of setting $m_1+m_2+m_3=0$, and multiplying by $(1-t^2)$, such that 
$\tr(\Phi)=0$ for the adjoint scalar $\Phi$. In other words
\begin{equation}
 \HS_{\sut}^{[a,b]}(t^2)= (1-t^2) \oint_{|z|=1} \frac{\diff z}{2 \pi \im z } 
\HS_{\ut}^{[a,b]}(t^2,z) \; .
\label{eqn:Reduction_to_SU3}
\end{equation}
As a consequence, the conformal dimension for $\sut$ itself is obtained 
from~\eqref{eqn:delta_U3_generic} via
\begin{equation}
 \Delta(m_1,m_2) \coloneqq \Delta(m_1,m_2,m_3) \big|_{m_3 =-m_1-m_2} \; .
 \label{eqn:delta_SU3_generic}
\end{equation}
The Weyl chamber is now characterised by $m_1 \geq \max\{m_2,-2m_2\}$. 
Multiplying~\eqref{eqn:dress_fct_U3} by $(1-t^2)$ and employing 
$m_3=-m_1-m_2$ results in the  classical dressing factors for $\sut$
\begin{equation}
 P_{\sut}(t^2,m_1,m_2) = \left\{
 \begin{matrix} \frac{1}{(1-t^2)^2} \; , & m_1 >\max\{m_2,-2m_2\} \; ,\\ 
                \frac{1}{(1-t^2)(1-t^4)} \; , & (m_1=m_2 ) \vee       
(m_1=-2m_2 )  \; , \\
\frac{1}{(1-t^4)(1-t^6)} \; , & m_1=m_2=0 \; .
                          \end{matrix} \right.
\label{eqn:dress_fct_SU3}
\end{equation}
%
% 
%%%%%%%%%%%%%%%%%%%%%%%%%%%%%%%%%%%%%%%%%%%%%%%%%%%%%%%%%%%%%%%%%%%%%%%% 
%%%%%%%%%%%%%%%%%%%%%%%%%%%%%%%%%%%%%%%%%%%%%%%%%%%%%%%%%%%%%%%%%%%%%%%%
%
\subsection{Hilbert basis}
\subsubsection{Fan and cones for \texorpdfstring{$\ut$}{U(3)}}
Following the ideas outline previously, 
$\Lambda_w(\widehat{\ut}) \slash 
\mathcal{W}_{\ut}$ can be described as a collection of semi-groups 
that originate from a fan.
Since this is our first $3$-dimensional example, we provide a detail 
description on how to obtain the fan. Consider the absolute values $|a m_1 +b 
m_2 +cm_2|$ in~\eqref{eqn:delta_SU3_generic} as \emph{Hesse normal form} for 
the hyperplanes
\begin{equation}
 \vec{n}\cdot \vec{m} \equiv \begin{pmatrix} a \\ b \\ c   \end{pmatrix}\cdot 
\begin{pmatrix} m_1 \\ m_2 \\ m_3   \end{pmatrix} =0
\end{equation}
which pass through the origin. Take all normal vectors $\vec{n}_j$, define the 
matrices $M_{i,j} = (\vec{n}_i,\vec{n}_j)^T$ (for $i<j$) and compute the null 
spaces (or kernel) $K_{i,j}\coloneqq\ker(M_{i,j})$. Linear algebra tell us 
that $\dim(K_{i,j}) \geq1$, but by the specific form\footnote{$\Delta$ is 
homogeneous and all hyperplanes pass through the origin; hence, no two 
hyperplanes can be parallel. This implies that no two normal vectors can be 
multiplies of each other.} of $\Delta$ we have the stronger condition 
$\rank(M_{i,j})=2$ for all $i<j$; thus, we always have $\dim(K_{i,j})=1$. Next, 
we select a basis vector $e_{i,j}$ of $K_{i,j}$ and check if $e_{i,j}$ or 
$-e_{i,j}$ intersect the Weyl-chamber. If it does, then it is going to be an 
edge for the fan and, more importantly, will turn out to be a ray generator 
(provided one defines $e_{i,j}$ via the intersection with the corresponding 
weight lattice). 
Now, one has to define all $3$-dimensional cones, merge them into a fan, and, 
lastly, compute the  Hilbert bases. The programs 
\texttt{Macaulay2} and \texttt{Sage} are convenient 
tools for such tasks.

As two examples, we consider the conformal 
dimension~\eqref{eqn:delta_SU3_generic} for $N_R=0$ and 
$N_R\neq0$ and preform the entire procedure. That is: firstly, compute the 
edges of the fan; secondly, define the all $3$-dimensional cones and; thirdly, 
compute the Hilbert bases.
\paragraph{Case $N_R=0$:} In this circumstance, we deduce the following edges
 \begin{align}
  \begin{pmatrix}1 \\ 0 \\ 0  \end{pmatrix} \, , \quad
  \begin{pmatrix}1 \\ 1 \\ 0  \end{pmatrix} \, , \quad
  \begin{pmatrix}1 \\ 1 \\ 1  \end{pmatrix} \, , \quad
  \begin{pmatrix}0 \\ 0 \\ -1  \end{pmatrix} \, , \quad
  \begin{pmatrix}0 \\ -1 \\ -1  \end{pmatrix} \, , \quad 
  \begin{pmatrix}-1 \\ -1 \\ -1  \end{pmatrix} \, .
 \end{align}
All these vectors are on the boundaries of the Weyl chamber.
The set of $3$-dimensional cones that generate the corresponding fan is given by
\begin{subequations}
 \begin{alignat}{2}
  C_1^{(3)} &= \cone\left\{ \colvec{1}{0}{0}, \colvec{1}{1}{0},\colvec{1}{1}{1} 
 \right\} , \;  &
  C_2^{(3)} &= \cone\left\{ \colvec{1}{0}{0}, 
\colvec{1}{1}{0},\colvec{0}{0}{-1} 
 \right\} ,\\
  C_3^{(3)} &= \cone\left\{ \colvec{1}{0}{0}, 
\colvec{0}{-1}{-1},\colvec{0}{0}{-1} 
 \right\} , \; &
  C_4^{(3)} &= \cone\left\{ \colvec{-1}{-1}{-1}, 
\colvec{0}{-1}{-1},\colvec{0}{0}{-1} 
 \right\} .
 \end{alignat}
\end{subequations}
A computation shows that all four cones are strictly convex, smooth, and 
simplicial. The Hilbert bases for the resulting semi-groups 
comprise solely the ray generators
\begin{subequations}
\label{eqn:Hilbert_basis_U3_NR=0}
  \begin{alignat}{2}
  \Hcal(S_1^{(3)}) &=\left\{ \colvec{1}{0}{0}, 
\colvec{1}{1}{0},\colvec{1}{1}{1} 
 \right\} , \;  &
  \Hcal(S_2^{(3)}) &= \left\{ \colvec{1}{0}{0}, 
\colvec{1}{1}{0},\colvec{0}{0}{-1} 
 \right\} ,\\
  \Hcal(S_3^{(3)}) &= \left\{ \colvec{1}{0}{0}, 
\colvec{0}{-1}{-1},\colvec{0}{0}{-1} 
 \right\} , \; &
  \Hcal(S_4^{(3)}) &= \left\{ \colvec{-1}{-1}{-1}, 
\colvec{0}{-1}{-1},\colvec{0}{0}{-1} 
 \right\} .
 \end{alignat}
\end{subequations}
From the above, we expect $6$ bare monopole operators plus their dressings for 
a generic theory with $N_R=0$. Since all ray generators lie at the boundary of 
the Weyl chamber, the residual gauge groups are $\ut$ for $\pm(1,1,1)$ and 
$\utwo \times \uo$ for the other four GNO-charges.
%
%%%%%%%%%%%%%%%%%%%%%%%%%%%%%%%% 
% 
\paragraph{Case $N_R\neq 0$:}  
Here, we compute the following edges:
\begin{subequations}
 \begin{alignat}{7}
  &\begin{pmatrix}1 \\ 0 \\ 0  \end{pmatrix} , \; &
  &\begin{pmatrix}1 \\ 1 \\ 0  \end{pmatrix} , \; &
  &\begin{pmatrix}1 \\ 1 \\ 1  \end{pmatrix} , \; &
  &\begin{pmatrix}2 \\ 1 \\ 0  \end{pmatrix} , \; &
&\begin{pmatrix}2 \\ 1 \\ 1  \end{pmatrix} , \; &
&\begin{pmatrix}2 \\ 2 \\ 1  \end{pmatrix}, \;&
&\begin{pmatrix}4 \\ 2 \\ 1  \end{pmatrix} , \\
 &\begin{pmatrix}0 \\ 0 \\ -1  \end{pmatrix}, \;&
 &\begin{pmatrix}0 \\ -1 \\ -1  \end{pmatrix} , \;&
  &\begin{pmatrix}-1 \\ -1 \\ -1  \end{pmatrix} , \;&
&\begin{pmatrix}0 \\ -1 \\ -2  \end{pmatrix} , \; &
&\begin{pmatrix}-1 \\ -1 \\ -2  \end{pmatrix} , \; &
&\begin{pmatrix}-1 \\ -2 \\ -2  \end{pmatrix} , \; &
&\begin{pmatrix}-1 \\ -2 \\ -4  \end{pmatrix} . 
 \end{alignat}
 \end{subequations}
 Now, we need to proceed and define all $3$-dimensional cones that constitute 
the fan and, in turn, will lead to the semi-groups that we wish to study.
\begin{subequations}
\label{eqs:Cones_U3_NR>0}
 \begin{alignat}{2}
  C_1^{(3)} &= \cone\left\{ \colvec{1}{0}{0},\colvec{2}{1}{0},\colvec{4}{2}{1} 
\right\} , \, &
C_2^{(3)} &= \cone\left\{ \colvec{4}{2}{1},\colvec{1}{0}{0},\colvec{2}{1}{1} 
\right\} , \\
  C_3^{(3)} &= \cone\left\{ \colvec{2}{2}{1},\colvec{1}{1}{0},\colvec{2}{1}{0} 
\right\} , \, &
C_4^{(3)} &= \cone\left\{ \colvec{2}{2}{1},\colvec{2}{1}{0},\colvec{4}{2}{1} 
\right\}  , \\
  C_5^{(3)} &= \cone\left\{ \colvec{2}{2}{1},\colvec{4}{2}{1},\colvec{2}{1}{1} 
\right\} , \, &
 C_6^{(3)} &= \cone\left\{ \colvec{2}{2}{1},\colvec{2}{1}{1},\colvec{1}{1}{1} 
\right\}  , \\
  C_7^{(3)} &= \cone\left\{ \colvec{0}{0}{-1},\colvec{1}{0}{0},\colvec{2}{1}{0} 
\right\} , \, &
 C_8^{(3)} &= \cone\left\{ \colvec{0}{0}{-1},\colvec{1}{1}{0},\colvec{2}{1}{0} 
\right\}  , \\
  C_9^{(3)} &= \cone\left\{ 
\colvec{0}{0}{-1},\colvec{0}{-1}{-2},\colvec{1}{0}{0} 
\right\}, \, &
C_{10}^{(3)} &= \cone\left\{ 
\colvec{0}{-1}{-2},\colvec{0}{-1}{-1},\colvec{1}{0}{0} 
\right\} , \\
  C_{11}^{(3)} &= \cone\left\{ 
\colvec{0}{0}{-1},\colvec{0}{-1}{-2},\colvec{-1}{-2}{-4} 
\right\} , \, &
 C_{12}^{(3)} &= \cone\left\{ 
\colvec{0}{0}{-1},\colvec{-1}{-2}{-4},\colvec{-1}{-1}{-2} 
\right\} , \\
  C_{13}^{(3)} &= \cone\left\{ 
\colvec{0}{-1}{-2},\colvec{-1}{-2}{-4},\colvec{0}{-1}{-1} 
\right\} , \, &
 C_{14}^{(3)} &= \cone\left\{ 
\colvec{0}{-1}{-1},\colvec{-1}{-2}{-4},\colvec{-1}{-2}{-2} 
\right\} , \\
  C_{15}^{(3)} &= \cone\left\{ 
\colvec{-1}{-2}{-4},\colvec{-1}{-2}{-2},\colvec{-1}{-1}{-2} 
\right\} , \, &
 C_{16}^{(3)} &= \cone\left\{ 
\colvec{-1}{-2}{-2},\colvec{-1}{-1}{-2},\colvec{-1}{-1}{-1} 
\right\}  . 
 \end{alignat}
\end{subequations}
All of the cones are strictly convex and simplicial, but only the cones $C_p$ 
for $p=1,2,3,6,\ldots,13,16$ are smooth.
Now, we compute the Hilbert bases for semi-groups $S_p^{(3)}$ for 
$p=1,2,\dots,16$ and obtain
\begin{subequations}
\label{eqs:Hilbert_basis_U3_NR>0}
 \begin{align}
  \Hcal(S_1^{(3)}) &= \left\{ 
\colvec{1}{0}{0},\colvec{2}{1}{0},\colvec{4}{2}{1} 
\right\} , \, &
  \Hcal(S_2^{(3)}) &= \left\{ 
\colvec{4}{2}{1},\colvec{1}{0}{0},\colvec{2}{1}{1} 
\right\}  , \\
  \Hcal(S_3^{(3)}) &= \left\{ 
\colvec{2}{2}{1},\colvec{1}{1}{0},\colvec{2}{1}{0} 
\right\} , \, &
 \Hcal(S_4^{(3)}) &= \left\{ 
\colvec{2}{2}{1},\colvec{2}{1}{0},\colvec{4}{2}{1},\colvec{3}{2}{1} 
\right\}  , \\
\Hcal(S_5^{(3)}) &= \left\{ 
\colvec{2}{2}{1},\colvec{4}{2}{1},\colvec{2}{1}{1},\colvec{3}{2}{1} 
\right\} , \, &
 \Hcal(S_6^{(3)}) &= \left\{ 
\colvec{2}{2}{1},\colvec{2}{1}{1},\colvec{1}{1}{1} 
\right\}  , \\
  \Hcal(S_7^{(3)}) &= \left\{ 
\colvec{0}{0}{-1},\colvec{1}{0}{0},\colvec{2}{1}{0} 
\right\} , \, &
 \Hcal(S_8^{(3)}) &= \left\{ 
\colvec{0}{0}{-1},\colvec{1}{1}{0},\colvec{2}{1}{0} 
\right\}  , \\
  \Hcal(S_9^{(3)}) &= \left\{ 
\colvec{0}{0}{-1},\colvec{0}{-1}{-2},\colvec{1}{0}{0} 
\right\} , \, &
\Hcal(S_{10}^{(3)}) &= \left\{ 
\colvec{0}{-1}{-2},\colvec{0}{-1}{-1},\colvec{1}{0}{0} 
\right\}  , \\
  \Hcal(S_{11}^{(3)}) &= \left\{ 
\colvec{0}{0}{-1},\colvec{0}{-1}{-2},\colvec{-1}{-2}{-4} 
\right\} , \, &
 \Hcal(S_{12}^{(3)}) &= \left\{ 
\colvec{0}{0}{-1},\colvec{-1}{-2}{-4},\colvec{-1}{-1}{-2} 
\right\}  , \\
  \Hcal(S_{13}^{(3)}) &= \left\{ 
\colvec{0}{-1}{-2},\colvec{-1}{-2}{-4},\colvec{0}{-1}{-1} 
\right\} , \, &
 \Hcal(S_{14}^{(3)}) &= \left\{ 
\colvec{0}{-1}{-1},\colvec{-1}{-2}{-4},\colvec{-1}{-2}{-2} ,\colvec{-1}{-2}{-3}
\right\}  , \\
  \Hcal(S_{15}^{(3)}) &= \left\{ 
\colvec{-1}{-2}{-4},\colvec{-1}{-2}{-2},\colvec{-1}{-1}{-2},\colvec{-1}{-2}{-3} 
\right\} , \, &
 \Hcal(S_{16}^{(3)}) &= \left\{ 
\colvec{-1}{-2}{-2},\colvec{-1}{-1}{-2},\colvec{-1}{-1}{-1} 
\right\}  . 
 \end{align}
\end{subequations}
We observe that there are four semi-groups $S_p^{}$ for $p=4,5,14,15$ for which 
the Hilbert bases exceeds the set of ray generators by an additional element.
Consequently, we expect $16$ bare monopoles plus their dressings for a generic 
theory with $N_R \neq0$. However, the dressings exhibit a much richer structure 
compared to $N_R=0$, because some minimal generators lie in the interior of 
the Weyl chamber. The residual gauge groups are $\ut$ for $\pm(1,1,1)$; $\utwo 
\times \uo$ for $(1,0,0)$, $(0,0,-1)$, $(1,1,0)$, $(0,-1,-1)$, $(2,1,1)$, 
$(-1,-1,-2)$, $(2,2,1)$, and $(-1,-2,-2)$; and $\uo^3$ for $(2,1,0)$, 
$(0,-1,-2)$,$(4,2,1)$, $(-1,-2,-4)$, $(3,2,1)$, and $(-1,-2,-3)$.
% 
% 
%%%%%%%%%%%%%%%%%%%%%%%%%%%%%%%%%%%%%%%%%%%%%%%%%%%%%%%%%%%%%%%%%%%%%%%%%%%% 
%%%%%%%%%%%%%%%%%%%%%%%%%%%%%%%%%%%%%%%%%%%%%%%%%%%%%%%%%%%%%%%%%%%%%%%%%%%%
% 
\subsubsection{Fan and cones for \texorpdfstring{$\sut$}{SU(3)}}
\begin{figure}[h]
\centering
\begin{subfigure}{0.485\textwidth}
\centering
    \begin{tikzpicture}
  \coordinate (Origin)   at (0,0);
  \coordinate (XAxisMin) at (-0.5,0);
  \coordinate (XAxisMax) at (4.5,0);
  \coordinate (YAxisMin) at (0,-2.5);
  \coordinate (YAxisMax) at (0,3.5);
  \draw [thin, gray,-latex] (XAxisMin) -- (XAxisMax);% Draw x axis
  \draw [thin, gray,-latex] (YAxisMin) -- (YAxisMax);% Draw y axis
  \draw (4.7,-0.2) node {$m_1$};
  \draw (-0.3,3.3) node {$m_2$};
  \foreach \x in {0,1,...,4}{% Two indices running over each
      \foreach \y in {-2,-1,...,3}{% node on the grid we have drawn 
        \node[draw,circle,inner sep=0.8pt,fill,black] at (\x,\y) {};
            % Places a dot at those points
            }
            }
    \draw[black,dashed,thick]  (Origin) -- (3.8,3.8);  
    \draw[black,dashed,thick]  (Origin) -- (1.2*4,-1.2*2);
    \draw[black,dashed,thick]  (Origin) -- (4.2,0);
% % 
 \draw[black,thick] (1,0) circle (4pt);
 \draw[black,thick] (1,1) circle (4pt);
 \draw[black,thick] (2,-1) circle (4pt);
\end{tikzpicture}
 \caption{$N_R=0$}
 \label{Fig:Hilbert_basis_SU3_NR=0}
\end{subfigure}
\begin{subfigure}{0.485\textwidth}
  \centering
    \begin{tikzpicture}
  \coordinate (Origin)   at (0,0);
  \coordinate (XAxisMin) at (-0.5,0);
  \coordinate (XAxisMax) at (4.5,0);
  \coordinate (YAxisMin) at (0,-2.5);
  \coordinate (YAxisMax) at (0,3.5);
  \draw [thin, gray,-latex] (XAxisMin) -- (XAxisMax);% Draw x axis
  \draw [thin, gray,-latex] (YAxisMin) -- (YAxisMax);% Draw y axis
  \draw (4.7,-0.2) node {$m_1$};
  \draw (-0.3,3.3) node {$m_2$};
  \foreach \x in {0,1,...,4}{% Two indices running over each
      \foreach \y in {-2,-1,...,3}{% node on the grid we have drawn 
        \node[draw,circle,inner sep=0.8pt,fill,black] at (\x,\y) {};
            % Places a dot at those points
            }
            }
    \draw[black,dashed,thick]  (Origin) -- (3.8,3.8);  
    \draw[black,dashed,thick]  (Origin) -- (1.2*4,-1.2*2);
    \draw[black,dashed,thick]  (Origin) -- (4.2,0);
    \draw[black,dashed,thick]  (Origin) -- (2.2*2,2.2*1);
    \draw[black,dashed,thick]  (Origin) -- (1.5*3,-1.5*1);
% % 
 \draw[black,thick] (1,0) circle (4pt);
 \draw[black,thick] (1,1) circle (4pt);
 \draw[black,thick] (2,-1) circle (4pt);
 \draw[black,thick] (3,-1) circle (4pt);
 \draw[black,thick] (2,1) circle (4pt);
\end{tikzpicture}
 \caption{$N_R\neq0$}
 \label{Fig:Hilbert_basis_SU3_NR>0}
\end{subfigure}
\caption{The semi-groups for $\sut$ and the corresponding ray generators 
(black circled points).}
\label{Fig:Hilber_basis_SU3}
\end{figure}
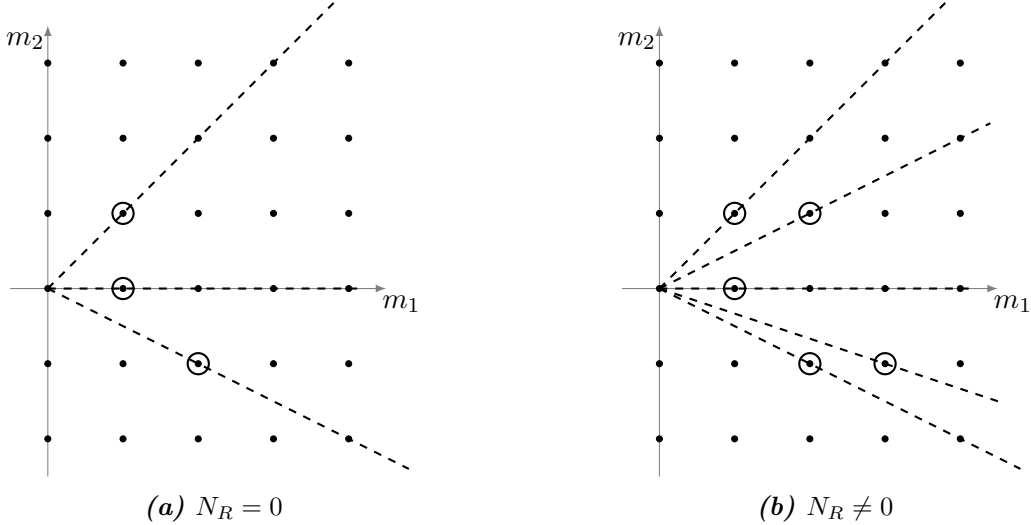

The conformal dimension~\eqref{eqn:delta_SU3_generic} divides the Weyl chamber 
of the GNO-dual into two different fans, depending on $N_R=0$ or $N_R\neq0$. 
\paragraph{Case $\boldsymbol{N_R=0}$:} 
For this situation, which is depicted in 
Fig.~\ref{Fig:Hilbert_basis_SU3_NR=0}, 
 there are three rays $~\sim |m_1|, |m_1-m_2|, |m_1+2m_2|$ present that 
intersect the Weyl chamber non-trivially. The corresponding fan is generated 
by two 
$2$-dimensional cones
 \begin{equation}
  C_1^{(2)}=\cone((2,-1),(1,0)) \und C_2^{(2)}=\cone((1,0),(1,1)) \; .
 \end{equation}
The Hilbert bases for the semi-groups, obtained by intersecting the cones with 
the weight lattice, are solely given by the ray generators, i.e.
 \begin{equation}
 \label{eqn:Hilbert_basis_SU3_NR=0}
  \Hcal(S_1^{(2)}) = \Big\{(2,-1),(1,0) \Big\} \und \Hcal(S_2^{(2)})= \Big\{ 
(1,0),(1,1) \Big\} \;. 
 \end{equation}
As a consequence, we expect three bare monopole operators (plus dressings) for 
a generic $N_R=0$ theory. The residual gauge group is $\su\times \uo$ for 
$(2,-1)$ and $(1,1)$, because these GNO-charges are at the boundary of the 
Weyl-chamber. In contrast, $(1,0)$ has residual gauge group $\uo^2$ as it lies 
in the interior of the dominant Weyl chamber. 
\paragraph{Case $\boldsymbol{N_R\neq 0}$:} 
For this circumstance, which is depicted in 
Fig.~\ref{Fig:Hilbert_basis_SU3_NR>0}, 
 there are two additional rays $~\sim  |m_1 -2m_2 |, |m_1+3m_2|$ 
present, compared to $N_R=0$, that intersect the Weyl chamber non-trivially. 
The corresponding fan is now generated by four $2$-dimensional cones
\begin{subequations}
 \begin{alignat}{2}
  C_{1-}^{(2)} &= \cone((2,-1),(3,-1)) \; ,\qquad  &
C_{1+}^{(2)} &= \cone((3,-1),(1,0)) \; , \\
  C_{2-}^{(2)} &= \cone((1,0),(2,1)) \; , \qquad &
  C_{2+}^{(2)} &= \cone((2,1),(1,1))  \; . 
 \end{alignat}
\end{subequations}
The Hilbert bases for the resulting semi-groups are given by the ray 
generators, i.e.
\begin{subequations}
\label{eqn:Hilbert_basis_SU3_NR>0}
\begin{alignat}{2}
  \Hcal(S_{1-}^{(2)}) &= \Big\{(2,-1),(3,-1)\Big\} \; ,\qquad  &
\Hcal(S_{1+}^{(2)}) &= \Big\{(3,-1),(1,0)\Big\} \; , \\
  \Hcal(S_{2-}^{(2)}) &= \Big\{(1,0),(2,1)\Big\} \; , \qquad &
  \Hcal(S_{2+}^{(2)}) &= \Big\{(2,1),(1,1)\Big\}  \; .
 \end{alignat}
\end{subequations}
Judging from the Hilbert bases, there are five bare monopole operators present 
in the generic case. The residual gauge group for $(1,0)$, $(3,-1)$, and $(2,1)$ 
is $\uo^2$, as they lie in the interior. For $(1,1)$ and $(2,-1)$ the residual 
gauge group is $\su \times \uo$, because these points lie at the boundary of 
the Weyl chamber.
% 
%%%%%%%%%%%%%%%%%%%%%%%%%%%%%%%%%%%%%%%%%%%%%%%%%%%%%%%%%%%%%%%% 
%%%%%%%%%%%%%%%%%%%%%%%%%%%%%%%%%%%%%%%%%%%%%%%%%%%%%%%%%%%%%%%%
% 
\subsection{Casimir invariance}
\subsubsection{Dressings for \texorpdfstring{$\ut$}{U(3)}}
Following the description of dressed monopole operators as 
in~\cite{Cremonesi:2013lqa}, we diagonalise the adjoint-valued scalar $\Phi$ 
along the moduli space, i.e.
\begin{equation}
 \diag \Phi = (\phi_1,\phi_2,\phi_3) \; .
\end{equation}
Moreover, the Casimir invariants of $\ut$ can then be written as $
 \Casi{j}= \tr(\Phi^j)= \sum_{l=1}^{3} (\phi_l)^j$ for $j=1,2,3$. We will now 
elaborate on the possible dressed monopole operators by means of the insights 
gained in Sec.~\ref{subsec:Dressings_as_HS} and App.~\ref{app:PL}.

To start with, for a monopole with GNO-charge such that $\Hh_{(m_1,m_2,m_3)}= 
\ut$ the dressings are described by 
\begin{equation}
 \frac{P_{\ut}(t,m_1,m_1,m_1)}{P_{\ut}(t,0)} -1 = 0 \; ,
\end{equation}
i.e.\ there are no dressings, because the Casimir invariants of the 
centraliser $\Hh_{(m_1,m_2,m_3)}$ are identical to those of $\G$, since the 
groups coincide. Prominent examples are the (bare) monopoles of GNO-charge
$\pm(1,1,1)$.

Next, a monopole of GNO-charge such that  $\Hh_{(m_1,m_2,m_3)}= \uo \times 
\utwo$ exhibit dressings governed by
\begin{equation}
 \frac{P_{\ut}(t,m_1,m_2,m_3)}{P_{\ut}(t,0)} -1 = 
\frac{(1-t^2)(1-t^4)(1-t^6)}{(1-t^2)^2(1-t^4)} -1 = t^2 + t^4 \; ,
\end{equation}
implying there to be exactly one dressing by a degree $2$ Casimir and one 
dressing by a degree $4$ Casimir. The two degree $2$ Casimir invariants of 
$\Hh_{(m_1,m_2,m_3)}$, one by $\uo$ and one by $\utwo$, are not both 
independent 
because there is the overall Casimir $\Casi{1}$ of $\ut$. Therefore, only one 
of them leads to an independent dressed monopole generator. The second dressing 
is then due to the second Casimir of $\utwo$. For example, the monopole of 
GNO-charge $(1,1,0)$, $(0,-1,-1)$, $(2,1,1)$, $(-1,-2,-2)$, $(2,2,1)$, and 
$(-1,-2,-2)$ exhibit these two dressings options.

Lastly, if the residual gauge group is $\Hh_{(m_1,m_2,m_3)} = \uo^3$ then the 
dressings are determined via
\begin{equation}
  \frac{P_{\ut}(t,m_1,m_2,m_3)}{P_{\ut}(t,0)} -1 = 
\frac{(1-t^2)(1-t^4)(1-t^6)}{(1-t^2)^3} -1 = 2t^2 + 2t^4 +t^6 \; . 
\end{equation}
Consequently, there are generically five dressings for each such bare monopole 
operator. Examples for this instance are $(2,1,0)$, $(0,-1,-2)$, $(3,2,1)$, 
$(-1,-2,-3)$, $(4,2,1)$, $(-1,-2,-4)$.
\subsubsection{Dressings for \texorpdfstring{$\sut$}{SU(3)}}
\label{subsec:Dress_SU3}
To determine the dressings, we take the adjoint scalar $\Phi$ and diagonalise 
it, keeping in mind that it now belongs to $\sut$, that is
\begin{equation}
 \diag \Phi = (\phi_1,\phi_2,-(\phi_1+\phi_2)) \; .
\end{equation}
While keeping in mind that each $\phi_i$ has dimension one, we can write down 
the dressings (in the dominant Weyl chamber):
$(1,0)$ can be dressed by two independent $\uo$-Casimir invariants, i.e.\ 
directly by $\phi_1$ and $\phi_2$
\begin{equation}
 V_{(1,0)}^{\mathrm{dress},(0,0)}\equiv (1,0) \longrightarrow 
\begin{cases} V_{(1,0)}^{\mathrm{dress},(1,0)} \equiv \phi_1 \ (1,0) \; ,\\
V_{(1,0)}^{\mathrm{dress},(0,1)} \equiv \phi_2 \ (1,0)   \; ,\end{cases}
\end{equation}
such that the dressings have conformal dimension $\Delta(1,0)+1$. 
Next, out of the three degree 2 combinations of $\phi_i$, only two of them are 
independent and we choose them to be
\begin{equation}
 V_{(1,0)}^{\mathrm{dress},(0,0)}\equiv (1,0) \longrightarrow 
\begin{cases} V_{(1,0)}^{\mathrm{dress},(2,0)} \equiv \phi_1^2 \ (1,0) \; 
,\\
V_{(1,0)}^{\mathrm{dress},(0,2)} \equiv \phi_2^2 \ (1,0) \; ,\end{cases}
\end{equation}
and these second order dressings have conformal dimension $\Delta(1,0)+2$.
Finally, one last dressing is possible 
\begin{equation}
  V_{(1,0)}^{\mathrm{dress},(0,0)}\equiv (1,0) \longrightarrow 
V_{(1,0)}^{\mathrm{dress},(3,0)+(0,3)} \equiv (\phi_1^3+ \phi_2^3) \ (1,0)  
\; ,
\end{equation}
having dimension $\Delta(1,0)+3$. Alternatively, we utilise App.~\ref{app:PL} 
and compute the number and degrees of the dressed monopole operators of 
magnetic charge $(1,0)$ via the quotient $ P_{\sut}(t^2,1,0) \slash  
P_{\sut}(t^2,0,0)=1+2t^2+2t^4+t^6$. 

For the two monopoles of GNO-charge $(1,1)$ and $(2,-1)$, the residual 
gauge group is $\su\times \uo$, 
i.e.\ the monopoles can be dressed by a degree one Casmir invariant of the 
$\uo$ 
and by a degree two Casimir invariant of the $\su$. These increase the 
dimensions by one and two, respectively. Consequently, we obtain
\begin{align}
 V_{(1,1)}^{\mathrm{dress},0}\equiv (1,1) \longrightarrow 
\begin{cases} V_{(1,1)}^{\mathrm{dress},\uo} \equiv (\phi_1+\phi_2) \ (1,1)
\; ,\\
V_{(1,1)}^{\mathrm{dress},\su} \equiv (\phi_1^2+\phi_2^2 ) \ (1,1)   \; 
,\end{cases}
\end{align}
and similarly
\begin{align}
 V_{(2,-1)}^{\mathrm{dress},0}\equiv (2,-1) \longrightarrow 
\begin{cases} V_{(2,-1)}^{\mathrm{dress},\uo} \equiv (\phi_1+\phi_2) \
(2,-1) \; ,\\
V_{(2,-1)}^{\mathrm{dress},\su} \equiv (\phi_1^2+\phi_2^2 ) \  (2,-1) 
\;. \end{cases}
\end{align}
Since the magnetic weights $(1,1)$, $(2,-1)$ lie at the boundary of the 
dominant Weyl chamber, we can derive the dressing behaviour via  $ 
P_{\sut}(t^2,(1,1)\text{ or }(2,-1)) \slash  P_{\sut}(t^2,0,0)=1+t^2+t^4$ and 
obtain agreement with our choice of generators.

The remaining monopoles of GNO-charge $(2,1)$ and $(3,-1)$ can be treated 
analogously to $(1,0)$ and we obtain
\begin{align}
 V_{(2,1)}^{\mathrm{dress},(0,0)}\equiv (2,1) &\longrightarrow 
\begin{cases} V_{(2,1)}^{\mathrm{dress},(1,0)} \equiv \phi_1 \ (2,1) \; ,\\
V_{(2,1)}^{\mathrm{dress},(0,1)} \equiv \phi_2 \ (2,1)   \; , \\
 V_{(2,1)}^{\mathrm{dress},(2,0)} \equiv \phi_1^2 \ (2,1) \; ,\\
V_{(2,1)}^{\mathrm{dress},(0,2)} \equiv \phi_2^2 \ (2,1)   \; , \\ 
V_{(2,1)}^{\mathrm{dress},(3,0)+(0,3)} \equiv (\phi_1^3+\phi_2^3) \ (2,1)   
\; ,  \end{cases}\\
 V_{(3,-1)}^{\mathrm{dress},(0,0)}\equiv (3,-1) &\longrightarrow 
\begin{cases} V_{(3,-1)}^{\mathrm{dress},(1,0)} \equiv \phi_1 \ (3,-1) \; 
,\\
V_{(3,-1)}^{\mathrm{dress},(0,1)} \equiv \phi_2 \ (3,-1)   \; , \\
 V_{(3,-1)}^{\mathrm{dress},(2,0)} \equiv \phi_1^2 \ (3,-1) \; ,\\
V_{(3,-1)}^{\mathrm{dress},(0,2)} \equiv \phi_2^2 \ (3,-1)   \; , \\ 
V_{(3,-1)}^{\mathrm{dress},(3,0)+(0,3)} \equiv (\phi_1^3+\phi_2^3) \ (3,-1) 
 \; .  \end{cases}
\end{align}
There can be circumstances in which not all dressings for the minimal 
generators determined by the Hilbert bases~\eqref{eqn:Hilbert_basis_SU3_NR>0} 
are truly independent. However, this will only occur for special 
configurations of $(N_F,N_A,F_R)$ and, therefore, is considered as 
``non-generic'' case.
% 
%%%%%%%%%%%%%%%%%%%%%%%%%%%%%%%%%%%%%%%%%%%%%%%%%%%%%%%%%%%%%%%%%%%%%%%% 
%%%%%%%%%%%%%%%%%%%%%%%%%%%%%%%%%%%%%%%%%%%%%%%%%%%%%%%%%%%%%%%%%%%%%%%%
%
\subsection{Category \texorpdfstring{$N_R =0$}{NR=0}}
% 
%%%%%%%%%%%%%%%%%%%%%%%%%%%%%%%%%%%%%%%%%%%%%%%%%%%%%%%%%%%%%%%%%%%%%%%%% 
% 
\subsubsection{\texorpdfstring{$N_F$}{NF} hypermultiplets in 
\texorpdfstring{$[1,0]$}{[1,0]} and \texorpdfstring{$N_A$}{NA} hypermultiplets 
in \texorpdfstring{$[1,1]$}{[1,1]}}
\paragraph{Intermediate step at $\boldsymbol{\ut}$}
The conformal dimension~\eqref{eqn:delta_U3_generic} reduces for $N_R=0$ to the 
following:
\begin{equation}
 \Delta(m_1,m_2,m_3)=\frac{N_F}{2} \sum_i |m_i| + (N_A-1) \sum_{i<j} |m_i-m_j| 
\; . \label{eqn:delta_U3_Rep10+11}
\end{equation}
The Hilbert series is then readily computed
\begin{subequations}
\label{eqn:HS_U3_Rep10+11}
\begin{align}
 &\HS_{\ut}^{[1,0]+[1,1]}(N_F,N_A,t,z)= \frac{R(N_F,N_A,t,z)}{P(N_F,N_A,t,z)} 
\;, \\
P(N_F,N_A,t,z) &= \prod_{j=1}^3 \left(1-t^{2j}\right) 
 \left(1-\tfrac{1}{z} t^{4 N_A+N_F-4}\right) 
\left(1-z t^{4 N_A+N_F-4}\right)
\label{eqn:HS_U3_Rep10+11_Den}\\*
&\qquad \quad \times 
 \left(1-\tfrac{1}{z^2} t^{4 N_A+2 N_F-4}\right)
\left(1-z^2 t^{4 N_A+2 N_F-4}\right) 
\left(1-\tfrac{1}{z^3} t^{3 N_F}\right) 
 \left(1-z^3 t^{3 N_F}\right) \; ,
 \notag \\
% 
%%%%%%%%%%%%%%%%%%%%%%%%%%%%%%%%%%%%%%%%%%
%
R(N_F,N_A,t,z)&=
1+t^{8 N_A+2 N_F-2}
 -t^{8 N_A+4 N_F-8} (1+2t^2+2t^4)
 +2 t^{8 N_A+6 N_F-8}(1-t^6)
 \label{eqn:HS_U3_Rep10+11_Num}\\
 &\qquad
 + t^{8 N_A+8 N_F-6}(2+2t^2+t^4)
 -t^{8 N_A+10 N_F-8}
 +t^{16 N_A+6 N_F-10} \notag \\
&\qquad
 -t^{16 N_A+12 N_F-10}
 -t^{6 N_F} \notag \\
&+\left(z+\frac{1}{z}\right) \bigg(
t^{4 N_A+N_F-2} (1+t^2)
+t^{4 N_A+7 N_F-4}
-t^{4 N_A+5 N_F-4} (1+t^2+t^4)
 \notag \\
&\phantom{+\left(z+\frac{1}{z}\right) \bigg( }
-t^{8 N_A+3 N_F-6}(1+t^2)
+t^{8 N_A+9 N_F-6}(1+t^2)
-t^{12 N_A + 5N_F-6}
 \notag \\
&\phantom{+\left(z+\frac{1}{z}\right) \bigg( }
+t^{12 N_A+7 N_F -10}(1+t^2+t^4)
-t^{12 N_A+11 N_F-10}(1+t^2) \bigg)
\notag\\
&+\left(z^2+\frac{1}{z^2}\right) \bigg(
 t^{4 N_A+2 N_F-2}
+t^{4 N_A+2 N_F}
-t^{4 N_A+4 N_F-4}(1+t^2+t^4)
+t^{4 N_A+8 N_F-4}
\notag \\
&\phantom{+\left(z^2+\frac{1}{z^2}\right)\bigg( }
-t^{12 N_A+4 N_F-6}
+t^{12 N_A+ 8 N_F-10}(1+t^2+t^4)
-t^{12 N_A+ 10 N_F-10}(1+t^2) \bigg)
\notag \\
&+\left(z^3+\frac{1}{z^3}\right) \bigg(
 t^{8 N_A+3 N_F-2}
-t^{8 N_A+5 N_F-6}(1+t^2+t^4)
\notag \\
&\phantom{+\left(z^3+\frac{1}{z^3}\right)\bigg( } 
+t^{8 N_A+7 N_F-8}(1+t^2+t^4)
-t^{8 N_A+9 N_F-8} \bigg) \; .
\notag
\end{align}

\end{subequations}
One can check that $R(N_F,N_A,t=1,z)=0$ and $\tfrac{\diff^n}{\diff 
t^n}R(N_F,N_A,t,z)|_{t=1,z=1}=0$ for $n=1,2$. 
Thus, the Hilbert series~\eqref{eqn:HS_U3_Rep10+11} has a pole of order $6$, 
which matches the dimension of the moduli space. Moreover, one computes the 
degree of the numerator~\eqref{eqn:HS_U3_Rep10+11_Num} to be 
$12 N_F +16 N_A -10$ and the degree of the 
denominator~\eqref{eqn:HS_U3_Rep10+11_Den} to be $12 N_F +16 N_A -4$, 
such that their difference equals the dimension of the moduli space.
The interpretation follows the results~\eqref{eqn:Hilbert_basis_U3_NR=0} 
obtained from the Hilbert bases and we summarise the minimal generators in 
Tab.~\ref{tab:Ops_U3_NR=0}.
\begin{table}[h]
\centering
 \begin{tabular}{c|c|c|c}
 \toprule
   \multicolumn{2}{c|}{$(m_1,m_2,m_3)$}  & 
$2\Delta(m_1,m_2,m_3)$ & $\Hh_{(m_1,m_2,m_3)}$ \\ \midrule
   $(1,0,0)$ & $(0,0,-1)$ & $N_F +4N_A -4$ & $\uo\times \utwo$ \\
  $(1,1,0)$ & $(0,-1,-1)$ & $2N_F +4N_A -4$ & $\uo\times \utwo$ \\
 $(1,1,1)$ & $(-1,-1,-1)$ & $3N_F$ & $\ut$ \\
 \bottomrule
 \end{tabular}
\caption{The monopole generators for a $\ut$ gauge theory with $N_R=0$ that 
together with the Casimir invariants generate the chiral ring.} 
\label{tab:Ops_U3_NR=0}
\end{table}
\paragraph{Reduction to $\boldsymbol{\sut}$}
Following the prescription~\eqref{eqn:Reduction_to_SU3}, we derive the 
following Hilbert series:
\begin{subequations}
\label{eqn:HS_SU3_Rep10+11}
\begin{align}
\HS_{\sut}^{[1,0]+[1,1]}(N_F,N_A,t)&= \frac{R(N_F,N_A,t)}{\left(1-t^4\right) 
\left(1-t^6\right)\left(1-t^{8 N_A+2 
N_F-8}\right) \left(1-t^{12 N_A+4 N_F-12}\right) } \; ,\\
R(N_F,N_A,t)&=1
+t^{8 N_A+2 N_F-6}(2+2t^2+t^4)
\label{eqn:HS_SU3_Rep10+11_Num}\\
& \qquad 
+t^{12 N_A+4 N_F-12}(1+2t^2+2t^4)
+t^{20 N_A+6 N_F-14} \notag \;.
\end{align}
\end{subequations}
An inspection yields that the numerator~\eqref{eqn:HS_SU3_Rep10+11_Num} is 
a palindromic polynomial of degree $20N_A+6N_F-14$; while the degree of the 
denominator is $20N_A+6N_F-10$. Thus, the difference in the degrees is 4, which 
equals the complex dimension of the moduli space. 
In addition, the Hilbert series~\eqref{eqn:HS_SU3_Rep10+11} has a pole of order 
four at $t\to 1$, which agrees with the dimension of Coulomb branch as well.

The minimal generators of~\eqref{eqn:Hilbert_basis_SU3_NR=0} are given by 
$V_{(1,0)}^{\mathrm{dress},(0,0)} $ with 
$2\Delta(1,0)=8 N_A+2 N_F-8$, and $V_{(1,1)}^{\mathrm{dress},0} $ and 
$V_{(2,-1)}^{\mathrm{dress},0}$ with $2\Delta(2,-1)=2\Delta(1,1)=12 N_A+4 
N_F-12 $. The dressed monopole operators are as described in 
Subsec.~\ref{subsec:Dress_SU3}.
% 
%%%%%%%%%%%%%%%%%%%%%%%%%%%%%%%%%%%%%%%%%%%%%%%%%%%%%%%%%%%%%%%
%
\subsubsection{\texorpdfstring{$N$}{N} hypermultiplets in 
\texorpdfstring{$[1,0]$}{[1,0]} 
representation}
Considering $N$ hypermultiplets in the fundamental representation is on extreme 
case of~\eqref{eqn:delta_U3_generic}, as $N_A=0=N_R$. We recall the results 
of~\cite{Cremonesi:2013lqa} and discuss them in the context of Hilbert bases 
for semi-groups.
\paragraph{Intermediate step at $\boldsymbol{\ut}$}
The Hilbert series has been computed to read
\begin{equation}
 \HS_{\ut}^{[1,0]}(N,t,z) = \prod_{j=1}^3 
\frac{1-t^{2N+2-2j}}{(1-t^{2j})(1-zt^{N+2-2j})(1-\tfrac{t^{N+2-2j}}{z})} \; .
 \end{equation}
Notably, it is a complete intersection in which the (bare and dressed) monopole 
operators of GNO-charge $(1,0,0)$ and $(0,0,-1)$ generate all other monopole 
operators. The to be expected minimal generators $(1,1,0)$, $(0,-1,-1)$, 
$(1,1,1)$, and $(-1,-1,-1)$ are now generated because
\begin{subequations}
\begin{align}
 V_{(1,1,0)}^{\mathrm{dress},0}&= V_{(1,0,0)}^{\mathrm{dress},1} 
+V_{(0,1,0)}^{\mathrm{dress},1} \; ,\\
 V_{(1,1,0)}^{\mathrm{dress},0}&= V_{(1,0,0)}^{\mathrm{dress},2} 
+V_{(0,1,0)}^{\mathrm{dress},2} +V_{(0,0,1)}^{\mathrm{dress},2}\; .
\end{align}
\end{subequations}
\paragraph{Reduction to $\boldsymbol{\sut}$}
The reduction leads to
\begin{equation}
\label{eqn:HS_SU3_Rep10_Hanany}
 \HS_{\sut}^{[1,0]}(N,t)= \frac{1+t^{2N-6} +2 t^{2N-4} +t^{2N-2} +t^{4N-8} 
}{(1-t^4)(1-t^6) (1-t^{2N-6})(1-t^{2N-8})} \; .
\end{equation}
Although the form of the Hilbert series~\eqref{eqn:HS_SU3_Rep10_Hanany} is 
suggestive: it has a pole of order $4$ for $t\to1$ and the numerator is 
palindromic, there is one drawback: no monopole operator of conformal dimension 
$(2N-6)$ exists. Therefore, we provide a equivalent rational function to 
emphasis the minimal generators:
\begin{equation}
\label{eqn:HS_SU3_Rep10_mod}
 \HS_{\sut}^{[1,0]}(N,t)= \frac{1
 + t^{2 N-6}(2+2t^2+t^4)
 +t^{4 N-12}(1+2t^2+2t^4)
 +t^{6 N-14}}{(1-t^4)(1-t^6) 
(1-t^{2N-8})(1-t^{4N-12})} \; .
\end{equation}
The equivalent form~\eqref{eqn:HS_SU3_Rep10_mod} still has a pole of order 
$4$ and a palindromic numerator. Moreover, the monopole generators are clearly 
visible, as we know the set of minimal 
generators~\eqref{eqn:Hilbert_basis_SU3_NR=0}, and can be summarise for 
completeness: $2\Delta(1,0)=2N-8$ and $2\Delta(1,1)=2\Delta(2,-1)=4N-12$.
\subsubsection{\texorpdfstring{$N$}{N} hypermultiplets in 
\texorpdfstring{$[1,1]$}{[1,1]} representation}
Investigating $N$ hypermultiplets in the adjoint representation is another 
extreme case of~\eqref{eqn:delta_U3_generic} as $N_F=0=N_R$. 
The conformal dimension in this circumstance reduces to
\begin{equation}
 \Delta(m_1,m_2,m_3)=(N-1) \sum_{i<j} | m_i-m_j| \; , 
\label{eqn:conformal_dim_11}
\end{equation}
and we notice that there is the shift symmetry $m_i \to m_i +a$ present. Due to 
this, the naive calculation of the $\ut$ Hilbert series is divergent, which we 
understand as follows:  Define overall $\uo$-charge $M\coloneqq m_1+m_2+m_3$, 
then the Hilbert series becomes
\begin{align}
 \HS_{\ut}^{(1,1)}= \sum_{M\in \mathbb{Z}} \sum_{m_1,m_2 \atop m_1\geq \max{(
m_2 , M-2m_2 )} } 
t^{2(N-1)(3m_1+3m_2-2M+|m_1-m_2|)} \ z^M \ P_{\ut}(t,m_1,m_2,m_3)  \; .
\end{align}
Since we want to use the $\ut$-calculation as an intermediate step to derive 
the $\sut$-case, the only meaningful choice to fix the shift-symmetry is 
$m_1+m_2+m_3=0$. But then
\begin{align}
 \HS_{\ut,\mathrm{fixed}}^{(1,1)}=  \sum_{m_1,m_2 \atop 
m_1\geq \max{(
m_2 ,-2m_2 )} } 
t^{2(N-1)(3m_1+3m_2+|m_1-m_2|)}  \ P_{\ut}(t,m_1,m_2,-m_1-m_2)  
\end{align}
and the transition to $\sut$ is simply
\begin{align}
  \HS_{\sut}^{(1,1)}&= (1-t^2) \int_{|z|=1} \frac{\diff z}{2 \pi z} 
\sum_{m_1,m_2 \atop m_1\geq \max{(
m_2 ,-2m_2 )} } 
t^{2(N-1)(3m_1+3m_2+|m_1-m_2|)}  \ P_{\ut}(t,m_1,m_2,-m_1-m_2) \notag \\
&= \sum_{m_1,m_2 \atop m_1\geq \max{(
m_2 ,-2m_2 )} } 
t^{2(N-1)(3m_1+3m_2+|m_1-m_2|)}  \ P_{\sut}(t,m_1,m_2) \; .
\end{align}
The computation then yields
\begin{equation}
\HS_{\sut}^{(1,1)}= \frac{1
+  t^{8 N-6} (2+2t^2+t^4)
+ t^{12 N-12} (1+2t^2+2t^4)
+ t^{20 N-14}}{\left(1-t^4\right) \left(1-t^6\right) 
\left(1-t^{8 N-8}\right) \left(1-t^{12 N-12}\right) } \; .
\label{eqn:HS_U3_Rep11}
\end{equation}
 We see that numerator of~\eqref{eqn:HS_U3_Rep11} is a palindromic 
polynomial of degree $20N-14$; while the degree of the denominator is 
$20N-10$. Hence, the difference in the degrees is 4, which coincides with the 
complex dimension of the moduli space. The same holds for the order of the pole 
of~\eqref{eqn:HS_U3_Rep11} at $t\to 1$.

The interpretation of the appearing monopole operators, and their dressings, is 
completely analogous to~\eqref{eqn:HS_SU3_Rep10+11} and reproduces the picture 
concluded from the Hilbert bases~\eqref{eqn:Hilbert_basis_U3_NR=0}. To be 
specific, $2\Delta(1,0)=8N-8$ and $2\Delta(1,1)=2\Delta(2,-1)=12N-12$.
% 
%%%%%%%%%%%%%%%%%%%%%%%%%%%%%%%%%%%%%%%%%%%%%%%%%%%%%%%%%%%%%%%%%%%%%%%% 
%%%%%%%%%%%%%%%%%%%%%%%%%%%%%%%%%%%%%%%%%%%%%%%%%%%%%%%%%%%%%%%%%%%%%%%%
%
\subsubsection{\texorpdfstring{$N$}{N} hypers in 
\texorpdfstring{$[3,0]$}{[3,0]} 
representation}
\paragraph{Intermediate step at $\boldsymbol{\ut}$}
The conformal dimension reads
\begin{equation}
\Delta(m_1,m_2,m_3)= \frac{3}{2} N \sum_i |m_i| +(N-1) \sum_{i<j} |m_i-m_j| \; .
\label{eqn:delta_U3_Rep30}
\end{equation}
We then obtain for $N>2$ the Hilbert series:
\begin{subequations}
\label{eqn:HS_U3_Rep30}
\begin{equation}
\HS_{\ut}^{[3,0]}(t,z)= \frac{R(N,t,z)}{P(N,t,z)} \; , 
\end{equation}
\begin{align}
P(N,t,z) &= 
\prod_{j=1}^3 \left(1-t^{2j}\right) 
 \left(1-\tfrac{1}{z}t^{7 N-4}\right) 
\left(1-z t^{7 N-4}\right) 
\left(1-\tfrac{1}{z^2} t^{10 N-4}\right) 
\left(1-z^2 t^{10 N-4}\right)
\label{eqn:HS_U3_Rep30_Den}\\*
&\qquad \times 
\left(1-\tfrac{1}{z^3} t^{9 N}\right) 
\left(1-z^3 t^{9 N}\right) \notag \; ,\\
R(N,t,z)&=  1+t^{14 N-2}-t^{18 N}-t^{20 N-8}-2 t^{20 N-6}-2 t^{20 N-4}+2 t^{26 
N-8}-2 t^{26 N-2} 
\label{eqn:HS_U3_Rep30_Num}\\
&\qquad +2 t^{32 N-6} +2 t^{32 N-4}+t^{32 N-2}+t^{34 N-10}-t^{38 N-8}-t^{52 
N-10}  \notag \\
&+(z+\tfrac{1}{z})\Big( 
t^{7 N-2}
+t^{7 N}
-t^{17 N-6}
-t^{17 N-4}
-t^{19 N-4}
-t^{19 N-2}
-t^{19 N}
+t^{25 N-4}\notag\\
&\phantom{(z+\tfrac{1}{z})}
-t^{27 N-6} 
+t^{33 N-10}
+t^{33 N-8}
+t^{33 N-6}
+t^{35 N-6}
+t^{35 N-4}
-t^{45 N-10}
-t^{45 N-8} \Big) \notag \\
&+(z^2+\tfrac{1}{z^2})\Big(
t^{10 N-2}
+t^{10 N}
-t^{16 N-4}
-t^{16 N-2}
-t^{16 N}
-t^{24 N-6}
+t^{28 N-4}
+t^{36 N-10} \notag\\
&\phantom{(z^2+\tfrac{1}{z^2})}+t^{36 N-8}+t^{36 N-6}-t^{42 N-10}-t^{42 N-8}
\Big) \notag \\
&+(z^3+\tfrac{1}{z^3})\Big(
t^{17 N-2}-t^{23 N-6}-t^{23 N-4}-t^{23 N-2}+t^{29 N-8}+t^{29 N-6}+t^{29 
N-4}-t^{35 N-8}\Big) \notag \; .
\end{align}
\end{subequations}
The Hilbert series~\eqref{eqn:HS_U3_Rep30} has a pole of order $6$ as $t\to1$, 
because $R(N,t=1,z)=0$ and $\tfrac{\diff^n}{\diff t^n}R(N,t,z)|_{t=1}=0$ for 
$n=1,2$. Therefore, the moduli space is $6$-dimensional. Also, the degree 
of~\eqref{eqn:HS_U3_Rep30_Num} is $52N-10$, while the degree 
of~\eqref{eqn:HS_U3_Rep30_Den} us $52N-4$; thus, the difference in degrees 
equals the dimension of the moduli space.%

As this example is merely a special case of~\eqref{eqn:HS_U3_Rep10+11}, we just 
summarise the minimal generators in Tab.~\ref{tab:Ops_U3_Rep30}.
\begin{table}[h]
\centering
 \begin{tabular}{c|c|c|c}
 \toprule
   \multicolumn{2}{c|}{$(m_1,m_2,m_3)$}  & 
$2\Delta(m_1,m_2,m_3)$ & $\Hh_{(m_1,m_2,m_3)}$ \\ \midrule
   $(1,0,0)$ & $(0,0,-1)$ & $7N-4$ & $\uo\times \utwo$ \\
  $(1,1,0)$ & $(0,-1,-1)$ & $10N-4$ & $\uo\times \utwo$ \\
 $(1,1,1)$ & $(-1,-1,-1)$ & $9N$ & $\ut$ \\
 \bottomrule
 \end{tabular}
\caption{The monopole generators for a $\ut$ gauge theory with matter 
transforming in $[3,0]$ that together with the Casimir invariants generate the 
chiral ring.} 
\label{tab:Ops_U3_Rep30}
\end{table}
% %
% 
\paragraph{Reduction to $\boldsymbol{\sut}$}
The Hilbert series reads
\begin{equation}
\label{eqn:HS_SU3_Rep30}
 \HS_{\sut}^{[3,0]}(t)=\frac{1+t^{14 N-6}(2+2t^2+t^4)
 +t^{24N-12}(1+2t^2+2t^4)+t^{38 N-14}}{\left(1-t^4\right) 
\left(1-t^6\right) \left(1-t^{14 N-8}\right) \left(1-t^{24 N-12}\right) } \; .
\end{equation}
It is apparent that the numerator of~\eqref{eqn:HS_SU3_Rep30} is a 
palindromic polynomial of degree $38N-14$; while the degree of the denominator 
is $38N-10$; hence, the difference in the degrees is $4$, which equals the 
complex 
dimension of the moduli space.

The structure of~\eqref{eqn:HS_SU3_Rep30} is merely a special case 
of~\eqref{eqn:HS_SU3_Rep10+11}, and the conformal dimensions of the minimal 
generators are $2\Delta(1,0)=14N-8$ and $2\Delta(1,1)=2\Delta(2,-1)=24N-12$.
% 
%%%%%%%%%%%%%%%%%%%%%%%%%%%%%%%%%%%%%%%%%%%%%%%%%%%%%%%%%%%%%%%%%%%%%%%%%%%
%%%%%%%%%%%%%%%%%%%%%%%%%%%%%%%%%%%%%%%%%%%%%%%%%%%%%%%%%%%%%%%%%%%%%%%%%%%
% 
\subsection{Category \texorpdfstring{$N_R\neq0$}{NR>0}}
% 
%%%%%%%%%%%%%%%%%%%%%%%%%%%%%%%%%%%%%%%%%%%%%%%%%%%%%%%%%%%%%%%%%%%%%%%%%%%
%%%%%%%%%%%%%%%%%%%%%%%%%%%%%%%%%%%%%%%%%%%%%%%%%%%%%%%%%%%%%%%%%%%%%%%%%%%
% 
\subsubsection{\texorpdfstring{$N_F$}{NF} hypers in 
\texorpdfstring{$[2,1]$}{[2,1]}, \texorpdfstring{$N_A$}{NA} hypers in 
\texorpdfstring{$[1,1]$}{[1,1]}, \texorpdfstring{$N_R$}{NR}
hypers in \texorpdfstring{$[2,1]$}{[2,1]} representation}
\paragraph{Intermediate step at $\boldsymbol{\ut}$}
The conformal dimension reads
\begin{align}
 2\Delta(m_1,m_2,m_3)=(4 N_R+N_A )\sum_{i=1}^3 \left| m_i\right|
&+N_R \sum_{i<j}\left(\left| 2 m_i-m_j\right| +\left| m_i-2 
m_j\right| \right) \\
&+2(N_A-1) \sum_{i<j}\left| m_i-m_j\right| \; . \notag
\end{align}
The Hilbert series reads
\begin{subequations}
\label{eqn:HS_U3_Rep10+11+21}
\begin{equation}
 \HS_{\ut}^{[1,0]+[1,1]+[2,1]}(t,z)= 
\frac{R(N_F,N_A,N_R,t,z)}{P(N_F,N_A,N_R,t,z)} \;, 
\end{equation}
with
\begin{align}
P(N_F,N_A,N_R,t,z)=\prod_{j=1}^3 \left(1-t^{2j}\right) 
&\left(1-\frac{t^{N_F+4 N_A+10 N_R-4}}{z}\right) 
\left(1-z t^{N_F+4 N_A+10 N_R-4}\right)  
\label{eqn:HS_U3_Rep10+11+21_Den}\\
&\times
\left(1-\frac{t^{2 N_F+4 N_A+16 N_R-4}}{z^2}\right) 
\left(1-z^2 t^{2 N_F+4 N_A+16 N_R-4}\right) \notag \\
&\times
\left(1-\frac{t^{3 N_F+18 N_R}}{z^3}\right) 
\left(1-z^3 t^{3 N_F+18 N_R}\right) \notag \\
&\times
\left(1-\frac{t^{3 N_F+8 N_A+24 N_R-8}}{z^3}\right)
\left(1-z^3 t^{3 N_F+8 N_A+24 N_R-8}\right) \notag \\
&\times
\left(1-\frac{t^{4 N_F+4 N_A+24 N_R-4}}{z^4}\right) 
\left(1-z^4 t^{4 N_F+4 N_A+24 N_R-4}\right) \notag \\
&\times
\left(1-\frac{t^{5 N_F+4 N_A+30 N_R-4}}{z^5}\right)
\left(1-z^5 t^{5 N_F+4 N_A+30 N_R-4}\right) \notag \\
&\times
\left(1-\frac{t^{7 N_F+12 N_A+46 N_R-12}}{z^7}\right) 
\left(1-z^7 t^{7 N_F+12 N_A+46 N_R-12}\right)\notag \; ,
\end{align}
\end{subequations}
and the numerator $R(N_F,N_A,N_R,t,z)$ is too long to be displayed, 
because it contains $28650$ monomials. We checked 
explicitly that $R(N_F,N_A,N_R,t=1,z) =0$ and $\tfrac{\diff^n}{\diff t^n} 
R(N_F,N_A,N_R,t,z)|_{t=1,z=1}=0$ for all $n=1,2\ldots, 10$. Therefore, the 
Hilbert series~\eqref{eqn:HS_U3_Rep10+11+21} has a pole of order $6$ at $t=1$, 
which equals the dimension of the moduli space. In addition, 
$R(N_F,N_A,N_R,t,z)$ is a polynomial of degree $50 N_F +72 N_A + 336 N_R -66$, 
while the denominator~\eqref{eqn:HS_U3_Rep10+11+21_Den} is of degree $50 N_F 
+72 
N_A + 336 N_R -60$. The difference in degrees reflects the dimension of the 
moduli space as well.

Following the analysis of the Hilbert bases~\eqref{eqn:Hilbert_basis_SU3_NR>0}, 
we identify the bare monopole operators and provide their conformal dimensions 
in Tab.~\ref{tab:Ops_U3_Rep10+11+21}.
\begin{table}[h]
\centering
 \begin{tabular}{c|c|c|c}
 \toprule
  \multicolumn{2}{c|}{$(m_1,m_2,m_3)$} & $2\Delta(m_1,m_2,m_3)$ & 
$\Hh_{(m_1,m_2,m_3)}$ \\ \midrule
  $(1,0,0)$ & $(0,0,-1)$ & $N_F+4 N_A+10 N_R-4$ & $\uo\times \utwo $ \\
  $(1,1,0)$ & $(0,-1,-1)$ & $2 N_F+4 N_A+16 N_R-4$ & $\uo\times \utwo $ 
\\
  $(1,1,1)$ & $(-1,-1,-1)$ & $3 N_F+18 N_R$ &  $\ut$ \\
  $(2,1,0)$ & $(0,-1,-2)$ & $3 N_F+8 N_A+24 N_R-8$ & $\uo^3$ \\
  $(2,1,1)$ & $(-1,-1,-2)$ & $4 N_F+4 N_A+24 N_R-4$ & $\uo \times \utwo $ 
\\
  $(2,2,1)$ & $(-1,-2,-2)$ & $5 N_F+4 N_A+30 N_R-4$ & $\uo \times \utwo $ 
\\
  $(3,2,1)$ & $(-1,-2,-3)$ & $6 N_F+8 N_A+38 N_R-8$ & $\uo^3$ \\
  $(4,2,1)$ & $(-1,-2,-4)$ & $7 N_F+12 N_A+46 N_R-12$ & $\uo^3$ \\
  \bottomrule
 \end{tabular}
\caption{The monopole generators for a $\ut$ gauge theory with a mixture of 
matter transforming in $[1,0]$, $[1,1]$, and $[2,1]$.} 
\label{tab:Ops_U3_Rep10+11+21}
\end{table}
 The result~\eqref{eqn:HS_U3_Rep10+11+21} has been tested against the 
independent calculations of the cases: $N$ hypermultiplets in $[1,0]$;
 $N_F$ hypermultiplets in $[1,0]$ together with $N_A$ hypermultiplets in 
$[1,1]$; and $N$ hypermultiplets in $[2,1]$. All the calculations agree.
\paragraph{Reduction to $\boldsymbol{\sut}$}
The Hilbert series for the $\sut$ theory reads
\begin{subequations}
\label{eqn:HS_SU3_Rep10+11+21}
\begin{equation}
\HS_{\sut}^{[1,0]+[1,1]+[2,1]}(N_F,N_A,N_R,t)= 
\frac{R(N_F,N_A,N_R,t)}{P(N_F,N_A,N_R,t)} \; ,
\end{equation}
\begin{align}
P(N_F,N_A,N_R,t)&=
\left(1-t^4\right) \left(1-t^6\right) 
\left(1-t^{2 N_F+8 N_A+20 N_R-8}\right) 
\label{eqn:HS_SU3_Rep10+11+21_Den}\\
&\qquad \times 
\left(1-t^{4 N_F+12 N_A+36 N_R-12}\right) 
\left(1-t^{6 N_F+20 N_A+54 N_R-20}\right) \; , \notag\\
R(N_F,N_A,N_R,t)&=1
+ t^{2 N_F+8 N_A+20 N_R-6} (2+2t^2+t^4)
\label{eqn:HS_SU3_Rep10+11+21_Num}
\\
&\qquad +t^{4 N_F+12 N_A+36 N_R-12}(1+2t^2+2t^4)
\notag \\
&\qquad +t^{6 N_F+20 N_A+54 N_R-20}(1+4t^2+4t^4+2t^6)
\notag \\
&\qquad - t^{6 N_F+20 N_A+56 N_R-20}(2+4t^2+4t^4+t^6)
\notag \\
&\qquad - t^{8 N_F+28 N_A+74 N_R-26}(2+2t^2+t^4)
\notag \\
&\qquad -t^{10 N_F+32 N_A+90 N_R-32}(1+2t^2+2t^4)
-t^{12 N_F+40 N_A+110 N_R-34}\notag \; .
\end{align}
\end{subequations}
Again, the numerator~\eqref{eqn:HS_SU3_Rep10+11+21_Num} is an anti-palindromic 
polynomial of degree $12 N_F+40 N_A+110 N_R-34$; while the 
denominator~\eqref{eqn:HS_SU3_Rep10+11+21_Den} is of degree $12 N_F+40 
N_A+110 N_R-30$, such that the difference is again $4$.

The minimal generators from~\eqref{eqn:Hilbert_basis_SU3_NR>0} are now realised 
with the following conformal dimensions: $2\Delta(1,0)=2 N_F+8 N_A+20 
N_R-8$, $2\Delta(1,1)=2\Delta(2,-1)= 4 N_F+12 N_A+36 N_R-12 $ , and 
$2\Delta(2,1)=2\Delta(3,-1)=6 N_F+20 N_A+54 N_R-20$. Moreover, the 
appearing dressed monopoles are as described in 
Subsec.~\ref{subsec:Dress_SU3}.
\paragraph{Remark}
The $\sut$ result~\eqref{eqn:HS_SU3_Rep10+11+21} has been tested against the 
independent calculations of the cases: $N$ hypermultiplets in $[1,0]$; $N$ 
hypermultiplets in $[1,1]$; $N_F$ hypermultiplets in $[1,0]$ together with 
$N_A$ hypermultiplets in $[1,1]$; and $N$ hypermultiplets in $[2,1]$. All the 
calculations agree.
\paragraph{Dressings of \texorpdfstring{$(2,1)$}{(2,1)} and 
\texorpdfstring{$(3,-1)$}{(3,-1)}}
From the generic analysis~\eqref{eqn:Hilbert_basis_SU3_NR>0} the bare monopoles 
of GNO-charges $(3,-1)$ and $(2,1)$ are necessary generators. However, not all 
of their dressings need to be independent generators, c.f.\ App.~\ref{app:PL}. 
\begin{itemize}
 \item $N_R=0$: $(2,1)$ and $(3,-1)$ are generated by 
$(1,0)$, $(1,1)$, and $(2,-1)$, which is the generic result 
of~\eqref{eqn:Hilbert_basis_SU3_NR=0}.
\item $N_R=1$: Here, $(2,1)$ and $(3,-1)$ are independent, but not all 
of their dressings, as we see
\begin{align}
 (2,1) = (1,1) + (1,0) \und \Delta(2,1) + 1 = \Delta(1,1) + 
\Delta(1,0) \; .
\end{align}
Hence, \emph{only one} of the degree one dressings 
$V_{(2,1)}^{\mathrm{dress},(1,0)}$, $V_{(2,1)}^{\mathrm{dress},(0,1)}$ is 
independent, while the other can be generated. (Same holds for $(3,-1)$.)
\item $N_R=2$: Here, $(2,1)$ and $(3,-1)$ are independent, but not all 
of their dressings, as we see
\begin{align}
 (2,1) = (1,1) + (1,0) \und \Delta(2,1) + 2 = \Delta(1,1) + 
\Delta(1,0) \; .
\end{align}
Hence, \emph{only one} of the degree two dressings 
$V_{(2,1)}^{\mathrm{dress},(2,0)}$, $V_{(2,1)}^{\mathrm{dress},(0,2)}$ is 
independent, while the other can be generated. However, \emph{both} degree one 
dressings $V_{(2,1)}^{\mathrm{dress},(1,0)}$, 
$V_{(2,1)}^{\mathrm{dress},(0,1)}$ are independent. (Same holds for 
$(3,-1)$.)
\item $N_R=3$: Here, $(2,1)$ and $(3,-1)$ are independent, but 
still not all of their dressings, as we see
\begin{align}
 (2,1) = (1,1) + (1,0) \und \Delta(2,1) + 3 = \Delta(1,1) + 
\Delta(1,0) \; .
\end{align}
Hence, the degree three dressing 
$V_{(2,1)}^{\mathrm{dress},(3,0)+(0,3)}$ is not 
independent. However, \emph{both} degree one 
dressings $V_{(2,1)}^{\mathrm{dress},(1,0)}$, 
$V_{(2,1)}^{\mathrm{dress},(0,1)}$ and \emph{both} degree two dressings 
$V_{(2,1)}^{\mathrm{dress},(2,0)}$, $V_{(2,1)}^{\mathrm{dress},(0,2)}$ 
are 
independent. (Same holds for $(3,-1)$.)
\item $N_R\geq4$: The bare and the all dressed monopoles corresponding to  
$(2,1)$ and $(3,-1)$ are independent.
\end{itemize}
%
%%%%%%%%%%%%%%%%%%%%%%%%%%%%%%%%%%%%%%%%%%%%%%%%%%%%%%%%%%%%%%%%%%%%%%%%%% 
% 
\subsubsection{\texorpdfstring{$N$}{N} hypers in 
\texorpdfstring{$[2,1]$}{[2,1]} 
representation}
\paragraph{Intermediate step at $\boldsymbol{\ut}$}
The conformal dimension reads
\begin{align}
 2\Delta(m_1,m_2,m_3)=4 N \sum_{i=1}^3 \left| m_i\right|
+N \sum_{i<j}\left(\left| 2 m_i-m_j\right| +\left| m_i-2 
m_j\right| \right) -2 \sum_{i<j}\left| m_i-m_j\right| \; .
\end{align}
From the calculations we obtain the Hilbert series
\begin{subequations}
\label{eqn:HS_U3_Rep21}
 \begin{equation}
 \HS_{\ut}^{[2,1]}(N,t,z)=\frac{R(N,t,z)}{P(N,t,z)}  \; ,
 \end{equation}
\begin{align}
 P(N,t,z)= \prod_{j=1}^3 \left(1-t^{2j}\right) 
 &\left(1-\frac{t^{10 N-4}}{z}\right) 
\left(1-z t^{10 N-4}\right)
\left(1-\frac{t^{16 N-4}}{z^2}\right)
\left(1-z^2 t^{16 N-4}\right) 
\label{eqn:HS_U3_Rep21_Den}\\
&\times 
\left(1-\frac{t^{18 N}}{z^3}\right)
\left(1-z^3 t^{18 N}\right) 
\left(1-\frac{t^{24 N-8}}{z^3}\right)
\left(1-z^3 t^{24 N-8}\right) \notag \\
& \times 
\left(1-\frac{t^{24 N-4}}{z^4}\right)
\left(1-z^4 t^{24 N-4}\right) 
\left(1-\frac{t^{30 N-4}}{z^5}\right)
\left(1-z^5 t^{30 N-4}\right) \notag \\
& \times 
\left(1-\frac{t^{46 N-12}}{z^7}\right)
\left(1-z^7 t^{46 N-12}\right) \notag \; ,
\end{align}
\end{subequations}
and the numerator $R(N,t,z)$ is with $13492$ monomials too long to be 
displayed. Nevertheless, we checked 
explicitly that $R(N,t=1,z) =0$ and $\tfrac{\diff^n}{\diff t^n} 
R(N,t,z)|_{t=1,z=1}=0$ for all $n=1,2\ldots, 10$. Therefore, the 
Hilbert series~\eqref{eqn:HS_U3_Rep21} has a pole of order $6$ at $t=1$, 
which equals the dimension of the moduli space. In addition, the degree of 
$R(N,t,z)$ is $296N-62$, while the denominator~\eqref{eqn:HS_U3_Rep21_Den} is 
of degree $296N-56$; therefore, the difference in degrees is again equal to the 
dimension of the moduli space.

The Hilbert series~\eqref{eqn:HS_U3_Rep21} appears as special case 
of~\eqref{eqn:HS_U3_Rep10+11+21} and as such the appearing monopole operators 
are the same. For completeness, we provide in Tab.~\ref{tab:Ops_U3_Rep21} the 
conformal dimensions of all 
minimal (bare) generators~\eqref{eqs:Hilbert_basis_U3_NR>0}.
\begin{table}[h]
\centering
 \begin{tabular}{c|c|c|c}
 \toprule
  \multicolumn{2}{c|}{$(m_1,m_2,m_3)$} & $2\Delta(m_1,m_2,m_3)$ & 
$\Hh_{(m_1,m_2,m_3)}$ \\ \midrule
  $(1,0,0)$ & $(0,0,-1)$ & $10N-4$ & $\uo\times \utwo $ \\
  $(1,1,0)$ & $(0,-1,-1)$ & $16N-4$ & $\uo\times \utwo $ \\
  $(1,1,1)$ & $(-1,-1,-1)$ & $18N$ &  $\ut$ \\
  $(2,1,0)$ & $(0,-1,-2)$ & $24N-8$ & $\uo^3$ \\
  $(2,1,1)$ & $(-1,-1,-2)$ & $24N-4 $ & $\uo \times \utwo $ \\
  $(2,2,1)$ & $(-1,-2,-2)$ & $30N-4 $ & $\uo \times \utwo $ \\
  $(3,2,1)$ & $(-1,-2,-3)$ & $38N-8 $ & $\uo^3$ \\
  $(4,2,1)$ & $(-1,-2,-4)$ & $46N-12 $ & $\uo^3$ \\
  \bottomrule
 \end{tabular}
\caption{The monopole generators for a $\ut$ gauge theory with matter 
transforming in $[2,1]$ that generate the chiral ring (together with the 
Casimir invariants).} 
\label{tab:Ops_U3_Rep21}
\end{table}
The GNO-charge $(3,2,1)$ is not apparent in the Hilbert series, but we 
know it to be present due to the analysis of the Hilbert 
bases~\eqref{eqs:Hilbert_basis_U3_NR>0}.
\paragraph{Reduction to $\boldsymbol{\sut}$}
After reduction~\eqref{eqn:Reduction_to_SU3} of~\eqref{eqn:HS_U3_Rep21} to 
$\sut$ we obtain the following Hilbert series:
\begin{subequations}
\label{eqn:HS_SU3_Rep21}
\begin{align}
HS_{\sut}^{(2,1)}&=
\frac{R(N,t)}{\left(1-t^4\right) 
\left(1-t^6\right) \left(1-t^{20 N-8}\right) \left(1-t^{36 N-12}\right) 
\left(1-t^{54 N-20}\right)} \; ,
\\
R(N,t)&=1
+ t^{20 N-6}(2+2t^2+t^4)
+t^{36 N-12}(1+2t^2+2t^4)
\label{eqn:HS_SU3_Rep21_Num}\\
&\quad
+t^{54 N-20} (1+4t^2+4t^4+2t^6)
-t^{56 N-20}(2+4t^2+4t^4+t^6)
\notag \\
&\quad
- t^{74 N-26}(2+2t^2+t^4)
-t^{90 N-32}(1+2t^2+2t^4)
-t^{110 N-34} \notag \; .
\end{align}
\end{subequations}
The numerator of~\eqref{eqn:HS_SU3_Rep21_Num} is an 
anti-palindromic polynomial of degree $110N-34$; while the numerator is of 
degree $110N-30$. Consequently, the difference in degree reflects the complex 
dimension of the moduli space.

The Hilbert series~\eqref{eqn:HS_SU3_Rep21} is merely a special case 
of~\eqref{eqn:HS_SU3_Rep10+11+21} and, thus, the appearing (bare and dressed) 
monopole operators are the same. For completeness we provide their conformal 
dimensions: $2\Delta(1,0)=20N-8$, $2\Delta(1,1)=2\Delta(2,-1)=36N-12$, and 
$2\Delta(2,1)=2\Delta(3,-1)= 54N-20$.
%%%%%%%%%%%%%%%%%%%%%%%%%%%%%%%%%%%%%%%%%%%%%%%%%%%%%%%%%%%%%%%%%%%%%%%%%%%%%%%%
  \section{Conclusions}
\label{sec:conclusions}
In this paper we introduced a geometric concept to identify and compute the set 
of bare and dressed monopole operators that are sufficient to describe the 
entire chiral ring $\C[\MCoulomb]$ of any $3$-dimensional $\Ncal=4$ gauge 
theory. The methods can be summarised as follows:
\begin{enumerate}
 \item The matter content together with the positive roots of the gauge group 
$\G$ define the conformal dimension, which in turn defines an arrangement of 
hyperplanes that divide the dominant Weyl chamber of $\GNOG$ into a fan.
 \item The intersection of the fan with the weight lattice of the GNO-dual 
group leads to a collection of affine semi-groups. All semi-groups are finitely 
generated and the unique, finite basis is called Hilbert basis.
\item The knowledge of the minimal generators, together with their properties 
$\su_R$-spin, residual gauge group $\Hh_m$, and topological charges $J(m)$,  is 
sufficient to explicitly sum and determine the Hilbert series as rational 
function.
\end{enumerate}
Utilising the fan and the Hilbert bases for each semi-group also allows to 
deduce the dressing behaviour of monopole operators. The number of dressed 
operators is determined by a ratio of orders of Weyl groups, while the degrees 
are determined by the ratio of the dressing factors associated to the GNO-charge 
$m$ divided by the dressing factor of the trivial monopole $m=0$. 

Most importantly, the entire procedure works for any rank of the gauge group, 
as indicated in Sec.~\ref{sec:SU3} for $\ut$. For the main part of the 
paper, we, however, have chosen to provide a comprehensive collection of rank 
two examples. 

Before closing, let us outline and comment on the approach to higher rank cases.
\begin{enumerate}[(a)]
 \item The gauge group $\G$ determines the GNO-dual group $\GNOG$ and the 
corresponding dominant Weyl chamber (or the product of several Weyl chambers). 
The Weyl chamber is understood as finite intersection of positive half-spaces 
$H^+_{\alpha} \subset \tfrak$, where $\alpha$ ranges over all simple roots of 
$\G$. (If $\G$ is a product, then the roots of one factor have to be embedded 
in a higher dimensional vector space.)
\item The \emph{relevant weights} $\mu_i$, as identified in 
Sec.~\ref{subsec:Hilbert basis}, define a finite set of cones via the 
intersection of all possible upper and lower half-spaces with the Weyl chamber. 
This step can, for instance, be implemented by means of the package 
\emph{Polyhedra} of \texttt{Macaulay2}.
\item Having defined all cones in \texttt{Macaulay2}, one computes the 
dimension and the Hilbert basis for each cone. Identifying all cones 
$C_p^{(\rank(G))}$ of the maximal dimension $\rank(\G)$ can typically reduce 
the number of cones one needs to consider.
\item Define the fan $F= 
\{C_p^{(\rank(G))} | p=1,\ldots,L\}$ generated by all top-dimensional cones in 
\texttt{Macaulay2}. This step is the computationally most demanding process so 
far.
\item Next, one employs the \emph{inclusion-exclusion principle} for each cone 
in the fan: that is the number of points in the (relative) interior
$\mathrm{Int}(S^{(p)})\coloneqq \mathrm{Relint}(C^{(p)}) \cap 
\Lambda_w(\GNOG)$ is given by 
\begin{subequations}
\begin{align}
 \#|\mathrm{Int}(S^{(p)})| &= |S^{(p)}| - \Bigg( \sum_{j=1}^{\kappa_p} 
|S_{j}^{(p-1)}| 
 -\sum_{1\leq i<j \leq \kappa_p} |S_{i}^{(p-1)} \cap S_{j}^{(p-1)}| \\*
 &\phantom{= |S^{(p)}| -}
 +\sum_{1\leq i<j<k \leq \kappa_p} |S_{i}^{(p-1)} \cap 
S_{j}^{(p-1)}\cap S_{k}^{(p-1)}|
- \ldots 
+(-1)^{\kappa_p -1} \left|\bigcap_{i=1}^{\kappa_p} S_{i}^{(p-1)} \right|
\Bigg) \notag \\
&\equiv|S^{(p)}| - |\partial S^{(p)} | \;,
\end{align}
\end{subequations}
where the $S_j^{(p-1)}$ for $j=1,\ldots,\kappa_p$ are the semi-groups resulting 
from the facets of $C^{(p)}$. Note that the last term $\bigcap_{i=1}^{\kappa_p} 
S_{i}^{(p-1)}$ equals the 
trivial semi-group, while the intermediate intersections give rise to all lower 
dimensional semi-groups contained in the boundary of $S^{(p)}$.
Then, the contribution for $\mathrm{Int}(S^{(p)})$ to the monopole formula is 
computed as follows:
\begin{subequations}
\begin{align}
 \HS(S^{(p)};t) &\coloneqq P_{\G}(t;S^{(p)}) \cdot \left[ \HilbS_{S^{(p)}}(t) 
-\HilbS_{\partial S^{(p)}}(t) \right]\; , \\
\HilbS_{S^{(p)}}(t) &\coloneqq   \sum_{m\in S^{(p)}} 
z^{J(m)} \, t^{\Delta(m)}  \; ,\\
 \HilbS_{\partial S^{(p)}}(t) &\coloneqq 
  \sum_{j=1}^{\kappa_p} 
\HilbS_{S_{j}^{(p-1)}}(t) 
 -\sum_{1\leq i<j \leq \kappa_p} \HilbS_{S_{i}^{(p-1)} \cap S_{j}^{(p-1)}}(t) 
\\*
 & \quad
 +\sum_{1\leq i<j<k \leq \kappa_p} \HilbS_{S_{i}^{(p-1)} \cap 
S_{j}^{(p-1)}\cap S_{k}^{(p-1)}}(t)
- \ldots 
+(-1)^{\kappa_p -1} \HilbS_{\bigcap_{i=1}^{\kappa_p} S_{i}^{(p-1)} }(t) \; . 
\notag
\end{align}
\end{subequations}
Each contribution $\HilbS_{S^{(p)}}(t)$ is evaluated as 
discussed in Sec.~\ref{subsec:summation_HS_unrefined} and 
\ref{subsec:summation_HS_refined}. Although this step is algorithmically 
simple, it can be computationally demanding. It is, however, crucial that the 
fan $F$ has been defined, in order to work with the correct faces of each cone 
and to sum over each cone in the fan only once.
\item Finally, one has to add all contributions 
\begin{equation}
 \HS(F;t)= \sum_{C \in F} \HS(S) \; .
\end{equation}
This last step is a simple sum, but to obtain the Hilbert series as a rational 
function in a desirable form can be cumbersome.
\end{enumerate}
% % 
Equipped with this procedure, we hope to report on Coulomb branches for higher 
rank gauge groups and quiver gauge theories in the future.
%%%%%%%%%%%%%%%%%%%%%%%%%%%%%%%%%%%%%%%%%%%%%%%%%%%%%%%%%%%%%%%% 
% 
\section*{Acknowledgements}
We thank Roger Bielawski, Simon Brandhorst, Stefano Cremonesi, Giulia Ferlito, 
Rudolph Kalveks, and Markus Röser for useful discussions.
A.~H.\ thanks the Institute für Theoretische Physik of the Leibniz Universität 
Hannover for hospitality. 
A.~H.\ is supported by STFC Consolidated Grant ST/J0003533/1, and EP- SRC 
Programme Grant EP/K034456/1.
M.~S.\ thanks the Theoretical Physics Group of the Imperial College London for 
hospitality.
M.~S.\ is supported by the DFG research training group GRK1463 ``Analysis, 
Geometry, and String Theory''. 
%%%%%%%%%%%%%%%%%%%%%%%%%%%%%%%%%%%%%%%%%%%%%%%%%%%%%%%%%%%%%%%%%%%%%%%%%%%%%%%%
  \begin{appendix}
    \section{Plethystic Logarithm}
\label{app:PL}
In this appendix we summarise the main properties of the plethystic logarithm. 
Starting with the definition, for a mulit-valued function $f(t_1,\ldots,t_m)$ 
with $f(0,\ldots,0)=1$, one defines
\begin{equation}
 \PL[f]\coloneqq \sum_{k=1}^{\infty} \frac{\mu(k)}{k} \log\left( 
f(t_1^k,\ldots,t_m^k) \right) \; ,
\end{equation}
where $\mu(k)$ denote the Möbius function~\cite{Benvenuti:2006qr}.
Some basic properties include
\begin{equation}
 \PL[f\cdot g] = \PL[f] + \PL[g] \und 
 \PL\left[ \frac{1}{\prod_n (1-t^n)^{a_n}}  \right] = \sum_n a_n \ t^n \; .
 \label{eqn:PL_properties}
\end{equation}
Now, we wish to compute the plethystic logarithm. Given a Hilbert series as 
rational function, i.e.\ of the form~\eqref{eqn:HS_generic_solved} 
or~\eqref{eqn:HS_generic_refined}, the denominator can be taken care of by 
means of~\eqref{eqn:PL_properties}, while the numerator is a polynomial with 
integer coefficients. In order to obtain an approximation of the PL, we employ 
the following two equivalent transformations for the numerator:
\begin{subequations}
\begin{align}
 \PL\left[ 1+ a t^n +\mathcal{O}(t^{n+1}) \right] &=
 \PL\left[ \frac{(1-t^n)^a \ \left( 1+ a t^n +\mathcal{O}(t^{n+1}) \right) 
}{(1-t^n)^a} \right] \notag \\
&= a t^n + \PL\left[1 +\mathcal{O}(t^{n+1})  \right] \; ,
\label{eqn:PL_mod_plus-term}\\
 \PL\left[ 1- a t^n +\mathcal{O}(t^{n+1}) \right] &=
 \PL\left[ \frac{(1-t^n)^a \ (1+t^n)^a \ \left( 1- a t^n +\mathcal{O}(t^{n+1}) 
\right) 
}{(1-t^{2n})^a} \right] \notag \\
&= -a t^n + a t^{2n} + \PL\left[1 +\mathcal{O}(t^{n+1})  \right] \; 
. \label{eqn:PL_mod_minus-term}
\end{align}
\end{subequations}
Now, we derive an approximation of the PL for a generic rank two 
gauge group in terms of $t^\Delta$. More precisely, consider the Hilbert basis 
$\{X_i\}$ then we provide an approximation of the PL up to second order, i.e.\ 
\begin{equation}
 \PL = \text{Casimir inv.} 
 + \left\{t^{\Delta(X_i)} \text{-terms} \right\}
 + \left\{t^{\Delta(X_i)+\Delta(X_j)} \text{-terms} \right\}
 +\mathcal{O}\left(t^{\Delta(X_i)+\Delta(X_j)+\Delta(X_k)}\right)
\end{equation}
Considering~\eqref{eqn:HS_generic_solved}, the numerator is denoted by $R(t)$, 
while the denominator $Q(t)$ is given by
\begin{equation}
 Q(t) = \prod_{i=1}^2 (1-t^{d_i}) \ \prod_{p=0}^L \left( 
1-t^{\Delta(x_p)}\right) \; ,
\end{equation}
with $d_i$ the degrees of the Casimir invariants. Then expand the numerator as 
follows:
\begin{align}
 R(t) = 1 
 &+ \sum_{q=0}^L \left( \frac{P_{\G}(t,x_q)}{P_{\G}(t,0)}-1\right) 
t^{\Delta(x_q)} 
+ \sum_{q=0}^L \sum_{s \in \mathrm{Int}(\mathcal{P}(C_q^{(2)}))} 
\frac{P_{\G}(t,s)}{P_{\G}(t,0)} t^{\Delta(s)} \\
&-\sum_{{{q,p=0}\atop{q\neq p}}}^L
\left(\frac{P_{\G}(t,x_q)}{P_{\G}(t,0)}-\frac{1}{2} \right) t^{\Delta(x_p) 
+\Delta(x_q)}
+\sum_{q=1}^L  \frac{P_{\G}(t,C_q^{(2)})}{P_{\G}(t,0)} t^{\Delta(x_{q-1}) 
+\Delta(x_q)} \notag \\
&-\sum_{q=1}^L \sum_{s \in \mathrm{Int}(\mathcal{P}(C_q^{(2)}))} 
\sum_{{{r=0}\atop{r\neq q{-}1,q}}}^L \frac{P_{\G}(t,s)}{P_{\G}(t,0)} 
t^{\Delta(s) + \Delta{(x_r)}} \notag \; .
\end{align}
Note that the appearing factor $\tfrac{1}{2}$ avoids double counting when 
changing summation $\sum_{q<p}$ to $\sum_{q\neq p}$. Still, the numerator is a 
polynomial with integer coefficients.
The PL then reads
\begin{equation}
 \PL\left[\HS_{\G}(t)\right] = \sum_{i=1}^2 t^{d_i} + \sum_{p=0}^L 
t^{\Delta(x_p)} 
+\PL\left[R(t)\right] \; .
\end{equation}
By step~\eqref{eqn:PL_mod_plus-term} we factor out the order $t^{\Delta(x_q)}$ 
and $t^{\Delta(s)}$ terms. However, this introduces further terms at order 
$t^{\Delta(x_q) +\Delta(s)}$ and so forth, which are given by
\begin{equation}
 -\left( \sum_{q=0}^L \left( \frac{P_{\G}(t,x_q)}{P_{\G}(t,0)}-1\right) 
t^{\Delta(x_q)} 
+ \sum_{q=1}^L \sum_{s \in \mathrm{Int}(\mathcal{P}(C_q^{(2)}))} 
\frac{P_{\G}(t,s)}{P_{\G}(t,0)} t^{\Delta(s)} \right)^2 \; .
\end{equation}
Subsequently factoring the terms of this order by means 
of~\eqref{eqn:PL_mod_minus-term}, one derives at the following expressing of the 
PL
\begin{align}
 \PL\left[\HS_{\G}(t)\right] = \sum_{i=1}^2 t^{d_i} &+ \sum_{q=0}^L 
\frac{P_{\G}(t,x_q)}{P_{\G}(t,0)} \ t^{\Delta(x_q)} 
+\sum_{q=1}^L \sum_{s \in \mathrm{Int}(\mathcal{P}(C_q^{(2)}))} 
\frac{P_{\G}(t,s)}{P_{\G}(t,0)}\  t^{\Delta(s)} 
\label{eqn:PL_truncated}\\
&-\sum_{{{q,p=0}\atop{q\neq p}}}^L
\left(\frac{P_{\G}(t,x_q)}{P_{\G}(t,0)}-\frac{1}{2} \right) t^{\Delta(x_p) 
+\Delta(x_q)}
+\sum_{q=1}^L  \frac{P_{\G}(t,C_q^{(2)})}{P_{\G}(t,0)} t^{\Delta(x_{q-1}) 
+\Delta(x_q)} \notag\\
&-\sum_{q=1}^L \sum_{s \in \mathrm{Int}(\mathcal{P}(C_q^{(2)}))} 
\sum_{{{r=0}\atop{r\neq q{-}1,q}}}^L \frac{P_{\G}(t,s)}{P_{\G}(t,0)} 
t^{\Delta(s) + \Delta{x_r}} \notag \\
&-\sum_{q,p=0}^{L}
\left( \frac{P_{\G}(t,x_q)}{P_{\G}(t,0)} -1 \right)
\left( \frac{P_{\G}(t,x_p)}{P_{\G}(t,0)} -1 \right) t^{\Delta(x_q)+\Delta(x_p)} 
\notag \\
&-2 \sum_{p=0}^{L} \sum_{q=1}^L
\left( \frac{P_{\G}(t,x_p)}{P_{\G}(t,0)} -1 \right)
  \sum_{s \in \mathrm{Int}(\mathcal{P}(C_q^{(2)}))} 
\frac{P_{\G}(t,s)}{P_{\G}(t,0)} \ t^{\Delta(x_p)+\Delta(s)} \notag \\
&-\sum_{q,p=1}^{L}  \sum_{s \in \mathrm{Int}(\mathcal{P}(C_p^{(2)}))}  \sum_{s' 
\in \mathrm{Int}(\mathcal{P}(C_q^{(2)}))} 
\frac{P_{\G}(t,s)}{P_{\G}(t,0)} \frac{P_{\G}(t,s')}{P_{\G}(t,0)}
t^{\Delta(s)+\Delta(s')} \notag \\
&+\PL\left[1 + 
\mathcal{O}\left(t^{\Delta(X_i)+\Delta(X_j)+\Delta(X_j)}\right)\right]\notag \; 
.
\end{align}
Strictly speaking, the truncation~\eqref{eqn:PL_truncated} is 
only meaningful if 
\begin{equation}
\begin{aligned}
&\max\{\Delta(X) \} +\max\{d_i | i=1,2\}  < \min\{\Delta(X)+\Delta(Y) \} = 2 
\cdot \min\{ \Delta(X)\} \\
&\for  X,Y= x_q \text{ or } s\; ,  s \in 
\mathrm{Int}(\mathcal{P}(C_p^{(2)})) , q =0,1,\ldots,l
\end{aligned}
\label{eqn:PL_condition}
\end{equation}
holds. Only in this case do the positive contributions, i.e.\ the generators, 
of the first line in~\eqref{eqn:PL_truncated} not mix with the negative 
contributions, i.e.\ first syzygies or relations, of the remaining lines. 
Moreover, the condition~\eqref{eqn:PL_condition} ensures that the remained 
$\mathcal{O}\left(t^{\Delta(X_i)+\Delta(X_j)+\Delta(X_k)}\right)$ does not 
spoil the truncation.

From the examples of Sec.~\ref{sec:U1xU1}-\ref{sec:SU3}, we see 
that~\eqref{eqn:PL_condition} is at most satisfied for scenarios with just a 
few generators, but not for elaborate cases. Nevertheless, there are some 
observations we summarise as follows:
\begin{itemize}
 \item The bare and dressed monopole operators associated to the GNO-charge $m$ 
are described by $\tfrac{P_\G(t,m)}{P_\G(t,0)} t^{\Delta(m)}$. In particular, 
we emphasis that the quotient of dressing factors provides information on the 
number and degrees of the dressed monopole operators.
\item The previous observation provides an upper bound on the number of dressed 
monopole operators associated to a magnetic weight $m$. In detail, the value of 
$\tfrac{P_\G(t,m)}{P_\G(t,0)}$ at $t=1$ equals the number of bare and 
dressed monopole operators associated to $m$. Let $\{d_i\}$ and $\{b_i\}$, for 
$i=1,\ldots,\rank(\G)$ denote the degree of the Casimir invariants for $\G$ and 
$\Hh_m$, respectively. Then
\begin{equation}
\begin{matrix} \text{\# dressed monopoles} \\
 +1 \text{ bare monopole}
\end{matrix} = 
 \lim_{t \to 1} \frac{P_\G(t,m)}{P_\G(t,0)} = \lim_{t\to 1} 
\frac{\prod_{i=1}^{\rank(\G)} \left(1-t^{d_i} \right) }{\prod_{j=1}^{\rank(\G)} 
\left(1-t^{b_j} \right)} = \frac{\prod_{i=1}^{\rank(\G)} 
d_i}{\prod_{j=1}^{\rank(\G)} b_j} = \frac{\left|\Wcal_{\G} 
\right|}{\left|\Wcal_{\Hh_{m}}\right| } \; ,
\end{equation}
where the last equality holds because the order of the Weyl group equals 
the product of the degrees of the Casimir invariants. Since $\Wcal_{\Hh_{m}} 
\subset \Wcal_{\G} $ is a subgroup of the finite group $\Wcal_{\G} $, 
Lagrange's theorem implies that $\tfrac{\left|\Wcal_{\G} 
\right|}{\left|\Wcal_{\Hh_{m}}\right| } \in \NN$ holds. 

The situation becomes obvious whenever $m$ belongs to the interior of the Weyl 
chamber, because $\Hh_m = \T$ and thus 
\begin{equation}
\left. \begin{matrix} \text{\# dressed monopoles} \\
 +1 \text{ bare monopole}
\end{matrix} \right|_{\text{interior of}\atop\text{Weyl chamber}}
=\left|\Wcal_{\G} \right|  \und 
 \frac{P_\G(t,m)}{P_\G(t,0)} =\prod_{i=1}^{\rank(\G)} \sum_{l_i=0}^{d_i -1} 
t^{l_i} \; .
\end{equation}
 \item The significance of the PL is limited, as, for instance, a positive 
contribution $\sim t^{\Delta(X_1)}$ can coincide with a negative contribution 
$\sim t^{\Delta(X_2)+\Delta(X_3)}$, but this does not necessarily imply that 
the object of degree $\Delta(X_1)$ can be generated by others. The situation 
becomes clearer if there exists an additional global symmetry $Z(\GNOG)$ on the 
moduli space. The truncated PL for~\eqref{eqn:HS_generic_refined} is obtained 
from~\eqref{eqn:PL_truncated} by the replacement
\begin{equation}
 t^{\Delta(X)} \mapsto \vec{z}^{\vec{J}(X)} \ t^{\Delta(X)} \; .
\end{equation}
Then the ``syzygy'' $\vec{z}^{\vec{J}(X_2+X_3)}t^{\Delta(X_2)+\Delta(X_3)}$ can 
cancel the ``generator'' $\vec{z}^{\vec{J}(X_1)}t^{\Delta(X_1)}$ only if the 
symmetry charges agree $\vec{z}^{\vec{J}(X_1)}=\vec{z}^{\vec{J}(X_2+X_3)}$, in 
addition to the $\su_R$ iso-spin.
\end{itemize}
Lastly, we illustrate the truncation with the two simplest examples:
\paragraph{Example: one simplicial cone}
For the Hilbert series~\eqref{eqn:Example_HS_simplicial} we obtain
\begin{align}
\label{eqn:Example_PL_simplicial}
 \PL=\sum_{i=1}^2 t^{d_i} &+ \frac{P_1(t)}{P_0(t)} \left(t^{\Delta(x_0)} + 
t^{\Delta(x_1)}\right) 
-\left(2 \frac{P_1(t)}{P_0(t)} -1 - \frac{P_2(t)}{P_0(t)} 
\right)  t^{\Delta(x_0)+\Delta(x_1)}  \\
&- \left(\frac{P_1(t)}{P_0(t)} \right)^2 
\left(t^{2\Delta(x_0)}+t^{2\Delta(x_1)} +2t^{\Delta(x_0)+\Delta(x_1)} \right) + 
\ldots \notag \; .
\end{align}
\paragraph{Example: one non-simplicial cone}
In contrast, for the Hilbert series~\eqref{eqn:Example_HS_non-simplicial} we 
arrive at
\begin{align}
\label{eqn:Example_PL_non-simplicial}
 \PL=\sum_{i=1}^2 t^{d_i} &+ \frac{P_1(t)}{P_0(t)} \left(t^{\Delta(x_0)} + 
t^{\Delta(x_1)}\right) 
+ \sum_{s \in 
\mathrm{Int}\mathcal{P}} \frac{P_2(t)}{P_0(t)} t^{\Delta(s)} \\
&-\left(2 \frac{P_1(t)}{P_0(t)} -1 - \frac{P_2(t)}{P_0(t)} 
\right)  t^{\Delta(x_0)+\Delta(x_1)}  \notag\\
&- \left(\frac{P_1(t)}{P_0(t)} \right)^2 
\left(t^{2\Delta(x_0)}+t^{2\Delta(x_1)} +2t^{\Delta(x_0)+\Delta(x_1)} \right)  
\notag \\
&-2\left(\frac{P_1(t)}{P_0(t)}-1 \right)\frac{P_2(t)}{P_0(t)} \sum_{s \in 
\mathrm{Int}\mathcal{P}} \left( t^{\Delta(s)+\Delta(x_0)}  
+t^{\Delta(s)+\Delta(x_1)} \right) \notag \\
&-\sum_{s \in 
\mathrm{Int}\mathcal{P}} \sum_{s' \in 
\mathrm{Int}\mathcal{P}} \left(\frac{P_2(t)}{P_0(t)} \right)^2 
t^{\Delta(s)+\Delta(s')} + \ldots \notag \; .
\end{align}

  \end{appendix}

 \bibliographystyle{JHEP}     % Zitierstil: alpha = [Nam88]
 {\footnotesize{\bibliography{references}}}

\end{document}